\documentclass[onecolumn,10pt]{article}%,nofootinbib,superscriptaddress]{article}
\usepackage{enumitem}
\usepackage[most]{tcolorbox}
\usepackage[T1]{fontenc}
\usepackage[latin9]{inputenc}
\usepackage{amsmath,braket,bm}
\usepackage{amssymb}
\usepackage{graphicx}
\usepackage{hyperref}
\def\oper{{\mathchoice{\rm 1\mskip-4mu l}{\rm 1\mskip-4mu l}
{\rm 1\mskip-4.5mu l}{\rm 1\mskip-5mu l}}}
\def\<{\langle}
\def\>{\rangle}
\newcommand{\ketbra}[2]{| #1 \rangle \langle #2 | }

\newcommand{\vac}{\mathrm{vac}}
\renewcommand{\i}{\mathrm{i}}
\newcommand{\e}{\mathrm{e}}
\newcommand{\g}{\mathrm{g}}
\newcommand{\W}[1]{W\!\left(#1\right)}
\newcommand{\Wdag}[1]{W^\dag\!\left(#1\right)}

\newtheorem{Theorem}{Theorem}

\newtheorem{Cor}{Corollary}
\newtheorem{Definition}{Definition}
\newtheorem{Remark}{Remark}
\newtheorem{Example}{Example}
\newtheorem{Proposition}{Proposition}

\usepackage{color}
\usepackage{soul}

\newcommand{\BH}{\mathcal{B}(\mathcal{H})}
\newcommand{\TTH}{\mathcal{T}(\mathcal{H})}

\newcommand{\LH}{{{\rm L}}(\mathcal{H})}
\newcommand{\Tr}{{\rm Tr}\,}
\newcommand{\HH}{\mathcal{H}}

\numberwithin{equation}{section}

\usepackage{amsmath}

\begin{document}

%\begin{frontmatter}

\title{{\bf Dynamical maps beyond Markovian regime\footnote{Dedicated to the memory of Professor Andrzej Kossakowski (1938-2021)} }}

\author{Dariusz Chru\'sci\'nski \\
%\email{darch@fizyka.umk.pl}
%\cortext[cor]{Corresponding author}
Institute of Physics, Faculty of Physics, Astronomy and
Informatics \\ Nicolaus Copernicus University \\
 Grudzi\c{a}dzka 5/7,
87--100 Toru\'n, Poland}

\date{}

\maketitle

\begin{abstract}
Quantum dynamical maps provide  suitable mathematical representation of quantum evolutions. When representing quantum states by density operators, the evident requirements for any dynamical map is positivity and trace-preservation. However, these properties are not consistent with quantum mechanics of composite systems. It is the very notion of complete positivity which provides a proper mathematical representation of quantum evolution and gives rise to the powerful generalization of unitary evolution of closed Hamiltonian systems. A prominent example of quantum evolution of an open system is a Markovian semigroup. In what follows, we analyze both the semigroups of positive and completely positive maps. In the latter case the dynamics is governed by the celebrated Gorini-Kossakowski-Lindblad-Sudarshan (GKLS) Master Equation.  Markovian semigroups, however, provide only an approximate description of  general quantum evolution. The main topic of our analysis are dynamical maps beyond this regime. Non-Markovian quantum evolution attracted a lot of attention in recent years and there is a vast literature dedicated to it. In this report we analyze quantum dynamics governed by time-local generators and/or non-local memory kernels. A special attention is devoted to the concept of {\em divisibility} which is often used as a definition of Markovianity. In particular, the concept of so called CP-divisibility (in contrast to P-divisibility) is widely accepted as a proper definition of quantum Markovianity.  We discuss a number of important physical implications of divisibility. We also briefly discuss the notion of Markovianity beyond the dynamical map, that is, when one has an access to the evolution of `system + environment'. The entire exposition is concentrated more on the general concepts and intricate connections between them than on studying particular systems. We illustrate the analyzed concepts by paradigmatic models of open quantum systems like the amplitude damping and phase damping models.
\end{abstract}

%\begin{keyword}
%open quantum systems \sep quantum Markovianity  \sep quantum dynamical maps \sep master equations \sep memory kernels
%\end{keyword}

%\journal{Physics Reports}
%\end{frontmatter}

\tableofcontents

%\addcontentsline{toc}{section}{List of abbreviations}

\section{Introduction} \label{intro}

Any real quantum system is never perfectly isolated and hence to determine its dynamics one  has to take into account the interaction between the system and its environment. Such analysis is  of fundamental importance for quantum physics and it defines the central objective of the theory of open quantum systems. The evolution of an isolated system is governed by the Schr\"odinger equation and, therefore, is  characterized by a family of unitary operators in the system's Hilbert space. It is no longer true when the system dynamics is influenced by the external world (environment, reservoir, thermal bath). The dynamics is no longer unitary and displays characteristic features induced by the interaction with environment, like dissipation and/or decoherence which cannot be properly described within the standard Hamiltonian formulation of quantum evolution. Theory of open quantum system has a rather long and interesting history. During the last two decades there has been an increasing interest in open quantum systems due to the rapid development of powerful experimental techniques enabling one to control quantum systems on the one hand, and the emergence of quantum information theory together with  modern quantum technologies such as quantum communication, cryptography, computation and an ever growing number of applications on the other. There are already several excellent monographs discussing evolution of open quantum systems  \cite{Open1,Open2,Open3,ALICKI,Open4,Open5,DEC1,DEC2,Carmichael,Carmichael-QR,Gardiner,Schaller,AF} (cf. also review papers \cite{Legget,Fabio-Roberto,Chemia}, and recent lecture notes \cite{LIDAR}, and the recent review on collision models in open quantum systems \cite{collisions}). They cover both mathematical tools and important applications (e.g. in quantum optics and solid state physics).
If the interaction between the system and environment is sufficiently weak and there is a large separation of characteristic time scales of system and environment  one derives a consistent dynamical equation (so-called master equation) for the evolution of the system's reduced density operator. Such Markovian approximation was studied by many people but the most general structure of the corresponding master equation was fully characterized in two seminal papers by Gorini-Kossakowski-Sudarshan \cite{GKS} and Lindblad \cite{L} --- so called GKLS master equation (see \cite{40-GKLS} for a brief history of these developments). Interestingly, GKLS master equation is based on an elegant mathematical concept of completely positive maps introduced by Stinespring \cite{Stinespring} (twenty years before GKLS papers \cite{GKS,L}). Nowadays completely positive maps turn out to define indispensable tool in modern quantum  physics \cite{QIT}. Actually, it is a quantum entanglement, a key feature of quantum physics, which requires the notion of complete positivity to properly describe operations on composite quantum systems. This provides another proof of the Wigner statement on {\em the unreasonable effectiveness of mathematics in the natural sciences} \cite{Wigner-ef}.

Markovian semigroups turned out to provide an effective description of evolution for several important systems in particular in quantum optics where the coupling between the system (atom) and the environment (electromagnetic field) is weak and the Markovian approximation is well justified.  However, recent development of new experimental techniques to control quantum systems and to produce new materials call for more refined approach which takes into account non-Markovian memory effects which are completely neglected in the  Markovian or memoryless approximation.
%A typical manifestation of a non-Markovian memory effect is an information backflow . As a result the system may for example partially or even %completely {\em recohere}. It can never happen for truly Markovian evolution.
Non-Markovian quantum dynamics attracted a lot of attention in recent years  and there is already a vast literature dealing both with several theoretical approaches and experimental realizations (cf. recent review articles \cite{NM1,NM2,NM3,NM4} and \cite{Piilo-I,Piilo-II} for a short introduction to the basic approaches to define and quantify quantum non-Markovianity and for recent experimental advances). It should be stressed that the very term {\em Markovian} is borrowed from the theory of classical stochastic processes \cite{Kampen}. Dealing with evolution of open quantum systems  {\em Markovian evolution} is often used in different contexts and actually several concept of (non)Markovian evolution are available in the current literature. In particular the review \cite{NM4} shows an intricate hierarchy of various notions and clearly indicates that quantum non-Markovianity is highly context-dependent.

%====================

The current report discusses quantum evolution of an open system beyond Markovian semigroup. In what follows we consider the evolution of an open system as a reduced dynamics of the  `system + environment'. They key object of interest is a {\em quantum dynamical map} --- a family of quantum channels --- which  fully characterizes the dynamical properties of a quantum system. Again, complete positivity turns out to play a crucial role in the entire analysis. We discuss both time-local master equations governed by time-local generators and memory kernel master equations of Nakajima-Zwanzig form. In particular we characterize properties  of admissible time-local generators and memory kernels giving rise to legitimate dynamical maps. An important property of the dynamical map is {\em divisibility}. The very concept of Markovianity is based on so-called CP-divisibility which requires that the dynamical map  may be represented as a composition of propagators and each propagator defines a quantum channel.

Clearly, these issues were already discussed by many authors. Our idea is to present these concepts in a unifying way revealing the difference between positivity and complete positivity and stressing how complete positivity usually allows to considerably simplify the entire analysis. We  discuss in parallel the classical and quantum cases showing what are the  universal properties of classical and quantum dynamics.
%Introducing the notion of dissipative generator it is shown that classical and quantum case correspond to distinct notion of dissipativity.
%We also reveal divisibility in the Schr\"odinger and Heisenberg picture and analyze what are the implications of distinct divisibility properties  %for the evolution of a quantum system.
The main goal of this report is to present a universal structure of dynamical maps (both quantum and classical) beyond Markovian semigroup.
Since several recent reviews \cite{NM1,NM2,NM3} already analyzed non-Markovian behaviour of many physical systems, here  instead of studying particular systems we propose to reveal unifying concepts like dissipativity of generators, contractivity and divisibility of maps. These are not only formal mathematical concepts but they are intimately linked to physical properties of quantum systems. In particular, we discuss distinct notions of divisibility  based on various degrees of positivity (from simple positivity to complete positivity). We also reveal intricate connections between time-local generators and non-local memory kernels. We stress that it is not a mathematical physics paper (see e.g. \cite{WCL5} for a more mathematically refined approach to open quantum systems). The mathematical side plays of course a significant role in the entire presentation. However, the goal is to eventually better understand the physical side and again, following Wigner, to show intricate connections between mathematical concepts and physical problems. To simplify the presentation we study only systems living in finite dimensional Hilbert spaces. A lot of concepts can be generalized to infinite dimensional case as well.

%===============================

The report is organized as follows: we begin in Section \ref{SEC-I} by introducing basic tools for describing the dynamics of open quantum systems. The primary objects of interest are positive and completely positive maps. A clear distinction between these two classes of maps is one of the main guiding principle throughout the paper. Both Schr\"odinger and Heisenberg pictures are discussed. The key property of positive trace-preserving maps (Schr\"odinger picture) or positive unital (Heisenberg picture) is contractivity w.r.t. an appropriate norm (trace norm in the Schr\"odinger picture and operator norm in the Heisenberg picture). In Section \ref{SEC-II} we summarize various different representations of completely positive maps and finally relate them to famous $A$ and $B$ matrix representations from a seminal paper by Sudarshan et. al. \cite{Sudarshan-1} published already sixty years ago (see also \cite{Sudarshan-1,Sudarshan-2} and the recent review \cite{Jagadish}).

%Actually, this is the notion of contractivity which allows for a unifying approach.

Equipped with basic mathematical formalism we analyze Markovian semigroup in Section \ref{SEC-IV}. Both semigroups of classical stochastic matrices, positive trace-preserving (PTP) maps, and eventually completely positive trace-preserving (CPTP) maps are considered. We present both Schr\"odinger picture and the dual Heisenberg picture. Spectral properties of generators and the quantum version of detailed balance condition are discussed.  Section \ref{APPROX} discusses various Markovian approximations leading to Markovian semigroup and section \ref{THERMO} analyzes how a semigroup dynamics links the basic laws of thermodynamics.

Section \ref{BEYOND} introduces a key concept of quantum dynamical map beyond Markovian semigroup. Basic properties of dynamical maps such as P- and CP-divisibility are discussed. In this report, following \cite{RHP} I call quantum evolution to be Markovian if it is represented by a CP-divisible dynamical map. This is the most natural intrinsic definition of Markovianity. However, as already stressed in \cite{NM4}, Markovianity is highly context-dependent and the authors of \cite{NM4} presented the intricate hierarchy of different concepts. The characteristic non-Markovian memory effects displayed by a quantum system, such as information backflow \cite{BLP}, are discussed and linked to the concept of divisibility. This section presents also several simple examples which serve as an illustration of the theoretical concepts. Section \ref{Paradigmatic} discusses two paradigmatic open system models: amplitude damping and phase damping.  Starting from the qubit scenario the multi-level scenario is analyzed and conditions for Markovianity are revealed.
Section \ref{IMPLICATIONS} analyzes instructive implications of Markovianity such as monotonicity of physically meaningful quantities. These include various entropic quantities based on the concept of relative entropy and distances between quantum states defined in terms of monotone Riemannian metrics. There are natural information quantities (e.g. mutual information, Fisher information, Wigner-Yanase-Dyson skew information, channel capacity) which display monotonic behaviour under divisible evolution. Finally, quantum entanglement and other correlations (like e.g. quantum discord) fit this scenario. Section \ref{Beyond-d} shows natural constructions of non-Markovian dynamical maps.

CP-divisibility provides an intrinsic characterization of Markovianity. Having access to the total system-environment unitary evolution the very concept of Markovianity may be considerably refined. In this case the property known as quantum regression formula provides the most natural definition which reduces to the standard definition of a classical stochastic process in the commutative (classical) scenario. Now, CP-divisibility provides only a necessary condition for quantum regression to hold and even a dynamical semigroup might violate quantum regression formula and hence display non-Markovian effects. These issues are discussed in section \ref{REGRESSION}.

Section \ref{SEC-MEMORY} characterizes physically admissible memory kernels. It is shown that a complete characterization requires an infinite hierarchy  of nontrivial conditions. Recalling classical semi-Markov evolution we characterize a subclass of admissible kernels giving rise to so called quantum semi-Markov dynamics. Finally, we propose a hybrid approach which uses both time-local generator and memory kernel. A proper engineering of these two mutually commuting generators allows to control complete positivity of the corresponding dynamical map.
Final conclusions are collected in Section \ref{CONCLUSIONS}.

Throughout the paper we set $\hbar =1$.

\section{Preliminaries: maps, channels and all that}   \label{SEC-I}

\subsection{Maps and quantum states}

Let $\mathcal{H}$ denote a finite dimensional Hilbert space and  $\LH$ a vector space of linear operators acting on $\mathcal{H}$. Majority of results can be generalized for infinite dimensional case as well. However, to avoid technical complications and keep the presentation as simple as possible I consider only finite level quantum systems. Quantum states are represented by density operators, i.e. $\rho \in \LH$ such that $\rho \geq 0$ and $\Tr \rho=1$. Positivity of $\rho$ means that for any vector $x \in \mathcal{H}$ one has $\<x|\rho|x\> \geq 0$. Equivalently, all eigenvalues of $\rho$ are nonnegative. Note, that fixing an orthonormal basis $\{|k\>\}_{k=1}^d$ in $\mathcal{H}$ any density operator gives rise to a probability distribution $p_k = \<k|\rho|k\>$. Hence $\rho$ encodes infinitely many classical probability distributions (for this reason sometimes it is called quantum probability). The formalism of density operators provides an appropriate tool for dealing with statistical mixtures:  given a statistical mixture $\{p_i,|\psi_i\>\}$  the corresponding density matrix is defined as $\rho = \sum_i p_i |\psi_i\>\<\psi_i|$.  A density matrix corresponds to a pure quantum state if there is only one element in this ensemble, i.e. $\rho = |\psi\>\<\psi|$. Two mixtures $\{p_i,|\psi_i\>\}$ and $\{q_i,|\phi_i\>\}$ are equivalent whenever $\sum_i p_i |\psi_i\>\<\psi_i|  = \sum_i q_i |\phi_i\>\<\phi_i|$. Density operators not only provide a natural generalization of state vectors from the Hilbert space. They are deeply rooted in the heart of quantum physics due to the intricate property of quantum entanglement: given a state vector of a composite $AB$ system $\psi_{AB} \in \mathcal{H}_A \otimes \mathcal{H}_B$ the reduced states of subsystems defined by a partial trace operation $\rho_A = {\rm Tr}_B |\psi_{AB}\>\<\psi_{AB}|$ and $\rho_B = {\rm Tr}_A |\psi_{AB}\>\<\psi_{AB}|$ are in general not pure.

A natural question arises how to transform density operators? In particular how to describe the time evolution $\rho \to \rho_t$? It turns out \cite{Gisin} that to exclude the possibility for superluminal communication different but equivalent initial mixtures have to stay equivalent for any $t > 0$. This implies
that only linear transformations are allowed. In particular such transformation $\Phi$ should map a convex combination  $p_1 \rho_1 + p_2 \rho_2$ into a convex combination $p_1 \Phi(\rho_1) + p_2 \Phi_2(\rho_2)$. Consider, therefore, a linear transformations $\Phi$

%\footnote{In general one considers transformations $\Phi : \mathcal{H}_1 \to \mathrm{L}(\mathcal{H}_2)$. In this paper, however, we are mainly interested i transforming states of the same system.}

\begin{equation}\label{}
  \Phi : \LH \to \mathrm{L}(\mathcal{H}') ,
\end{equation}
mapping states living in $\HH$ into states living in $\HH'$.  Clearly, it should preserve Hermiticity, that is, $\Phi(X)^\dagger = \Phi(X^\dagger)$, and the trace $\Tr \Phi(X) = \Tr X$. Finally, since $\rho$ is a positive operator $\Phi$ should preserve positivity meaning that if $X \geq 0$ then $\Phi(X) \geq 0$. One calls such $\Phi$ a  positive map.  Hence, any  map transforming states into states is positive and trace-preserving (PTP).

Surprisingly, this class of maps is not consistent with principles of quantum mechanics.  Suppose that $\Phi_1 : \mathrm{L}(\mathcal{H}_1) \to \mathrm{L}(\mathcal{H}'_1)$ and  $\Phi_2 : \mathrm{L}(\mathcal{H}_2) \to \mathrm{L}(\mathcal{H}'_2)$ are two positive and trace-preserving maps. Consider two composite systems living in $\HH_1 \otimes \HH_2$ and $\HH'_1 \otimes \HH'_2$, respectively. It is, therefore, natural to consider a tensor product map

\begin{equation}\label{}
  \Phi_1 \otimes \Phi_2 : \mathrm{L}(\HH_1 \otimes \HH_2) \to \mathrm{L}(\HH'_1 \otimes \HH'_2) .
\end{equation}
It turns out that even if $\Phi_1$ and $\Phi_2$ are positive the tensor product map $\Phi_1 \otimes \Phi_2$ need not be positive. Hence, one can not use $\Phi_1 \otimes \Phi_2$ to safely transform states of a composite quantum systems. What goes wrong? Note, that if a state $\rho$ is separable, that is, it can be represented via the following convex combination \cite{QIT,HHHH}

\begin{equation}\label{}
  \rho = \sum_k p_k \rho^{(1)}_k \otimes \rho^{(2)}_k ,
\end{equation}
where $p_k > 0$, $\sum_k p_k = 1$, and $\rho^{(1)}_k$ and $\rho^{(2)}_k$ are states of $S_1$ living in $\HH_1$ and $S_2$ living in $\HH_2$, respectively, then

\begin{equation}\label{}
  (\Phi_1 \otimes \Phi_2)(\rho) = \sum_k p_k \Phi_1(\rho^{(1)}_k) \otimes \Phi_2(\rho^{(2)}_k) ,
\end{equation}
is a legitimate separable state of $S'_1 + S'_2$. Hence, it is clear that if something goes wrong, i.e. the map $\Phi_1 \otimes \Phi_2$ is not positive, it is because of quantum entanglement. This observation clearly shows that one needs a more refined class of physically admissible maps. Suppose we couple a system $S$ to another system living in $\mathbb{C}^k$ ($k$-dimensional ancilla). Denote by $M_k(\mathbb{C})$ a set of $k \times k$ complex matrices and let ${\rm id}_k :  M_k(\mathbb{C}) \to M_k(\mathbb{C})$ denote an identity map.

\begin{tcolorbox}
\begin{Definition} A linear map $\Phi : \LH \to \mathrm{L}(\mathcal{H}')$ is $k$-positive if

\begin{equation}\label{}
  {\rm id}_k \otimes \Phi : M_k(\mathbb{C}) \otimes \LH \ \to\   M_k(\mathbb{C}) \otimes \mathrm{L}(\HH')
\end{equation}
is positive. $\Phi$ is completely positive (CP) if it is $k$-positive for all $k=1,2,3,\ldots $.
\end{Definition}
\end{tcolorbox}
The very notion of completely positive map was introduced by Stinespring in 1955 \cite{Stinespring} (cf. detailed exposition in \cite{Paulsen,Stormer-63,Stormer,Bhatia}). The above definition of completely positive maps can be generalized for infinite dimensional case as follows \cite{Paulsen}: consider a linear map $\Phi : \mathfrak{A} \to \BH$, where $\mathfrak{A}$ denotes  an infinite dimensional $C^*$-algebra  and $\BH$ denotes the vector space of bounded operators acting on $\HH$. The map $\Phi$ is  completely positive if  the following condition is satisfied

\begin{equation}\label{}
  \sum_{i,j=1}^N \< x_i | \Phi(a^*_i a_j)| x_j \> \geq 0 ,
\end{equation}
for any $a_1,\ldots,a_N \in \mathfrak{A}$, and $x_1,\ldots,x_n \in \HH$, and $N=1,2,\ldots$. Twenty years after Stinespring paper  \cite{Stinespring} this  beautiful mathematical concept turned out to play distinguished  role in characterizing dynamics of quantum systems.

%How to understand the action of $ {\rm id}_k \otimes \Phi $? Note that any element  $X \in M_k(\mathbb{C}) \otimes \LH$ may be represented as %follows
%
%\begin{equation}\label{}
%  X = \sum_{i,j=1}^k |i\>\<j| \otimes X_{ij} \ , \ \ \ X_{ij} \in \LH ,
%\end{equation}
%where $\{|1\>,\ldots,|k\>\}$ is a conical basis in $\mathbb{C}^k$. One defines
%
%\begin{equation}\label{}
%  ({\rm id}_k \otimes \Phi)(X) := \sum_{i,j=1}^k |i\>\<j| \otimes \Phi(X_{ij}) .
%\end{equation}
%
%\begin{Example} Consider the simplest example when $\HH_1=\HH_2 = \mathbb{C}^2$ and $\HH_1'=\HH_2'=\mathbb{C}^2$. Fix a basis $|0\>,|1\>$ and define maximally entangled state $P^+ = |\psi^+\>\<\psi^+|$, where $|\psi^+\> = 1/\sqrt{2}\sum_{i=0}^{1}|i \otimes i\>$.

%\begin{equation}\label{}
%  P^+ = \frac 12 \left( \begin{array}{cc|cc}
%                  1 & 0 & 0 & 1 \\
%                  0 & 0 & 0 & 0 \\ \hline
%                  0 & 0 & 0 & 0 \\
%                  1 & 0 & 0 & 1
%                \end{array} \right) .
%\end{equation}
%One finds for $\Phi=T$ (transposition in the basis $\{|0\>,|1\>\}$)

%\begin{equation}\label{}
% ({\rm id} \otimes T)(P^+) = \frac 12 \left( \begin{array}{cc|cc}
%                  1 & 0 & 0 & 0 \\
%                  0 & 0 & 1 & 0 \\ \hline
%                  0 & 1 & 0 & 0 \\
%                  0 & 0 & 0 & 1
%                \end{array} \right) ,
%\end{equation}
%which is no longer positive (has one negative eigenvalue) and hence ${\rm id} \otimes T$ is not a positive map.
%\end{Example}
%
%
%Completely positive maps were introduced by Stinespring \cite{Stinespring}. \cite{Kadison,Pauslen,Stormer,Wolf}.
The above definition of $k$-positive map is rather formal. What about a physical meaning of $k$-positive maps? Consider a composite system in $\HH \otimes \HH'$. Any state vector $\psi$ gives rise to the Schmidt decomposition

\begin{equation}\label{}
  |\psi\> = \sum_{i=1}^r s_i |e_i\> \otimes |f_i\> ,
\end{equation}
where the Schmidt rank ${\rm SR}(|\psi\>) = r$ satisfies $1 \leq r \leq d_m = \min\{d,d'\}$ ($d={\rm dim}\, \HH$ and $d'={\rm dim}\, \HH$). This concept may be easily generalized for density operators \cite{Terhal}: given $\rho$ one defines its Schmidt number

\begin{equation}\label{}
  {\rm SN}(\rho) = \min_{\{p_k,\psi_k\}} \max_k {\rm SR}(\psi_k)
\end{equation}
where one minimizes over all pure state decompositions $\rho = \sum_k p_k |\psi_k\>\<\psi_k|$. If $\rho = |\psi\>\<\psi|$, then ${\rm SN}(\rho) = {\rm SR}(\psi)$.

\begin{Proposition} The map $\Phi : \LH \to \mathrm{L}(\HH)$ is $k$-positive if and only if

\begin{equation}\label{}
   ({\rm id} \otimes \Phi)(\rho) \geq 0 ,
\end{equation}
for any $\rho \in \mathrm{L}(\HH \otimes \HH)$  with ${\rm SN}(\rho) \leq k$.
\end{Proposition}
Hence, if the map $\Phi : \LH \to \mathrm{L}(\mathcal{H})$ is $k$-positive, then the operation ${\rm id} \otimes \Phi$ safely transforms entangled states $\rho$ in $\HH \otimes\HH$ into states in $\HH \otimes \HH'$ provided ${\rm SN}(\rho) \leq k$. It shows that classification of entangled states is closely related to classification of positive maps. Denote by $\mathcal{P}_k$ a set of $k$-positive maps $\LH  \to  \mathrm{L}(\HH')$.  It is evident that $\mathcal{P}_k \subset \mathcal{P}_l$ whenever $k < l$. Interestingly, it was shown  by Choi \cite{Choi-75} that if the map is $d$-positive ($d = {\rm dim}\HH$), then it is completely positive. It gives rise to the following chain of inclusions

%\begin{tcolorbox}
\begin{equation}\label{k-hierarchy}
\mbox{positive maps} =  \mathcal{P}_1 \supset \mathcal{P}_2 \supset \ldots \supset \mathcal{P}_{d} = \mbox{CP maps} .
\end{equation}
%\end{tcolorbox}

\begin{Example} Consider a map $\Phi : \LH \to \LH$ defined as follows

\begin{equation}\label{Phi-k}
  \Phi_p(X) = p\, \oper_d {\rm Tr} X - X ,
\end{equation}
with $p > 0$. Choi showed \cite{Choi-72} that $\Phi_p$ is $k$-positive but not $(k+1)$-positive if and only if $k \leq p < k+1$. In particular $\Phi$ is positive if $p \geq 1$. In entanglement theory the map $\Phi_1$ is called reduction map and plays important role in classifying states of composite systems \cite{HHHH,Guhne,TOPICAL}.
\end{Example}
Completely positive maps in finite dimensional case are fully characterized due to the following remarkable result \cite{Choi-75}

\begin{tcolorbox}
\begin{Theorem}  A linear map $\Phi : \LH \to \mathrm{L}(\HH)$ is completely positive if and only if

\begin{equation}\label{}
  ({\rm id}_d \otimes \Phi)(P^+_d) \geq 0 ,
\end{equation}
where $P^+_d$ is a maximally entangled state in $\HH \otimes \HH$ ($d = {\rm dim}\,\HH$).
\end{Theorem}
\end{tcolorbox}
It means that to check for complete positivity one needs to check the spectrum of a single operator $({\rm id}_d \otimes \Phi)(P^+_d)$. In other words, complete positivity of $\Phi$ is equivalent to positivity of the extended map ${\rm id}_d \otimes \Phi$ and remarkably positivity of  ${\rm id}_d \otimes \Phi$ is guaranteed by positivity on $P^+_d$. Such spectral property is no longer true for positive maps which are not completely positive and it makes the analysis of positive maps highly nontrivial. Completely positive trace-preserving (CPTP) map  $\Phi : \LH \to \mathrm{L}(\HH)$ is called a quantum channel.

\begin{Example} To illustrate the difference between completely positive and positive maps consider three simple examples of qubit maps using well known Bloch representation of a qubit density operator

\begin{equation}\label{}
  \rho = \frac 12 \left( \oper + \sum_{k=1}^3 x_k \sigma_k \right) ,
\end{equation}
where the Bloch vector $\mathbf{x}=(x_1,x_2,x_3)$ satisfies $|\mathbf{x}|\leq 1$ (Bloch ball). Let $\mathbf{x}'=(x_1',x_2',x_3')$ be a Bloch vector corresponding to $\rho'=\Phi(\rho)$. Consider the following three maps

\begin{eqnarray}
  \Phi_1 &:& \ \ \mathbf{x}\ \to\ (x_1',x_2',x_3') = (x_1,-x_2,x_3) \\
    \Phi_2 &:& \ \ \mathbf{x}\ \to\ (x_1',x_2',x_3') = -(x_1,x_2,x_3)\\
      \Phi_3 &:& \ \ \mathbf{x}\ \to\ (x_1',x_2',x_3') = (-x_1,x_2,-x_3)
\end{eqnarray}
Clearly all three maps are positive due to $|\mathbf{x}|=|\mathbf{x}'|$. However, only $\Phi_3$ is completely positive. Note, that

\begin{equation}\label{}
  \Phi_1(\rho) = \rho^T \ , \ \ \ \Phi_2(\rho) = \oper_2 {\rm Tr}\rho - \rho ,
\end{equation}
where $\rho^T$ denotes a transposition, and for the map  $\Phi_3 = \Phi_1 \Phi_2$ one finds

\begin{equation}\label{}
  \Phi_3(\rho) = \frac 12 {\rm Tr}\rho - \rho^T = \sigma_2 \rho \sigma_2 ,
\end{equation}
which represents unitary quantum channel (cf. Figure 1). This example clearly shows that complete positivity is fundamentally different from the standard positivity which is already guaranteed by the fact that $\mathbf{x}'$ belongs to the Bloch ball.
A qubit map transforming Bloch ball into itself is necessarily positive but needs not be completely positive. Interestingly, a composition of two maps which are not completely positive may lead do completely positive one.
\end{Example}

\begin{center}
\begin{figure} \label{}
\hspace*{.4cm} \includegraphics[width=4cm]{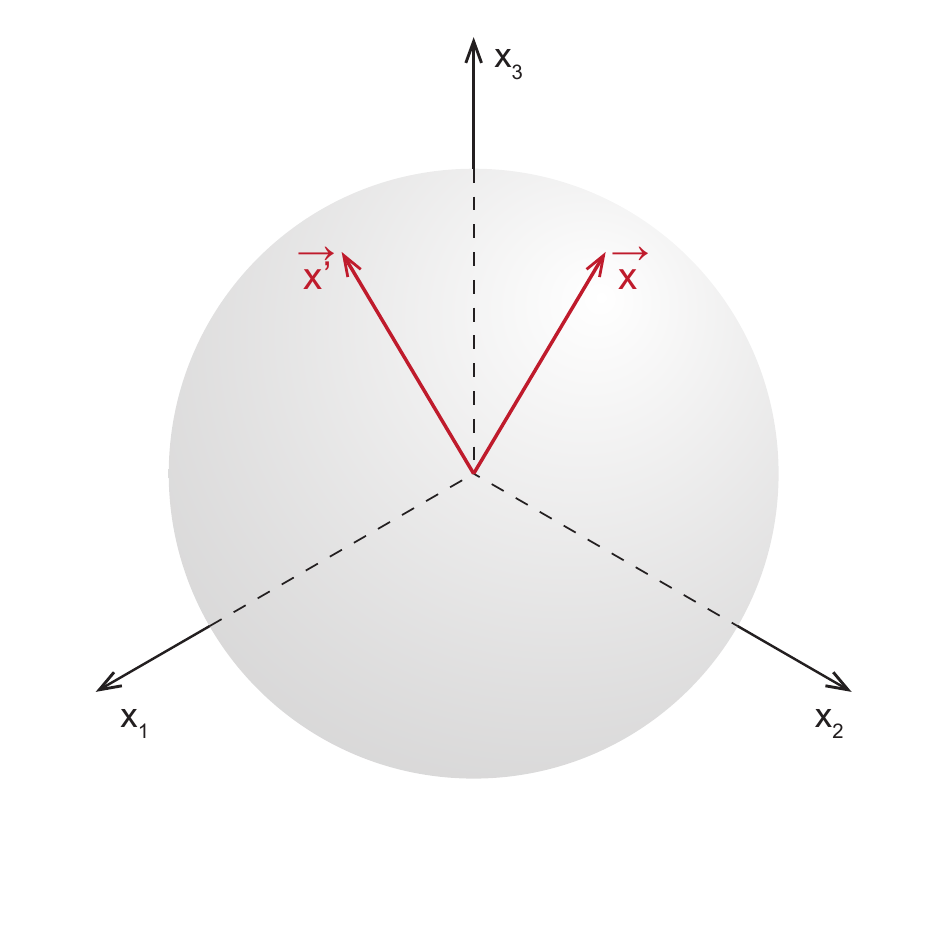} \hspace{.4cm} \includegraphics[width=4cm]{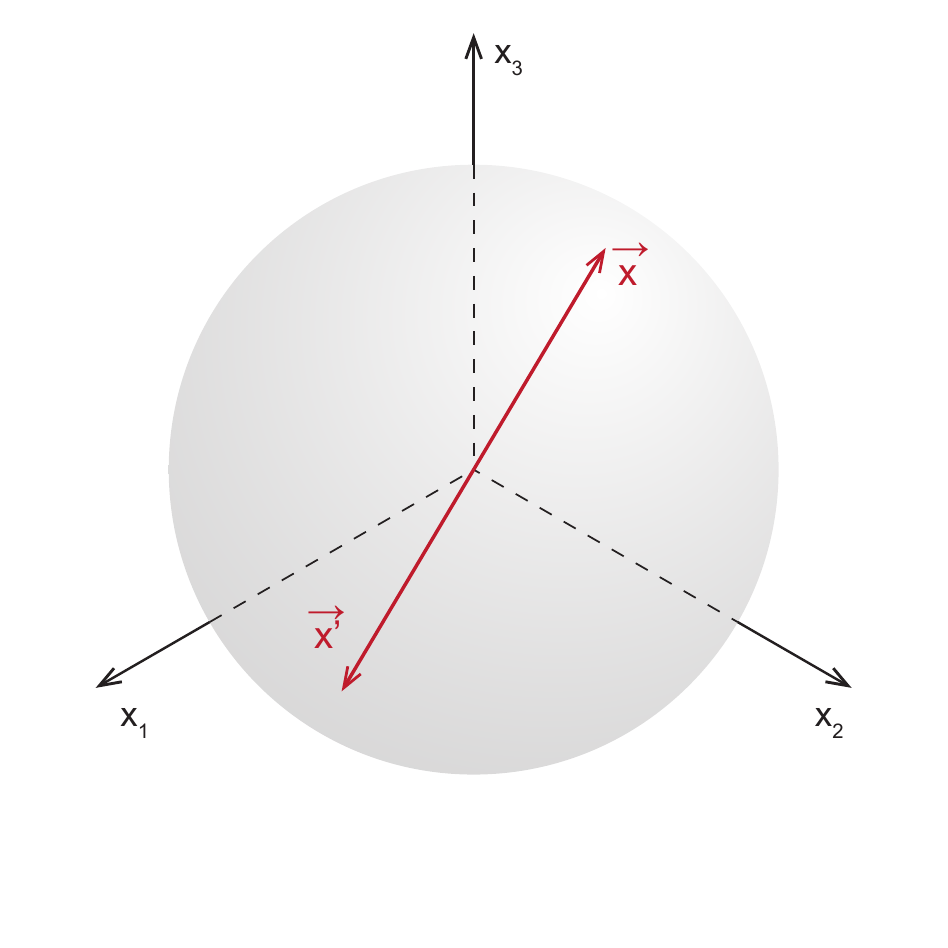} \hspace{.4cm}  \includegraphics[width=4cm]{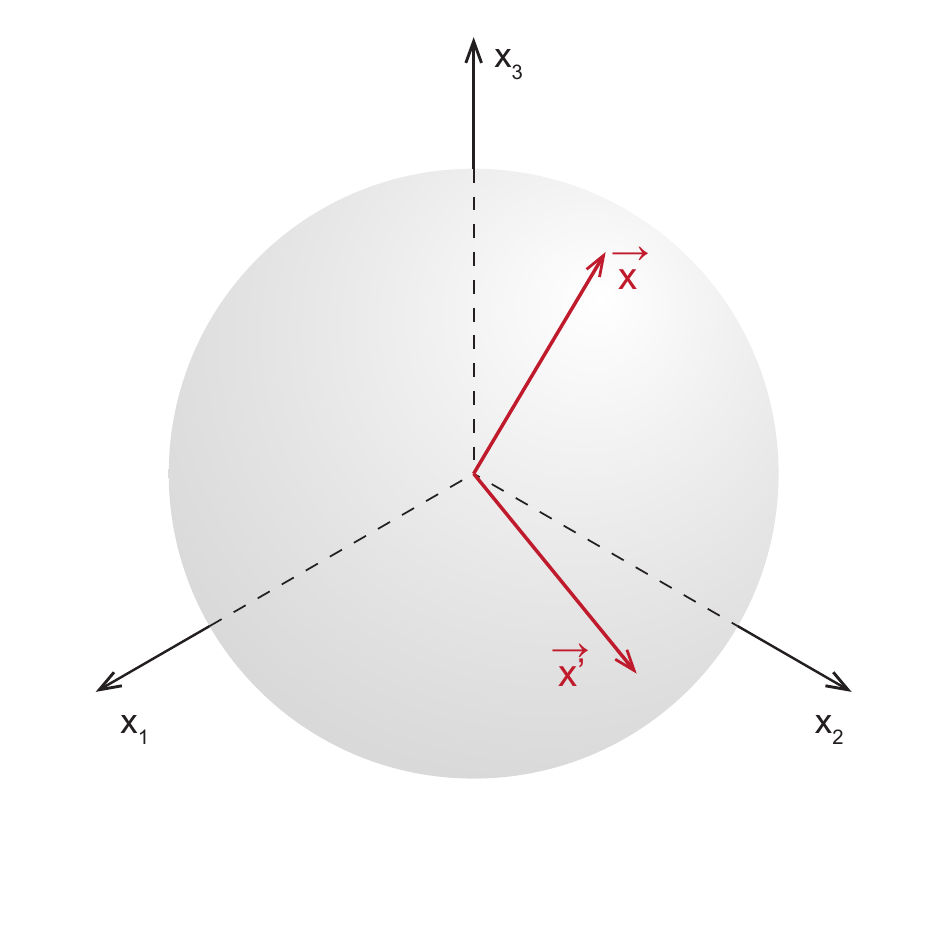}
\caption{(Color online) Graphical representation of the action of $\Phi_1$ (left), $\Phi_2$ (middle), and $\Phi_3$ (right). }
\end{figure}
\end{center}

%\begin{tcolorbox}

%\end{tcolorbox}

For any linear map $\Phi : \LH \to \LH$ one defines a dual map $\Phi^\ddag : \LH \to \LH$ via

\begin{equation}\label{}
  (\Phi^\ddag(X),Y)_{\rm HS} := (X,\Phi(Y))_{\rm HS} ,
\end{equation}
for all $X,Y \in \LH$, and $(X,Y)_{\rm HS} = {\rm Tr}(X^\dagger Y)$
denotes the Hilbert-Schmidt inner product. The vector space $\LH$ equipped with the Hilbert-Schmidt inner product becomes a Hilbert space. It is clear that $\Phi^{\ddag \ddag} =\Phi$.

\begin{tcolorbox}
\begin{Proposition} One has the following relations

\begin{enumerate}
  \item $\Phi$ is Hermiticity preserving if and only if $\Phi^\ddag$ is Hermiticity preserving,
  \item $\Phi$ is trace-preserving if and only if $\Phi^\ddag$ is unital, i.e. $\Phi^\ddag(\oper)=\oper$,
  \item $\Phi$ is positive if and only if $\Phi^\ddag$ is positive,
  \item $\Phi$ is completely positive if and only if $\Phi^\ddag$ is completely positive.
\end{enumerate}

\end{Proposition}
\end{tcolorbox}

%\subsection{Positive maps vs. stochastic matrices}

A real $d \times d$ matrix $T_{ij}$ is column stochastic if $T_{ij} \geq 0$ and $\sum_{i=1}^d T_{ij} = 1$. It is row stochastic  if $T_{ij} \geq 0$ and $\sum_{j=1}^d T_{ij} = 1$. Finally, it is doubly stochastic (or bi-stochastic) if it is both column and row stochastic. Consider a positive trace-preserving map $\Phi : \LH \to \LH$. Fixing a basis $\{|k\>\}_{k=1}^d$ in $\HH$ one defines a matrix

\begin{equation}\label{}
  T_{ij} := {\rm Tr}(|i\>\<i| \Phi(|j\>\<j|)) = \<i|  \Phi(|j\>\<j|)| i\> .
\end{equation}
Clearly, $T_{ij}$ is column stochastic. Similarly, one may define a row stochastic matrix using a unital positive map $\Phi^\ddag$

\begin{equation}\label{}
  S_{ij} := {\rm Tr}(|i\>\<i| \Phi^\ddag(|j\>\<j|)) .
\end{equation}
Note, that $S_{ij} = T_{ji}$. This way any positive trace-preserving map encodes an infinite set of column stochastic matrices. Similarly, any density operator $\rho$  encodes an infinite set of probability distributions $p_k = \<k|\rho|k\>$.

\subsection{Positive maps and quantum entanglement}

A positive map $\Phi : \LH \to \LH$ which is positive but not completely positive serves as a universal tool for detecting quantum entanglement. One has the following fundamental result \cite{HHHH}

\begin{Theorem} A state $\rho$ living in $\mathcal{H}_1 \otimes \mathcal{H}_2$ is entangled if and only if there exists a positive map  $\Phi : \mathrm{L}(\HH_2) \to \mathrm{L}(\HH_3)$ such that

\begin{equation}\label{Ent}
  ({\rm id} \otimes \Phi)(\rho) \ngeq 0.
\end{equation}
\end{Theorem}
Note, that (\ref{Ent}) implies the existence of $|\psi\> \in \HH_1 \otimes \HH_3$ such that

\begin{equation}\label{}
  \< \psi |  ({\rm id} \otimes \Phi)(\rho) | \psi\> < 0 ,
\end{equation}
and clearly $|\psi\>$ is entangled. Similarly, one has

\begin{Cor} A state $\rho$ living in $\mathcal{H}_1 \otimes \mathcal{H}_2$ is entangled with ${\rm SN}(\rho) > k$ if and only if there exists a $k$-positive  map  $\Phi : \mathrm{L}(\HH_2) \to \mathrm{L}(\HH_3)$ such that

\begin{equation}\label{Ent-k}
  ({\rm id} \otimes \Phi)(\rho) \ngeq 0.
\end{equation}
\end{Cor}
For bipartite systems living in $\mathbb{C}^2 \otimes \mathbb{C}^2$ or $\mathbb{C}^2 \otimes \mathbb{C}^3$ a state is separable if and only if it is PPT (positive under partial transposition ${\rm id} \otimes T$). Hence, in this simple scenario a single transposition map decides about separability. It is no longer true for other bipartite systems due to so called bound entangled states \cite{HHHH,Guhne,TOPICAL}. This concept gives rise to the following

\begin{Definition} A positive map $\Phi : \LH \to \mathrm{L}(\mathcal{H}')$ is decomposable if
  \begin{equation}\label{DEC}
    \Phi = \Phi_1 + T\circ \Phi_2 ,
  \end{equation}
where $\Phi_1$ and $\Phi_2$ are CP ($T$ is a transposition map). A map which can not be represented as (\ref{DEC}) is called non-decomposable.
\end{Definition}
All positive maps $M_k(\mathbb{C}) \to M_\ell(\mathbb{C})$ with $k\cdot \ell \leq 6$ are decomposable \cite{Woronowicz}. Unfortunately, there is no a general method to construct non-decomposable maps (cf. \cite{TOPICAL}). If a state $\rho$ is PPT, then $({\rm id}\otimes \Phi)(\rho) \geq 0$ for any decomposable positive map $\Phi$ and hence one needs non-decomposable maps  in order to witness bound entangled states \cite{HHHH}.

%Another important tool for entanglement detection is an {\em entanglement witness}, that is, a Hermitian operator $W \in \mathrm{L}(\HH_1 \otimes \HH_2)$ such that $\< \psi|W|\psi\> \geq 0$ for all product vectors $|\psi\> \in \HH_1 \otimes \HH_2$. Note, that any positive map $\Phi$ gives rise to an entanglement witness defined as follows

%\begin{equation}\label{}
%  W = ({\rm id} \otimes \Phi)(P^+) ,
%\end{equation}
%where $P^+$ is a maximally entangled state in

\subsection{Contractive properties of positive maps and stochastic matrices}

A key property of trace-preserving positive maps is contractivity. In order to analyze this property one needs an appropriate norm in $\LH$. Introducing a trace norm  $\|X\|_1 = {\rm Tr}|X|$ one endows the vector space $\LH$ with a structure of a Banach space $\mathcal{T}(\mathcal{H}) =(\LH,\|\ \|_1)$. Note, that $\rho \geq 0$ if and only if ${\rm Tr}\rho = \|\rho\|_1$.

\begin{tcolorbox}
\begin{Proposition} \label{PRO-I} Let $\Phi : \TTH \to \TTH$ be a trace-preserving and Hermiticity preserving map. $\Phi$ is positive if and only if

\begin{equation}\label{}
  \| \Phi(X) \|_1 \leq \|X\|_1 ,
\end{equation}
for all Hermitian operators $X^\dagger=X \in \TTH$.
\end{Proposition}
\end{tcolorbox}

Similarly, the operator norm $\|X\|_\infty$  endows $\LH$ with a structure of an  algebra of bounded operators $\BH =(\LH,\| \ \|_\infty)$. In fact it is a $C^*$-algebra and  $\|XY\|_\infty \leq \|X\|_\infty \|Y\|_\infty$ for any $X,Y \in \BH$.

\begin{tcolorbox}
\begin{Proposition} \label{PRO-II} Let $\Psi : \BH \to \BH$ be a Hermiticity preserving unital map. $\Psi$ is positive if and only if

\begin{equation}\label{}
  \| \Phi(X) \|_\infty \leq \|X\|_\infty ,
\end{equation}
for all  operators $X \in \BH$.
\end{Proposition}
\end{tcolorbox}
For the proof of the above Propositions cf. \cite{Paulsen,Bhatia} (actually, Proposition \ref{PRO-I} was originally proposed by Kossakowski \cite{Kossakowski1972}). These results show that essentially positivity is nothing but contractivity w.r.t. an appropriate norm.
In what follows we call a positive trace-preserving  map $\Phi : \TTH \to \TTH$ a Schr\"odinger picture of the map and its dual $\Phi^\ddag : \BH \to \BH$ the  corresponding Heisenberg picture.

Similar contractivity properties hold for column and row stochastic matrices. Introducing the following norms in $\mathbb{R}^d$: $\ell_1$-norm

\begin{equation}\label{ell-1}
  \|\mathbf{x}\|_1 = \sum_{i=1}^{d} |x_i| ,
\end{equation}
where $\mathbf{x} =(x_1,\ldots,x_d)$, and $\ell_\infty$-norm

\begin{equation}\label{ell-infty}
  \|\mathbf{x}\|_\infty = \max_i |x_i| ,
\end{equation}
one defines a real Banach space $\mathcal{T}_d = (\mathbb{R}^d,\| \ \|_1)$ and a commutative algebra $\mathcal{B}_d=(\mathbb{R}^d,\| \ \|_\infty)$.

\begin{tcolorbox}
\begin{Proposition} Let $T_{ij}$ be a real $d \times d $ matrix such that $\sum_i T_{ij}=1$. Then $T_{ij}$ is column stochastic if and only if
\begin{equation}\label{}
  \| T \mathbf{x} \|_1 \leq \|\mathbf{x}\|_1 ,
\end{equation}
for all $\mathbf{x} \in \mathcal{T}_d$. Similarly, let $S_{ij}$ be a real $d \times d $ matrix such that $\sum_j S_{ij}=1$. Then $S_{ij}$ is row stochastic if and only if
\begin{equation}\label{}
  \| S \mathbf{x} \|_\infty \leq \|\mathbf{x}\|_\infty ,
\end{equation}
for all $\mathbf{x} \in \mathcal{B}_d$.
\end{Proposition}
\end{tcolorbox}

\subsection{Spectral properties of positive maps}

Celebrated Perron-Frobenius theorem \cite{Matrices-1,Matrices-2} states that a spectrum $ T \mathbf{x} =\lambda \mathbf{x}$ of any column stochastic matrix satisfies the following properties:

%\begin{equation}\label{}
%  T \mathbf{x} =\lambda \mathbf{x} , %\ \ \mathbf{x} \in \mathbb{R}^d \ \  ???????\ \ \mathbb{C}^d
%\end{equation}

\begin{enumerate}
  \item there is a leading eigenvalue $\lambda_0=1$, and the corresponding eigenvector $\mathbf{x}$ satisfies $x_i \geq 0$,
  \item all other eigenvalues belong to the unit disc in the complex plane, $|\lambda_k|\leq 1$ for $k=1,\ldots,d-1$,
  \item the spectrum is symmetric w.r.t. real line.
\end{enumerate}
Interestingly, a spectrum of a linear positive trace-preserving map $\Phi : \LH \to \LH$ enjoys similar properties. Consider a spectral problem

\begin{equation}\label{}
  \Phi(X_\alpha) = \lambda_\alpha X_\alpha \ , \ \ X_\alpha \in \LH .
\end{equation}
One has the following quantum version of Perron-Frobenius theorem \cite{Perron-1,Perron-2}

\begin{tcolorbox}
\begin{Proposition} A spectrum of any trace-preserving positive map $\Phi :\LH \to \LH$ satisfies the following properties

\begin{enumerate}
  \item there is a leading eigenvalue $\lambda_0=1$, and the corresponding eigenvector $X_0 \geq 0$,
  \item all other eigenvalues belong to the unit disc in the complex plane, $|\lambda_k|\leq 1$ for $k=1,\ldots,d^2-1$,
  \item the spectrum is symmetric w.r.t. real line.
\end{enumerate}

\end{Proposition}
\end{tcolorbox}
Normalizing $X_0$ one defines a steady state $\rho_{\rm ss} = X_0/{\rm Tr}X_0$ of the map. Note that a dual map has exactly the same spectrum, that is, $\Phi^\ddag(Y_\alpha) = \lambda_\alpha^* Y_\alpha$, with $Y_0 = \oper$.  Recall, that $\Phi$ is irreducible if the inequality $\Phi(P) < \lambda P$, where $P$ is a Hermitian projector and $\lambda > 0$, implies that $P \in \{0,\oper\}$. For irreducible positive trace-preserving maps the steady state $\rho_{\rm ss}$ is unique and faithfull, i.e. $\rho_{\rm ss} > 0$.

Eigenvalue $\lambda$ is called peripheral if $|\lambda|=1$. If $\Phi$ is a irreducible CPTP map then all peripheral eigenvalues are non degenerated  and

\begin{equation}\label{per}
  \lambda_k = e^{2\pi i k/m} , \ \ \ k=0,1,\ldots,m-1,
\end{equation}
for some $m \leq d^2$ (cf. \cite{Perron-1,WOLF}). Irreducible positive trace-preserving maps enjoy the following  ergodic property:

\begin{Proposition} If $\Phi$ is an irreducible positive trace-preserving map then

\begin{equation}\label{}
  \lim_{N \to \infty} \frac 1N\, \sum_{n=1}^N T^n(\rho) = \rho_{\rm ss} .
\end{equation}
 for any initial $\rho \in \TTH$.
\end{Proposition}
For more detailed exposition of spectral properties of positive maps cf.  \cite{WOLF}.

\section{How to represent quantum maps} \label{SEC-II}

\subsection{Stinespring representation}

Consider a CPTP map $\Phi : \LH \to \mathrm{L}(\HH')$. There exists a Hilbert space $\mathcal{H}_E$ and an isometry $W : \HH \to \HH' \otimes \HH_E$ (i.e. $W^\dagger W =\oper_{\HH}$)  such that

\begin{equation}\label{ST}
  \Phi(\rho) = {\rm Tr}_E ( W \rho W^\dagger) .
\end{equation}
The dual map $\Phi^\ddag : \mathrm{L}(\HH') \to \LH$ reads as follows

\begin{equation}\label{}
  \Phi^\ddag(X) = W^\dagger(X \otimes \oper_E)W ,
\end{equation}
and satisfies $\Phi^\ddag(\oper_{\HH'}) = \oper_{\HH}$. One calls (\ref{ST}) a Stinespring representation of $\Phi$.

\begin{Remark}
Actually, in his seminal paper \cite{Stinespring} Stinespring provided the following representation in a more general scenario: a linear map $\Phi :\mathfrak{A} \to \BH$ is completely positive if there exists a Hilbert space $\mathcal{K}$, representation $\pi : \mathfrak{A} \to \mathcal{B}(\mathcal{K})$ of a $C^*$-algebra $\mathfrak{A}$, and a linear operator $V : \mathcal{K} \to \mathcal{H}$ such that

\begin{equation}\label{}
  \Phi(a) = V \pi(a) V^\dagger ,
\end{equation}
for any $a \in \mathfrak{A}$.
\end{Remark}

\subsection{Environmental  representation}

For  any CP map $\Phi : \LH \to \mathrm{L}(\HH')$ there exists an auxiliary Hilbert space $\mathcal{H}_E$ and a linear operator

\begin{equation}\label{}
  \mathbb{V} : \HH \otimes \mathcal{H}_E \to \HH' \otimes \mathcal{H}_E ,
\end{equation}
such that

\begin{equation}\label{Stine}
  \Phi(\rho) = \Tr_E (\mathbb{V} \rho \otimes |\psi\>\<\psi| \mathbb{V}^\dagger) ,
\end{equation}
for some state vector $|\psi\> \in \HH_E$. One calls (\ref{Stine}) an environmental representation of $\Phi$. Again,  $\Phi$ is trace-preserving if $\mathbb{V}$ satisfies additional condition $\mathbb{V}^\dagger \mathbb{V} = \oper$. Indeed,

$$   \Tr \Phi(\rho) = \Tr (\mathbb{V} \rho \otimes |\psi\>\<\psi| \mathbb{V}^\dagger) = \Tr (\mathbb{V}^\dagger \mathbb{V} \rho \otimes |\psi\>\<\psi| ) = \Tr \rho , $$
whenever $\mathbb{V}^\dagger \mathbb{V} = \oper$, that is, $\mathbb{V}$ is an isometry. It should be stressed that the above representation is not unique. In the special case when $\HH'=\HH$,  $\mathbb{V}$ is not only an isometry but it is a unitary operator. The dual map reads as follows

\begin{equation}\label{}
  \Phi^\ddag(X) = {\rm Tr}_E (\mathbb{V}^\dagger(X \otimes \oper_E)\mathbb{V}) .
\end{equation}

% One usually interpret $\mathcal{K} = \mathcal{H}_E$ as an environmental Hilbert space  and eventually arrives at the familiar environmental %representation of the quantum channel
%
%\begin{equation}\label{Stine-U}
%  \Phi(\rho) = \Tr_{\rm E} (U \rho \otimes |\psi\>\<\psi| U^\dagger) .
%\end{equation}

\subsection{Operator sum representation}

If $\HH$ and $\HH'$ are finite dimensional, then one can always find finite dimensional environmental Hilbert space  $\mathcal{H}_E$ to provide the corresponding  representations (\ref{ST}) or (\ref{Stine}). Let $|\psi\> = |0\>,|1\>,\ldots,|D\>$ be an orthonormal basis in $\mathcal{K}$ and let

\begin{equation}\label{}
  \mathbb{V} = \sum_{i,j=0}^D V_{ij} \otimes |i\>\<j| \ ; \ \ \ V_{ij} : \HH \to \HH' .
\end{equation}
Observe that condition $\mathbb{V}^\dagger \mathbb{V} = \oper$ implies

\begin{equation}\label{VV}
  \sum_{i=0}^D V^\dagger_{ik} V_{ij} = \delta_{kj} \oper .
\end{equation}
Now, using (\ref{Stine}) one gets

\begin{equation}\label{}
  \Phi(\rho) = \Tr_\mathcal{K} (\mathbb{V} \rho \otimes |\psi\>\<\psi| \mathbb{V}^\dagger) = \sum_{i,j=0}^D \sum_{k,l=0}^D V_{ij} \rho V^\dagger_{kl} \Tr(|i\>\<j|0\>\<0|l\>\<k|) = \sum_{i=0}^D V_{i0} \rho V_{i0}^\dagger ,
\end{equation}
and finally introducing a set of Kraus operators $K_i := V_{i0}$, one arrives at the following Kraus or operator sum representation \cite{Kraus-71,KRAUS}

\begin{equation}\label{Kraus}
  \Phi(\rho) = \sum_{i=0}^D K_{i} \rho K_{i}^\dagger ,
\end{equation}
where due to (\ref{VV}) Kraus operators satisfy

\begin{equation}\label{}
  \sum_{i=0}^D K_i^\dagger K_i =\oper .
\end{equation}

\subsection{Channel state duality}

The space of linear maps $\Phi : \LH \to \mathrm{L}(\HH')$ can be identified with $\mathrm{L}(\HH \otimes\HH')$. Indeed, they have the same dimension $(dd')^2$ and hence they are isomorphic. Of course one may design infinitely many isomorphisms between space of maps and $\LH \to \mathrm{L}(\HH')$ and space of operators $\mathrm{L}(\HH \otimes\HH')$. However, there is one special isomorphism which plays a distinguished role in what follows.  Fixing an orthonormal basis $\{|1\>,\ldots,|d\>\}$ in $\HH$ one defines so called Choi operator $\mathbf{C}_\Phi \in \mathrm{L}(\HH \otimes \mathcal{H}')$

\begin{equation}\label{CJ}
  \mathbf{C}_\Phi := ({\rm id} \otimes \Phi)(|\psi^+\>\<\psi^+|) =  \frac 1d \sum_{i,j=1}^d |i\>\<j| \otimes \Phi(|i\>\<j|) ,
\end{equation}
where $|\psi^+\> \in \HH \otimes \HH$ denotes maximally entangled vector $|\psi^+\> =  \sum_{i=1}^d |i \otimes i\>/\sqrt{d}$.  Conversely, given an operator $\mathbf{C} \in \mathrm{L}(\HH \otimes \mathcal{H}')$ one defines a linear map

\begin{equation}\label{}
  \Phi_\mathbf{C}(\rho) := d \, \Tr_{\HH'} (\mathbf{C} \rho^T \otimes \oper_{\HH'}) .
\end{equation}
The correspondence $\Phi \leftrightarrow \mathbf{C}$ is called {\em Choi-Jamio{\l}kowski isomorphism} \cite{Choi-75,Jam} (cf. \cite{Karol-d, KAROL,duality,duality-2} for more details). The most important property of the above correspondence was proved by Choi \cite{Choi-75}

\begin{Proposition} The map $\Phi$ is completely positive if and only if its Choi operator $\mathbf{C}_\Phi \geq 0$.
Moreover, it  defines a quantum channel (CPTP)  if additionally
\begin{equation}\label{CId}
  \Tr_{\HH'} \mathbf{C}_\Phi = \frac{\oper}{d} .
\end{equation}
\end{Proposition}
Indeed, using (\ref{CJ}) one finds

$$  \Tr_{\HH'} \mathbf{C}_\Phi = \frac{1}{d} \sum_{i,j=1} |i\>\<j| \Tr \Phi(|i\>\<j|) = \frac 1d \sum_{i=1}^d |i\>\<i| = \frac 1d \oper ,$$
due to $\Tr \Phi(|i\>\<j| = \Tr |i\>\<j| = \delta_{ij}$. Hence, there is 1-1 correspondence between quantum channels $\LH \to \mathrm{L}(\HH')$ and quantum states in $\HH \otimes \HH'$ satisfying (\ref{CId}). This is why it is often called {\em channel state duality} \cite{KAROL}.

It should be stressed that $\mathbf{C}_\Phi$ defined in (\ref{CJ}) does depend upon the basis $\{|1\>,\ldots,|d\>\}$. However, both Hermiticity $\mathbf{C}_\Phi^\dagger = \mathbf{C}_\Phi$ and positivity $\mathbf{C}_\Phi \geq 0$ are basis independent. Interestingly, a basis independent isomorphism was proposed by de Pillis \cite{Pillis}

\begin{equation}\label{Pillis}
  \mathbf{P}_\Phi =  \frac 1d \sum_{i,j=1}^d |i\>\<j| \otimes \Phi(|j\>\<i|) = (T \otimes {\rm id}) \mathbf{C}_\Phi .
\end{equation}
He showed \cite{Pillis} that $\Phi$ is Hermiticity preserving if and only if $\mathbf{P}_\Phi$ is a Hermitian operator (partial transposition preserves Hermiticity). Moreover, he observed that if $\mathbf{P}_\Phi  \geq 0$, then $\Phi$ is a positive map. Using (\ref{Pillis}) (and not (\ref{CJ})!) Jamio{\l}kowski proved \cite{Jam} that $\Phi$ is positive if and only if $\mathbf{P}_\Phi$ satisfies the following property

\begin{equation}\label{psi-psi}
  \< \psi \otimes \psi'| \mathbf{P}_\Phi |\psi \otimes \psi'\> \geq 0 ,
\end{equation}
for all $\psi \in \HH$ and $\psi' \in \HH'$. Actually, the above property does not depend whether one uses $\mathbf{P}_\Phi$ or $\mathbf{C}_\Phi$. Hence, both assignments could be used for characterizing positive maps in terms of operators in $\mathrm{L}(\HH \otimes \HH')$. However, concerning complete positivity (\ref{CJ}) is much more suitable. Similar correspondence was also considered by Arveson \cite{Arveson}. Interestingly, a Choi operator $\mathbf{C}_\Phi$ can be used to characterize $k$-positivity as well.

\begin{Proposition} The map $\Phi : \LH \to \mathrm{L}(\mathcal{H}')$ is $k$-positive if and only if
\begin{equation}\label{psi-psi-k}
  \< \Psi| \mathbf{C}_\Phi |\Psi\> \geq 0 ,
\end{equation}
for all $\Psi \in \HH \otimes \HH'$ such that the Schmidt rank of $\Psi$ is not larger than $k$.
\end{Proposition}
This property turns out to play a key role in entanglement theory (cf. \cite{TOPICAL} for more details).

\begin{tcolorbox}
Consider a linear  map $\Phi : \LH \to \mathrm{L}(\mathcal{H}')$.

\begin{itemize}

\item $\Phi$ preserves Hermiticity if and only if  $\mathbf{C}_\Phi$ is Hermitian (equivalently $\mathbf{P}_\Phi$ is Hermitian),

  \item $\Phi$ is positive if and only if $\< \psi \otimes \psi'|\mathbf{C}_\Phi| \psi \otimes \psi'\> = \< \psi^* \otimes \psi'|\mathbf{P}_\Phi| \psi^* \otimes \psi'\> \geq 0$ for any $\psi \in \HH$ and $\psi' \in \HH'$,

  \item $\Phi$ is $k$-positive if and only if  $\< \Psi| \mathbf{C}_\Phi |\Psi\> \geq 0$ for all $\Psi \in \HH \otimes \HH'$ with ${\rm SR}(\Psi) \leq k$,
  \item $\Phi$ is completely positive if and only if $\mathbf{C}_\Phi \geq 0$.
\end{itemize}

\end{tcolorbox}
The above discussion can be generalized for infinite dimensional case as well (cf. \cite{Stormer,Paulsen,Holevo-duality}). The original correspondence $\Phi \leftrightarrow \mathbf{C}_\Phi$ uses maximally entangled state which is not well defined in the infinite dimensional case. Note, however, that one may reformulate the isomorphism between maps and operators in $\mathrm{L}(\HH \otimes \HH')$ as follows: given a map $\Phi$ one defines a linear functional $\widetilde{\Phi} : \mathrm{L}(\HH \otimes \HH') \to \mathbb{C}$ via

\begin{equation}\label{}
  \widetilde{\Phi}(X \otimes Y) = {\rm Tr}(\Phi(X) Y^T)  .
\end{equation}
One proves \cite{Stormer,Paulsen} that

\begin{itemize}
  \item $\Phi$ is positive if and only if $\widetilde{\Phi}$ is positive on $\mathrm{L}_+(\HH) \otimes \mathrm{L}_+(\HH')$,
  \item $\Phi$ is completely positive if and only if $\widetilde{\Phi}$ is positive on $\mathrm{L}_+(\HH \otimes \HH')$,
\end{itemize}
where $\mathrm{L}_+(\HH)$ denotes (a convex cone) of positive operators in $\LH$. This construction may be generalized for infinite dimensional case.

\subsection{Vectorisation}

Channel state duality provides an elegant characterization of complete positivity of $\Phi$ in terms of the Choi operator $\mathbf{C}_\Phi$. However, this representation is not well suited when one composes maps. Having two maps $\Phi_1$ and $\Phi_2$ and the corresponding Choi operators $\mathbf{C}_1$ and $\mathbf{C}_2$ one may ask about Choi operator $\mathbf{C}$ for $\Phi_2 \circ \Phi_1$. It turns out that $\mathbf{C}$ is a quite nontrivial function of $\mathbf{C}_1$ and $\mathbf{C}_2$. To find a representation of $\Phi$ compatible with composition of maps let us apply well known  vectorisation procedure \cite{Watrous,Gilchrist}. It assigns to any operator $X \in \LH$ a vector $|X\>\!\> \in \mathcal{H} \otimes \mathcal{H}$

\begin{equation}\label{VEC}
  X \to |X\>\!\> =\sum_{i,j=1}^d X_{ij} |i \otimes j\> ,
\end{equation}
where $X_{ij} = \<i|X|j\>$. The vectorization of an $n \times n$ matrix $X$ is a column vector obtained by stacking the rows of the matrix $X$ on top of one another

\begin{equation}\label{}
  |X\>\!\> = [X_{11},\ldots,X_{1n},X_{21},\ldots,X_{2n},\ldots,X_{n1},\ldots,X_{nn}]^{\rm T} \ .
\end{equation}
Using this operation one may assign to any linear map $\Phi : \LH \to \mathrm{L}(\mathcal{H}')$ a super-operator $\widehat{\Phi} : \mathcal{H} \otimes \mathcal{H} \to \mathcal{H}' \otimes \mathcal{H}'$ defined by

\begin{equation}\label{}
  \widehat{\Phi}|X\>\!\> := |\Phi(X)\>\!\> .
\end{equation}
One easily finds \cite{Watrous,Gilchrist} that

\begin{equation}\label{}
  |A X B^\dagger\>\!\> = A \otimes {B^*}  |X\>\!\> ,
\end{equation}
for any $A,B : \HH \to \HH'$. This assignment enjoys a fundamental property: if $\Phi_1 : \mathrm{L}(\mathcal{H}_1) \to \mathrm{L}(\mathcal{H}_2)$ and $\Phi_2 : \mathrm{L}(\mathcal{H}_2) \to \mathrm{L}(\mathcal{H}_3)$, then

\begin{equation}\label{}
  \widehat{\Phi_2 \Phi}_1 = \widehat{\Phi}_2 \widehat{\Phi}_1 .
\end{equation}
Some authors use another convention and define

\begin{equation}\label{}
  X \to |X\>\!\>\!\> =\sum_{i,j=1}^d X_{ij} |j \otimes i\> ,
\end{equation}
that is,  vectorization of an $n \times n$ matrix $X$ is a column vector obtained by stacking the columns of the matrix $X$ on top of one another

\begin{equation}\label{}
  |X\>\!\>\!\> = [X_{11},\ldots,X_{n1},X_{12},\ldots,X_{n2},\ldots,X_{1n},\ldots,X_{nn}]^{\rm T} \ .
\end{equation}
One finds

\begin{equation}\label{}
  |A X B^\dagger\>\!\>\!\> = B^* \otimes {A}  |X\>\!\>\!\> .
\end{equation}
In what follows we use (\ref{VEC}).

\begin{tcolorbox}
\begin{Proposition} For a quantum channel having a Kraus representation $\Phi(\rho) = \sum_i K_i \rho K_i^\dagger$ one has
  \begin{equation}\label{}
  \widehat{\Phi} = \sum_i K_i \otimes K_i^* ,
\end{equation}
and
\begin{equation}\label{}
  \mathbf{C}_\Phi = \sum_i |K_i \>\!\> \<\!\< K_i | .
\end{equation}
\end{Proposition}
\end{tcolorbox}

\subsection{Bloch representation}

Bloch representation of qubit states enables one to treat quantum states as vectors in $\mathbb{R}^3$ and hence provides very intuitive picture of qubit states. Recall, that any qubit density operator may be represented via $ \rho = \frac 12 (\oper + \sum_{k=1}^3 x_k \sigma_k )$,
where $(\sigma_1,\sigma_3,\sigma_3)$ are Pauli matrices. Now,  $\rho \geq 0$ if and only if $|\mathbf{x}|\leq 1$, where $\mathbf{x}=(x_1,x_2,x_3)$. Consider a qubit map $\Phi$ and define $4 \times 4 $ real matrix

\begin{equation}\label{}
  \Phi_{\alpha\beta} = \frac 12 \Tr( \sigma_\alpha \Phi(\sigma_\beta)) ,\ \ \ \alpha,\beta=0,1,2,3,
\end{equation}
with $\sigma_0=\oper$. Trace-preservation implies that $\Phi_{\alpha\beta}$ has the following structure

\begin{equation}\label{Delta}
  \Phi_{\alpha\beta} = \left( \begin{array}{c|c} 1 & 0 \\ \hline  \mathbf{r} & \Delta \end{array} \right) ,
\end{equation}
{where $\mathbf{r} \in \mathbb{R}^3$, $\Delta$ is a $3\times 3$ real matrix}, and the map $\rho \to \Phi(\rho)$ in the Bloch representation is realized via the following affine transformation

\begin{equation}\label{affine}
  \mathbf{x} \to {\Delta} \mathbf{x} + \mathbf{r} .
\end{equation}
This representation can be immediately generalized for an arbitrary quantum map $\Phi : \LH \to \mathrm{L}(\mathcal{H}')$. Let $F_\alpha$ ($\alpha=0,1,\ldots,d^2-1$) be an orthonormal Hermitian basis in $\LH$, i.e. $\Tr(F_\alpha F_\beta) = \delta_{\alpha\beta}$, together with $F_0 = \oper/\sqrt{d}$. Similarly, let $F'_\mu$ ($\mu=0,1,\ldots,d'^2-1$) define orthonormal Hermitian basis in $\mathrm{L}(\HH')$ such that $F'_0 = \oper/\sqrt{d'}$. Any state in $\HH$ can be represented as follows

\begin{equation}\label{}
  \rho = \frac 1d \left(\oper + \sum_{\alpha=1}^{d^2-1} x_\alpha F_\alpha \right)
\end{equation}
where the generalized Bloch vector $\mathbf{x} =(x_1,\ldots,x_{d^2-1}) \in \mathbb{R}^{d^2-1}$. Let us define $d'^2 \times d^2$ real matrix

\begin{equation}\label{}
  \Phi_{\alpha\beta} =  \Tr( F'_\alpha \Phi(F_\beta)) .
\end{equation}
Again, $ \Phi_{\alpha\beta} $ has the same structure as (\ref{Delta}), where now $\Delta$ is a $(d'^2-1) \times (d^2-1)$ real matrix, and hence the generalized Bloch vector $\mathbf{x}$ transforms according to the affine transformation (\ref{affine}). Unfortunately, for $d>2$ the geometric structure of admissible Bloch vector is highly nontrivial (cf. \cite{KAROL,Beppe}). Bloch representation is well suited if one considers composition of channels: if $\Phi_1 \leftrightarrow \{\Delta_1,\mathbf{r}_1\}$ and  $\Phi_2 \leftrightarrow \{\Delta_2,\mathbf{r}_2\}$, then

$$   \Phi_2 \circ \Phi_1 \ \leftrightarrow\  \{ \Delta_2\, \Delta_1,\mathbf{r}_2 + \Delta_2 \mathbf{r}_1 \} . $$
Interestingly, there exists a simple relation between the super-operator $\widehat{\Phi}$ and the matrix $\Phi_{\alpha\beta}$:

%\begin{tcolorbox}
\begin{Proposition} For any map $\Phi : \LH \to \mathrm{L}(\mathcal{H}')$ one has

\begin{equation}\label{}
  \Phi_{\alpha\beta}= \<\!\< F'_\alpha|\widehat{\Phi}|F_\beta \>\!\> .
\end{equation}
\end{Proposition}
%\end{tcolorbox}

\subsection{Sudarshan $A$ and $B$ representations}

The problem of characterization of admissible maps transforming quantum states of $S$ into quantum states of $S'$ was initiated in a seminal paper of Sudarshan and collaborators \cite{Sudarshan-1} (see also \cite{Sudarshan-2,Sudarshan-3}). Since, the map is linear it is clear that it should be possible to find appropriate matrix representation. Authors of \cite{Sudarshan-1} proposed two representations:

\begin{equation}\label{}
  \rho'_{i'j'} = \sum_{i,j=1}^d \mathcal{A}_{i'j';ij} \rho_{ij} ,
\end{equation}
where $\mathcal{A}_{i'j';ij}$ is a $d'^2 \times d^2$ complex matrix.  Sudarshan $A$-matrix enjoys the following properties:

\begin{enumerate}
  \item $\mathcal{A}_{j'i';ji} = \mathcal{A}^*_{i'j';ij} \ \ \ (\mbox{Hermiticity-preservation})$,
  \item $\sum_{i'=1}^{d'} \mathcal{A}_{i'i';ij} = 1 \ \ \ (\mbox{trace-preservation})$ .
\end{enumerate}
However, is it not clear how to characterize positivity condition for the map $\Phi$ transforming $\rho \to \rho'=\Phi(\rho)$ in terms of $A$-matrix. To provide characterisation of positivity Sudarshan et al.considered  another matrix representation, so called $B$-matrix, defined as follows

\begin{equation}\label{}
  \mathcal{B}_{i'i;j'j} := \mathcal{A}_{i'j';ij} ,
\end{equation}
that is, $B$-matrix is related to $A$-matrix by a simple reshuffling of indices (cf. \cite{KAROL} for more details). Now, in terms of $B$-matrix the basic properties of the map $\Phi$ read as follows:

\begin{enumerate}
  \item $\mathcal{B}_{j'j;i'i} = \mathcal{B}^*_{i'i;j'j} \ \ \ (\mbox{Hermiticity-preservation})$,
  \item $\sum_{i'=1}^{d'} \mathcal{B}_{i'i;i'j} = \delta_{ij} \ \ \ (\mbox{trace-preservation})$.
\end{enumerate}
Note, that $B_{i'i;j'j}$ is a square $dd' \times dd'$ complex matrix and condition $\mathcal{B}_{j'j;i'i} = \mathcal{B}^*_{i'i;j'j}$ says that this matrix is Hermitian.

\begin{tcolorbox}
\begin{Proposition} One has the following correspondence
\begin{equation}\label{}
  \mathcal{A}_{i'j';ij} = \<i' \otimes j'|  \widehat{\Phi}| i \otimes j\> ,
\end{equation}
that is, Sudarshan $\mathcal{A}$-matrix is nothing but the matrix of the super-operator $\widehat{\Phi}$ in the basis $|i \otimes j\>$ in $\HH \otimes \HH$ and $|i' \otimes j'\>$ in $\HH' \otimes \HH'$. Moreover

\begin{equation}\label{}
  \mathcal{B}_{i'i;j'j} = \<i' \otimes i|  \mathbf{C}_\Phi | j' \otimes j\> ,
\end{equation}
that is, Sudarshan $\mathcal{B}$-matrix is nothing but the matrix of the  Choi operator $\mathbf{C}_{\Phi}$ in the basis $|i' \otimes i\>$ in $\HH' \otimes \HH$.
\end{Proposition}
\end{tcolorbox}
It is, therefore, clear, that Sudarshan contribution already provides all essential representations of quantum maps. How to characterize positivity of the original map? Sudarshan et al.came to the following natural conclusions: since preservation of Hermiticity by the map corresponds to Hermicity of the $B$-matrix it is natural to expect that positivity of the map corresponds to positivity of the $B$-matrix. To stress the fact that $B$-matrix encodes the fundamental property of the map, i.e. preservation of Hermiticity and positivity authors of \cite{Sudarshan-1} called $B$-matrix a {\em dynamical matrix}. As we shall see, however, positivity of $B$ is only sufficient but not necessary for positivity of $\Phi$. Interestingly, the authors of \cite{Sudarshan-1} derived operator sum representation for the map $\Phi$ ten years before Kraus without using any notion of complete positivity. Note, however, that positivity of $B$-matrix is equivalent to complete positivity of the original map $\Phi$ (cf. the recent review \cite{Jagadish} and \cite{40-GKLS}).

%\subsection{Superchannels}

%====================================================

\section{Mathematical structure of Markovian semigroups}   \label{SEC-IV}

%In this section we develop a general structure of Markovian semigroups. The central role is played by the property of {\em contractivity} w.r.t. %an appropriate norm. It gives rise to a key concept of {\em dissipative} generator. This structure is universal. It works in the classical case %(semigroups of stochastic matrices) and quantum case (semigroups of PTP and CPTP maps).

\subsection{Group of unitary evolutions}

Recall, that  Schr\"odinger evolution

\begin{equation}\label{}
  i|\dot{\psi}_t\> = H |{\psi}_t\> ,
\end{equation}
gives rise to a 1-parameter family of unitary operators $U_t = e^{-iHt}$ for any $t \in \mathbb{R}$. Clearly, they define a 1-parameter group

\begin{equation}\label{}
  U_t U_s = U_s U_t = U_{t+s} \ ,  \ \ U_0 = \oper .
\end{equation}
Similarly,  Schr\"odinger evolution of a density operator governed by the von Neumann equation

\begin{equation}\label{}
  \dot{\rho}_t = -i[H,\rho_t] ,
\end{equation}
is represented by a family of maps $\mathbb{U}_t : \LH \to \LH$ defined by

\begin{equation}\label{}
  \mathbb{U}_t(\rho) = U_t \rho U_t^\dagger .
\end{equation}
Again, a family of maps $\{\mathbb{U}_t\}_{t \in \mathbb{R}}$ defines a 1-parameter group

\begin{equation}\label{}
  \mathbb{U}_t \mathbb{U}_s = \mathbb{U}_s \mathbb{U}_t = \mathbb{U}_{t+s} \ ,  \ \ \mathbb{U}_0 = {\rm id} .
\end{equation}
Note, that $\mathbb{U}_t$ is CPTP for all $t \in \mathbb{R}$. Moreover the dual map $\mathbb{U}^\ddag_t$ satisfies

\begin{equation}\label{}
  \mathbb{U}_t^\ddag = \mathbb{U}_t^{-1} = \mathbb{U}_{-t} ,
\end{equation}
which shows that $\mathbb{U}_t$ is unitary w.r.t. Hilbert-Schmidt inner product. Unitary maps $\mathbb{U}_t$ satisfy

\begin{equation}\label{U11}
  \| \mathbb{U}_t(X) \|_1 = \|X\|_1
\end{equation}
for any $X \in \LH$. This follows form the unitary invariance of the trace norm $\| UXV\|_1 = \| X\|_1$ for any unitaries $U,V$ \cite{Matrices-2}.
In particular for any pair of initial density operators $\rho_1$ and $\rho_2$ one has

\begin{equation}\label{}
  \| \mathbb{U}_t(\rho_1 - \rho_2) \|_1 = \|\rho_1 - \rho_2\|_1 ,
\end{equation}
 which shows that the trace distance $\| \rho_1 - \rho_2\|_1$ does not change in time during the Schr\"odinger evolution.  Similarly, in the Heisenberg picture

\begin{equation}\label{U-op}
  \| \mathbb{U}^\ddag_t(X) \|_\infty = \| X \|_\infty  .
\end{equation}
Such unitary evolution is a special case of much more general scenario.

\subsection{Semigroups of positive trace-preserving maps}    \label{DIS}

Consider a family of linear maps $\Lambda_t : \LH \to \LH$ for $t \geq 0$.

\begin{tcolorbox}
\begin{Definition} The family $\{ \Lambda_t\}_{t\geq 0}$ defines a semigroup of positive trace-preserving maps if

\begin{enumerate}

\item $\Lambda_t$ is a positive trace-preserving map for all $t \geq 0$,

\item $\Lambda_t$ is a continuous function of $t$,

\item $\Lambda_{t=0} = {\rm id}$,

\item $\Lambda_t \Lambda_s = \Lambda_{t+s}$ for all $t,s \geq 0$,

\end{enumerate}

\end{Definition}
\end{tcolorbox}
Note, that equivalently we may call $\{ \Lambda_t\}_{t\geq 0}$ a semigroup of contractive maps in $\TTH$, that is,

\begin{enumerate}

\item $\Lambda_t$ is a contractive map for all $t \geq 0$

\begin{equation}\label{Lambda-1}
 \| \Lambda_t(X)\|_1 \leq \|X\|_1 ,
\end{equation}
for all Hermitian operators $X$, and

\item $\| \Lambda_t(X)\|_1 = \|X\|_1$ for all  $X\geq 0$.

\end{enumerate}
In the dual (Heisenberg) picture one has

\begin{tcolorbox}
\begin{Definition} The family $\{ \Lambda^\ddag_t\}_{t\geq 0}$ defines a semigroup of positive unital maps if

\begin{enumerate}

\item $\Lambda^\ddag_t$ is a positive unital map for all $t \geq 0$,

\item $\Lambda^\ddag_t$ is a continuous function of $t$,

\item $\Lambda^\ddag_{t=0} = {\rm id}$,

\item $\Lambda^\ddag_t \Lambda^\ddag_s = \Lambda^\ddag_{t+s}$ for all $t,s \geq 0$,

\end{enumerate}

\end{Definition}
\end{tcolorbox}
Such semigroup satisfies the following contractivity property

\begin{equation}\label{Lambda-op}
  \| \Lambda_t^\ddag(X)\|_\infty \leq \|X\|_\infty ,
\end{equation}
for all operators $X$. Note, that (\ref{U11}) and (\ref{U-op}) are special  cases of much more general properties (\ref{Lambda-1}) and (\ref{Lambda-op}), respectively. The price we pay for this generalization is that $\{\Lambda_t\}_{t \geq 0}$ is no longer 1-parameter group and it is defined only for $t \geq 0$. It does not mean that $\Lambda_t$ is not invertible (as a linear map). The inverse $\Lambda_t^{-1}$ does exist, however, it does not define a positive map and hence violates (\ref{Lambda-op}).

Actually a semigroup  of  positive trace-preserving maps is an example of a contraction semigroup acting on a Banach space $(\mathcal{B},\|\  \|)$, i.e. a vector space $\mathcal{B}$ equipped with a norm $\|X\|$ for $X \in \mathcal{B}$. In our case $\mathcal{B}$ is a Banach space of trace-class operators equipped with a trace norm. One calls $\Phi_t : \mathcal{B} \to \mathcal{B}$ a contraction semigroup if $\| \Phi_t(X)\| \leq \|X\|$ for any $X \in \mathcal{B}$ and for all $t \geq 0$. One proves \cite{Yosida,Engel,Reed} that $\{\Phi_t\}_{t\geq 0}$ satisfies

\begin{equation}\label{}
  \frac{d}{dt} \Phi_t = L \Phi_t , \ \ \ \ \Phi_0 = {\rm id} ,
\end{equation}
with the generator $L : \mathcal{B} \to \mathcal{B}$ defined by

\begin{equation}\label{}
  L := \frac{d}{dt} \Phi_t\Big|_{t=0} .
\end{equation}
The corresponding (unique) solution reads $\Phi_t = e^{L t}$. A generator of a contraction semigroup is called {\em dissipative} \cite{Lumer-Phillips}.

Consider now a PPT semigroup $\{\Lambda_t\}_{t \geq 0}$ in  $\TTH$. Denote by $\mathcal{L} : \LH \to \LH$, the corresponding generator. Introducing a time-dependent density operator $\rho_t = \Lambda_t(\rho)$ one obtains the following dynamical equation

\begin{equation}\label{}
  \frac{d}{dt}\rho_t = \mathcal{L}(\rho_t) \ , \ \ \ \rho_{t=0}=\rho .
\end{equation}
Note, that $\Lambda_t$ is trace-preserving if and only if $\mathcal{L}$ annihilates the trace
${\rm Tr}\mathcal{L}(X) =0$  for any $X$. Equivalently, a dual generator $\mathcal{L}^\ddag$ annihilates identity operator $\mathcal{L}^\ddag(\oper)=0$. Kossakowski  \cite{Kossakowski1972} found the following necessary and sufficient conditions for the generator of positive trace-preserving (contractive) semigroups:

%\begin{tcolorbox}
\begin{Theorem}  \label{TH-KOS} A linear map $\mathcal{L} : \TTH \to \TTH$ such that $\mathcal{L}$ annihilates the trace, generates a semigroup of positive trace-preserving maps $\{\Lambda_t\}_{t\geq 0}$  if and only if

\begin{equation}\label{QLP}
{\rm Tr}(Q \mathcal{L}(P))\geq 0 ,
\end{equation}
for any mutually orthogonal rank-1 projectors $P \perp Q$. Equivalently

\begin{equation}\label{PPP}
  P^\perp \mathcal{L}(P) P^\perp \geq 0,
\end{equation}
for any rank-1 projector $P$ ($P^\perp = \oper - P$ stands for a complementary projector).
\end{Theorem}
%\end{tcolorbox}
One calls \cite{Evans-1979,Evans-1977} a map $\mathcal{L} : \LH \to \LH$ satisfying (\ref{QLP}) {\em conditionally positive}. Hence $\mathcal{L}$ generates a semigroup of positive maps if and only if it is conditionally positive, or, equivalently dissipative.

\subsection{Classical stochastic semigroups}

A classical counterpart of a positive trace-preserving semigroup is provided by a
%\subsubsection{Schr\"odinger picture --- evolution of states}
a semigroup of $d\times d$ column stochastic matrices $\{ T(t) \}_{t\geq 0}$ satisfying   $ T(t+s) = T(t) T(s)$.
Recall that for any $t \geq 0$ one has

\begin{equation}\label{T-ell-1}
  \| T(t) \mathbf{x} \|_1 \leq \| \mathbf{x}\|_1 ,
\end{equation}
where $\| \mathbf{x}\|_1$ is $\ell_1$-norm defined in (\ref{ell-1}). A semigroup $T(t)$ satisfies
\begin{equation}\label{TLT}
  \frac{d}{dt} T(t) = L T(t) \ ;\ \ \ T(0) = \oper ,
\end{equation}
with a generator $L$ being $d\times d$  real matrix. Equivalently, one has the following equation for a state of the classical stochastic system represented by a probability vector $\mathbf{p} \in \mathbb{R}^d_+$

\begin{equation}\label{pLp}
  \frac{d}{dt} \mathbf{p}(t) = L \mathbf{p}(t) \ ,
\end{equation}
where $\mathbf{p}(t) = T(t)\mathbf{p}_0$. It is well known \cite{Kampen} that $L$ generates a semigroup of stochastic matrices if and only if

\begin{itemize}
  \item $\sum_i L_{ij}=0$,
  \item $L_{ij} \geq 0$, for $i\neq j$.
\end{itemize}
One often calls such $L$ a Kolmogorov generator \cite{Kampen}.  Any Kolmogorov generator  $L_{ij}$ can be represented via

\begin{equation}\label{Kolmogorov-L}
  L_{ij}= W_{ij} - \delta_{ij} W_j \ , \ \ \ W_j = \sum_{k=1}^d W_{kj} ,
\end{equation}
where $W_{ij} \geq 0$ for $i \neq j$ are interpreted as transition rates from a state `$j$' to `$i$'. If $\mathbf{p}_0 \in \mathcal{T}_d$ is an initial probability vector, then (\ref{pLp}) reproduces well know classical Pauli rate equation \cite{Kampen} for $\mathbf{p}(t) = T(t) \mathbf{p}_0$

%\begin{tcolorbox}
\begin{equation}\label{PAULI}
  \frac{dp_i(t)}{dt} = \sum_{j=1}^d \Big[ W_{ij} p_j(t) - W_{ji} p_i(t) \Big] \ , \ \ \ i=1,\ldots,d .
\end{equation}
%\end{tcolorbox}
Consider now a generator $\mathcal{L} : \LH \to \LH$ of a positive trace-preserving semigroup. Let $\{e_1,\ldots,e_d\}$ be an arbitrary orthonormal basis in $\HH$ and define

\begin{equation}\label{}
  L_{ij} = {\rm Tr}(P_i \mathcal{L}(P_j)) ,
\end{equation}
where $P_i = |e_i\>\<e_i|$. One has

$$  \sum_i  L_{ij} = {\rm Tr}\Big(\sum_i P_i \mathcal{L}(P_j) \Big) =   {\rm Tr}\mathcal{L}(P_j) = 0 , $$
since $\mathcal{L}$ annihilates the trace, and $\sum_i P_i = \oper$. Moreover for $ i \neq j$

$$   L_{ij} = {\rm Tr}(P_i \mathcal{L}(P_j)) \geq 0 ,   $$
since $P_i \perp P_j$. Hence, ${L}_{ij}$ defines a generator of a column stochastic semigroup.

\begin{Cor} A map $\mathcal{L} : \LH \to \LH$ defines  a generator of positive trace-preserving semigroup if and only if $L_{ij} = {\rm Tr}(P_i \mathcal{L}(P_j))$ defines a generator of a column stochastic semigroup for any  orthonormal basis in $\HH$.  It shows that $\mathcal{L}$ encodes information about infinitely many classical Kolmogorov generators. Let us recall that in the same way a density operator $\rho$ encodes information about infinitely many probability distributions $p_k = \<e_k|\rho |e_k\>$.
\end{Cor}

A similar analysis can be provided for a semigroup of row stochastic matrices. If $T(t)$ is a semigroup of column stochastic matrices, then $S(t) = T^{\rm T}(t)$ defines a semigroup of row stochastic matrices. In analogy with quantum mechanics we may call $T(t)$ a Schr\"odinger picture and $S(t)$ a Heisenberg  picture of the classical stochastic evolution. For any classical observable represented by a real vector  $\mathbf{x} \in \mathbf{R}^d$, and a probability vector  $\mathbf{p} \in \mathbb{R}^d_+$ one has

\begin{equation}\label{}
  (\mathbf{x},T(t) \mathbf{p}) = (T^{\rm T}(t)\mathbf{x},\mathbf{p}) ,
\end{equation}
that is, $\mathbf{x}(t) = T^{\rm T}(t) \mathbf{x}$ defines a (Heisenberg picture) evolution of $\mathbf{x}$. It is clear that

\begin{equation}\label{pLp-1}
  \frac{d}{dt} \mathbf{x}(t) = L^{\rm T} \mathbf{x}(t) \ ,
\end{equation}
or, in a more explicit form

\begin{equation}\label{PAULI-x}
  \frac{dx_i(t)}{dt} = \sum_{j=1}^d \Big[ W_{ji} x_j(t) - W_{ji} x_i(t) \Big] \ , \ \ \ i=1,\ldots,d .
\end{equation}

\subsection{From positive to completely positive semigroups --- GKS generator}

A semigroup of positive trace-preserving maps is uniquely characterized by the corresponding dissipative generator $\mathcal{L}$ satisfying (\ref{QLP}). This elegant condition is however not constructive, i.e. there is no general method which enables to construct such generators. Interestingly, one can find a general structure for generators if one considers a special subclass of positive trace-preserving semigroups.

\begin{Definition} Let $\{\Lambda_t\}_{t \geq 0}$ be a semigroup of positive trace-preserving maps. One calls $\{\Lambda_t\}_{t \geq 0}$

\begin{enumerate}
  \item a semigroup of $k$-positive trace-preserving maps if and only if $\Lambda_t$ is $k$-positive,
  \item a semigroup of completely positive trace-preserving maps if and only if $\Lambda_t$ is  completely positive.
\end{enumerate}
\end{Definition}
The corresponding generator of a $k$-positive semigroup is called $k$-dissipative, and that for completely positive semigroup is called {\em completely dissipative}.

To find a suitable representation of the generator $\mathcal{L}$ let us fix an orthonormal basis $\{F_\alpha\}_{\alpha=0}^{d^2-1}$ in $\LH$ such that $F_0 = \oper/\sqrt{d}$.  In particular ${\rm Tr}F_k=0$ for $k=1,\ldots,d^2-1$. Now, since $\mathcal{L}$ preserves Hermiticity one has

\begin{equation}\label{}
  \mathcal{L}(\rho) = \sum_{\alpha,\beta=0}^{d^2-1} C_{\alpha\beta} F_\alpha \rho F_\beta^\dagger ,
\end{equation}
with a Hermitian matrix $C_{\alpha\beta}$. Defining a Hermitian operator

\begin{equation}\label{}
  H = \frac{i}{2\sqrt{d}} \sum_{k=1}^{d^2-1} \Big( C_{k0} F_k - C_{0k} F^\dagger_k \Big) ,
\end{equation}
simple algebra gives rise to

%\begin{tcolorbox}

\begin{equation}\label{L-canonical}
  \mathcal{L}(\rho) = -i[H,\rho] + \sum_{k,l=1}^{d^2-1} C_{kl} \left( F_k \rho F_l^\dagger - \frac 12 \{ F_l^\dagger F_k,\rho\} \right) ,
\end{equation}
%\end{tcolorbox}
 where $\{A,B\} := AB+BA$. This is the canonical form of any $\mathcal{L}$ which preserves Hermiticity and the trace: $H$ is an arbitrary Hermitian operator, and $C_{kl}$ is an arbitrary $(d^2-1)\times (d^2-1)$ Hermitian matrix. The additional condition (\ref{QLP}) provides additional constraints for the matrix $C_{kl}$ which is however not constructive. The situation considerably simplifies if $\mathcal{L}$ is a generator of a completely positive trace-preserving semigroup. A seminal result of Gorini, Kossakowski and Sudarshan states \cite{GKS}

\begin{Theorem} \label{TH-GKS} $\mathcal{L}$ generates a semigroup of CPTP maps  if and only if the matrix $C_{kl}$ is positive definite.
\end{Theorem}
Equivalently,  the following map

\begin{equation}\label{}
  \Phi(\rho) = \sum_{k,l=1}^{d^2-1} C_{kl}(t)  F_k \rho F_l^\dagger ,
\end{equation}
is completely positive. Note, that $\sum_{k,l=1}^{d^2-1} C_{kl}(t)   F_l^\dagger F_k = \Phi^\ddag(\oper)$ and hence the canonical form can be rewritten in the compact form as follows

\begin{tcolorbox}
\begin{equation}\label{R1}
  \mathcal{L}(\rho) = - i[H,\rho] +  \Phi(\rho) - \frac 12 \{ \Phi^\ddag(\oper),\rho \}  ,
\end{equation}
and similarly for the dual generator (Heisenberg picture)

\begin{equation}\label{R2}
  \mathcal{L}^\ddag(X) = i[H,X] +  \Phi^\ddag(\rho) - \frac 12 \{ \Phi^\ddag(\oper),X\} .
\end{equation}
\end{tcolorbox}
 Note, however, that this representation in highly non unique. Indeed,  introducing a new map

\begin{equation}\label{}
  \Phi'(\rho) = \Phi(\rho) + \mathbf{K}\rho + \rho\, \mathbf{K}^\dagger ,
\end{equation}
with an arbitrary operator $\mathbf{K} = A + iB$ ($A,B$ Hermitian), one finds

\begin{equation}\label{}
  \mathcal{L}(\rho) = - i[H',\rho] + \Phi'(\rho) - \frac 12 \{ {\Phi'}^\ddag(\oper),\rho\} ,
\end{equation}
with $H'=H + B$. Since the matrix $C_{kl}$ is Hermitian it can be diagonalized by a unitary matrix $U_{mn}$, that is, $C_{kl} = \sum_{mn} U_{km} \gamma_m U^*_{lm}$, and hence it leads to the so-called diagonal representation

\begin{equation}\label{L-DIAG}
  \mathcal{L}(\rho) = -i[H,\rho] + \sum_{k=1}^{d^2-1} \gamma_k \left( L_k \rho L_k^\dagger - \frac 12 \{ L_k^\dagger L_k,\rho\} \right) ,
\end{equation}
with positive  rates $\gamma_k$, and $L_k = \sum_m U_{km}F_m$. Note, that new  operators $L_k$ are traceless and mutually orthogonal $\Tr (L^\dagger_k L_l) = \delta_{kl}$. The formula (\ref{L-DIAG}) is often presented in the following equivalent form

\begin{equation}\label{L-DIAGa}
  \mathcal{L}(\rho) = -i[H,\rho] + \frac 12 \sum_{k=1}^{d^2-1} \gamma_k \left( [L_k \rho, L_k^\dagger] + [L_k, \rho L_k^\dagger] \} \right) ,
\end{equation}
and when all $L_k$ are Hermitian, then

\begin{equation}\label{L-DIAGb}
  \mathcal{L}(\rho) = -i[H,\rho] - \frac 12 \sum_{k=1}^{d^2-1} \gamma_k  [L_k,[L_k, \rho]] .
\end{equation}
In analogy to (\ref{PPP}) Theorem \ref{TH-GKS} can be reformulated as follows \cite{Evans-1977}

%\begin{tcolorbox}
\begin{Proposition} $\mathcal{L}$ generates a semigroup of completely positive maps if and only if

\begin{equation}\label{CCP}
 (\oper -  P^+_d) [{\rm id}_d \otimes \mathcal{L}](P^+_d) (\oper - P^+_d) \geq 0,
\end{equation}
where $P^+_d$ stands for maximally entangled projector in $\mathbb{C}^d \otimes \mathbb{C}^d$.
\end{Proposition}
%\end{tcolorbox}
One calls such $\mathcal{L}$  conditionally completely positive \cite{Evans-1977}. To illustrate the difference between generators of positive and completely positive semigroups let us consider the following

\begin{Example} Consider a qubit generator (already in the diagonal canonical form)

\begin{equation}\label{Pauli}
  \mathcal{L}(\rho) = \frac 12 \sum_{k=1}^3 \gamma_k( \sigma_k \rho \sigma_k - \rho) ,
\end{equation}
with $\gamma_k \in \mathbb{R}$. Clearly, $\mathcal{L}$ generates CPTP evolution if and only if all $\gamma_k \geq 0$. Now, $\mathcal{L}$ generates PTP evolution if and only if condition (\ref{PPP}) holds, with $\Phi(\rho) = \frac 12 \sum_{k=1}^3 \gamma_k \sigma_k \rho \sigma_k$. Taking the following rank-1 projectors:

$$   P=  \begin{bmatrix}      1 & 0 \\
                              0 & 0
                            \end{bmatrix}  , \ \ P' = \frac 12 \begin{bmatrix}
                              1 & 1 \\
                              1 & 1
                            \end{bmatrix} \ , \ \ P'' = \frac 12 \begin{bmatrix}
                              1 & -i \\
                              i & 1
                            \end{bmatrix} , $$
one obtains
\begin{equation}\label{}
  {\rm Tr}[(\oper-P) \Phi(P)] = \gamma_1 + \gamma_2 \ , \    {\rm Tr}[(\oper-P') \Phi(P')] = \gamma_2 + \gamma_3  \ , \  {\rm Tr}[(\oper-P'') \Phi(P'')] = \gamma_1 + \gamma_3 .
\end{equation}
%and hence $\gamma_1 + \gamma_2 \geq 0$. Similarly one finds $\gamma_1 + \gamma_3 \geq 0$ and $\gamma_2 + \gamma_3 \geq 0$.
Interestingly, these conditions are also sufficient for $\mathcal{L}$ to generate a positive trace-preserving semigroup.  It should be stressed that the map $\Phi(\rho) = \sum_{k=1}^3 \gamma_k\sigma_k \rho \sigma_k$ is completely positive if and only if all $\gamma_k \geq 0$. However, it is never positive whenever one of $\gamma_k$ is negative. Nevertheless, whenever

\begin{equation}\label{}
  \gamma_i + \gamma_j \geq \  0 , \ \ \ i \neq j ,
\end{equation}
then $e^{t \mathcal{L}}$ defines a family of positive maps.

\end{Example}

\begin{Example}  Consider the qubit evolution governed by following generator

\begin{equation}\label{L+-}
  \mathcal{L}(\rho) = - \frac{\omega}{2} [\sigma_z,\rho] + \frac{\gamma_+}{2} \mathcal{L}_+(\rho) + \frac{\gamma_-}{2} \mathcal{L}_-(\rho) + \frac{\gamma_z}{2} \mathcal{L}_z(\rho) ,
\end{equation}
where

\begin{equation}\label{}
\mathcal{L}_\pm(\rho) = \sigma_\pm \rho \sigma_\mp - \frac 12 \{ \sigma_\mp\sigma_\pm,\rho\} \ , \ \ \mathcal{L}_z(\rho) = \sigma_z\rho \sigma_z -\rho ,
\end{equation}
and $\sigma_{\pm}=\frac{1}{2}(\sigma_{x}\pm i\sigma_{y})$.  This generator is already in the diagonal form. It generates completely positive evolution if and only if $\gamma_\pm \geq 0$ and $\gamma_z \geq 0$. The evolution of the density operator is represented as follows \cite{ALICKI}

\begin{equation}\label{}
 \rho = \left( \begin{array}{cc} \rho_{00} & \rho_{01} \\ \rho_{10} & \rho_{11} \end{array} \right)  \ \to \ \Lambda_t(\rho)  =
 \left( \begin{array}{cc} 1-P_e(t) & C(t) \rho_{01} \\ C^*(t) \rho_{10} & P_e(t) \end{array} \right) ,
\end{equation}
where

\begin{equation}\label{}
  P_e(t) = e^{- \frac 12 (\gamma_+ +\gamma_-)t} ( G(t) + P_e(0) ) , \ \ \ C(t) = e^{i (\omega - (\gamma_+ +\gamma_-)/4 - \gamma_z)t} ,
\end{equation}
together with

\begin{equation}\label{}
  %\gamma = \frac 12 (\gamma_+ +\gamma_-) \ , \ \ \
  G(t) = \frac{\gamma_+}{\gamma_++\gamma_-} \left( 1-e^{- \frac 12 (\gamma_+ +\gamma_-)t} \right) .
\end{equation}
Interestingly, positive evolution does not require positivity of $\gamma_z$. One finds \cite{Sergey} the following necessary and sufficient conditions for $\mathcal{L}$:

\begin{equation}\label{}
\gamma_\pm \geq 0 , \ \ \ \sqrt{\gamma_+\gamma_-} + 2 \gamma_z \geq 0 .
\end{equation}
This clearly shows that complete positivity is much more restrictive than positivity.
\end{Example}

\subsection{Quantum semigroup in the Heisenberg picture -- Lindblad approach}

In the Heisenberg picture the characterization of generator is essentially the same as in the Schr\"odinger picture. Consider a linear map $\mathcal{L}^\ddag : \BH \to \BH$ such that $\mathcal{L}^\ddag(\oper) =0$.

%\begin{tcolorbox}

\begin{itemize}
  \item $\mathcal{L}^\ddag$ generates a semigroup of positive unital maps if and only if $P^\perp \mathcal{L}^\ddag(P) P^\perp \geq 0$ for any rank-1 projector $P$ in $\mathcal{H}$,

%  \item $\mathcal{L}^\ddag$ generates a semigroup of $k$-positive unital maps if and only if $P^\perp ({\rm id}_k \otimes \mathcal{L}^\ddag(P)] P^\perp %\geq 0$ for any rank-1 projector $P$ in $\mathbb{C}^k \otimes \mathcal{H}$,

  \item $\mathcal{L}^\ddag$ generates a semigroup of completely positive unital (CPU) maps if and only if $P_+^\perp ({\rm id}_d \otimes \mathcal{L}^\ddag(P_+)] P_+^\perp \geq 0$, where $P_+$ denotes a maximally entangled projector in $\HH \otimes \HH$.
\end{itemize}
%\end{tcolorbox}
Interestingly in the Heisenberg picture one can exploit another family of semigroups satisfying  Schwarz inequality.   Let us recall that a unital map $\Phi : \LH \to \LH$ satisfies Schwarz inequality \cite{Paulsen,Stormer} if

\begin{equation}\label{KS}
  \Phi(X^\dagger X) \geq \Phi(X^\dagger)\Phi(X) ,
\end{equation}
for all $X \in \LH$.  Any unital Schwarz map, i.e. a map satisfying (\ref{KS}), is evidently positive. However, the converse needs not be true (a counter example is provided e.g. by transposition). It was shown by Kadison \cite{Kadison1,Paulsen,Stormer} that any positive unital map satisfies (\ref{KS}) for Hermitian $X$, that is,

\begin{equation}\label{KS-H}
  \Phi(X^2) \geq \Phi(X)\Phi(X) ,
\end{equation}
for all $X = X^\dagger \in \LH$. This observation was then further generalized by Choi \cite{Choi-KS1,Choi-KS2} who proved that for any positive unital map one has

\begin{equation}\label{}
   \Phi(X^\dagger X) \geq \Phi(X^\dagger)\Phi(X) \  , \ \ \  \Phi(X^\dagger X) \geq \Phi(X)\Phi(X^\dagger) ,
\end{equation}
for all normal operators.

Let $\mathcal{S}_k$ denote a set of unital maps $\Phi : \LH \to \LH$ such that  ${\rm id}_k \otimes \Phi$ is a Schwarz map. One has the following hierarchy of inclusions

%\begin{tcolorbox}
\begin{equation}\label{KS-hierarchy}
   \mathcal{S}_1 \supset \mathcal{S}_2 \supset \ldots \supset  \mathcal{S}_d = \mbox{CPU maps} .
\end{equation}
%\end{tcolorbox}
Denote by $\mathcal{P}_k^U$ a set of unital $k$-positive maps. One may prove that

\begin{equation}\label{}
  \mathcal{P}_k^U \supset \mathcal{S}_k \supset \mathcal{P}_{k+1}^U ,
\end{equation}
which generalizes the well known result $\mathcal{P}_1^U \supset \mathcal{S}_1 \supset \mathcal{P}_2^U$. Hence merging (\ref{k-hierarchy}) and (\ref{KS-hierarchy})  one arrives at the following refined hierarchy

%\begin{tcolorbox}
\begin{equation}\label{KS-hierarchy-1}
  \mathcal{P}_1^U \supset \mathcal{S}_1 \supset \mathcal{P}_2^U \supset \ldots \supset \mathcal{P}_d^U = \mathcal{S}_d = \mbox{CPU maps} .
\end{equation}
%\end{tcolorbox}
In particular, Schwarz maps interpolate between positive and 2-positive unital maps.

In his seminal paper \cite{L} Lindblad analyzed the structure of semigroups $\{\Lambda_t^\ddag\}_{t \geq 0}$ consisting of maps satisfying (\ref{KS}).

% belonging to $\mathcal{S}_k$ for $k=1,2,\ldots$.

%\begin{tcolorbox}
\begin{Theorem} A unital semigroup $\Lambda^\ddag_t = e^{\mathcal{L}^\ddag t}$ satisfies (\ref{KS}) for all $t \geq 0$ if and only if $\mathcal{L}^\ddag(\oper)=0$ and

\begin{equation}\label{KS-L}
  \mathcal{L}^\ddag(X^\dagger X) \geq \mathcal{L}^\ddag(X^\dagger) X + X^\dagger \mathcal{L}^\ddag(X) ,
\end{equation}
for all $X \in \LH$.
\end{Theorem}
%\end{tcolorbox}
Indeed,  the proof immediately  follows from the Schwarz inequality

\begin{equation}\label{}
  e^{\mathcal{L}^\ddag t}(X^\dagger X) \geq  e^{\mathcal{L}^\ddag t}(X^\dagger)   e^{\mathcal{L}^\ddag t}(X) .
\end{equation}
Time derivatives at $t=0$ implies (\ref{KS-L}). Lindblad \cite{L} called $\mathcal{L}^\ddag$ satisfying (\ref{KS-L}) a dissipative generator. It should be stressed that this is a different notion of dissipativity that the one discussed in Section \ref{DIS}. Actually,
if one restricts (\ref{KS-L}) to Hermitian operators one arrives at the following

%\begin{tcolorbox}
\begin{Cor}  $\mathcal{L}^\ddag$ generates a semigroup $\{\Lambda_t^\ddag\}_{t\geq 0}$ of positive unital maps if and only if $\mathcal{L}^\ddag(\oper)=0$ and

\begin{equation}\label{KS-LH}
  \mathcal{L}^\ddag(X^2) \geq \mathcal{L}^\ddag(X) X + X \mathcal{L}^\ddag(X) ,
\end{equation}
for all $X =X^\dagger \in \LH$.
\end{Cor}
%\end{tcolorbox}

%To summarise, $e^{t \mathcal{L}^\ddag}$ is a semigroup of positive unital maps if $\mathcal{L}^\ddag$ satisfies (\ref{KS-LH}) and it satisfies %Schwarz inequality if additionally $\mathcal{L}^\ddag$ satisfies (\ref{KS-L}).

%To link this concept of dissipativity to the one considered in section \ref{SEC-diss} let $\varrho > 0$ be an invariant state for $\mathcal{L}$, %that is, $\mathcal{L}(\varrho)=0$. One obviously has ${\rm Tr}(\varrho \mathcal{L}^\ddag(X^\dagger X)) = {\rm Tr}(\mathcal{L}(\varrho) X^\dagger %X)=0$, and hence

%\begin{equation}\label{}
%  0 \geq {\rm Tr}(\varrho \mathcal{L}^\ddag(X^\dagger) X) + {\rm Tr}(\varrho X^\dagger \mathcal{L}^\ddag(X)) ,
%\end{equation}
%which can be rewritten as follows

%\begin{equation}\label{DISS-I}
%  {\rm Re}\, (X,\mathcal{L}^\ddag(X))_\varrho \leq 0 ,
%\end{equation}
%where we introduced the following inner product

%\begin{equation}\label{rho-product}
%  (X,\mathcal{L}^\ddag(X))_\varrho := {\rm Tr}(\varrho X^\dagger Y) ,
%\end{equation}
%for any $X,Y\in \LH$. Note, that condition (\ref{DISS-I}) is a special case of the general dissipativity condition (\ref{zLz}).

%It  implies the following dissipativity condition

%\begin{equation}\label{DISS-II}
%   (X,\mathcal{L}^\ddag(X))_\varrho \leq 0 ,
%\end{equation}
%for all  $X =X^\dagger \in \LH$. Note, that condition (\ref{DISS-II}) is a special case of the general dissipativity condition (\ref{Diss-1}).
%Evidently, condition (\ref{DISS-I}) is much stronger than (\ref{DISS-II}).

Interestingly, a generator of positive semigroup can be equivalently characterized by the following

\begin{Proposition}\cite{Evans-1979} $\mathcal{L} : \TTH \to \TTH$ with $\mathcal{L}^\ddag(\oper)=0$ generates a semigroup of positive trace-preserving maps if and only if
\begin{equation}\label{KS-U}
   \mathcal{L}^\ddag(U^\dagger) U + U^\dagger \mathcal{L}^\ddag(U) \leq 0 ,
\end{equation}
for all unitary operators $U$ in $\LH$.
\end{Proposition}

\begin{Example} Consider once again the qubit generator (\ref{Pauli}) for which one has $\mathcal{L}^\ddag = \mathcal{L}$. One easily computes

\begin{equation}\label{}
  \mathcal{L}(\sigma_1) = -[\gamma_2 + \gamma_3]\sigma_1 \ , \ \  \mathcal{L}(\sigma_2) = -[\gamma_1 + \gamma_3]\sigma_3 \ , \ \  \mathcal{L}(\sigma_3) = -[\gamma_1 + \gamma_2]\sigma_3 \ ,
\end{equation}
and hence condition (\ref{KS-U}) reproduces $\gamma_i + \gamma_j \geq 0$ for $i\neq j$.

\end{Example}

\subsection{Quantum detailed balance}

Consider a semigroup of column stochastic matrices $\{T(t)\}_{t \geq 0}$. It possesses at least one stationary state

 \begin{equation}\label{}
  T(t) \mathbf{p}^{\rm ss} = \mathbf{p}^{\rm ss} ,
\end{equation}
for all $t \geq 0$. The corresponding generator

$$ L_{ij} = W_{ij} - \delta_{ij} \sum_{k=1}^d W_{kj} , $$
with $W_{ij} \geq 0$ for $i\neq j$,  satisfies detailed balance w.r.t. $\mathbf{p}^{\rm ss}$ if \cite{Kampen}

\begin{equation}\label{CDB}
  W_{ij} p^{\rm ss}_j = W_{ji} p^{\rm ss}_i ,
\end{equation}
for all $i,j=1,\ldots,d$. Interestingly, the above condition may be reformulated as follows

\begin{Proposition} $L$ satisfies detailed balance w.r.t. $\mathbf{p}^{\rm ss}$ if only if

\begin{equation}\label{}
  (L^{\rm T}\mathbf{x},\mathbf{y})_{\rm ss} =  (\mathbf{x},L^{\rm T}\mathbf{y})_{\rm ss} ,
\end{equation}
where the inner product $(\mathbf{x},\mathbf{y})_{\rm ss}$ is defined as follows

\begin{equation}\label{()p}
  (\mathbf{x},\mathbf{y})_{\rm ss} := \sum_{i=1}^d p_i^{\rm ss} x_i y_i ,
\end{equation}
for any $\mathbf{x},\mathbf{y} \in \mathbb{R}^d$.
\end{Proposition}
The above result shows that $L$ satisfies detailed balance w.r.t. $\mathbf{p}^{\rm ss}$ if $L^{\rm T}$ (i.e. the dual generator) is Hermitian w.r.t. inner product  (\ref{()p}). This property may be considered as an equivalent definition of the classical detailed balance.

Consider now a quantum counterpart, i.e. a  semigroup of CPTP maps $\{\Lambda_t\}_{t\geq 0}$. Similarly to the classical scenario it possesses at least one  steady state

\begin{equation}\label{}
  \Lambda_t(\rho_{\rm ss}) = \rho_{\rm ss} ,
\end{equation}
for all $t \geq 0$. Let us assume that $\rho_{\rm ss} > 0$ and introduce a quantum analog of (\ref{()p})

\begin{equation}\label{rho-product}
  (X,Y)_{\rm ss} := {\rm Tr}(\rho_{\rm ss} X^\dagger Y) ,
\end{equation}
for any $X,Y\in \LH$. The quantum analog of the detailed balance condition is provided by the following definition \cite{Alicki-DB,Gorini-DB,Gorini-DB2,ALICKI,Fagnola}

\begin{tcolorbox}
\begin{Definition} A GKLS generator $\mathcal{L}$ satisfies the detailed balance condition with respect to $\rho_{\rm ss}$ iff

\begin{itemize}
  \item $\mathcal{L}(\rho_{\rm ss}) = 0$,

\item there exists a Lindblad representation $\mathcal{L}^\ddag(X) = i[H,X] + \Phi^\ddag(X) - \frac 12 \{\Phi^\ddag(\oper),X\}$ such that

\begin{equation}\label{rho-dual}
  (X,\Phi^\ddag(Y))_{\rm ss} = (\Phi^\ddag(X),Y)_{\rm ss} ,
\end{equation}
for all $X,Y \in \LH$.
\end{itemize}

\end{Definition}
\end{tcolorbox}
Condition (\ref{rho-dual}) shows that the map $\Phi^\ddag$ is self-dual with respect to $(\ , \ )_{\rm ss}$. For any map $\Psi : \LH \to \LH$ one defines its dual $\widetilde{\Psi}$ w.r.t. (\ref{rho-product}) via

\begin{equation}\label{}
  (\widetilde{\Psi}(X),Y)_{\rm ss} := (X,\Psi(Y))_{\rm ss} ,
\end{equation}
 and easily finds
\begin{equation}\label{tilde}
  \widetilde{\Psi}(X) = \Psi^\ddag(X \rho_{\rm ss})\rho_{\rm ss}^{-1} .
\end{equation}
Clearly, if $\rho_{\rm ss}=\oper/d$, one has $\widetilde{\Psi} = \Psi^\ddag$. However, in general $\widetilde{\Psi}$ and $\Psi^\ddag$ are different maps. The quantum detailed balance implies

\begin{equation}\label{}
  \Phi^\ddag(X) = \widetilde{\Phi^\ddag}(X) = \Phi(X \rho_{\rm ss}) \rho_{\rm ss}^{-1} ,
\end{equation}
or, equivalently

\begin{equation}\label{}
  \Phi^\ddag(X) \rho_{\rm ss} =  \Phi(X \rho_{\rm ss}) .
\end{equation}
If $\mathcal{L}$ satisfies quantum detailed balance condition w.r.t. $\rho_{\rm ss}$, then

\begin{equation}\label{}
\mathcal{L}_0^\ddag(X) := \frac 12 \left(  \mathcal{L}^\ddag(X) - \widetilde{\mathcal{L}^\ddag}(X) \right)  = i[H,X] ,
\end{equation}
for some $H^\dagger=H \in \LH$. Moreover

\begin{equation}\label{}
  \mathcal{D}^\ddag := \frac 12 \left( \mathcal{L}^\ddag + \widetilde{\mathcal{L}^\ddag} \right) ,
\end{equation}
satisfies $\widetilde{\mathcal{D}^\ddag} = \mathcal{D}^\ddag$.

Let us define a linear map $\Delta : \LH \to \LH$ by the following formula

\begin{equation}\label{}
  \Delta(X) = \rho_{\rm ss} X \rho_{\rm ss}^{-1} .
\end{equation}
One has the following
\begin{Proposition}  If $\mathcal{L}$ satisfies quantum detailed balance condition w.r.t. $ \rho_{\rm ss}$, then

\begin{equation}\label{commutative}
  [\mathcal{L}^\ddag,\Delta]=  [\mathcal{L},\Delta] = 0 .
\end{equation}
\end{Proposition}
Indeed, $\widetilde{\mathcal{L}^\ddag}$ is a generator of CP semigroup and hence preserves Hermiticity. One has, therefore,

\begin{equation}\label{}
  \widetilde{\mathcal{L}^\ddag}(X) = \mathcal{L}(X \rho_{\rm ss})\rho_{\rm ss}^{-1} \ , \ \ \   \widetilde{\mathcal{L}^\ddag}(X) =
  \rho_{\rm ss}^{-1} \mathcal{L}(\rho_{\rm ss}X) ,
\end{equation}
and hence

\begin{equation}\label{}
  \widetilde{\mathcal{L}^\ddag}(\rho_{\rm ss} X\rho_{\rm ss}^{-1}) = \mathcal{L}(\rho_{\rm ss}X )\rho_{\rm ss}^{-1} =  \rho_{\rm ss}\widetilde{\mathcal{L}^\ddag}( X)\rho_{\rm ss}^{-1} .
\end{equation}

%In particular one has $[H,\varrho]=[\Phi^\ddag(\oper),\varrho]=0$ and hence $\mathcal{D}(\varrho)=0$.

Let $\rho_{\rm ss} = \sum_k p^{\rm ss}_k |k\>\<k|$  be a spectral decomposition of $\rho_{\rm ss}$. Let us assume that the spectrum of $\rho_{\rm ss}$ is not degenerate.  The quantum detailed balance condition implies $[H,\rho_{\rm ss}]=0$ and $\mathcal{D}(\rho_{\rm ss})=0$. One finds the following characterization of the dissipative part \cite{Gorini-DB2}

\begin{equation}\label{}
  \mathcal{D}^\ddag(X) = \sum_{i,i'}\sum_{j,j'} C_{ii',jj'} \Big( |i\>\<i'|X|j'\>\<j| - \frac 12 \delta_{i'j'} \{ |i\>\<j|,X\} \Big) ,
\end{equation}
where the matrix $C_{ii',jj'}$ is positive definite and satisfies the following condition

\begin{equation}\label{}
  C_{ii',jj'} p^{\rm ss}_j = C_{j'j,i'i} p^{\rm ss}_{j'} .
\end{equation}

%\begin{Proposition}  If $\mathcal{L}$ satisfies quantum detailed balance condition w.r.t. $\rho_{\rm ss}$, then

%\begin{equation}\label{commutative}
%  [\mathcal{L}^\ddag,\Delta]=  [\mathcal{L}^\ddag_0,\Delta]=  [\mathcal{D}^\ddag,\Delta] = [\Phi^\ddag,\Delta]= 0 ,
%\end{equation}
%where  $\Delta : \LH \to \LH$ is defined via

%\begin{equation}\label{}
%  \Delta(X) := \rho_{\rm ss} X \rho_{\rm ss}^{-1}  .
%\end{equation}
%In particular one has $[H,\rho_{\rm ss}]=[\Phi^\ddag(\oper),\rho_{\rm ss}]=0$ and hence $\mathcal{D}(\rho_{\rm ss})=0$.
%\end{Proposition}
%Commutativity conditions (\ref{commutative}) have  interesting implications  \cite{Fagnola}

Finally, let us link the quantum detailed balance condition to the classical one (\ref{CDB}). Assuming $\rho_{\rm ss} = \sum_{k=1}^d p_k^{\rm ss} P_k $ (with $P_k= |k\>\<k|$)  let us define a classical Kolmogorov generator $L_{ij} = {\rm Tr}(P_i \mathcal{L}(P_j) )$. One has for $i \neq j$

\begin{equation}\label{}
  W_{ij} = {\rm Tr}(P_i \mathcal{L}(P_j) ) = {\rm Tr}(P_i \Phi(P_j) ) = {\rm Tr}(\Phi^\ddag(P_i) P_j ) = {\rm Tr}(\Phi(P_i \rho_{\rm ss})\rho_{\rm ss}^{-1} P_j ) ,
\end{equation}
where we used $\Phi^\ddag(X) = \widetilde{\Phi^\ddag}(X) = \Phi(X\rho_{\rm ss})\rho_{\rm ss}^{-1}$. Finally using $P_i \rho_{\rm ss} = p_i^{\rm ss} P_i$, one obtains

\begin{equation}\label{}
  W_{ij} = \frac{p^{\rm ss}_i}{p^{\rm ss}_j} {\rm Tr}(\Phi(P_i) P_j ) =  \frac{p^{\rm ss}_i}{p^{\rm ss}_j}  W_{ji} ,
\end{equation}
which recovers a classical condition (\ref{CDB}).

Assuming that $\rho_{\rm ss} = \sum_{k=1}^d p_k^{\rm ss} |k\>\<k|$ has a non-degenerate spectrum the structure of detailed balance generator reads as follows

\begin{equation}\label{}
  \mathcal{L}(\rho) = -i[H,\rho] + \sum_{k\neq l} W_{kl} \Big( |k\>\<l| \rho |l\>\<k| - \frac 12 \{ |l\>\<l|,\rho\} \Big)  + \sum_{k,l} D_{kl} \Big(|k\>\<k| \rho |l\>\<l| - \delta_{kl}    \{|l\>\<l|,\rho\} \Big) ,
\end{equation}
where $  W_{ij}{p^{\rm ss}_j} = W_{ji}{p^{\rm ss}_i}$, $H =  \sum_{k=1}^d E_k |k\>\<k|$, and $[D_{kl}]$ is a positive definite matrix (actually, it is assumed that no `accidental' degeneracies can occur, i.e. $p^{\rm ss}_i/p^{\rm ss}_j \neq p^{\rm ss}_k/p^{\rm ss}_l$. For the most general structure of GKLS generator satisfying quantum detailed balance cf. \cite{Alicki-DB}.

\subsection{Spectral properties  and relaxation rates}

%\subsubsection{Spectra of generators}

Consider a semigroup of CPTP maps $\{\Lambda_t\}_{t\geq 0}$. It possesses at least one stationary state (or steady state) $  \Lambda_t(\rho_{\rm ss}) = \rho_{\rm ss}$ for all $t \geq 0$. It turns out that a set of fixed points of the map, i.e. operators $X \in \LH$ satisfying $\Lambda_t(X)=X$, is spanned by stationary states. Let $\mathcal{L} : \LH \to \LH$ be the corresponding generator and let $\ell_\alpha$ be  eigenvalues of $\mathcal{L}$

\begin{equation}\label{ell}
  \mathcal{L}(X_\alpha) = \ell_\alpha X_\alpha ,
\end{equation}
for $\alpha=0,1,\ldots,d^2-1$.  Clearly, time-dependent eigenvalues of $\Lambda_t$ read $\lambda_\alpha(t) = e^{t\ell_\alpha}$.
Spectral properties of $\mathcal{L}$ are summarized as follows

%Since $\mathcal{L}$ does preserve Hermiticity it implies that $\mathcal{L}(X^\dagger_\alpha) = \ell^*_\alpha X_\alpha^\dagger$, that is, if %$\ell_\alpha$ is complex, then $\ell^*_\alpha$ is also an eigenvalue.

%\begin{tcolorbox}

\begin{enumerate}
\item the spectrum is symmetric w.r.t. real line,
  \item all eigenvalues $\ell_\alpha$ belong to the left half-plane $\mathbb{C}_- = \{ z\in \mathbb{C}\, |\, {\rm Re}\, z \leq 0\}$,
  \item $\ell_0=0$  and the corresponding eigenvector $X_0\geq 0$ and hence $\rho_{\rm ss} := X_0/\Tr X_0$ defines a steady state. If $\ell_0$ is a simple eigenvalue (not degenerated), then $\rho_{\rm ss} > 0$ and the corresponding semigroup is relaxing, i.e. for any initial state $\rho$ one has $e^{\mathcal{L} t}(\rho) \to \rho_{\rm ss}$ as $t \to \infty$.
 % \item `peripheral' eigenvalues (i.e. ${\rm Re}\, \ell_k =0$) are semisimple (the corresponding Jordan blocks are 1-dimensional),
\end{enumerate}

%\end{tcolorbox}
The fact that the spectrum is symmetric w.r.t. real line follows from Hermiticity-preserving property, that is, $\mathcal{L}(X^\dagger) = [\mathcal{L}(X)]^\dagger$. Hence, if $\mathcal{L}(X) = \ell X$ then  $\mathcal{L}(X^\dagger) = \ell^* X^\dagger$. Note, that the condition ${\rm Re}\ell \leq 0$ protects the spectrum of the quantum channel $e^{t \mathcal{L}}$ to escape the unit disk in the complex plane.  Otherwise, the evolution `explodes' due to $|e^{t\ell}|\to \infty$ as $t \to \infty$.
%One defines a set of relaxation rates
%\begin{equation}\label{}
%  \Gamma_\alpha := {\rm Re}\, \ell_\alpha \ , \ \ \ \alpha=1,\ldots,d^2-1 .
%\end{equation}
%Clearly $_\alpha \geq 0$ for all $\alpha=1,\ldots,d^2-1$.
Defining the corresponding super-operator $\widehat{\mathcal{L}} : \mathrm{L}(\HH \otimes \HH) \to \mathrm{L}(\HH \otimes \HH)$ via $\widehat{\mathcal{L}} |X\rangle\!\rangle = |\mathcal{L}(X) \rangle\!\rangle $, the eigenvalue problem (\ref{ell}) is equivalent to

\begin{equation}\label{}
  \widehat{\mathcal{L}} |X_\alpha\rangle\!\rangle = \ell_\alpha  |X_\alpha\rangle\!\rangle .
\end{equation}
Now, if $\mathcal{L}$ is a diagonal form of GKLS generator
\begin{equation}\label{L-diag}
  \mathcal{L}(\rho) = -i[H,\rho] + \sum_k \gamma_k \Big(L_k \rho L_k^\dagger - \frac 12 \{ L_k^\dagger L_k,\rho\} \Big) ,
\end{equation}
then the corresponding super-operator $ \widehat{\mathcal{L}}$ has the following form

\begin{equation}\label{}
  \widehat{\mathcal{L}} = \sum_k \gamma_k  L_k \otimes \overline{L}_k -  \left( \mathbf{C} \otimes \oper + \oper \otimes \overline{\mathbf{C}} \right) ,
\end{equation}
where  $\,   \mathbf{C} = i H + \frac 12 \sum_k \gamma_k L_k^\dagger L_k$. Spectral properties of $\mathcal{L}$ were analyzed by several authors \cite{WOLF,Albert,Heide-1,Heide-2,Heide-3}

\begin{tcolorbox}
\begin{Proposition} Any GKLS generator $\mathcal{L}$ has the following spectral representation

\begin{equation}\label{}
  \mathcal{L} = \mathcal{L}_ \varphi + \sum_{{\rm Re}\ell_k < 0} (\ell_k \mathcal{P}_k + \mathcal{N}_k) ,
\end{equation}
where $\mathcal{P}_k$ are projectors satisfying $ \mathcal{P}_k \mathcal{P}_\ell = \delta_{k\ell} \mathcal{P}_k$,
%\begin{equation}\label{}
%  \mathcal{P}_k \mathcal{P}_\ell = \delta_{k\ell} \mathcal{P}_k ,
%\end{equation}
and $\mathcal{N}_k$ are nilpotent maps, i.e. $\mathcal{N}_k^{n_k} = 0$ for some $n_k > 0$. Moreover,
%
%\begin{equation}\label{}
 $  \mathcal{P}_k \mathcal{N}_\ell = \mathcal{N}_\ell  \mathcal{P}_k   = \delta_{k\ell} \mathcal{N}_k$.
%\end{equation}
The peripheral part defined via

\begin{equation}\label{}
  \mathcal{L}_ \varphi = \sum_{{\rm Re}\ell_k = 0} \ell_k \mathcal{P}_k ,
\end{equation}
satisfies $\,   \mathcal{L}_ \varphi = \mathcal{L} \mathcal{P}_ \varphi = \mathcal{P}_ \varphi \mathcal{L}$,  where
%
%\begin{equation}\label{}
 $ \mathcal{P}_ \varphi =  \sum_{{\rm Re}\ell_k = 0}  \mathcal{P}_k$
%\end{equation}
is a projection onto the peripheral part of the spectrum.

\end{Proposition}
\end{tcolorbox}
Note, that $\mathcal{L}(\rho) = -i[H,\rho]$ is purely peripheral. It generates a group of unitary maps $\mathbb{U}_t = e^{\mathcal{L}t}$. Actually, for any GKLS generator $ e^{\mathcal{L}_\varphi t}\mathcal{P}_\varphi$ is CPTP for any $t \in \mathbb{R}$ \cite{Facchi}. Majority of examples considered in the literature analyze generators $\mathcal{L}$ with semi-simple eigenvalues only (1-dimensional Jordan blocks). The following example  illustrates a non-trivial Jordan structure   \cite{Yuasa}.

\begin{Example} Consider the following qubit generator

\begin{equation}\label{}
  \mathcal{L}(\rho) = -i \omega [\sigma_x,\rho] + \gamma(\sigma_z \rho \sigma_z - \rho) .
\end{equation}
One finds

\begin{equation}\label{}
 \widehat{\mathcal{L}} = \left(
\begin{array}{cccc}
 -\gamma -i \omega & 0 & 0 & \gamma \\
 0 & -\gamma & \gamma & 0 \\
 0 & \gamma & -\gamma & 0 \\
 \gamma & 0 & 0 & -\gamma + i \omega  \\
\end{array}
\right) ,
\end{equation}
with the corresponding eigenvalues:

$$ \left\{0,-2 \gamma,-\gamma-\sqrt{\gamma^2-\omega^2},-\gamma + \sqrt{\gamma^2-\omega^2} \right\}  .$$
In the generic case $\gamma \neq \omega$ all eigenvalues are simple and the semigroup $e^{\mathcal{L}t}$ is relaxing to a maximally mixed qubit state. For $\gamma=\omega$ a doubly degenerate $\ell_2=\ell_3= -\gamma$ is not semisimple. Indeed, the corresponding Jordan decomposition has the following form

\begin{equation}\label{}
  \widehat{\mathcal{L}} = \widehat{\mathcal{S}}\widehat{\mathcal{L}}\widehat{\mathcal{S}}^{-1}
\end{equation}
with

\begin{equation}\label{}
  \widehat{\mathcal{S}} = \left(
\begin{array}{cccc}
 0 & 0 & -i & 1/\gamma \\
 1 & -1 & 0 & 0 \\
 1 & 1 & 0 & 0 \\
 0 & 0 & 1 & 0 \\
\end{array}
\right) \ ,\ \ \ \ \ \  \widehat{\mathcal{J}} =  \left(
\begin{array}{cccc}
 0 & 0 & 0 & 0 \\
 0& -2\gamma & 0 & 0  \\
 0 & 0& -\gamma & 1  \\
 0 & 0 & 0&   -\gamma  \\

\end{array}
\right) .
\end{equation}
$\widehat{\mathcal{J}} $ has a single $2 \times 2$ Jordan block. However, the corresponding semigroup is still relaxing since $\ell_0=0$ is simple.
\end{Example}
Recently several authors studied spectral properties of random GKLS generators. Using Random Matrix Theory techniques interesting universal properties of such spectra were derived \cite{random1,random2,random3,random4,random5,random6}.

%\subsubsection{Relaxation rates}

If $\widehat{\mathcal{L}}$ is diagonalizable, then

\begin{equation}\label{}
  \widehat{\mathcal{L}} |X\>\!\> = \sum_{\alpha=0}^{d^2-1} e^{-i \omega_\alpha t} e^{-\Gamma_\alpha t} |X_\alpha\>\!\> \<\!\< Y_\alpha| X \>\!\>
\end{equation}
where  $ \widehat{\mathcal{L}} |X_\alpha\>\!\> = \ell_\alpha  |X_\alpha\>\!\>$, and $ \widehat{\mathcal{L}^\ddag} |Y_\alpha\>\!\> = \ell^*_\alpha  |Y_\alpha\>\!\>$, together with

\begin{equation}\label{}
  \ell_\alpha = - i \omega_\alpha - \Gamma_\alpha .
\end{equation}
where $ \Gamma_\alpha := - {\rm Re}\, \ell_\alpha \geq 0$ being a set of relaxation rates.
Complete positivity is guaranteed by positivity of  $\gamma_k$ in the diagonal GKSL representation

\begin{equation}\label{L-diag-1}
  \mathcal{L}(\rho) = -i[H,\rho] + \sum_{k=1}^{d^2-1} \gamma_k \left( L_k \rho L_k^\dagger - \frac 12 \{ L_k^\dagger L_k,\rho\} \right) .
\end{equation}
It should be stressed, however, that $\gamma_k$ are not directly measured in the experiment and moreover they are not uniquely defined since the representation (\ref{L-diag-1}) is not unique. The quantities which are directly measured are relaxation rates $\Gamma_\alpha$. It is, therefore, clear that  complete positivity has to imply some additional constraints for relaxation rates.

\begin{Example} Consider a qubit generator

\begin{equation}  \label{PAULI-L}
  \mathcal{L}(\rho) = \frac 12 \sum_{k=1}^3 \gamma_k(\sigma_k \rho \sigma_k - \rho) .
\end{equation}
One finds for the relaxation rates

\begin{equation}\label{}
  \Gamma_1 = \gamma_2 + \gamma_3 \ , \ \ \ \ \ \Gamma_2 = \gamma_3 + \gamma_1 \ , \ \ \ \ \ \Gamma_3 = \gamma_1 + \gamma_2 \ .
\end{equation}
Now, if $\mathcal{L}$ generates a semigroup of positive maps, then $\Gamma_k \geq 0$ for $k=1,2,3$. However, it is not sufficient for complete positivity which requires $\gamma_k \geq 0$. Complete positivity is guaranteed by

\begin{equation}\label{}
  \Gamma_1 + \Gamma_2  \geq  \Gamma_3  , \ \ \  \Gamma_2 + \Gamma_3  \geq  \Gamma_1  , \ \ \ \Gamma_3 + \Gamma_1  \geq  \Gamma_2  .
\end{equation}
Interestingly, by defining $\Gamma := \Gamma_1 + \Gamma_2 + \Gamma_3$, the above conditions can be rewritten as follows

\begin{equation}\label{2GG}
  2 \Gamma_k \leq \Gamma \ , \ \ \ k=1,2,3 ,
\end{equation}
or, equivalently, defining relative rates $R_k := \frac{\Gamma_k}{\Gamma}$ one finds
\begin{equation}\label{R1/2}
  R_k \leq \frac 12  \ , \ \ \ k=1,2,3 .
\end{equation}
\end{Example}

\begin{Example} For a qubit generator considered in Example \ref{L+-} it is well known \cite{GKS} that complete positivity implies the following  relation between relaxation times $T_k := 1/\Gamma_k$
\begin{equation}\label{TT}
 2\, T_{\rm L} \geq  T_{\rm T} ,
\end{equation}
where the longitudinal rate $ \Gamma_{\rm L}$ and the transversal rate  $\Gamma_{\rm T}$ read as follows

\begin{equation}\label{GLGT}
 \Gamma_{\rm L} = \Gamma_3= \gamma_+ + \gamma_-  \ , \ \ \ \
 \Gamma_{\rm T} = \Gamma_1 = \Gamma_2 = \frac{1}{2} (\gamma_+ + \gamma_- ) + 2\gamma_z
\end{equation}
Condition (\ref{TT}) was experimentally demonstrated to be true \cite{ALICKI,LT,LT2}. Violation of (\ref{TT}) shows that the generator does not provide legitimate CPTP evolution.
\end{Example}
The above examples illustrate the following general result \cite{GK}

%\begin{tcolorbox}
\begin{Theorem} For any qubit GKLS generator the relative relaxation rates satisfy (\ref{R1/2}).
\end{Theorem}
%\end{tcolorbox}
In \cite{Gen-2020} this result was partially generalized for $d > 2$. Consider a GKLS generator such that $\mathcal{L}(\rho_{\rm ss}) =0$ with $\rho_{\rm ss} > 0$, and let $\mathcal{L}^\ddag(Y_\alpha) = \ell_\alpha Y_\alpha$.

%\begin{tcolorbox}
\begin{Proposition} \label{GAMMA}  The relaxation rates $\Gamma_\alpha$ can be represented as follows
\begin{equation}\label{!a}
   \Gamma_\alpha  = \frac{1}{2 \| Y_\alpha\|^2_{\rm ss}}  \sum_k \gamma_k \| [L_k,Y_\alpha] \|^2_{\rm ss} ,
\end{equation}
where $\|X\|^2_{\rm ss} = (X,X)_{\rm ss} = {\rm Tr}(\rho_{\rm ss} X^\dagger X)$.
\end{Proposition}
%\end{tcolorbox}

%\begin{tcolorbox}
\begin{Theorem} \label{TH-Gen}
  For unital CPTP semigroup the relaxation rates satisfy
\begin{equation}\label{CON}
  \Gamma_\alpha \leq  \frac 1d \Gamma ,
\end{equation}
or, equivalently, the relative rates satisfy $R_\alpha \leq \frac 1d$ for $\alpha = 1,\ldots,d^2-1$.
\end{Theorem}
%\end{tcolorbox}
Proof: inserting $\rho_{\rm ss} = \oper/d$ into  (\ref{!a}) one arrives at

\begin{equation}\label{!!}
   \Gamma_\alpha  = \frac{1}{2 \| Y_\alpha\|_{\rm HS}^2}  \sum_k \gamma_k \| [L_k,Y_\alpha] \|_{\rm HS}^2 ,
\end{equation}
where now $\|A\|_{\rm HS}^2 = {\rm Tr}(A^\dagger A)$ is a Hilbert-Schmidt norm. To prove (\ref{CON}) we use the following inequality \cite{BW}

\begin{equation}\label{BW}
  \| [A,B]\|_{\rm HS}^2 \leq 2 \| A \|_{\rm HS}^2 \|B \|_{\rm HS}^2 ,
\end{equation}
which implies $\Gamma_\alpha   \leq  \sum_k \gamma_k \| L_k\|_{\rm HS}^2$. Assuming the following normalization $\| L_k \|_{\rm HS}^2  =1$ {as well as the condition $\Tr L_k = 0$}, one shows that $\sum_k \gamma_k = \frac 1 d \sum_\alpha \Gamma_\alpha$ and hence  (\ref{CON}) follows. \hfill $\Box$

\begin{Remark} Note, that  if one can prove the following  generalization of (\ref{BW})

\begin{equation}\label{BW-gen}
  \| [A,B]\|^2_{\rm ss} \leq 2 \| A \|_{\rm HS}^2 \|B \|_{\rm ss}^2 ,
\end{equation}
then (\ref{CON}) holds for arbitrary GKLS generator satisfying $\rho_{\rm ss} > 0$. Simple numerical analysis shows that condition (\ref{BW-gen}) is not true.
\end{Remark}
Theorem \ref{TH-Gen} can be generalized as follows

%\begin{tcolorbox}
\begin{Proposition} Let $\mathcal{L}$ be a GKLS generator with the corresponding invariant state $\rho_{\rm ss} > 0$. If the $\rho_{\rm ss}$-dual generator $\widetilde{\mathcal{L}^\ddag}$ has a GKLS form, then relaxation rates of $\mathcal{L}$ satisfy (\ref{CON}).
\end{Proposition}
%\end{tcolorbox}
Proof: recall, that

\begin{equation}\label{}
  \widetilde{\mathcal{L}^\ddag}(X) = \mathcal{L}(X \rho_{\rm ss}) \rho_{\rm ss}^{-1} ,
\end{equation}
and hence if $\mathcal{L}(X_\alpha) = \ell_\alpha X_\alpha$, then

\begin{equation}\label{}
   \widetilde{\mathcal{L}^\ddag}(X_\alpha \rho_{\rm ss}^{-1}) = \mathcal{L}(X) \rho_{\rm ss}^{-1} = \ell_\alpha X_\alpha \rho_{\rm ss}^{-1} ,
\end{equation}
that is, both $\mathcal{L}$ and $ \widetilde{\mathcal{L}^\ddag}$ have the same spectra. Moreover,

\begin{equation}\label{}
   \widetilde{\mathcal{L}^\ddag}(\oper) = \mathcal{L}(\rho_{\rm ss}) \rho_{\rm ss}^{-1} = 0 ,
\end{equation}
and hence, if $ \widetilde{\mathcal{L}^\ddag}$ is of GKLS form, then Theorem \ref{TH-Gen} implies (\ref{CON}). \hfill $\Box$

%Interestingly, one has the following result \cite{Franco}

\begin{Cor} If $\mathcal{L}$ is a GKLS generator with the corresponding invariant state $\rho_{\rm ss} > 0$  commuting with the operator $\Delta$, then (\ref{CON}) holds. In particular  if $\mathcal{L}$ satisfies quantum detailed balance, then conditions (\ref{CON}) hold.
\end{Cor}

%\begin{Example}
%  Covariant ??????????????????
%\end{Example}

\begin{Remark} In \cite{Gen-2020} it is conjectured that condition (\ref{CON}) is satisfied for any GKLS generator. This conjecture is strongly supported by numerical analysis. Another constraints for relaxation rates were also derived in \cite{rates1,rates2,rates3}.
\end{Remark}

\subsection{Historical remarks}

The problem of characterising physically admissible transformations of quantum states and admissible generators of quantum evolution has a quite interesting history (cf. \cite{40-GKLS} for a recent historical account). Sudarshan et al.\cite{Sudarshan-1,Sudarshan-2,Sudarshan-3} formulated necessary and sufficient conditions for admissible transformation $\Phi : \LH \to \LH$ in terms of $A$ and $B$-matrices. By admissible they meant that $\Phi$ defines a positive map. Discussing $B$-matrix they say ``It immediately follows that $B$ is Hermitian and positive
semidefinite''. Now, $B$ matrix is nothing but a Choi matrix corresponding to $\Phi$ and hence positivity of $B$ is equivalent to complete positivity of $\Phi$. Not surprisingly they derived operator sum representation of the map $\Phi$ ten years before Kraus (formulae (32) and (33) in \cite{Sudarshan-1}). Positivity of $B$ is sufficient but not necessary for positivity of $\Phi$.

A similar positivity argument was used by Belavin et al.\cite{Belavin} (they did not cite \cite{Sudarshan-1}). They studied the master equation with a single rate $\gamma > 0$ and then proposed generalization $\gamma \to \gamma_{ij}$ (eq. (14) in \cite{Belavin}). Again the claim is that the Hermitian matrix $\gamma_{ij}$ is positive definite. However, positivity of $\gamma_{ij}$ enforces completely positive evolution of the density operator which is much stronger than positivity \cite{Belavin}.

Interestingly, Bausch \cite{Bausch} citing \cite{Sudarshan-1} noticed that positivity of $B$-matrix is not necessary for positivity of  the evolution. He derived diagonal form of the master equation (eq. (2.11) in \cite{Bausch}) and remarked that there is no need that all rates are positive. Neither Belavin et al.nor Bausch knew the notion of complete positivity which appeared in physics literature with papers of Ludwig, Haag, Kraus, Hellwig and others \cite{CP-1,CP-2,CP-3,CP-4,Kraus-71}. Ten years after Bausch two seminal papers of Gorini et al.\cite{GKS} and Lindblad \cite{L} fully exploited the notion of complete positivity. In the same year, however, there was a paper of Franke \cite{Franke} who followed Bausch observation. Interestingly, Franke cited \cite{Sudarshan-1,Sudarshan-2,CP-3,CP-4} but did not use the very concept of complete positivity. Instead, he derived the following form of master equation (eq. (5.3) in \cite{Franke})

\begin{equation}\label{}
  \dot{\rho} = - i [H,\rho] +  \sum_{k=1}^N (A_k \rho A_k^\dagger - \frac 12 \{A_k^\dagger A_k, \rho\}) +  \sum_{l=1}^{N'}  ((B_l \rho B_l^\dagger)^T - \frac 12 \{B_l^\dagger B_l, \rho\}) ,
\end{equation}
with arbitrary operators $A_k, B_l \in \LH$. He claims that ``{\em for $d = 2$ this equation will be the most general master equation that does not contain a "memory". For $d\geq 7$, this is not so.}" Note, that this equation has exactly the Lindblad form with

\begin{equation}\label{}
  \Phi(\rho) = \sum_{k=1}^N A_k \rho A_k^\dagger + \sum_{l=1}^{N'}  (B_l \rho B_l^\dagger)^T .
\end{equation}
Such map is positive and decomposable. It is well known that in the qubit case ($d=2$) all positive maps are of this form \cite{Woronowicz}. However, already for $d>2$ there are positive not decomposable maps (cf. \cite{TOPICAL}). Again, even in the qubit case Franke generator gives only a sufficient condition for positive evolution. The necessary and sufficient condition is represented by (\ref{PPP}) which does not require positivity of $\Phi$. Anyway, Franke paper shows that even  for qubit evolution the positivity condition is by no means trivial (recently, authors of \cite{FGKLS}  proposed to call GKLS the  FGKLS master equation. We stress, however, that Franke paper provides only a particular class of generators giving rise to PTP evolution contrary to GKLS generator which is universal for CPTP evolutions).

The celebrated result of Gorini et al.and Lindblad did not spread very fast. In 1984  Banks, Susskind and Peskin  \cite{Banks} motivated by some problems in quantum field theory posed the question about the dynamical equation for the density operator which enables the transition from pure to mixed states in the course of time. They were completely unaware of \cite{GKS} and \cite{L}. Without mentioning complete positivity (no citation to Kraus or other paper) they essentially derived GKSL-like master equation (eqs. (7-9) in \cite{Banks}) and finally came to the following problem: ``{\em We still need to implement the requirement that $\rho$ remains positive}". And then conclude ``{\em We do not know
what conditions are necessary to insure these properties, but we can state some simple sufficient conditions}". The reader immediately guesses what these sufficient conditions tell. They are nothing but conditions ensuring complete positivity. In a recent paper \cite{Barbara} an interesting  relationship between the classic magnetic resonance density matrix relaxation theories of Bloch and Hubbard and the modern Lindbladian master equation methods is reviewed.

This short historical review clearly shows the prominent role of complete positivity in deriving necessary and sufficient conditions for physically admissible generator of quantum dynamical semigroup.

\section{Markovian approximations in open quantum systems}   \label{APPROX}

The previous Section discusses a number of mathematical properties of Markovian semigroups of positive and completely positive trace preserving semigroups. Here we briefly recall how Markovian semigroup appears as an evolution of open quantum system. This is one of the central issue of the theory of open quantum systems and has been discussed in great details by many authors \cite{Open1,Open2,ALICKI,AF,Fabio-Roberto}. A semigroup of CPTP maps $\{ \Lambda_t = e^{\mathcal{L} t} \}_{t \geq 0}$ is a special example of a much more general scenario.  Any quantum evolution may be represented by a family of linear maps $\{\Lambda_{t,t_0}\}_{t\geq t_0}$  mapping an initial quantum state $\rho$ at time $t_0$ to a quantum state at time $t > t_0$.

\begin{tcolorbox}
\begin{Definition} One calls a family $\{ \Lambda_{t,t_0}\}_{t \geq t_0}$ of maps  ($t \geq t_0$)  a quantum dynamical map if

\begin{itemize}
  \item $\Lambda_{t,t_0}$ is completely positive and trace preserving for $t \geq t_0$,
  \item $\Lambda_{t_0,t_0} = {\rm id}$.
\end{itemize}
$\Lambda_{t,t_0}$ is time homogeneous if $\Lambda_{t,t_0} = \Lambda_{t-t_0}$. In this case one usually sets $t_0=0$.

\end{Definition}
\end{tcolorbox}
The simplest example is provided  by a unitary evolution represented by $\{\mathbb{U}_{t,t_0}\}_{t \in \mathbb{R}}$. It is defined for all $t \in \mathbb{R}$ meaning that the evolution is reversible. Obviously, $\mathbb{U}_{t,t_0}^{-1} = \mathbb{U}_{t_0,t}$. It is no longer true for a general dynamical map $\{ \Lambda_{t,t_0}\}_{t \geq t_0}$ which is defined for $t \geq t_0$ only. Even if the map $\Lambda_{t,t_0}$ is invertible (as a linear operator) the inverse needs not be CPTP. Actually, $\Lambda_{t,t_0}^{-1}$ is CPTP if and only if $\Lambda_{t,t_0}$ is unitary, i.e. $\Lambda_{t,t_0} = \mathbb{U}_{t,t_0}$.

\subsection{Reduced evolution}    \label{Reduced}

The physical origin of quantum dynamical map is based on an idea of reduced evolution. Consider a quantum system $S$ interacting with an environment $E$. The total $SE$ Hamiltonian $\mathbf{H}$ has the following form

\begin{equation}\label{}
 \mathbf{H} =  H_0 +  \lambda H_{\rm int} \ , \ \ \ \ H_0 = H_S \otimes \oper_E + \oper_S \otimes H_E ,
\end{equation}
where we introduced a coupling constant $\lambda$. Let us assume  that initially $S$ and $E$ are not correlated, i.e. $
  \rho_{SE} = \rho_S \otimes \rho_E$. One finds the unitary evolution of the composite state

\begin{equation}\label{}
  \rho_{SE} \ \longrightarrow \ \rho_{SE}(t) := U^{SE}_{t-t_0} \rho_S \otimes \rho_E U^{SE \dagger}_{t-t_0} =: \mathbb{U}^{SE}_{t-t_0}( \rho _S \otimes\rho_E) ,
\end{equation}
where $U^{SE}_t = e^{-i \mathbf{H} t}$. Till now, the composite $SE$ system is closed and its evolution is unitary. One {\em opens} the system tracing out all environmental degrees of freedom.  Assuming, that the initial environmental state is invariant under the evolution generated by $H_E$

\begin{equation}\label{}
  \mathbb{U}^E_{t-t_0}(\rho_E) = e^{-i H_E (t-t_0)} \rho_E e^{i H_E (t-t_0)} = \rho_E ,
\end{equation}
and taking into account  the fact that the total SE Hamiltonian $\mathbf{H}$ is time independent the operation of partial trace gives  rise to the time homogeneous dynamical map  $\Lambda_{t} : \TTH \to \TTH$:

\begin{equation}\label{reduced}
   \rho_S \to \Lambda_{t}(\rho_S) := {\rm Tr}_E \, \mathbb{U}^{SE}_{t}( \rho_S \otimes\rho_E) ,
   %\Big( U^{SE}_{t-t_0} \rho \otimes \rho_E U^{SE \dagger}_{t-t_0}  \Big)  ,
\end{equation}
where we already fixed $t_0=0$. The above map is by construction  CPTP.  One calls the evolution $\rho_S \to \Lambda_{t}(\rho_S)$ the {\em reduced evolution} of the system. Note, that $\Lambda_{t}$ is a composition of three completely positive maps

\begin{equation}\label{}
  \Lambda_{t} = {\rm Tr}_E \circ \mathbb{U}^{SE}_{t}\circ \mathcal{A} ,
\end{equation}
where $\mathcal{A} : \mathcal{T}(\mathcal{H}_S) \to \mathcal{T}(\mathcal{H}_S\otimes \mathcal{H}_E)$ defines so-called assignment map defined via

\begin{equation}\label{}
  \mathcal{A}(\rho_S) := \rho_S \otimes \rho_E .
\end{equation}
One may easily define a dynamical map in the Heisenberg picture introducing a dual map $\mathcal{E} = \mathcal{A}^\ddag : \mathcal{B}(\mathcal{H}_S\otimes \mathcal{H}_E) \to \mathcal{B}(\mathcal{H}_S)$ via

\begin{equation}\label{}
  {\rm Tr}_{S+E}[ \mathcal{A}(\rho_S) X_{SE}] = {\rm Tr}_S[\rho_S \mathcal{E}(X_{SE})] ,
\end{equation}
for any $X_{SE} \in \mathcal{B}(\mathcal{H}_S \otimes \mathcal{H}_E)$. Note, that $\mathcal{E}(\oper_S \otimes\oper_E) = \oper_S$ and $
  \mathcal{E}(X_S \otimes Y_E) = X_S\, {\rm Tr}(Y_E \rho_E)$. Finally, one finds
\begin{equation}\label{}
  \Lambda_{t}^\ddag(X_S) = \mathcal{E}(\mathbb{U}^\ddag_{t}(X_S \otimes \oper_E)) = {\rm Tr}_E [(\oper_S \otimes \rho_E) \mathbb{U}^\ddag_{t}(X_S \otimes \oper_E)] .
\end{equation}
How to find the dynamical equation governing the evolution of the reduced density operator $\Lambda_{t}(\rho)$? This problem was already analyzed by several authors (cf. monographs \cite{Open1,Open2,Open3,Schaller,ALICKI,Open4,Haake} and in what follows we skip the detailed derivation and point out only the most essential steps. Representing an interaction Hamiltonian

\begin{equation}\label{}
 H_{\rm int} =  \sum_k A_k \otimes B_k ,
\end{equation}
one shows that the reduced time homogeneous evolution satisfies the following Nakajima-Zwanzig master equation \cite{Nakajima,Zwanzig,Haake}

\begin{equation}\label{GME}
  \partial_t \Lambda_{t} = \mathcal{L}_{\rm eff} \Lambda_{t} + \lambda^2 \int_{t}^t \mathcal{K}^{\rm NZ}_{t-\tau} \Lambda_{\tau} d\tau \ , \ \ \ \Lambda_0 = {\rm id} ,
\end{equation}
where   $\mathcal{L}_{\rm eff}: \TTH \to \TTH$ is defined by

\begin{equation}\label{}
  \mathcal{L}_{\rm eff}(\rho_S) = - i[H_{\rm eff},\rho_S] ,
\end{equation}
together with an effective Hamiltonian

\begin{equation}\label{}
  H_{\rm eff} = H_S + \lambda \sum_\alpha A_\alpha {\rm Tr}(B_\alpha \rho_E) .
\end{equation}
The linear operator $\mathcal{K}^{\rm NZ}_t : \TTH \to \TTH$ is usually called a Nakjima-Zwanzig memory kernel. It is a highly nontrivial function of the total Hamiltonian $\mathbf{H}$ and the initial environmental state $\rho_E$. To represent the structure of the kernel $\mathcal{K}^{\rm NZ}_t$ let us define the following  projection operator

\begin{equation}\label{}
  \mathcal{P}(\rho_{SE}) := {\rm Tr}_E \, \rho_{SE} \otimes \rho_E ,
\end{equation}
and let $Q := {\rm id} - \mathcal{P}$ denote a complementary projection. The reduced evolution of the system $S$ satisfies the following compact relation

\begin{equation}\label{}
  \mathcal{A}\circ \Lambda_{t} = \mathcal{P}\circ \mathbb{U}_{t} \circ \mathcal{A} ,
\end{equation}
for any $t \geq 0$. Introducing

\begin{equation}\label{}
  \mathbf{L}(\rho_{SE}) := - i[\mathbf{H},\rho_{SE}] \ , \ \ \  \mathbf{L}_{\rm int}(\rho_{SE}) := - i[H_{\rm int},\rho_{SE}] ,
\end{equation}
one finds  the following structure of the memory kernel

\begin{equation}\label{}
  \mathcal{K}^{\rm NZ}_t(\rho_S) = {\rm Tr}_E \Big( \mathbf{L}_{\rm int} Q \, e^{t Q \mathbf{L} Q}\, Q \mathbf{L}_{\rm int} (\rho_ \otimes \rho_E) \Big) .
\end{equation}
It should be clear, however, that this highly nontrivial formula for  $ \mathcal{K}^{\rm NZ}_t$ is in practice rather untractable.
%Moreover, to compute this kernel one has to know the complete solution of the `system + environment' evolution.
Hence, in order to find a tractable description of the system's evolution one  tries to find a suitable approximation which allows to simply the structure of $ \mathcal{K}^{\rm NZ}_t$.

Finding a physically consistent approximation of (\ref{GME}) is a very delicate issue. Any legitimate approximation should provide CPTP evolution for $\Lambda_t$. In most cases one assumes that the system-environment interaction is weak enough such that a second order approximation in the coupling constant $\lambda$ is sufficient. Such weak coupling approximation works perfectly in several physical contexts. These include in particular quantum optical systems but also nuclear magnetic resonance and molecular systems.
In the weak coupling regime one considers two key approximations:

\begin{enumerate}
  \item {\it Born approximation} which essentially neglects all correlations which might developed between the system and the environment, that is, the time evolved total state $\rho_{SE}(t) \approx \rho_S(t) \otimes \rho_E$,

 \item {\it Markov approximation} which is based on the assumption of separation of characteristic time scales $\tau_S$ and $\tau_E$ of the system and environment, that is, $\tau_E \ll \tau_S$. It essentially means that environment degrees of freedom evolve much faster that the corresponding degrees of freedom of the system.
\end{enumerate}

\subsection{Bloch-Redfield master equation}

Born and Markov approximations allow to derive the following master equation for the system density operator

\begin{equation}\label{}
  \dot{\rho}_S(t) = \mathcal{L}_{\rm eff}(\rho_S(t)) + \lambda^2\, \mathbf{L}_{\rm BR}(\rho_S(t)) ,
\end{equation}
where the Bloch-Redfield generator $\mathbf{L}_{\rm BR}$ reads as follows \cite{Open1,Redfield,Redfield-2,Bloch,Fabio-Roberto,LIDAR}

\begin{equation}\label{BR-1}
  \mathbf{L}_{\rm BR}(\rho_S) = \int_0^\infty d\tau\, \mathbb{U}^S_{-\tau}\, {\rm Tr}_E \Big\{ [H_{\rm int}, \mathbb{U}^S_{\tau} \otimes \mathbb{U}^E_{\tau}[H_{\rm int},\rho_S \otimes \rho_E]] \Big\} .
\end{equation}
To find the canonical GKLS form of the  Bloch-Redfield generator  (\ref{BR-1}) one considers a spectral resolution of the system Hamiltonian

\begin{equation}\label{}
  H_S = \sum_{k=1}^d \epsilon_k P_k ,
\end{equation}
and defines

\begin{equation}\label{}
  A_\alpha(\omega) = \sum_{\epsilon_m - \epsilon_n = \omega} P_m A_\alpha P_n ,
\end{equation}
where the above sum is performed over all Bohr frequencies   $\omega = \epsilon_m - \epsilon_n$. Note that

\begin{equation}\label{}
  [A_\alpha(\omega),H_S] = \omega A_\alpha(\omega) ,
\end{equation}
together with
\begin{equation}\label{}
  A_\alpha^\dagger(\omega) = A_\alpha(-\omega) ,  \ \ \ \ \ \sum_\omega A_\alpha(\omega) = A_\alpha .
\end{equation}
Simple algebra leads to the following form

\begin{equation}\label{BR-3}
   \mathbf{L}_{\rm BR}(\rho_S) = -i[\rho_S,\Delta H] + \sum_{\alpha,\beta} \sum_{\omega\omega'} \gamma_{\alpha\beta}(\omega,\omega') \Big( A_\beta(\omega')\rho_S A_\alpha^\dagger(\omega) - \frac 12 \{ A_\alpha^\dagger(\omega)A_\beta(\omega'),\rho_S \} \Big) ,
\end{equation}
where the Lamb shift correction

\begin{equation}\label{}
  H_{\rm LS} =  \sum_{\alpha,\beta} \sum_{\omega,\omega' } s_{\alpha\beta}(\omega,\omega') A^\dagger_\alpha(\omega) A_\beta(\omega') ,
\end{equation}
and

\begin{equation}\label{}
  s_{\alpha\beta}(\omega,\omega') = \tilde{h}_{\alpha\beta}(\omega') + \tilde{h}^*_{\beta\alpha}(\omega) \ , \ \ \ \gamma_{\alpha\beta}(\omega,\omega') = \frac{1}{2i} \Big( \tilde{h}_{\alpha\beta}(\omega') - \tilde{h}^*_{\beta\alpha}(\omega) \Big) ,
\end{equation}
with

\begin{equation}\label{}
  \tilde{h}_{\alpha\beta}(\omega) = \int_0^\infty d\tau\, e^{i\omega \tau} h_{\alpha\beta}(\tau) ,
\end{equation}
where we introduced the environment two-point correlation function

\begin{equation}\label{}
  h_{\alpha\beta}(\tau) = {\rm Tr}[ B_\alpha(t) B_\beta \rho_E] .
\end{equation}
 Interestingly, if $\rho_E$ is a thermal state of the environment $\rho_E = e^{-\beta H_E}/Z_E$ at the inverse temperature $\beta$, then  $h_{\alpha\beta}$ satisfies the celebrated Kubo-Martin-Schwinger (KMS) condition \cite{KMS-1,KMS-2,KMS-3}

\begin{equation}\label{KMS}
  h_{\mu\nu}(t) = h_{\nu\mu}(-t-i\beta) .
\end{equation}
Unfortunately, the Bloch-Redfield master equation has a serious drawback. The corresponding semigroup

\begin{equation}\label{}
  \Lambda^{\rm BR}_t = e^{(\mathcal{L}_{\rm eff} + \lambda^2 \mathbf{L}_{\rm BR})t} ,
\end{equation}
is by construction trace-preserving. However, it is not completely positive. Actually, even positivity is not guaranteed. This issue was extensively discussed in the literature \cite{Fabio-Roberto,Dumcke,Pollard,Fabio-JPA,Whitney,FG,Walter,Davidovic,Anton,Anders1,Winczewski,FDR}.

\subsection{Secular approximation and Davies generator}

The problem of violation of positivity by the Bloch-Redfield evolution was rigorously analyzed by Davies \cite{Davies1,Davies2,Davies3} (see also \cite{Pule} and \cite{SPOHN} for the review).  Davies approach based on the rigorous van Hove limit \cite{Hove} gives rise to the following limiting semigroup in the interaction picture w.r.t. to $H_S \otimes \oper_E + \oper_S \otimes H_E$:

%\begin{tcolorbox}
\begin{equation}\label{}
  e^{t\mathbf{L}_{\rm D}} : = \lim_{\lambda \to 0} \mathbb{U}^S_{-t/\lambda^2} \Lambda_{t/\lambda^2} ,
\end{equation}
%\end{tcolorbox}
where $\Lambda_t(\rho) = {\rm Tr}_E \mathbb{U}_t(\rho \otimes \rho_E)$ is an exact (without any approximation) dynamical map (for a rigorous derivation of a GKLS master equation  cf. also recent papers \cite{WCL1,WCL2,WCL3} and \cite{WCL5}). With appropriate conditions upon two-point environmental correlation functions Davies proved \cite{Davies1} that for a fixed (but arbitrary) rescaled time $\tau = \lambda^2 t $

\begin{equation}\label{Davies!}
  \lim_{\lambda \to 0}  \| \tilde{\rho}_S(\tau) - \tilde{\rho}^{\rm D}_S(\tau) \|_1 = 0 ,
\end{equation}
where $\tilde{\rho}_\tau$ and $\tilde{\rho}^{\rm D}_\tau$ represents the true reduced  evolution (in the interaction picture) and the one governed by the Davies generator, respectively.  Recently, Merkli \cite{Merkli} provided interesting improvement of Davies result showing that

\begin{equation}\label{Merkli!}
   \| \tilde{\rho}_S(\tau) - \tilde{\rho}^{\rm D}_S(\tau) \|_1 \leq C \lambda^2 ,
\end{equation}
is valid for all times $t \geq 0$. It turns out \cite{Davies1,Dumcke} that the Davies generator is related to the Bloch-Redfield genereator via the following ergodic average

\begin{equation}\label{}
  \mathbf{L}_{\rm D} = \lim_{T \to \infty} \frac{1}{2T} \int_{-T}^T dt \mathbb{U}^S_{-t} \mathbf{L}_{\rm BR} \mathbb{U}^S_t ,
\end{equation}
and it corresponds to a secular approximation

\begin{equation}\label{}
   s_{\alpha\beta}(\omega,\omega') \approx  \   s_{\alpha\beta}(\omega) \delta_{\omega,\omega'} \ ,\ \ \gamma_{\alpha\beta}(\omega,\omega') \approx  \gamma_{\alpha\beta}(\omega,\omega) \delta_{\omega,\omega'} ,
\end{equation}
with

\begin{equation}\label{s-g}
  s_{\alpha\beta}(\omega) = s_{\alpha\beta}(\omega,\omega) \ , \ \ \ \gamma_{\alpha\beta}(\omega) = \gamma_{\alpha\beta}(\omega,\omega) .
\end{equation}
These quantities are related via \cite{ALICKI}

\begin{equation}\label{}
  s_{\alpha\beta}(\omega) = \mathcal{P} \frac{1}{2\pi} \int_{-\infty}^\infty \frac{ \gamma_{\alpha\beta}(x)}{\omega - x} .
\end{equation}
It eventually gives rise to the following canonical form of the generator

\begin{equation}\label{D-1}
   \mathbf{D}_{\rm D}(\rho_S) = -i[\rho_S,\Delta H] + \sum_{\alpha,\beta} \sum_{\omega} \gamma_{\alpha\beta}(\omega) \Big( A_\beta(\omega)\rho_S A_\alpha^\dagger(\omega) - \frac 12 \{ A_\alpha^\dagger(\omega)A_\beta(\omega),\rho_S \} \Big) ,
\end{equation}
where now the Lamb shift correction  reads as follows

\begin{equation}\label{}
  H_{\rm LS} =  \sum_{\alpha,\beta} \sum_{\omega } s_{\alpha\beta}(\omega) A^\dagger_\alpha(\omega) A_\beta(\omega) .
\end{equation}
Using Bochner theorem one proves \cite{Davies1,ALICKI,Open1} that the matrix $\gamma_{\alpha\beta}(\omega)$ is positive definite (for all Bohr frequencies $\omega$) and hence (\ref{D-1}) provides the canonical form of GKLS generator.
The master equation (in the Schr\"odinger picture) has therefore the following form

\begin{equation}\label{}
  \dot{\rho}_S(t) = -i[H_{\rm eff},\rho_S(t)] + \lambda^2 \, \mathbf{L}_{\rm D}(\rho_S(t)) .
\end{equation}
If the environmental state is thermal, i.e. satisfies KMS condition (\ref{KMS}), then $\mathcal{L}_{\rm eff}+  \lambda^2 \mathbf{L}_{\rm D}$ satisfies quantum detailed balance condition

\begin{equation}\label{}
\gamma_{\mu\nu}(-\omega) = e^{- \beta \omega} \gamma_{\nu\mu}(\omega) .
\end{equation}
Hence, for an initial thermal state of the bath the dissipative part the Davies generator has the following structure

%\begin{tcolorbox}
\begin{eqnarray}\label{Davies-th}
 \mathcal{L}_D(\rho_S) &=&  \frac{\lambda^2}{2} \sum_{\mu,\nu} \sum_{\omega \geq 0}  \tilde{h}_{\mu\nu}(\omega) \Big\{  [A_\mu(\omega)\rho_S,A^\dagger_\nu(\omega)] + [A_\mu(\omega),\rho_S A^\dagger_\nu(\omega)]   \nonumber \\
 &+&  e^{-\beta \omega} \Big( [A^\dagger_\nu(\omega)\rho_S,A_\mu(\omega)] + [A^\dagger_\nu(\omega),\rho_S A_\mu(\omega)] \Big) \Big\}
\end{eqnarray}
%\end{tcolorbox}
\noindent where we used $A_\alpha(-\omega) = A_\alpha^\dagger(\omega)$. The most important properties of the Davies generator $\mathcal{L}_H + \mathcal{L}_D$ are summarized as follows:

\begin{tcolorbox}

\begin{itemize}
  \item Hamiltonian part $\mathcal{L}_H$ and dissipative part $\mathcal{L}_D$ commute,

  \item if $\rho_E = e^{-\beta H_E}/{\mathcal{Z}_E}$ is a thermal state with a partition function $\mathcal{Z}_E = {\rm Tr} e^{-\beta H_E}$, then the asymptotic state of the system is also thermal at the inverse temperature $\beta$, and $\rho_{\rm ss} = e^{-\beta H_S}/{\mathcal{Z}_S}$, with $\mathcal{Z}_S = {\rm Tr} e^{-\beta H_S}$,

  \item the Davies generator  satisfies quantum detailed balance condition w.r.t.  $\rho_{ss}$,

  \item the following covariance property holds

\begin{equation}\label{}
  e^{- iH_S t} \mathcal{L}(\rho_S)  e^{i H_S t} = \mathcal{L}( e^{- iH_S t} \rho_S  e^{iH_S t}) .
\end{equation}

\end{itemize}

\end{tcolorbox}
\noindent Weak coupling limit for a time dependent system Hamiltonian was generalized by Davies and Spohn \cite{D-S}. This generalization is essential for thermodynamics of open quantum systems \cite{Spohn-Lebowitz}. Recently, weak coupling limit was derived for periodically driven systems \cite{Periodic-1,Periodic-2} in connection to quantum heat engines.

\subsection{Strong coupling limit}

There exists a complementary method to derive Markovian master equation -- so called singular coupling limit \cite{ALICKI,Open1}. It corresponds to a physical scenario when decay time of the correlation functions of the environment $\tau_E \to 0$.
%It implies that the environment two-point correlation functions $h_{\mu\nu}(t) \to h_{\mu\nu}\delta(t)$.
Interestingly, it was observed in \cite{Palmer} that formally the corresponding semigroup can be obtained in a weak coupling limit of the following total Hamiltonian

\begin{equation}\label{}
  H = \lambda^2 H_S \otimes \oper_E + \oper_S \otimes H_E + \lambda H_{\rm int} ,
\end{equation}
cf. also \cite{Accardi-singular}. In particular if the interaction Hamiltonian $H_{\rm int} = \sum_\alpha A_\alpha \otimes B_\alpha$ with Hermitian $A_\alpha$ and $B_\alpha$, then the corresponding generator reads

\begin{equation}\label{}
  \mathcal{L}(\rho_S) = -i[H_{\rm eff},\rho_S] + \frac 12 \sum_{\alpha,\beta} \gamma_{\alpha\beta} \Big( [A_\alpha,\rho_S A_\beta] + [A_\alpha\rho_S, A_\beta] \Big) ,
\end{equation}
where

\begin{equation}\label{}
  H_{\rm eff} = H_S + \sum_{\alpha,\beta} s_{\alpha\beta} A_\beta A_\alpha ,
\end{equation}
and $s_{\alpha\beta}$ and $\gamma_{\alpha\beta}$ are defined in terms of two-point correlation function via

\begin{equation}\label{}
  \int_{0}^{\infty} h_{\alpha\beta}(t) dt = \frac 12 \gamma_{\alpha\beta} + \frac i2 s_{\alpha\beta} .
\end{equation}
The Hermitian matrix $\gamma_{\alpha\beta}$ is positive definite.  The Markovian semigroup in the strong coupling limit was applied for description of transport properties of nanomaterials \cite{strong}.

%Another derivation of Markovian semigroup uses so called stochastic limit developed by Accardi et al.\cite{ACCARDI}. Recently, an interesting %proposal which unifies weak coupling and singular coupling limit was proposed in \cite{Anton-2}.

\subsection{Coarse-grained master equation}

Secular approximation (well known as rotating wave approximation in quantum optics) gives rise to a Davies generator which has  legitimate GKLS structure. However, this approximation is not always physically justified and the corresponding Davies semigroup might substantially deviates from the true reduced evolution \cite{Fabio-2,Fabio-3}. An interesting alternative approach to Markovian master equation is based on a coarse-graining procedure \cite{Lidar-2}. Consider an interaction picture total unitary evolution

\begin{equation}\label{}
  \tilde{U}(t) = \mathcal{T} \exp\left( - i \lambda \int_0^t \tilde{H}_{\rm int}(\tau) d\tau \right) ,
\end{equation}
where $\mathcal{T}$ denotes a chronological product, and  $\tilde{H}_{\rm int}(t) = \mathbb{U}^S_{-t} \otimes \mathbb{U}^E_{-t} H_{\rm int} \mathbb{U}^S_{t} \otimes \mathbb{U}^E_{t}$. The reduce system evolution (interaction picture) reads

\begin{equation}\label{}
  \tilde{\rho}_S(t)  = {\rm Tr}_E \Big(  \tilde{U}(t) \rho_S(0) \otimes \rho_E(0)  \tilde{U}^\dagger(t) \Big) ,
\end{equation}
and hence gives rise to the following dynamical map

\begin{equation}\label{}
  \tilde{\rho}_S(t) = \tilde{\Lambda}_t(\rho_S(0)) = \sum_{\alpha,\beta=0}^{d^2-1} \chi_{\alpha\beta}(t) F_\alpha\rho_S(0) F_\beta^\dagger ,
\end{equation}
with $F_0 = \oper$, and $\chi_{\alpha\beta}(t)$ is a positive definite matrix such that $\chi_{\alpha\beta}(0) = \delta_{\alpha\beta}$. Trace-preservation condition is realized via

\begin{equation}\label{}
  \sum_{\alpha,\beta=0}^{d^2-1} \chi_{\alpha\beta}(t) F_\beta^\dagger F_\alpha = \oper ,
\end{equation}
for all $t \geq 0$. One finds therefore

\begin{eqnarray}\label{}
  \tilde{\rho}_S(0) &=& \frac 12 \sum_{\alpha,\beta=0}^{d^2-1} \chi_{\alpha\beta}(t) \{ F_\beta^\dagger F_\alpha,\tilde{\rho}_S(0) \}  = \chi_{00}(t) \tilde{\rho}_S(0)  \\ &+& \frac 12 \sum_{\alpha=1}^{d^2-1} \Big(  \chi_{\alpha 0}(t) \{ F_\alpha, \tilde{\rho}_S(0) \} +  \chi_{0 \alpha }(t) \{ \tilde{\rho}_S(0),F_\alpha^\dagger \} \Big) + \frac 12 \sum_{\alpha,\beta=1}^{d^2-1} \chi_{\alpha\beta}(t) \{ F_\beta^\dagger F_\alpha,\tilde{\rho}_S(0) \} , \nonumber
\end{eqnarray}
which eventually gives rise to the following equation \cite{Lidar-2}

\begin{equation}\label{CG-1}
  \frac{d}{dt}  \tilde{\rho}_S(t) = - i[ \dot{S}(t),  \tilde{\rho}_S(0)] +  \frac 12 \sum_{\alpha,\beta=1}^{d^2-1} \dot{\chi}_{\alpha\beta}(t) \Big(  [F_\alpha, \tilde{\rho}_S(0) F_\beta^\dagger ] +  [F_\alpha \tilde{\rho}_S(0), F_\beta^\dagger ] \Big) ,
\end{equation}
with

\begin{equation}\label{}
  S(t) = \frac{i}{2} \sum_{\alpha=1}^{d^2-1}\Big( \chi_{\alpha 0}(t) F_\alpha - \chi_{0 \alpha}(t) F^\dagger_\alpha \Big) .
\end{equation}
Note, that (\ref{CG-1}) does not still define the dynamical equation for $ \tilde{\rho}_S(t)$. Indeed, the rhs of (\ref{CG-1}) includes the initial state $\rho_S(0)$ and not the current one $ \tilde{\rho}_S(t)$. To find the Markovian master equation one has to introduce Markovian-like  approximation. Let us introduce a coarse-graining time scale $\Delta t$ satisfying $\tau_E \ll \Delta t \ll \tau_S$ and assume that after each period $\Delta t$ the environment effectively resets \cite{Lidar-2,Lidar-3}. This Markovian approximation gives rise to the following dynamical equation

\begin{equation}\label{CG-2}
  \frac{d}{dt}\tilde{\rho}_S(t) = - i[ \<\dot{S}\>,  \tilde{\rho}_S(t)] +  \frac 12 \sum_{\alpha,\beta=1}^{d^2-1} \<\dot{\chi}_{\alpha\beta}\> \Big(  [F_\alpha, \tilde{\rho}_S(t) F_\beta^\dagger ] +  [F_\alpha \tilde{\rho}_S(t), F_\beta^\dagger ] \Big) ,
\end{equation}
where

\begin{equation}\label{}
  \< X \> = \frac{1}{\Delta t} \int_0^{\Delta t} X(t) dt .
\end{equation}
One proves \cite{Lidar-2} that the matrix $\<\dot{\chi}_{\alpha\beta}\>$ is positive definite and hence $ \mathbf{L}_{\Delta t}$ has a GKLS form.
Note, that one still has a freedom to decide about the parameter $\Delta t$.
It is argued \cite{Lidar-3} that optimizing `$\Delta t$'  the solution to (\ref{CG-2}) is closer to the exact solution than the one provided by the Davies secular master equation.

\subsection{Refined weak coupling limit}   \label{sub-refined}

A similar approach to that based on the time coarse-graining was proposed in \cite{Brandes-1,Brandes-2,Fabio-2,Fabio-3,Angel-refined}. Starting from

\begin{equation}\label{RWC-1}
  \frac{d}{dt}  \tilde{\rho}_S(t) = - i\lambda {\rm Tr}_E[ \tilde{H}_{\rm int}(t),  \tilde{\rho}_{SE}(t)]  ,
\end{equation}
which implies

\begin{equation}\label{}
  \tilde{\rho}_S(\Delta t) = \tilde{\rho}_S(0) - \frac{\lambda^2}{2} \mathcal{T}  \int_0^{\Delta t} dt_1 \int_0^{\Delta t} dt_2 {\rm Tr}_E [\tilde{H}_{\rm int}(t_1),[\tilde{H}_{\rm int}(t_2),\rho_S(0) \otimes \rho_E]] + O(\lambda^3) ,
\end{equation}
where for simplicity we assumed ${\rm Tr}(B_\alpha \rho_E)=0$. Interestingly, the `$\lambda^2$' term has already a GKLS form

%Assuming $H_{\rm int} = \sum_\alpha A_\alpha \otimes B_\alpha$ one finds (in the Schr\"odinger picture)

%\begin{equation}\label{CG-2}
%  \frac{d}{dt}  {\rho}_S(t) = - i[ H_S,  {\rho}_S] +  \lambda^2 \mathbf{L}_{\Delta t} (\rho_S)  ,
%\end{equation}
%where the coarse-grained generator has the following form

\begin{equation}\label{}
  \mathbf{L}_{\Delta t} (\rho_S) = -i [ \delta H,\rho_S] + \sum_{\omega,\omega'} \sum_{\alpha,\beta} \gamma_{\alpha\beta}(\omega,\omega';\Delta t)\Big( A_\beta(\omega') \rho_S A_\alpha^\dagger(\omega) - \frac 12 \{  A_\alpha^\dagger(\omega)  A_\beta(\omega') ,\rho_S\} \Big) ,
\end{equation}
with the Hamiltonian correction $\delta H = \sum_{\omega,\omega'} \sum_{\alpha,\beta} s_{\alpha\beta}(\omega,\omega';\Delta t)  A^\dagger_\alpha(\omega) A_\beta(\omega)$ together with

\begin{equation}\label{}
 s_{\alpha\beta}(\omega,\omega';\Delta t) = \frac{1}{2i \Delta t} \int_0^{\Delta t} dt_1 \int_0^{\Delta t} dt_2 {\rm sign}(t_1-t_2)\, dt_2 e^{i(\omega t_1 - \omega' t_2)} h_{\alpha\beta}(t_1-t_2) ,
\end{equation}
and

\begin{equation}\label{}
 \gamma_{\alpha\beta}(\omega,\omega';\Delta t) = \frac{1}{\Delta t}\int_0^{\Delta t} dt_1 \int_0^{\Delta t} dt_2 \, e^{i(\omega t_1 - \omega' t_2)} h_{\alpha\beta}(t_1-t_2) ,
\end{equation}
and $h_{\alpha\beta}(t)$ is an environmental two-point correlation function. Hence, one may approximate the exact solution via the following Markovian semigroup (in the Schr\"odinger picture)

\begin{equation}\label{RWC-2}
  \dot{\rho}_S(t) = -i[H_S,\rho_S(t)] + \frac{\lambda^2}{\Delta t}  \mathbf{L}_{\Delta t}(\rho_S(t)) .
\end{equation}
Contrary to the Davies weak coupling master equation which provides a good approximation of the accurate evolution for large `$t$', the above master equation provides a good approximation for small times. One calls it {\em refined weak coupling limit} \cite{Angel-refined}.  Interestingly, in the limit $\Delta t \to \infty$ one recovers the Davies generator

\begin{equation}\label{}
  \lim_{\Delta t \to \infty} \mathbf{L}_{\Delta t} = \mathbf{L}_{\rm D} .
\end{equation}
A similar generator can be derived via so called cumulant expansion \cite{Alicki-1989}: let

\begin{equation}\label{}
  \tilde{\rho}_S(t)  = \tilde{\Lambda}_t(\rho_S(0)) ,
\end{equation}
and represent the dynamical map via the following cumulant expansion (cf. a pedagogical exposition in \cite{Piotr})

\begin{equation}\label{}
  \tilde{\Lambda}_{\Delta t} = \exp\Big( \sum_{n=1}^\infty \lambda^n K^{(n)}(\Delta t) \Big) = {\rm id} + \lambda K^{(1)}  + \lambda^2 \Big(  K^{(2)}(\Delta t) + \frac 12 (K^{(1)}(\Delta t))^2 \Big) + O(\lambda^3) .
\end{equation}
Assuming again ${\rm Tr}(B_\alpha \rho_E)=0$ one finds $K^{(1)}  = 0$, and hence

\begin{equation}\label{}
  \tilde{\Lambda}_{\Delta t} =  {\rm id} + \lambda^2   K^{(2)}(\Delta t)  + O(\lambda^3) .
\end{equation}
It should be not surprising that

\begin{equation}\label{}
   K^{(2)}(\Delta t) = \mathbf{L}_{\Delta t} ,
\end{equation}
that is, the second cumulant perfectly recovers the generator derived in the refined weak coupling limit.

\section{Markovian semigroups and nonequilibrium thermodynamics}   \label{THERMO}

We now address the question how Markovian dynamical semigroup derived in the weak coupling limit fits the law of quantum thermodynamics.  Quantum thermodynamics attracts in recent years a lot of attention and there is a vast body of literature (cf. e.g. \cite{QT0,QT1,QT2,QT3,QT4,QT5,QT6,QT7}). Here we show how Markovian semigroups (classical and quantum) fit the main principles of nonequilibrium quantum thermodynamics.

\subsection{Classical scenario}

Let us start with a classical scenario and consider a Markovian semigroup governed by the Pauli rate equation \cite{Open1,Kampen}

\begin{equation}\label{PAULI-2}
  \dot{p}_i(t)  = \sum_{j=1}^d \Big( W_{ij} p_j(t) - W_{ji} p_i(t) \Big) \ , \ \ \ i=1,\ldots,d ,
\end{equation}
with $W_{ij} \geq 0$. We briefly review Schnakenberg's \cite{Schnakerberg} approach to entropy production for the processes governed by (\ref{PAULI-2}). Suppose that  $k$th state has an energy $E_k$ and hence $\<E\> = \sum_k E_k p_k$ gives rise to

\begin{equation}\label{balance}
  \frac{d}{dt} \<E\> = \sum_k \frac{\partial{E_k}}{\partial t} p_k + \sum_k E_k \dot{p}_k = \<  \frac{\partial{E}}{\partial t} \> + \mathcal{J} ,
\end{equation}
where the heat flux (heat current) $\mathcal{J}$ is defined via $\mathcal{J} =  \sum_k E_k \dot{p}_k$, and $\<  \frac{\partial{E}}{\partial t} \>$ may be interpreted as an external power. The energy balance (\ref{balance}) represents the first law of thermodynamics. To analyze the second law one computes the time derivative of the Shannon entropy $S(\mathbf{p}) = - \sum_{i=1}^d p_i \ln p_i$

\begin{equation}\label{}
  \dot{S}(t) = \frac 12 \sum_{i,j=1}^d \Big( W_{ij} p_j(t) - W_{ji} p_i(t)\Big) \ln \frac{p_j(t)}{p_i(t)} ,
\end{equation}
which  may be decomposed as follows

\begin{equation}\label{}
  \dot{S}(t) = \dot{\Sigma}(t) + \dot{\Phi}(t) ,
\end{equation}
where

\begin{equation}\label{s1}
  \dot{\Sigma}(t) = \frac 12 \sum_{i,j=1}^d \Big( W_{ij} p_j(t) - W_{ji} p_i(t) \Big) \ln \frac{ W_{ij}p_j(t)}{W_{ji}p_i(t)} ,
\end{equation}
is interpreted as the entropy production rate  \cite{Schnakerberg}, and

\begin{equation}\label{s2}
  \dot{\Phi}(t) = \frac 12 \sum_{i,j=1}^d (W_{ij} p_j(t) - W_{ji} p_i(t)) \ln \frac{ W_{ij}}{W_{ji}} ,
\end{equation}
is interpreted as the entropy flux rate. Using $(a-b) \ln (a/b) \geq 0$ for non-negative $a,b$ one immediately  proves that the entropy production rate (\ref{s1}) satisfies

\begin{equation}\label{}
  \dot{\Sigma}(t) \geq 0 ,
\end{equation}
which definitely shows that stochastic evolution governed by the Pauli rate equation (\ref{PAULI-2}) is compatible with the second law of thermodynamics. These definitions are justified by the corresponding formulae one obtains assuming that Kolmogorov generator satisfies the detailed balance condition

%represents a thermal equilibrium

\begin{equation}\label{}
  W_{ij} \overline{p}_j = W_{ji} \overline{p}_i ,
\end{equation}
with $\overline{\mathbf{p}}$ being a steady state of the classical evolution. One easily finds that (\ref{s1}) reduces to

\begin{equation}\label{s1a}
  \dot{\Sigma}(t) = - \frac{d}{dt} S(\mathbf{p}(t) |\!| \overline{\mathbf{p}}) ,
\end{equation}
where

\begin{equation}\label{}
  S(\mathbf{p} |\!| {\mathbf{q}}) = \sum_{i=1}^d p_i \ln\frac{p_i}{q_i} ,
\end{equation}
stands for the relative entropy or Kullback-Leibler divergence. The entropy flux rate simplifies to

\begin{equation}\label{s2a}
  \dot{\Phi} = \sum_{i=1}^d  \dot{p}_i \ln \overline{p}_i ,
\end{equation}
and when $\overline{\mathbf{p}}$ defines a thermal equilibrium, i.e. $\overline{p}_i = e^{-\beta E_i}/Z$, one finds

\begin{equation}\label{s2b}
  \dot{\Phi}(t) = -\beta \sum_{i=1}^d E_i(t) \dot{p}_i(t) =  -\beta \mathcal{J}(t) .
\end{equation}

This approach may be generalized for multiple reservoirs. Such scenario enables one to consider nonequilibrium steady states which are essential for many applications like e.g.  engines or refrigerators  \cite{QT0,QT3}. Assuming that different reservoirs do not interfere one has

\begin{equation}\label{Wnu}
  W_{ij} = \sum_\nu W^{(\nu)}_{ij} ,
\end{equation}
where $ W^{(\nu)}_{ij}$ represent rates corresponding to reservoir $\nu$. If moreover each reservoir is thermal at $T_\nu = 1/\beta_\nu$, then

\begin{equation}\label{WW-T}
  \frac{ W^{(\nu)}_{ij}}{W^{(\nu)}_{ji}} = e^{-\beta_\nu (E_i - E_j)} .
\end{equation}
The first law may me formulated as

\begin{equation}\label{balance2}
  \frac{d}{dt} \<E\> = \sum_k \frac{\partial{E_k}}{\partial t} p_k + \sum_k E_k \dot{p}_k = \<  \frac{\partial{E}}{\partial t} \> + \sum_\nu \mathcal{J}_\nu ,
\end{equation}
where $\mathcal{J}_\nu$ is a heat current to the $\nu$th bath (environment), i.e.

\begin{equation}\label{}
  \mathcal{J}_\nu(t) =   \sum_{i,j=1}^d E_i(t) \Big( W^{(\nu)}_{ij} p_j(t) - W^{(\nu)}_{ji} p_i(t) \Big) .
\end{equation}
Based on (\ref{Wnu}) one identifies the entropy production rate as follows \cite{Esposito-2010,Esposito-2015}

\begin{equation}\label{snu}
  \dot{\Sigma}(t) = \frac 12 \sum_{i,j=1}^d \sum_\nu \Big( W_{ij}^{(\nu)} p_j(t) - W_{ji}^{(\nu)} p_i(t)\Big) \ln \frac{ W_{ij}^{(\nu)}p_j(t)}{W_{ji}^{(\nu)}p_i(t)}  .
\end{equation}
Again, whenever (\ref{WW-T}) holds, then

\begin{equation}\label{snu}
  \dot{S}(t) = \dot{\Sigma}(t) + \sum_\nu \beta_\nu \mathcal{J}_\nu(t) .
\end{equation}
In the stationary state $\rho_{\rm ss}$ one has $\dot{S}=0$, and the second law reduces to

\begin{equation}\label{}
  \sum_\nu \beta_\nu \mathcal{J}_\nu \leq 0 .
\end{equation}

\subsection{Quantum semigroup}   \label{SUB-QS}

Consider now a Markovian semigroup in the weak coupling limit governed by Davies GKLS generator $\mathcal{L}$. Let $\rho_\beta = e^{- \beta H_S}/{\rm Tr}e^{- \beta H_S}$ be an equilibrium thermal state. Actually, the derivation of Davies weak coupling limit generator can be extended to describe driven systems with time dependent Hamiltonians $H_S(t)$  \cite{D-S}. The simplest scenario is the adiabatic case with a slowly varying $H_S(t)$, e.g. $H_S(t) = H_S + h(t)$. The quantum version of the first law reads as follows

\begin{equation}\label{}
  \frac{d}{dt} E(t) = \mathcal{P}(t) + \mathcal{J}(t) ,
\end{equation}
where the internal system's energy reads

\begin{equation}\label{}
  E(t) = {\rm Tr}(H_S(t) \rho_S(t)) ,
\end{equation}
together with the power $\mathcal{P}(t)$ and the heat current $\mathcal{J}(t)$

\begin{equation}\label{}
  \mathcal{P}(t) = {\rm Tr}(\frac{\partial H_S(t)}{\partial t} \rho_S(t)) , \ \ \ \ \mathcal{J}(t) =  {\rm Tr}(H_S(t) \dot{\rho}_S(t)) = {\rm Tr}(\rho_S(t) \mathcal{L}^\ddag(H_S(t))) .
\end{equation}
Following \cite{Spohn-Lebowitz} one defines the entropy production rate

\begin{equation}\label{}
  \dot{\Sigma}(t) = - \frac{d}{dt} S( e^{t\mathcal{L}}(\rho)|\!|\rho_\beta)
\end{equation}
where $S(\rho|\!|\sigma)$ denotes the relative entropy

\begin{equation}\label{Relative-S}
  S(\rho|\!|\rho') = \left\{ \begin{array}{ll} {\rm Tr}[\rho(\log\rho - \log\rho'])\ , & \ \mbox{if}\ {\rm supp}\, \rho \subseteq {\rm supp}\, \rho' \\ + \infty \ , & \ \mbox{otherwise} \end{array} \right. \ .
\end{equation}
Lindblad showed \cite{Lindblad} that $S(\rho|\!|\sigma) $ satisfies the data processing inequality \cite{QIT,Wilde,Hayashi}:  for any quantum channel $\mathcal{E} : \LH \to \LH$

\begin{equation}\label{mon-S}
   S(\rho|\!|\sigma) \geq  S(\mathcal{E}(\rho)|\!|\mathcal{E}(\sigma)) .
\end{equation}
Actually Uhlmann proved \cite{Uhlmann} that  (\ref{mon-S}) holds for any trace-preserving map $\mathcal{E}$ such that its dual $\mathcal{E}^\ddag$ is a Schwarz map. In particular, it holds for any trace-preserving 2-positive maps  \cite{Petz-RE}. Recently this result was generalized in \cite{Reeb} to arbitrary positive trace-preserving maps. Monotonicity property (\ref{mon-S})  implies that $\dot{\Sigma} \geq 0$ which is the statement of the second law of thermodynamics. One finds

\begin{equation}\label{pr-1}
  \dot{\Sigma}(t) = -{\rm Tr}( \mathcal{L}(\rho_t) \ln \rho_t)  - \beta \, {\rm Tr}(\rho_t \mathcal{L}^\ddag(H_S)) = \dot{S}(t) - \beta \mathcal{J}(t) .
\end{equation}
Positivity of $\dot{\Sigma}(t)$ follows from the Spohn inequality \cite{Spohn-Lebowitz}

\begin{equation}\label{}
  {\rm Tr}( \mathcal{L}(\rho)(\ln \rho - \ln \rho_{\rm ss}) \leq 0 ,
\end{equation}
where $\rho_{\rm ss}$ is a steady state.

Consider now a multiple reservoir case

\begin{equation}\label{LLL}
  \mathcal{L} = \mathcal{L}_1 + \ldots + \mathcal{L}_N ,
\end{equation}
and let $\rho_j = e^{-\beta_j H_S}/Z_j$ be a thermal state such that each pair $(\rho_j,\mathcal{L}_j)$ satisfies quantum detailed balance.
Following \cite{Spohn-Lebowitz} one defines the entropy production rate

\begin{equation}\label{}
  \dot{\Sigma}(t) = \sum_j \dot{\Sigma}_j(t)  ,
\end{equation}
with

\begin{equation}\label{pr-k}
  \dot{\Sigma}_k(t) = - {\rm Tr}( \mathcal{L}_k(\rho_t) \ln \rho_t) - \beta_k \, \rho  {\rm Tr}(\rho_t \mathcal{L}_k^\ddag(H_S) ) .
\end{equation}
It provides a natural generalization of (\ref{pr-1}). The entropy flux reads as follows

\begin{equation}\label{}
  \dot{\Phi}(t) = \sum_k \beta_k \mathcal{J}_k(t) ,
\end{equation}
where the heat flux to $k$th reservoir reads $\mathcal{J}_k(t) = {\rm Tr}(\rho_t \mathcal{L}_k^\ddag(H_S))$. Clearly,

\begin{equation}\label{}
  \dot{S}(t) = \dot{\Sigma}(t) - \sum_j \beta_j \mathcal{J}_j(t) .
\end{equation}
If $\rho_{\rm ss}$ is a steady state, then $ \dot{S}(\rho_{\rm ss})=0$ and hence the entropy production rate reads

\begin{equation}\label{}
   \dot{\Sigma}_{\rm ss} =  \sum_j \beta_j {\rm Tr}(\rho_{\rm ss} \mathcal{L}_j^\ddag(H_S)) ,
\end{equation}
that is, there is a constant entropy production rate in the nonequilibrium steady state  (NESS).

\subsection{Large deviation principle and full counting statistics}

Consider again a system weakly coupled to $N$ thermal reservoirs governed by  (\ref{LLL}). Introducing a counting (vector) field $\chi=(\chi_1,\ldots,\chi_N) \in \mathbb{R}^N$ one defines

\begin{equation}\label{}
  \mathcal{L}_{(\chi)}(\rho) = \sum_j \mathcal{L}_j(\rho \rho_j^{\chi_j}) \rho_j^{-\chi_j} .
\end{equation}
Note, that $\mathcal{L}_{(\chi=0)}=\mathcal{L}$, and

\begin{equation}\label{}
  \mathcal{L}_{(\chi=1)}(\rho) = \sum_j \widetilde{\mathcal{L}_j^\ddag}(\rho)  ,
\end{equation}
where $\widetilde{\mathcal{L}^\ddag}_j$ is the dual of $\mathcal{L}_j^\ddag$ w.r.t. inner product $(A,B)_j = {\rm Tr}(\rho_j A^\dagger B)$ (cf. (\ref{tilde})).

\begin{Proposition} If each pair $(\mathcal{L}_j,\rho_j)$ satisfies quantum detailed balance, then for any $\chi \in \mathbb{R}^N$ a family $\{e^{t  \mathcal{L}_{(\chi)}} \}_{t \geq 0}$ defines a semigroup of completely positive maps.
\end{Proposition}
Clearly, being completely positive $e^{t  \mathcal{L}_{(\chi)}}$ is not trace-preserving (unless $\chi=0$). The following formula

\begin{equation}\label{pt-rho}
  {\rm Tr}\Big( e^{t  \mathcal{L}_{(\chi)}}(\rho) \Big) = \int_{\mathbb{R}^N} e^{-t x\cdot \chi} p^t_\rho(x) dx
\end{equation}
where $x \cdot \chi = \sum_k x_k \chi_k$, defines a probability measure $p^t_\rho(x)$ on $\mathbb{R}^N$. If $\{ e^{t \mathcal{L}}\}_{t \geq 0}$ is relaxing to a unique steady state $\rho_{\rm ss}$, then $\{e^{t  \mathcal{L}_{(\chi)}} \}_{t \geq 0}$ is also relaxing to $\rho_{\rm ss}$ for any $\chi \in \mathbb{R}^N$. Moreover,

\begin{equation}\label{}
  \lim_{t \to \infty} \frac 1t \log  {\rm Tr}\Big( e^{t  \mathcal{L}_{(\chi)}}(\rho) \Big) = \lambda(\chi) ,
\end{equation}
where

\begin{equation}\label{}
  \lambda(\chi) = \max \{ {\rm Re} \lambda \, | \, \lambda \in {\rm sp}( \mathcal{L}_{(\chi)}) \} .
\end{equation}
Hence, $\lambda(\chi)$  is a simple eigenvalue of $\mathcal{L}_{(\chi)}$ and this operator has no other eigenvalues on the
line ${\rm Re} z = \lambda(\chi)$. Clearly, $\lambda(0)=0$. Let us define the Legendre-Fenchel transform of $\lambda(\chi)$

\begin{equation}\label{}
  I(x) = - \sup_{\chi \in \mathbb{R}^N} (x \cdot \chi - \lambda(\chi)) .
\end{equation}

%\begin{tcolorbox}
\begin{Proposition}
The probability distribution $p^t_\rho(x)$ defined in (\ref{pt-rho}) satisfies the Large Deviation Principle \cite{LDP} with the rate function $I(x)$, that is, for any subset $A \in \mathbb{R}^N$ one has

\begin{equation}\label{}
  - \inf_{x \in {\rm int}(A)} I(x) \leq \lim_{t\to \infty}\inf \frac 1t \log P^t_\rho(A)\leq \lim_{t\to \infty}\sup \frac 1t \log P^t_\rho(A) \leq  - \inf_{x \in {\rm cl}(A)} I(x) ,
\end{equation}
where int$(A)$ and cl$(A)$ denote the interior and the closure of the subset $A$, and $P^t_\rho(A) = \int_A p^t_\rho(x)dx$.

\end{Proposition}
%\end{tcolorbox}
Interestingly, $p^t_\rho(x)$ defines a probability distribution of $x_k = \Delta s_k/t$, i.e. the mean rate of entropy transport from the system initially in the state $\rho$ to the $k$th reservoir in the time interval $[0,t]$. Using full counting statistics approach \cite{FCS}
one obtains the following formula for the first moment of the distribution of $\Delta s_k$

\begin{equation}\label{}
  \< \Delta s_k \>_t = -  \frac{\partial}{\partial \chi_k} G_\rho(\chi,t)\Big|_{\chi=0} ,
\end{equation}
where
\begin{equation}\label{}
  G_\rho(\chi,t) = \ln  {\rm Tr}\Big( e^{t  \mathcal{L}_{(\chi)}}(\rho) \Big) ,
\end{equation}
is the cumulant generating function. The technique of full counting statistics provides a powerful method to characterize the dynamics of the open system using statistics of appropriate measurements. The detailed presentation of this technique is beyond a scope of this report (cf. e.g. \cite{FCS}).

\section{Quantum dynamical maps beyond Markovian semigroup}   \label{BEYOND}
              %Divisibility vs. Markovianity} \label{SEC-VI}

%This section discusses an important concept of divisible dynamical maps. Very often the evolution is considered to be Markovian whenever it is %represented by CP-divisible dynamical map. However, there are weaker notion of divisibility, like e.g. P-divisibility, which also display %interesting properties both from mathematical and physical point of view.

%We discuss divisibility in the Schr\"odinger and Heisenberg picture. In the latter case one may introduce an interesting concept of Schwarz %divisibility. Our analysis is illustrated by simple but instructive examples.

%\subsection{Schr\"odinger divisibility}

\subsection{Divisible dynamical maps}

Consider a general dynamical map $\Lambda_t : \TTH \to \TTH$ which is not necessarily a semigroup.

\begin{tcolorbox}
\begin{Definition}
A quantum dynamical map $\left\{ \Lambda_{t}\right\}_{t\geq 0}$ is called divisible if for any $t \geq s$

\begin{equation}\label{DIV}
  \Lambda_t = V_{t,s} \Lambda_s ,
\end{equation}
and $V_{t,s} : \TTH \to \TTH$. Moreover,

\begin{enumerate}
  \item $\{\Lambda_t\}_{t\geq 0}$ is called $k$-divisible if $V_{t,s}$ is $k$-positive and trace-preserving.
  \item In particular it is called CP-divisible if $V_{t,s}$ is CPTP, and
  \item P-divisible if $V_{t,s}$ is PTP.
\end{enumerate}
\end{Definition}
\end{tcolorbox}
This concept provides a direct generalization of a semigroup property for which one has   $V_{t,s} = e^{(t-s)\mathcal{L}}$, and hence a semigroup is obviously CP-divisible.  Note, that any invertible map $\Lambda_t$ is always divisible. Indeed, the corresponding  propagator $V_{t,s}$ can be uniquely defined by $V_{t,s} = \Lambda_t \Lambda_s^{-1}$. Moreover, the propagator $V_{t,s}$ satisfies a local composition law

\begin{equation}\label{CL}
  V_{t,u} V_{u,s} = V_{t,s} ,
\end{equation}
for $t \geq u \geq s$. In this paper following \cite{Wolf-Mar,Wolf-Mar2,Wolf-Mar3,RHP} we adopt the following

%\begin{tcolorbox}

\begin{Definition} \label{MARKOV} Quantum evolution represented by a dynamical map $\{\Lambda_t\}_{t\geq 0}$ is Markovian if and only if  the map is CP-divisible.
%Similarly, $\left\{ \Lambda_{t,t_0}\right\}_{t\geq t_0}$ is Markovian if it is CP-divisible for all $t_0$.
\end{Definition}

%\end{tcolorbox}
This definition uses only intrinsic properties of the system's evolution and does not refer to the total `system + environment' dynamics.
%In what follows to simplify the presentation we consider time homogeneous case. However, the same analysis applies for time non homogeneous case %as well.
Note, that  Definition \ref{MARKOV} when applied to a classical dynamical map represented by a family of stochastic matrices $\{T(t)\}_{t \geq 0}$ implies that it is Markovian if

\begin{equation}\label{TTT}
  T(t) = T(t,s) T(s)  \ , \ \ \  t \geq s ,
\end{equation}
and $T(t,s)$ is a stochastic matrix.  The matrix elements $T_{ij}(t,s) \equiv p(i,t|j,s)$ can be interpreted as conditional probability.
%  or 2-point correlation function.

Another related concept is a Markovianity of a single quantum channel $\mathcal{E} : \LH \to \LH$. Authors of  \cite{Wolf-Mar2} call a channel $\mathcal{E}$ Markovian if $\mathcal{E} = e^{\mathcal{L}}$ for some GKLS generator $\mathcal{L}$. Interestingly, the problem of  deciding Markovianity of $\mathcal{E}$ turns out to be NP-hard \cite{Wolf-Mar3}. Actually, similar problem for stochastic matrices (so called embedding problem) --- given $T$ decide whether  ${T}= e^K$ for some Kolmogorov generator $K$ --- is NP-hard as well (cf.  recent papers \cite{Mario,Puchala} see also \cite{Sergey-2017}).

%\begin{Proposition}\label{PRO-Koss} A linear trace-preserving and Hermiticity-preserving map $\Phi : \LH \to \LH$ is positive if and only if

%\begin{equation}\label{CONTR}
%  \| \Phi(X) \|_1 \leq \| X\|_1 ,
%\end{equation}
%foe all Hermitian $X \in \LH$.
%\end{Proposition}

Divisibility of $\{\Lambda_t\}_{t \geq 0}$ is characterized by the following monotonicity property \cite{Angel,Sabrina}

%\begin{tcolorbox}
\begin{Theorem} \label{TH-Div-k}  Let us assume that $\{\Lambda_t\}_{t\geq 0}$ is an invertible dynamical map. Then $\Lambda_t$ is $k$-divisible if and only if
\begin{equation}\label{kDIV}
  \frac{d}{dt} \| [{\rm id}_k \otimes \Lambda_t](X) \|_1 \leq  0 ,
\end{equation}
for any Hermitian $X \in M_k(\mathbb{C}) \otimes \mathcal{T}(\mathcal{H})$.
\end{Theorem}
%\end{tcolorbox}
Proof: invertibility guarantees the existence of a propagator $ V_{t+\epsilon,t}$ for any $\epsilon > 0$. Hence

\begin{eqnarray*}\label{}
    \frac{d}{dt} \| [{\rm id}_k \otimes \Lambda_t](X) \|_1 &=& \lim_{\epsilon \to 0+} \frac{1}{\epsilon} \Big(  \| [{\rm id}_k \otimes \Lambda_{t+\epsilon}](X) \|_1 -  \| [{\rm id}_k \otimes \Lambda_t](X) \|_1  \Big) \\ &=& \lim_{\epsilon \to 0+} \frac{1}{\epsilon}
  \Big(  \| [{\rm id}_k \otimes V_{t+\epsilon,t} \Lambda_{t}](X) \|_1 -  \| [{\rm id}_k \otimes \Lambda_t](X) \|_1  \Big) \\
  &= &   \lim_{\epsilon \to 0+} \frac{1}{\epsilon}  \Big(  \| [{\rm id}_k \otimes V_{t+\epsilon,t}](Y)  \|_1 -  \| Y \|_1  \Big) ,
\end{eqnarray*}
with $   Y =  [{\rm id}_k \otimes \Lambda_t](X)$.  Now, assuming $k$-divisibility the propagator ${\rm id}_k \otimes V_{t+\epsilon,t}$ is PTP and hence $ \| [{\rm id}_k \otimes V_{t+\epsilon,t}](Y) \|_1 \leq  \| Y \|_1$ which implies (\ref{kDIV}). Conversely, if (\ref{kDIV}) holds for any Hermitian $X$, then due to Proposition \ref{PRO-I}
${\rm id}_k \otimes V_{t+\epsilon,t}$ is positive since (due to invertibility)  $ \| [{\rm id}_k \otimes V_{t+\epsilon,t}](Y) \|_1 \leq  \| Y \|_1$ for all $Y \in M_k(\mathbb{C}) \otimes \LH$. Hence, $\Lambda_t$ is $k$-divisible. \hfill $\Box$

\begin{Cor} An invertible dynamical map $\{\Lambda_t\}_{t\geq 0}$ is P-divisible if and only if

\begin{equation}\label{P-div}
   \frac{d}{dt} \|  \Lambda_t(X) \|_1 \leq  0 ,
\end{equation}
for any Hermitian $X \in \mathcal{T}(\mathcal{H})$.
\end{Cor}

Note, that if the dynamical map $\{\Lambda_t\}_{t \geq 0}$ satisfies commutativity condition $[\Lambda_t,\Lambda_s]=0$, then the corresponding eigenvectors of $\Lambda_t$

\begin{equation}\label{}
  \Lambda_t(X_\alpha) = \lambda_\alpha(t) X_\alpha ,
\end{equation}
are time independent and  hence
\begin{equation}\label{}
   \frac{d}{dt} \|  \Lambda_t(X_\alpha) \|_1 =   \frac{d}{dt} |  \lambda_\alpha(t)| \, \| X_\alpha \|_1 .
\end{equation}
It is, therefore, clear that condition (\ref{P-div}) implies

\begin{equation}\label{dt-lambda}
  \frac{d}{dt} |  \lambda_\alpha(t)|  \leq 0 , \ \ \ \ \alpha=1,\ldots,d^2-1 .
\end{equation}
In the case of Markovian semigroup  one has $\lambda_\alpha(t) = e^{-(i\omega_\alpha + \Gamma_\alpha)t}$ and hence

\begin{equation}\label{dt-lambda}
  \frac{d}{dt} |  \lambda_\alpha(t)|  = -\Gamma_\alpha e^{- \Gamma_\alpha t} \leq 0 ,
\end{equation}
due to $\Gamma_\alpha \geq 0$.

\begin{figure} \label{}
\begin{center}
%\hspace*{.4cm}
\includegraphics[width=7cm]{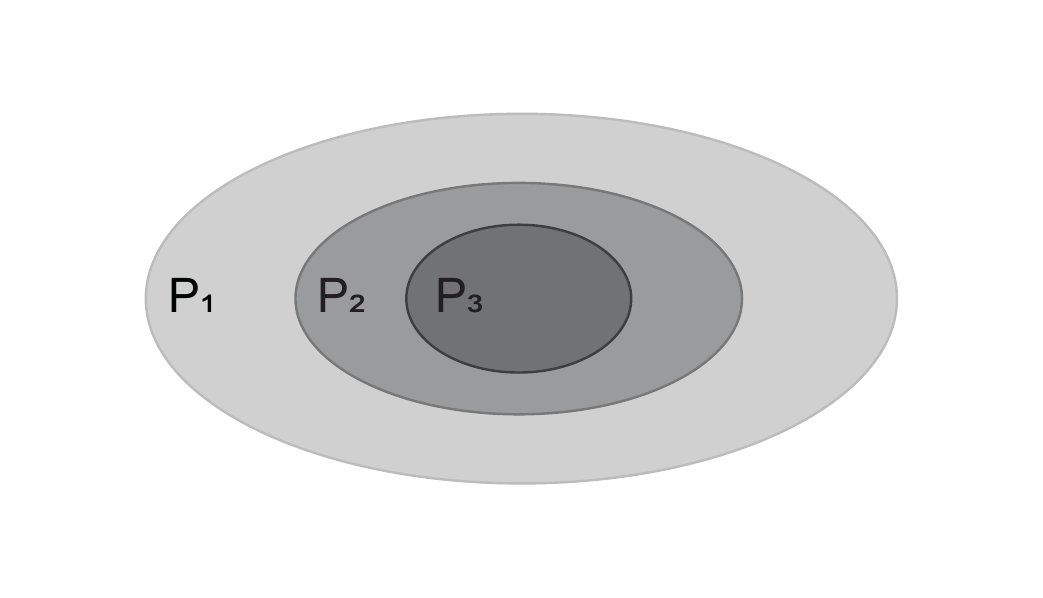}  \hspace{.5cm}  \includegraphics[width=7cm]{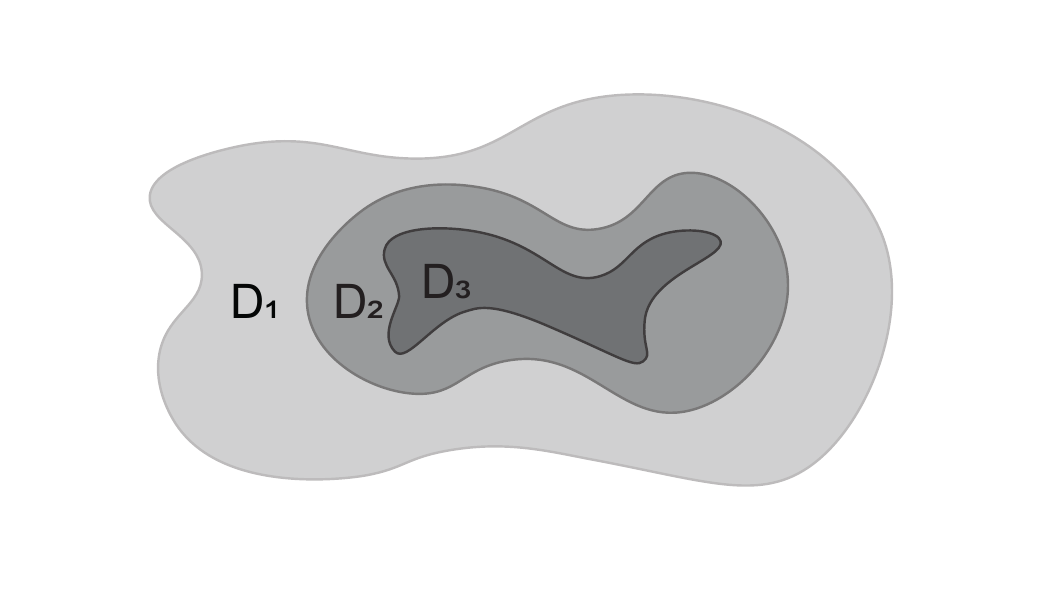}
\caption{(Color online) Left panel: convex sets $\mathcal{P}_k$. Right panel: not convex sets $\mathcal{D}_k$.  } \label{2ab}
\end{center}
\end{figure}

\begin{Cor} An invertible classical dynamical $\{T(t)\}_{t \geq 0}$ is P-divisible if and only if

\begin{equation}\label{}
  \frac{d}{dt} \| T(t) \mathbf{x} \|_1 \leq 0 ,
\end{equation}
for any vector $\mathbf{x} \in \mathbb{R}^d$.

%The classical version of BLP condition reads
%\begin{equation}\label{}
%  \frac{d}{dt} \| T(t)( \mathbf{p}_1 - \mathbf{p}_2) \|_1 \leq 0 ,
%\end{equation}
%where $\mathbf{p}_1$ and $\mathbf{p}_2$ are probability distributions.

\end{Cor}
Denoting by $\mathcal{D}_k$ a set of $k$-divisible maps one has the following hierarchy of inclusions

$$   {\rm Markovian} = \mathcal{D}_d \subset \mathcal{D}_{d-1} \subset \ldots \subset \mathcal{D}_1 = \mbox{P-divisible} \subset \mbox{all maps}
. $$
This hierarchy is an analog of (\ref{k-hierarchy}). There is, however, an important difference: a set $\mathcal{P}_k$ of $k$-positive maps is convex. It is no longer true for a set of $k$-divisible maps $\mathcal{D}_k$ (cf. Figure \ref{2ab}).

In \cite{Sabrina} a concept of non-Markovianity degree (NMD) was introduced: CP-divisible maps have NMD$=0$, and $(d-k)$-divisible maps have NMD$=k$. Dynamical maps which are at least P-divisible were called weakly non-Markovian and these which are not even P-divisible --- essentially non-Markovian. These  concepts were experimentally tested in \cite{Sciarrino} (cf. also \cite{Budini-MAX}). Another equivalent characterization of CP-divisibility is provided in \cite{BOGNA}

%\begin{tcolorbox}
\begin{Theorem} \label{B} Let us assume that $\{\Lambda_t\}_{t \geq 0}$ is an invertible dynamical map. Then $\{\Lambda_t\}_{t \geq 0}$ is CP-divisible if and only if
\begin{equation}\label{Bogna}
  \frac{d}{dt} \| [{\rm id}_{d+1} \otimes \Lambda_t]({\rho}_1-{\rho}_2)\|_1 \leq  0 ,
\end{equation}
for any pair of density operators ${\rho}_1$, ${\rho}_2$ in $\mathbb{C}^{d+1}  \otimes \mathcal{H}$.
\end{Theorem}
%\end{tcolorbox}
Hence,  enlarging the dimension of the ancilla $d \rightarrow d+1$,  one can restrict oneself to traceless operators only.  Actually, there is an intriguing relation between P- and CP-divisibility \cite{Fabio}

\begin{Proposition} A dynamical map $\{\Lambda_t\}_{t \geq 0}$ is CP-divisible if and only if $\{\Lambda_t \otimes \Lambda_t\}_{t \geq 0}$ is P-divisible.
\end{Proposition}
If $\{\Lambda_t\}_{t \geq 0}$ is a Markovian semigroup then this result shows that $e^{t \mathcal{L}}$ is CPTP if and only if $e^{t \mathcal{L}}\otimes e^{t \mathcal{L}}$ is PTP \cite{Fabio-2}.

\subsection{Divisibility vs. time-local generator}

Suppose that a dynamical map $\{\Lambda_t\}_{t \geq 0}$ satisfies time-local master equation $\dot{\Lambda}_t = \mathcal{L}_t \Lambda_t$.
Note, that given $\{\Lambda_t\}_{t \geq 0}$ one can formally find the corresponding time-local generator $\mathcal{L}_t= \dot{\Lambda}_t \Lambda_t^{-1}$ \cite{Fulinski,Ful-Kra,Shibata,PRL-2010}. A natural question arises: what are the properties of $\mathcal{L}_t$ which guarantee that $\Lambda_t$ is $k$-divisible? Let us consider invertible dynamical maps for which  the generator $\mathcal{L}_t= \dot{\Lambda}_t \Lambda_t^{-1}$ is regular.

\begin{Proposition} A regular time-local generator $\mathcal{L}_t$ gives rise to $k$-divisible dynamical map if and only if

\begin{equation}\label{k-Lt}
  P_\perp ([{\rm id}_k \otimes \mathcal{L}_t](P)) P_\perp \geq 0 ,
\end{equation}
for any rank-1 projector $P$ in $\mathbb{C}^k \otimes \mathcal{H}$ ($P_\perp$ is an orthogonal complement of $P$).
\end{Proposition}
Hence a regular time-local generator $\mathcal{L}_t$ gives rise to P-divisible dynamical map if and only if

\begin{equation}\label{P-Lt}
  P_\perp \mathcal{L}_t(P) P_\perp \geq 0 ,
\end{equation}
for any rank-1 projector $P$ in $\mathcal{H}$. Condition (\ref{P-Lt}) was used to provide stochastic unraveling of time-local master equation \cite{Diosi1988} (see also \cite{Caiaffa2017}). Condition (\ref{k-Lt}) considerably simplifies for CP-divisibility. One  proves

\begin{Proposition} A regular time-local generator $\mathcal{L}_t$ gives rise to CP-divisible dynamical map if and only if

\begin{equation}\label{d-Lt}
  P^+_\perp ([{\rm id}_d \otimes \mathcal{L}_t](P^+)) P^+_\perp \geq 0 ,
\end{equation}
where $P^+$ is a maximally entangled projector in  $\mathbb{C}^d \otimes \mathcal{H}$.
\end{Proposition}
Hence, to guarantee CP-divisiblity (i.e. Markovianity) it is sufficient to satisfy (\ref{k-Lt}) just for a single projector $P^+$. This is an analog of Choi condition for complete positivity.

\begin{Cor} A regular generator $\mathcal{L}_t$ gives rise to CP-divisible evolution if and only if it has a GKLS form for all $t \geq 0$

\begin{equation}\label{L-DIAGt}
  \mathcal{L}_t(\rho) = - i[H(t),\rho] + \sum_{k=1}^{d^2-1} \gamma_k(t) \Big( L_k(t) \rho L_k^\dagger(t) - \frac 12\{ L_k^\dagger(t)L_k(t),\rho\} \Big) ,
\end{equation}
with $\gamma_k(t) \geq 0$ for all $t \geq 0$.
\end{Cor}
It shows that essentially CP-divisible dynamical maps are solutions of time-local master equation with time-dependent generator $ \mathcal{L}_t$ having a GKLS form for any $t \geq 0$. The corresponding propagator $V_{t,s}$ reads as follows

\begin{equation}\label{}
  V_{t,s} = \mathcal{T} \exp\left( \int_s^t \mathcal{L}_\tau d\tau \right) ,
\end{equation}
and by construction it does satisfy the local composition low (\ref{CL}).

\subsection{Distinguishability of states and information back flow}   \label{SUB-BLP}

Given two states $\rho_1$ and $\rho_2$ one defines their natural distance \cite{Holevo,Ziman-Teiko}

\begin{equation}\label{D12}
  D(\rho_1,\rho_2) = \frac 12 \|\rho_1 - \rho_2\|_1 ,
\end{equation}
and calls $D(\rho_1,\rho_2)$  distinguishability of $\rho_1$ and $\rho_2$. Such definition implies that $0 \leq  D(\rho_1,\rho_2) \leq 1$. In particular $ D(\rho_1,\rho_2)=0$ if and only if $\rho_1=\rho_2$ and $ D(\rho_1,\rho_2)=1$ if and only if $\rho_1 \perp \rho_2$ (i.e. $\rho_1$ and $\rho_2$ are orthogonally supported) and in this case these two states are perfectly distinguishable. If $\{\Lambda_t\}_{t \geq 0}$ is Markovian, then

\begin{equation}\label{BLP}
  \sigma(\rho_1,\rho_2;t) =\frac{d}{dt} D(\Lambda_t(\rho_1),\Lambda_t(\rho_2)) \leq 0 ,
\end{equation}
for any pair of initial states $\rho_1$ and $\rho_2$. The above condition was proposed in a seminal paper \cite{BLP} as a condition for Markovianity of the evolution represented by the map $\{\Lambda_t\}_{t\geq 0}$. Breuer, Laine and Piilo \cite{BLP} interpreted  a violation of
(\ref{BLP}) as an information backflow from the environment into the system, that is, whenever $ \sigma(\rho_1,\rho_2;t) < 0$ then the system looses information and the information flows from the system to the environment. However, whenever $ \sigma(\rho_1,\rho_2;t) >0$ the information flows back from the environment into the open system. This suggestive interpretation provides a powerful tool in characterizing non-Markovian memory effects. In what follows we call the condition (\ref{BLP}) derived by Breuer, Laine and Piilo  the BLP condition. Equivalently,  BLP condition may be rephrased as (\ref{P-div}) for all Hermitian traceless operators $X$ which shows that it is weaker than P-divisibility condition (\ref{P-div}) which has to be satisfied for all Hermitian $X$. Violation of BLP condition provides a clear sign of non-Markovian memory effects \cite{NM2}.

Note, that if the pair of states $\{\rho_1,\rho_2\}$ is provided with the corresponding probabilities $\{p_1,p_2\}$, respectively, such that $p_1 + p_2=1$, then the distinguishability of such pair is defined as \cite{Helstrom}

\begin{equation}\label{}
  D(\{\rho_1,p_1\},\{\rho_2,p_2\}) =  \|p_1\rho_1 - p_2\rho_2\|_1 ,
\end{equation}
which recovers (\ref{D12}) if $p_1=p_2=1/2$. If the dynamical map is invertible and

\begin{equation}\label{}
  \frac{d}{dt} \|  \Lambda_t(p_1\rho_1 - p_2 \rho_2) \|_1  \leq 0 ,
\end{equation}
for all pairs $\{\rho_1,p_1\},\{\rho_2,p_2\}$, then $\{\Lambda_t\}_{t \geq 0}$ is P-divisible \cite{Angel}. Hence, P-divisibility is restored if we generalize a discrimination scenario for the biased case $p_1 \neq p_2$.

Consider now the unitary evolution of the `system + environment' and the corresponding reduced system's dynamics

\begin{equation}\label{}
  \Lambda_t(\rho_S) = {\rm Tr}_E (U_t \rho_S \otimes \rho_E U_t^\dagger) .
\end{equation}
Given two initial state $\rho_S$ and $\sigma_S$ suppose that distinguishability is not monotonic, i.e. for some $t > s$ one has

\begin{equation}\label{DD>}
   D(\Lambda_t(\rho_S),\Lambda_t(\sigma_S)) - D(\Lambda_s(\rho_S),\Lambda_s(\sigma_S)) > 0 ,
\end{equation}
which provides a clear sign of information backflow. One may ask whether there exists a natural upper bound for the l.h.s. of (\ref{DD>}). Interestingly, one finds  \cite{Bassano-2021} the following result

\begin{eqnarray}\label{}
 && D(\Lambda_t(\rho_S),\Lambda_t(\sigma_S)) - D(\Lambda_s(\rho_S),\Lambda_s(\sigma_S)) \nonumber \\
 && \leq D(\rho_E(s),\sigma_E(s)) + D(\rho_{SE}(s),\rho_S(s) \otimes \rho_E(s)) + D(\sigma_{SE}(s),\sigma_S(s) \otimes \sigma_E(s)) ,
\end{eqnarray}
where $\rho_{SE}(t) = U_t  \rho_S \otimes \rho_E U_t^\dagger$, $\sigma_{SE}(t) = U_t  \sigma_S \otimes \rho_E U_t^\dagger$, and similarly for $\rho_{SE}(s)$ and $\sigma_{SE}(s) $. Hence the gain of information from times `$s$' to time `$t$' is upper bounded by the sum of three contributions at time `$s$': the distance between reduced environmental states  $D(\rho_E(s),\sigma_E(s))$, and the distances between the total states $\rho_{SE}(s)$ and $\sigma_{SE}(s)$ and the corresponding products of marginals (reduced states) $\rho_S(s) \otimes \rho_E(s)$ and $\sigma_S(s) \otimes \sigma_E(s)$, respectively.

\subsection{Non-Markovianity measures}

How to measure the departure from Markovianity? There are several measures of non-Markovianity proposed in the literature (cf. \cite{NM1,NM2,NM3} for the detailed review). All these measures are based on distinct concept of Markovianity. Since in this report we concentrate on the notion of Markovianity based on CP-divisibility let us recall the measure already proposed by Rivas, Huelga and Plenio \cite{RHP} (henceforth the RHP measure). It is defined as follows: let

\begin{equation}\label{}
  g(t) := \lim_{\epsilon \to 0+} \frac{ \| [{\rm id}_d \otimes V_{t+\epsilon,t}](|\psi^+_d\>\<\psi^+_d|)\|_1 - 1}{\epsilon} ,
\end{equation}
where $|\psi^+_d\>$ stands for a maximally entangled state in $\HH \otimes \HH$ (and $d = {\rm dim}\ \HH$). Clearly, whenever $g(t) > 0$, then the propagator $V_{t+\epsilon,t}$ is not CP and hence the evolution is non-Markovian (not CP-divisible). Finally, one defines \cite{RHP}

\begin{equation}\label{}
  \mathcal{N}_{\rm RHP} = \int_0^\infty g(t) dt .
\end{equation}
Sometimes one uses a normalized version $ \mathcal{N}_{\rm RHP}/(1+ \mathcal{N}_{\rm RHP} )$ which tends to one when $\mathcal{N}_{\rm RHP}$ is infinite. This measure was computed for several examples of quantum dynamical maps (cf. e.g. \cite{rhp-1,rhp-2,rhp-3,rhp-4}). Assuming that $\Lambda_t$ satisfies time-local mater equation CP-divisibility is controlled by the properties of the time-local generator. One immediately finds \cite{RHP}

\begin{equation}\label{}
  g(t) := \lim_{\epsilon \to 0+} \frac{ \| [{\rm id}_d \otimes ({\rm id}_d + \epsilon \mathcal{L}_{t})](|\psi^+_d\>\<\psi^+_d|)\|_1 - 1}{\epsilon} .
\end{equation}
As was shown in \cite{Erika} the quantity $g(t)$ is directly related to the transition rates $\gamma_k(t)$ in the canonical diagonal representation (\ref{L-DIAGt}): define

\begin{equation}\label{}
  f_k(t) = \max \{-\gamma_k(t),0\} ,
\end{equation}
and let $f(t) = \sum_{k=1}^{d^2-1} f_k(t)$. It turns out \cite{Erika} that $f(t) = \frac d2 g(t)$.

According to (\ref{d-Lt})  CP-divisibility is equivalent to the following condition

\begin{equation}\label{CCP-1}
 \mathbb{L}(t) := (\oper -  P^+_d) [{\rm id}_d \otimes \mathcal{L}_t](P^+_d) (\oper - P^+_d) \geq 0 ,
\end{equation}
for any $t \geq 0$. Now, if this condition is violated, then

\begin{equation}\label{}
   \mathbb{L}(t) =  \mathbb{L}_+(t) -  \mathbb{L}_-(t) := | \mathbb{L}(t)| - (| \mathbb{L}(t)| -  \mathbb{L}(t)) ,
\end{equation}
and $ \mathbb{L}_\pm(t) \geq 0$. Whenever $ \mathbb{L}_-(t) \neq 0$ for some $t$ then the evolution is non-Markovian. One finds the following formula which does not depend on a particular representation of the generator \cite{Erika}

\begin{equation}\label{}
  g(t) = \frac 1d \, {\rm Tr}\,   \mathbb{L}_-(t) .
\end{equation}
Similar measure may be introduced to  quantify the departure  from $k$-divisibility. It was originally introduced in \cite{Sabrina}. Here we propose a slight modification which reproduces RHP measure if $k=d$.
Any regular generator $\mathcal{L}_t$ gives rise to $k$-divisible evolution if and only if (\ref{k-Lt}) is satisfied. Let

\begin{equation}\label{}
  \mathbb{L}^{(k)}(P,t) :=  P_\perp ([{\rm id}_k \otimes \mathcal{L}_t](P)) P_\perp .
\end{equation}
One has again the following decomposition

\begin{equation}\label{}
  \mathbb{L}^{(k)}(P,t) = \mathbb{L}_+^{(k)}(P,t) - \mathbb{L}_-^{(k)}(P,t) := | \mathbb{L}^{(k)}(P,t) | - ( | \mathbb{L}^{(k)}(P,t) | - \mathbb{L}^{(k)}(P,t) ) ,
\end{equation}
and finally introducing

\begin{equation}\label{}
  g^{(k)}(P,t) = \frac 1k \, {\rm Tr}\,   \mathbb{L}_-^{(k)}(P,t) ,
\end{equation}
one defines

\begin{equation}\label{}
  \mathcal{N}^{(k)} = \sup_P \int_0^\infty  g^{(k)}(P,t) dt ,
\end{equation}
where the supremum is over all rank-1 projectors in $\mathbb{C}^k \otimes \HH$.

\begin{Cor} A dynamical map $\{ \Lambda_t \}_{t \geq 0}$ is $k$-divisible if and only if $ \mathcal{N}^{(k)} = 0$.
\end{Cor}
One has the following hierarchy of measures

\begin{equation}\label{}
   \mathcal{N}^{(1)} \leq  \mathcal{N}^{(2)} \leq \ldots \leq  \mathcal{N}^{(d-1)} \leq  \mathcal{N}^{(d)} ,
\end{equation}
and the evolution is essentially non-Markovian \cite{Sabrina} if $\mathcal{N}^{(1)}> 0$, i.e. the dynamical map $\{\Lambda_t\}_{t \geq 0}$ is not even P-divisible.

 Another very popular measure (so called BLP measure) was proposed Breuer, Laine and Piilo \cite{BLP}. Strictly speaking, when we adopted the definition of Markovianity as a CP-divisibility of the corresponding dynamical map, the  BLP measure is not a measure of non-Markovianity but a measure of information backflow. It is defined as follows \cite{BLP}

\begin{equation}\label{}
  \mathcal{N}_{\rm BLP} = \max_{\rho_1,\rho_2} \int_{\sigma < 0} \sigma(\rho_1,\rho_2;t) dt
\end{equation}
where the information flow $\sigma(\rho_1,\rho_2;t)$ is defined in (\ref{BLP}),
and the supremum is over all pairs $\{\rho_1,\rho_2\}$ of system's density operators. Note, however, that even if $\mathcal{N}_{\rm BLP}=0$ the evolution might be still non-Markovian, i.e. not CP-divisible. It is, therefore, clear that $\mathcal{N}_{\rm BLP} =0$ defines a necessary condition for Markovianity. The problem of an optimal pair $\{\rho_1,\rho_2\}$ was analyzed in \cite{BLP-optimal,Liu}. It turns out \cite{Liu} that

\begin{equation}\label{}
  \mathcal{N}_{\rm BLP} = \max_{\rho \in \partial U(\rho_0) } \int_{\tilde{\sigma} < 0} \tilde{\sigma}(\rho,\rho_0;t) dt ,
\end{equation}
where

\begin{equation}\label{}
  \tilde{\sigma}(\rho,\rho_0;t)  = \frac{\sigma(\rho,\rho_0;t)}{ \| \rho-\rho_0\| }  ,
\end{equation}
and $\rho_0$ is a fixed state in the interior of the subset of states enclosed by an arbitrary closed surface $\partial U(\rho_0) $.

%This measure was generalized in \cite{Sabrina} to witness the violation of $k$-divisibility

\subsection{Examples: qubit evolution}

For a qubit evolution  the hierarchy of $k$-divisible evolutions reduces to

$$   {\rm Markovian} = \mathcal{D}_2  \subset  \mathcal{D}_{1}  \subset \mathcal{D}_{\rm BLP}  \subset \mbox{all qubit dynamical maps} . $$
Using the Bloch representation

\begin{equation}\label{}
  \rho = \frac 12 (\oper_2 + \mathbf{x} \cdot\mbox{\boldmath$\sigma$} ) ,
\end{equation}
one finds for the trace norm of $\Delta\rho =\rho_1 - \rho_2 = \frac 12 (\mathbf{x}_1-\mathbf{x}_2) \cdot\mbox{\boldmath$\sigma$}  $

\begin{equation}\label{}
  \| \Delta \rho \|_1^2 = ({\rm Tr} |\Delta\rho |)^2 =  |\Delta \mathbf{x}|^2 ,
\end{equation}
where $\Delta \mathbf{x} = \mathbf{x}_1-\mathbf{x}_2$. The evolution of the Bloch vector is governed by the following equation

\begin{equation}\label{}
 \dot{ \mathbf{x}}(t) = D(t) \mathbf{x}(t) + \mathbf{r}(t) ,
\end{equation}
where the real $3 \times 3$ matrix $D(t)$ is defined via

\begin{equation}\label{}
  D_{kl}(t) := \frac 12 {\rm Tr}(\sigma_k \mathcal{L}_t(\sigma_\ell)) ,
\end{equation}
and the drift vector reads

\begin{equation}\label{}
  \mathbf{r}(t) := \frac 12  {\rm Tr}(\mbox{\boldmath$\sigma$} \mathcal{L}_t(\oper))  = \frac 12 {\rm Tr} \mathcal{L}_t^\ddag( \mbox{\boldmath$\sigma$}) .
\end{equation}
One finds therefore

\begin{equation}\label{}
 \frac{d}{dt} \| \Delta\rho_t \|_1^2 = \frac {d}{dt}  |\Delta \mathbf{x}(t)|^2 =  \Delta \mathbf{x}^T(t) (D(t) + D^T(t))  \Delta \mathbf{x}(t) ,
\end{equation}
and hence BLP condition (\ref{BLP}) is equivalent to

\begin{equation}\label{D(t)}
  D(t) + D^T(t) \leq 0 ,
\end{equation}
for all $t \geq 0$.
% This condition is a special case of dissipativity condition (\ref{Diss-1}).
    In what follows we analyze two important examples of qubit evolution: commutative Pauli-like dynamics and non-commutative phase covariant dynamics.

\subsubsection{Pauli channel-like dynamics}

 Consider a qubit evolution governed by

\begin{equation}\label{sss}
  \mathcal{L}_t(\rho) = \frac 12 \sum_{k=1}^3 \gamma_k(t) ( \sigma_k \rho \sigma_k - \rho) .
\end{equation}
Note, that $\mathcal{L}_t$ is commutative $\mathcal{L}_t \mathcal{L}_s = \mathcal{L}_s \mathcal{L}_t$, and self-dual $\mathcal{L}_t^\ddag = \mathcal{L}_t$. The corresponding evolution is represented by the following dynamical map

\begin{equation}\label{Pauli-t}
  \Lambda_t(\rho) = \sum_{\alpha=0}^{3} p_\alpha(t) \sigma_\alpha \rho \sigma_\alpha ,
\end{equation}
where $p_\alpha(t)$ have the following form

\begin{equation}\label{}
  \left( \begin{array}{c} p_0(t) \\ p_1(t) \\ p_2(t) \\ p_3(t) \end{array} \right) =  \frac 14 \left( \begin{array}{cccc} 1 & 1 & 1 & 1 \\ 1 & 1 & - 1 & -1 \\ 1 & -1 & 1 & -1 \\ 1& -1&-1& 1 \end{array} \right) \left( \begin{array}{c} 1 \\ \lambda_1(t) \\ \lambda_2(t) \\ \lambda_3(t) \end{array} \right) ,
\end{equation}
together with

\begin{equation}\label{}
  \lambda_1(t) = e^{-\Gamma_2(t) - \Gamma_3(t)} \ , \ \ \lambda_2(t) = e^{-\Gamma_3(t) - \Gamma_1(t)} \ , \ \ \lambda_3(t) = e^{-\Gamma_1(t) - \Gamma_2(t)} \ ,
\end{equation}
and $\Gamma_k(t) = \int_0^t \gamma_k(\tau)d\tau$ for $k=1,2,3$. Moreover $\lambda_k(t)$ are eigenvalues of the dynamical map $\,\Lambda_t(\sigma_k) = \lambda_k(t) \sigma_k$. Clearly, the map (\ref{Pauli-t}) is a family of time-dependent Pauli channels provided that $p_\alpha(t) \geq 0$ for $t \geq 0$. This condition implies nontrivial constraints for the transition rates $\gamma_k(t)$. Due to commutativity one finds the following representation of the dynamical map

\begin{equation}\label{}
  \Lambda_t = e^{ \int_0^t \mathcal{L}_\tau d\tau } = e^{\frac 12 \Gamma_1(t) \mathcal{L}_1} \,e^{\frac 12 \Gamma_2(t) \mathcal{L}_2} \, e^{\frac 12 \Gamma_3(t) \mathcal{L}_3} ,
\end{equation}
with $\mathcal{L}_k(\rho) = \sigma_k \rho\sigma_k - \rho$. What are conditions for $\gamma_k(t)$ which guarantee that $\{\Lambda_t\}_{t \geq 0}$ is CPTP? If $\int_0^t \mathcal{L}_\tau d\tau $ is a GKLS generator for all $t \geq 0$, then $\{\Lambda_t\}_{t \geq 0}$ is a legitimate dynamical map. Note, however, that the above condition is only sufficient but not necessary. It may happen that $\{\Lambda_t\}_{t\geq 0}$ is CPTP even if temporally $\int_0^t \mathcal{L}_\tau d\tau $ fails to be a GKLS generator. In general to decide whether $\mathcal{L}_t$ is an admissible generator one has to check for positivity of all $p_\alpha(t)$ for all $t \geq 0$. Clearly, the evolution is CP-divisible if and only if all $\gamma_k(t) \geq 0$. Now, to analyze BLP condition (\ref{BLP}) one needs to analyze the corresponding matrix $D(t)$

\begin{equation}\label{}
  D(t) =  \left( \begin{array}{ccc} \gamma_1(t) - \gamma(t) & 0 & 0 \\ 0 & \gamma_2(t) - \gamma(t) & 0 \\ 0 & 0 & \gamma_3(t) - \gamma(t)\end{array} \right) ,
\end{equation}
with $\gamma(t)=\gamma_1(t) +\gamma_2(t)+\gamma_3(t)$. Clearly,  BLP condition states that $\gamma(t) - \gamma_k(t)\geq 0$, or equivalently

\begin{equation}\label{ggg}
  \gamma_1(t) + \gamma_2(t) \geq 0 , \ \  \gamma_2(t) + \gamma_3(t) \geq 0 , \ \  \gamma_1(t) + \gamma_3(t) \geq 0 ,
\end{equation}
for all $t\geq 0$ \cite{Filip-PLA,Filip-PRA,Erika}. Interestingly, in this case BLP condition coincides with P-divisibility. It is evident that in order  to satisfy (\ref{ggg}) at most one $\gamma_k(t)$ can be negative for a given moment $t$. An interesting example of qubit evolution was proposed in \cite{Erika} for which $\gamma_1=\gamma_2=1$ and $\gamma_3(t) = -{\rm tanh}\, t$. Due to the fact that one of the rates is always negative (for $t>0$) the corresponding evolution is called {\em eternally   non-Markovian} \cite{Erika}. One easily finds the representation (\ref{Pauli-t}) with

\begin{equation}\label{ppp-Erika}
  p_0(t) = \frac{1}{2}(1+ e^{-2t}) \ , \ p_1(t)=p_2(t) =  \frac{1}{4}(1 - e^{-2t})\ ,\ p_3(t)=0 .
\end{equation}
Interestingly, being non-Markovian it is perfectly P-divisible. Moreover, the map $e^{\frac 12 \Gamma_3(t) \mathcal{L}_3}$ is never CP due to $\Gamma_3(t) < 0$ for $t > 0$. Still the composition of three maps $e^{\frac 12 \Gamma_1(t) \mathcal{L}_1} \,e^{\frac 12 \Gamma_2(t) \mathcal{L}_2} \, e^{\frac 12 \Gamma_3(t) \mathcal{L}_3}$ is perfectly CP.

\subsubsection{Phase covariant dynamics}

Consider the qubit evolution governed by following time-local generator

\begin{equation}\label{L+-t}
  \mathcal{L}_t(\rho)  = - \frac{\omega(t)}{2} [\sigma_z,\rho] + \frac{\gamma_+(t)}{2} \mathcal{L}_+(\rho)  + \frac{\gamma_-(t)}{2} \mathcal{L}_-(\rho)  + \frac{\gamma_z(t)}{2} \mathcal{L}_z(\rho)  ,
\end{equation}
which is a time dependent version of (\ref{L+-}).   This generator enjoys the following covariance property

\begin{equation}\label{}
  U_\mathbf{x} \mathcal{L}(\rho) U^\dagger_\mathbf{x} = \mathcal{L}( U_\mathbf{x} \rho U^\dagger_\mathbf{x} )  ,
\end{equation}
where $U_\mathbf{x}$ is a diagonal unitary matrix

\begin{equation}\label{}
  U_\mathbf{x} = e^{i x_0} |0\>\<0| + e^{i x_1} |1\>\<1| ,
\end{equation}
with $\mathbf{x} =(x_0,x_1) \in \mathbb{R}^2$. The evolution of the density operator is represented as follows

\begin{equation}\label{}
 \rho = \left( \begin{array}{cc} \rho_{00} & \rho_{01} \\ \rho_{10} & \rho_{11} \end{array} \right)  \ \to \ \Lambda_t(\rho)  =
 \left( \begin{array}{cc} 1-P_e(t) & C(t) \rho_{01} \\ C^*(t) \rho_{10} & P_e(t) \end{array} \right) ,
\end{equation}
where the population of the excited state $P_e(t)$ and the coherence $C(t)$ read \cite{Maniscalco-PRA,Maniscalco-NJP,OSID-2014,Sergey,Francesco-NJP,Anita-2020}

\begin{equation}\label{}
  P_e(t) = e^{-\Gamma(t)} ( G(t) + P_e(0) ) , \ \ \ C(t) = e^{i \Omega(t) - \Gamma(t)/2 - \Gamma_z(t)} ,
\end{equation}
together with

\begin{equation}\label{}
  \Gamma(t) = \frac 12 \int_0^t ( \gamma_+(\tau) +\gamma_-(\tau)) d\tau \ , \ \ \ \Gamma_z(t) = \int_0^t  \gamma_z(\tau) d\tau ,
\end{equation}
and

\begin{equation}\label{}
  G(t) = \frac 12 \int_0^t e^{\Gamma(\tau)} \gamma_+(\tau) d\tau \ , \ \ \ \Omega(t) = \int_0^t  \omega(\tau)d\tau .
\end{equation}
Clearly, the corresponding evolution is CP-divisible if all rates are non-negative $\gamma_\pm(t) \geq 0$ and $\gamma_z(t) \geq 0 $. Beyond CP-divisibility the problem is non trivial since the complete positivity of $\{\Lambda_t\}_{t \geq 0}$ is not guaranteed if we relax positivity of $\gamma_\pm(t)$ and $\gamma_z(t)$.  The corresponding dynamical map can be represented as follows

\begin{equation}\label{}
  \Lambda_t(\rho) = \sum_{i,j=0}^1 T_{ij}(t) |i\>\<j| \rho |j\>\<i| + \sum_{i\neq j} C_{ij}(t) |i\>\<i| \rho |j\>\<j| ,
\end{equation}
where

\begin{equation}\label{}
   T_{ij}(t) = \left( \begin{array}{cc} 1 - e^{-\Gamma(t)}G(t) & 1 -  e^{-\Gamma(t)}(1+G(t)) \\  e^{-\Gamma(t)}G(t) &  e^{-\Gamma(t)}(1+G(t)) \end{array} \right)  ,
\end{equation}
and $\,  C_{12}(t) = C(t) =  C_{21}^*(t)$.  The map  $\{\Lambda_t\}_{t \geq 0}$ is completely positive if and only if the following matrix

\begin{equation}\label{CC}
    \left( \begin{array}{cc} 1 - e^{-\Gamma(t)}G(t) & C(t) \\  C^*(t) &  e^{-\Gamma(t)}(1+G(t)) \end{array} \right) \geq 0 ,
\end{equation}
for all $t \geq 0 $. Note, that condition (\ref{CC}) provides quite nontrivial constraints for the rates $\gamma_\pm(t)$ and $\gamma_z(t)$.

\begin{tcolorbox}
\begin{Proposition} Suppose that condition (\ref{CC}) is satisfied. Then the dynamical map $\{\Lambda_t\}_{t \geq 0}$

\begin{itemize}
		%\item is CP-divisible if and only if $\gamma_\pm(t) \geq 0$ and $\gamma_z(t) \geq 0$,
		\item is P-divisible iff
		\begin{equation}\label{ggg-1}
		\gamma_\pm(t) \geq 0 , \ \ \ \sqrt{\gamma_+(t)\gamma_-(t)} + 2 \gamma_z(t) \geq 0 ,
		\end{equation}
for all $t \geq 0$,
		\item satisfies BLP condition iff
		\begin{equation}\label{ggg-2}
		\gamma_+(t) + \gamma_-(t)  \geq 0 , \ \ \ \gamma_+(t) + \gamma_-(t) + 4 \gamma_z(t)  \geq 0 ,
		\end{equation}	
for all $t \geq 0$.
		%\item satisfies a  geometric criterion (\ref{GEO}) if and only if 	
		%\begin{equation}\label{}
		%\gamma_+(t) + \gamma_-(t) + 2 \gamma_z(t) \geq 0 ,
		%\end{equation}
	\end{itemize}
% for all $t \geq 0$.

\end{Proposition}
\end{tcolorbox}
Condition (\ref{ggg-1}) (recently derived in \cite{Sergey}) for P-divisibility is indeed more restrictive than BLP condition (\ref{ggg-2}). Condition (\ref{ggg-2}) immediately follows from  (\ref{D(t)}). Indeed, one easily finds for the matrix $D(t)$:

\begin{equation}\label{}
  D(t) =  - \left( \begin{array}{ccc} \frac 14 [\gamma_+(t) + \gamma_-(t)]  + \gamma_z(t)) & \omega(t) & 0 \\ -\omega(t) & \frac 14[\gamma_+(t) + \gamma_-(t)]  + \gamma_z(t) & 0 \\ 0 & 0 & \frac 12[\gamma_+(t) + \gamma_-(t)] \end{array} \right) ,
\end{equation}
and hence $D(t) + D^T(t) \leq 0$ is equivalent to (\ref{ggg-2}).

\subsection{Example: qudit evolution}

%Two examples of qubit dynamics can be generalized for $d > 2$.

%Note, that Pauli channel defined in terms of Pauli matrices can be essentially generalized in two ways: either by a collection of unitary %operators defining an orthonormal basis in $\LH$, or by

\subsubsection{From Pauli channels to Weyl-like dynamics}

Recall that a set of Pauli matrices $\{\sigma_\alpha\}_{\alpha=0}^3$ defines an orthonormal basis in $M_2(\mathbb{C})$. Moreover, each Pauli matrix $\sigma_\alpha$ is both Hermitian and unitary. In higher dimensions there is no operator basis which is both unitary and Hermitian. One may generalize Pauli matrices to a set of $d^2$ unitary Weyl operators defined as follows

\begin{equation}\label{Weyl}
  U_{k\ell} = \sum_{m=0}^{d-1} \omega_d^{\ell m}|k+m\>\<\ m| ,
\end{equation}
with $\omega_d = e^{2\pi i/d}$.  Note, that $U_{00}=\oper$ and

\begin{equation}\label{}
  {\rm Tr}(U_{k\ell} U^\dagger_{mn}) = d \delta_{km} \delta_{\ell n} .
\end{equation}
Moreover

\begin{equation}\label{}
  U_{mn} U_{k\ell} = \omega_d^{nk} U_{m+k,n+\ell} \ ,\ \ \ U_{k\ell}^\dagger = \omega_d^{k \ell} U_{-k,-\ell} .
\end{equation}
Let us introduce a single index $(k,\ell) \to \alpha := k d + \ell$ (mod $d$) $(\alpha = 0,1,\ldots,d^2-1$). Define a time-local generator

\begin{equation}\label{LtUU}
  \mathcal{L}_t(\rho)  = \sum_{\alpha=1}^{d^2-1} \gamma_\alpha(t)\left( U_\alpha \rho U^\dagger_\alpha -  \rho \right) ,
\end{equation}
which is a direct generalization of (\ref{sss}) for $d > 2$. It gives rise to the following dynamical map

\begin{equation}\label{}
  \Lambda_t(\rho) = \sum_{\alpha=0}^{d^2-1} p_\alpha(t) \, U_\alpha \rho U^\dagger_\alpha ,
\end{equation}
where

\begin{equation}\label{}
  p_{\alpha}(t) = \frac{1}{d^2} \sum_{\beta=0}^{d-1} H_{\alpha\beta}   \lambda_{\beta}(t) ,
\end{equation}
and $H_{\alpha\beta}$ is a (complex) Hadamard matrix  $H_{\alpha\beta} = \omega_d^{\alpha \times \beta}$, with $\alpha \times \beta = (k,\ell) \times (m,n) := {kn - \ell m}$, and

\begin{equation}\label{}
  \lambda_\alpha(t) = \exp\left( \sum_{\beta=1}^{d^2-1} H_{\alpha\beta} \Gamma_\beta(t) \right) \ , \ \ \ \Gamma_\beta(t) = \int_0^t \gamma_\beta(\tau) d\tau .
\end{equation}
Clearly, the generator (\ref{LtUU}) is physically legitimate if and only if $p_\alpha(t) \geq 0$ for $\alpha=0,1,\ldots,d^2-1$ and $t \geq 0$. Again, this condition provides a collection of non-trivial constraints for the transition rates $\gamma_\alpha(t)$. The evolution is CP-divisible if and only if all rates $\gamma_\alpha(t) \geq 0$. However, it is much more difficult to characterize P-divisibility.

%\begin{tcolorbox}
\begin{Proposition} Suppose that at most $d-1$ rates $\gamma_\alpha(t)$ can be negative at each $t \geq 0$. If $\gamma_\alpha(t) < 0$ for $\alpha > d^2-d$, and

\begin{equation}\label{ggg-Weyl}
  \gamma_\beta(t) \geq \sum_{\alpha > d^2-d} |\gamma_\alpha(t)|  ,
\end{equation}
for all $\beta \leq d^2-d$, then $\{\Lambda_t\}_{t\geq 0}$ is P-divisible.
\end{Proposition}
%\end{tcolorbox}
For the proof cf. \cite{Filip-PRA}. Note, that condition (\ref{ggg-Weyl}) becomes also necessary if $d=2$. Indeed,  assuming that $\gamma_3(t) < 0$, this condition reduces to  $\,  \gamma_k(t) \geq |\gamma_3(t)|\ , (k=1,2)$, which reproduces (\ref{ggg}). For $d>2$ we were able to provide only a sufficient condition.

\subsubsection{Generalized Pauli-like dynamics}   \label{SUB-Pauli}

Another generalization of the qubit generator is based on the observation that three eigen-basis of $\{\sigma_1,\sigma_2,\sigma_3\}$ are mutually unbiased. Recall that two orthonormal basis $\{e_k\}_{k=1}^d$ and $\{f_\ell\}_{k=1}^d$ are mutually unbiased if

\begin{equation}\label{}
  | \< e_k|f_\ell\>|^2 = \frac 1d ,
\end{equation}
for any pair $k,\ell$. It is well known \cite{MUB-1} that the number $N(d)$ of mutually unbiased bases (MUBs) in $\mathbb{C}^d$ is bounded by $N(d) \leq d + 1$  \cite{MUB-1,MUB-2,MUB-3}. If $d = p^r$
with $p$ being a prime number, one has exactly $N(d) = d + 1$. It was shown \cite{MUB-4} that in any dimension there exist at least three MUBs.
Suppose that the system Hilbert space $\HH$ allows for a maximal $d+1$ number of MUBs $\{|i_\alpha\>\}_{i=0}^{d-1}$ with $\alpha=1,\ldots,N(d)=d+1$. Construct the corresponding rank-1 projectors $P^\alpha_i :=|i_\alpha\>\<i_\alpha|$. Now, for each $\alpha$ define the following unitary operator

\begin{equation}\label{}
  U_\alpha = \sum_{k=0}^{d-1} \omega_d^{k} P^\alpha_k ,
\end{equation}
and the following completely positive map

\begin{equation}\label{}
  \mathbb{U}_\alpha(\rho) = \sum_{k=0}^{d-1} U_\alpha^k \rho U_\alpha^{k\dagger} .
\end{equation}
Note, that for $d=2$ these maps recover unitary channels $\sigma_\alpha \rho \sigma_\alpha$. For $d>2$ the map $\mathbb{U}_\alpha$ is no longer trace-preserving being a sum of unitary channels. However, one may define the following quantum channel

\begin{equation}\label{Phi-a}
  \Phi_\alpha(\rho) = \sum_{k=0}^{d-1} P^\alpha_k \rho P^\alpha_k ,
\end{equation}
such that $\,  \mathbb{U}_\alpha = d \Phi_\alpha - {\rm id}$. A family of channels $\Phi_\alpha$ enjoy the following properties

\begin{eqnarray*}
  \Phi_\alpha \Phi_\alpha = \Phi_\alpha  \ , \ \ \
  \Phi_\alpha \Phi_\beta = \Phi_\beta \Phi_\alpha = \Phi_0 \ , \ \ \ (\alpha \neq \beta) ,
\end{eqnarray*}
where $\Phi_0(\rho) = \frac 1d \oper {\rm Tr}\rho$ denotes a totally depolarising channel.

%\begin{tcolorbox}
\begin{Definition} The following map

\begin{equation}\label{GPC}
  \Lambda(\rho) = p_0\, {\rm id} +  \sum_{\alpha=1}^{d+1} p_\alpha\, \mathbb{S}_\alpha  ,
\end{equation}
where $(p_0,p_1,\ldots,p_{d+1})$ is a probability distribution, and $\, \mathbb{S}_\alpha = \frac{1}{d-1} \mathbb{U}_\alpha\,$ is called a generalized Pauli channel \cite{Kasia-2016,Ruskai-GP,Petz-GP}.
\end{Definition}
%\end{tcolorbox}
For $d=2$ the above definition reproduces a Pauli channel with $\mathbb{S}_k(\rho) = \sigma_k \rho \sigma_k$. Define the following time-local generator

\begin{equation}\label{GPC-L}
  \mathcal{L}_t = \frac 1d \sum_{\alpha=1}^{d^2-1} \gamma_\alpha(t) \Big( \mathbb{U}_\alpha - [d-1]{\rm id}\Big) = \sum_{\alpha=1}^{d^2-1} \gamma_\alpha(t) \Big( \Phi_\alpha - {\rm id} \Big) .
\end{equation}
Clearly, for $d=2$ it again reduces to (\ref{sss}). Due to commutativity $[\mathbb{S}_\alpha,\mathbb{S}_\beta]=0$, the corresponding evolution is a composition of $d+1$ maps

\begin{equation}\label{}
  \Lambda_t =  \Lambda^{(1)}_t \circ \ldots \circ \Lambda^{(d+1)}_t ,
\end{equation}
where each single map is defined via

\begin{equation}\label{}
  \Lambda^{(\alpha)}_t = e^{-\Gamma_\alpha(t)} {\rm id} + (1- e^{-\Gamma_\alpha(t)})\Phi_\alpha ,
\end{equation}
and $\Gamma_\alpha(t) = \int_0^t \gamma_\alpha(\tau)d\tau$. One finds \cite{Kasia-2016}  that $\Lambda_t$ defines a time dependent generalized Pauli map (\ref{GPC}) with

\begin{equation}\label{}
  p_0(t) = \frac{1}{d^2} \Big( 1 + (d-1)[\lambda_1(t) + \ldots +\lambda_{d+1}(t)]\Big) ,
\end{equation}
and

\begin{equation}\label{}
  p_\alpha(t) = \frac{d-1}{d^2} \Big( 1 + (d-1)\lambda_\alpha(t) - \sum_{\beta \neq \alpha}\lambda_\beta(t) \Big) ,
\end{equation}
where $  \lambda_\alpha(t) = \exp( \Gamma_\alpha(t) - \Gamma(t))$, and $\Gamma(t) = \Gamma_1(t) + \ldots + \Gamma_{d+1}(t)$. It is clear that a time-local generator (\ref{GPC-L}) generates legitimate dynamical map $\{\Lambda_t\}_{t\geq 0}$ if and only if $p_0(t) \geq 0$ and $p_\alpha(t) \geq 0$ for all $t \geq 0$. Moreover, $\{\Lambda_t\}_{t\geq 0}$ is CP-divisible if and only if $\gamma_\alpha(t) \geq 0$.
Suppose now that at each time at most single $\gamma_\alpha(t)$  can be  negative.

\begin{Proposition} Assume that $\gamma_{d^2-1}(t) < 0$. If

\begin{equation}\label{gg-Gen}
  \gamma_\alpha(t) \geq |\gamma_{d^2-1}(t)| ,
\end{equation}
for all $\alpha < d^2-1$, then   $\{\Lambda_t\}_{t\geq 0}$ is P-divisible.
\end{Proposition}
For the proof cf. \cite{Kasia-JMP}. Note, that for $d=2$ condition (\ref{gg-Gen}) reduces again to $\, \gamma_k(t) \geq |\gamma_3(t)|\ , \ (k=1,2)\,$ which reproduces (\ref{ggg}) which is necessary and sufficient for P-divisibility.

\subsubsection{Covariant dynamics}  \label{SEC-COV}

To generalize phase-covariant qubit evolution consider a map $\Phi : \LH \to \LH$ which satisfies the following covariance property

\begin{equation}\label{}
  \mathbb{U}_\mathbf{x} \, \Phi = \Phi \, \mathbb{U}_\mathbf{x} ,
\end{equation}
where $\mathbb{U}_\mathbf{x}(\rho) = U_\mathbf{x} \rho U_\mathbf{x}^\dagger$ with

\begin{equation}\label{}
  U_\mathbf{x} = \sum_{k=0}^{d-1} e^{i x_k} |k\>\<k| ,
\end{equation}
and $\mathbf{x}=(x_0,\ldots,x_{d-1}) \in \mathbb{R}^d$.

%\begin{tcolorbox}
\begin{Proposition} A map $\Phi$ is covariant w.r.t. maximal commutative subgroup of $U(d)$ if and only if

\begin{equation}\label{}
  \Phi(\rho) = \sum_{k\neq \ell =0}^{d-1} a_{k\ell} |k\>\<\ell| \rho |\ell\>\<k| + \sum_{k,\ell =0}^{d-1} b_{k\ell} |k\>\<k| \rho |\ell\>\<\ell| .
\end{equation}
$\Phi$ is Hermiticity preserving if and only if $a_{k\ell} \in \mathbb{R}$ and $d \times d$ complex matrix  $[b_{k\ell}]$  is Hermitian. Moreover, $\Phi$ is completely positive if $a_{kl} \geq 0$ and the matrix $[b_{kl}]$ is positive definite.
\end{Proposition}
%\end{tcolorbox}
Consider a covariant time-local generator

\begin{equation}\label{L-covariant}
  \mathcal{L}_t = \mathcal{L}^{0}_t + \mathcal{L}^{\rm class}_t  + \mathcal{L}^{\rm dec}_t
\end{equation}
where $\mathcal{L}^{0}_t(\rho) = -i[H(t),\rho]$, with $H(t) = \sum_{k=0}^{d-1} \epsilon_k(t) |k\>\<k|$. Moreover,  the classical generator $\mathcal{L}^{\rm class}_t$ and the dephasing (decoherence) generator $\mathcal{L}^{\rm dec}_t $ read
\begin{eqnarray}
 \mathcal{L}^{\rm class}_t(\rho) &=&  \sum_{k\neq \ell =0}^{d-1} t_{k\ell}(t) \Big( |k\>\<\ell| \rho |\ell\>\<k|  -\frac 12   \{ |\ell\>\<\ell|,\rho\} \Big) , \\  \mathcal{L}^{\rm dec}_t(\rho) &=& \sum_{k,\ell =0}^{d-1}  d_{k\ell} \Big( |k\>\<k| \rho |\ell\>\<\ell| - \frac 12  \{ |\ell\>\<\ell|,\rho\} \Big) .
\end{eqnarray}
Note that for $i \neq j$ one has

\begin{equation}\label{}
  \mathcal{L}_t(|i\>\<j|) = \ell_{ij}(t) |i\>\<j| ,
\end{equation}
with

\begin{equation}\label{}
  \ell_{ij}(t) = -\Big( i\omega_{ij}(t) + \frac{b_i(t)+b_j(t)}{2} - d_{ij}(t) + \frac{d_{ii}(t) + d_{jj}(t)}{2} \Big) ,
\end{equation}
where $\omega_{ij}(t) = \epsilon_i(t) - \epsilon_j(t)$, together with $b_j(t) = \sum_k t_{kj}(t)$. A necessary condition for P-divisibility
${\rm Re}\, \ell_{ij}(t) \leq 0$ implies

\begin{equation}\label{bbdd}
  b_i(t) + b_j(t) + d_{ii}(t) + d_{jj}(t) - 2 {\rm Re}\, d_{ij}(t) \geq 0 ,
\end{equation}
for all $t \geq 0$. Moreover,

\begin{equation}\label{}
  {\rm Tr}(|i\>\<i| \mathcal{L}_t(|j\>\<j|) = t_{ij}(t) ,
\end{equation}
and hence $t_{ij}(t) \geq 0$.  For $d=2$ the condition (\ref{bbdd}) reduces to

\begin{equation}\label{}
  \gamma_+(t) + \gamma_-(t) + 4\gamma_z(t) \geq 0 ,
\end{equation}
where $\gamma_+= b_0$ and $\gamma_-=b_1$. As we already observed this condition  is necessary but not sufficient for P-divisibility. Actually, together with $\gamma_+(t) + \gamma_-(t) \geq 0$, it is sufficient for BLP condition (\ref{BLP}).

\subsection{Divisibility of non-invertible dynamical maps}

%\begin{figure} \label{}
%\begin{center}
%\hspace*{.4cm}
%\includegraphics[width=9cm]{rys-cross.pdf}
%\caption{(Color online) Due to non-invertibility of the dynamical map two trajectories starting from two different initial states $\rho_1$ and %$\rho_2$ do cross at some moment of time  $t > 0$. } \label{FIG-cross}
%\end{center}
%\end{figure}
What happens if a dynamical map $\{\Lambda_t\}_{t \geq 0}$ is not invertible? Note, that if $\Lambda_t$ is invertible, i.e. the inverse $\Lambda_t^{-1}$ exists for all $t >0$, then if $\rho_1 \neq \rho_2$ the two trajectories $\Lambda_t(\rho_1)$ and $\Lambda_t(\rho_1)$ never cross (at finite time). It is no longer true if the map is not invertible.  Recall that invertibility is essential for divisibility since it does guarantee the very existence of propagators $V_{t,s} = \Lambda_t \Lambda_s^{-1}$. Quantum evolution of a qubit represented by non-invertible dynamical maps was already considered in \cite{Cresser2,Cresser3,non-1,non-2,Francesco-NJP,Anita-2020}. Hence, non-invertible dynamical maps may indeed represent the evolution of a real system. For non-invertible maps the divisibility is no longer guarantied. Actually, one proves \cite{PRL-2018} the following

\begin{Proposition} A dynamical  map $\{\Lambda_t\}_{t\geq 0}$ is divisible if and only if

\begin{equation}\label{KK}
  {\rm Ker}(\Lambda_s) \subseteq {\rm Ker}(\Lambda_t) ,
\end{equation}
for any $t > s$.
\end{Proposition}
Condition (\ref{KK}) means that the kernel of the map does not decrease in the course of time. Clearly, if the map is invertible then ${\rm Ker}(\Lambda_t)=0$ and (\ref{KK}) is trivially satisfied. Now, if $\Lambda_s$ is not invertible, then

\begin{itemize}
  \item $V_{t,s}$ is uniquely defined only on the image of $\Lambda_s$,
  \item the extension of $V_{t,s}$ to $\LH$ is not unique,
  \item $V_{t,s}$ is always trace-preserving on the image of $\Lambda_s$,
  \item even if $V_{t,s}$ is positive (or CP) on the image of $\Lambda_s$ there needs not exist a  positive (or CP) extension to $\LH$.
\end{itemize}
Recall that for invertible maps one has the following characterization

%\begin{tcolorbox}
\begin{equation}\label{}
  \{\Lambda_t\}_{t \geq 0} \ \mbox{is P-divisible} \ \Longleftrightarrow\ \ \frac{d}{dt} \| \Lambda_t(X)\|_1 \leq 0 ,
\end{equation}
and

\begin{equation}\label{}
  \{\Lambda_t\}_{t \geq 0} \ \mbox{is CP-divisible} \ \Longleftrightarrow\ \ \frac{d}{dt} \| [{\rm id} \otimes \Lambda_t](X)\|_1 \leq 0 .
\end{equation}
%\end{tcolorbox}
P- or CP-divisibility always implies monotonicity of the trace norm. The key property of invertible maps is that the converse is also true, that is,  monotonicity of the trace norm implies P- or CP-divisibility.

\begin{Proposition} Consider a linear trace-preserving map $\Phi : \mathcal{M} \to \LH$, where $\mathcal{M}$ is a linear subspace of $\LH$. If $\|\Phi(X)\|_1 \leq \|X\|_1$ for $X\in \mathcal{M}$, then $\Phi$ is a positive map.
\end{Proposition}
Note that if $\mathcal{M} = \LH$, then one recovers Proposition \ref{PRO-I}. Consider now a dynamical map $\{\Lambda_t\}_{t \geq 0}$  which is not invertible at $t=s$ but satisfies $\partial_t  \| \Lambda_t(X)\|_1 \leq 0$ for all $t \geq 0$. It means that the corresponding propagator

\begin{equation}\label{}
  V_{t,s} : {\rm Im} \Lambda_s \to \LH ,
\end{equation}
is positive and trace-preserving on the image of $\Lambda_s$. Similarly, if the dynamical map satisfies $\partial_t  \| [{\rm id}\otimes \Lambda_t](X)\|_1 \leq 0$ for all $t \geq 0$, then $V_{t,s}$ is CPTP on the image of $\Lambda_s$. Could we extend $V_{t,s}$ from the image of $\Lambda_s$ to $\LH$ keeping (complete) positivity and trace-preservation? The problem of CP extensions is well studied in mathematical literature \cite{Paulsen}. The key result was derived by Arveson \cite{Arveson}

\begin{Proposition} Let $\Phi : S \to \BH$ be a CP unital map, where $S$ is an operator system in $\BH$, i.e. if $X \in S$, then $X^\dagger \in X$, and $\oper \in S$. Then there exists (not unique) CP unital extension $\widetilde{\Phi} : \BH \to \BH$.
\end{Proposition}
Actually, if $S$ contains a strictly positive operator $X>0$ and $\Phi : S \to \BH$ is CP, then  there exists a CP  extension $\widetilde{\Phi} : \BH \to \BH$ \cite{Teiko}. It should be stressed that there is no similar result for positive maps \cite{Paulsen,Stormer,Bhatia}. It shows again a crucial difference between the notion of positivity and complete positivity. Another interesting result was provided in \cite{Jencova}

\begin{Proposition} Consider a CP map $\Phi : \mathcal{M} \to \BH$, where $\mathcal{M}$ is spanned by positive operators (e.g. density operators). Then $\Phi$ can be extended to a CP map $\widetilde{\Phi} : \BH \to \BH$.
\end{Proposition}
Using these results one proves \cite{PRL-2018}

\begin{Theorem} Let $\{\Lambda_t\}_{t \geq 0}$ be a dynamical map satisfying $\partial_t  \| [{\rm id}\otimes \Lambda_t](X)\|_1 \leq 0$ for all $t \geq 0$. Then there exists a family of completely positive propagators $\widetilde{V}_{t,s} : \LH \to \LH$.
\end{Theorem}
Note, however, that $\widetilde{V}_{t,s}$ being an extension of $V_{t,s}$ needs not be trace-preserving. Trace-preservation is guaranteed only on the image of $\Lambda_s$ where $\widetilde{V}_{t,s}$ coincides with $V_{t,s}$. Consider now a special class of maps which satisfy

\begin{equation}\label{ImIm}
  {\rm Im} \Lambda_t \subset {\rm Im} \Lambda_s ,
\end{equation}
for $t > s$. One calls $\{\Lambda_t\}_{t \geq 0}$ an {\em image nonicreasing dynamical map} \cite{PRL-2018}.

\begin{Theorem} An image non-increasing dynamical map $\{\Lambda_t\}_{t \geq 0}$ satisfying $\partial_t \| [{\rm id} \otimes \Lambda_t](X)\|_1 \leq 0$ is CP-divisible.
\end{Theorem}

\begin{Example} Two simple examples of image non-increasing dynamical map:

\begin{enumerate}
  \item $\Lambda_t$ is normal, i.e. $\Lambda_t \Lambda_t^\ddag = \Lambda_t^\ddag \Lambda_t$,
  \item $\Lambda_t$ is diagonalizable and commutative, that is,
\begin{equation}\label{}
  \widehat{\Lambda}_t = \sum_\alpha \lambda_\alpha(t) |X_\alpha\>\!\> \<\!\< Y_\alpha| ,
\end{equation}
and due to commutativity right and left eigenvectors $|X_\alpha\>\!\>,|Y_\alpha\>\!\> $ are time independent.
\end{enumerate}

\end{Example}
Actually, we are not aware of any example of a dynamical map  satisfying $\partial_t \| [{\rm id} \otimes \Lambda_t](X)\|_1 \leq 0$ which is not CP-divisible. It would be interesting to find such example or to show that $\partial_t \| [{\rm id} \otimes \Lambda_t](X)\|_1 \leq 0$ is equivalent to CP-divisibility even if the map is not invertible. Interestingly in the qubit case  ($d=2$) one proves \cite{CC19} the following general result

\begin{Theorem} A qubit dynamical map $\{\Lambda_t\}_{t \geq 0}$ is CP-divisible if and only if  $\partial_t \| [{\rm id} \otimes \Lambda_t](X)\|_1 \leq 0$.
\end{Theorem}
One may wonder how to construct an extension of the propagator $V_{t,s}$ from the image of $\Lambda_s$ to the entire $\LH$. In a recent paper \cite{Ujan} a simple method based on generalized inverse was proposed.

\section{Paradigmatic models of open quantum systems}   \label{Paradigmatic}

In this section we briefly review well known models of open quantum systems and analyze divisibility and Markovianity of the corresponding dynamical maps.

\subsection{Qubit amplitude damping evolution}

Consider a paradigmatic model of a qubit decay \cite{Open1,Open2,Open3,Legget} defined by the following Hamiltonian

\begin{equation}\label{eq:H}
		\mathbf{H}=\omega_{0}|{\e}\>\<{\e}|\otimes\oper_{\rm E}+\oper_{\rm S}\otimes H_{\rm E} +H_{\rm int},
	\end{equation}
where $H_{\rm E} = \int \mathrm{d}\omega\;\omega\,b^\dag_\omega b_\omega$, and the interaction term reads
	\begin{equation}\label{eq:hint_amplitude}
		H_{\rm int}= \sigma_-  \otimes b^\dagger(f) + \sigma_+ \otimes b(f^*) ,
%
%\otimes\int \mathrm{d}\omega f(\omega)b^\dag(\omega)+ \sigma_+ \otimes\int \mathrm{d}\omega f^*(\omega)b(\omega) .
	\end{equation}
where $b^\dagger(f) = \int \mathrm{d}\omega f(\omega)b^\dag_\omega$ and $b(f^*) = \int \mathrm{d}\omega f^*(\omega)b_\omega$. In this section $|\e\>$ stands for the excited state and $|\g\>$ for the ground state.  Creation and annihilation operators $b^\dag_\omega$, $b_\omega$ satisfy the canonical commutation relations  $[b_\omega,b_{\omega'}]=0$ and $[b_\omega,b^\dag_{\omega'}]= \delta(\omega-\omega')$, and $f(\omega)$ denotes a form-factor. Since the total Hamiltonian commutes with the operator of the  total number of excitations  one can easily find  the evolution $U_t = \e^{-\i t\mathbf{H}}$ in the single-excitation sector~\cite{Open1}. Assuming $|\Psi(0)\> = |\e\> \otimes |{\rm vac}\>$, where $|\vac\>$ denotes the vacuum state of the boson field, one finds
	
\begin{equation}\label{}
		|\Psi(t)\> = a(t) |\e\> \otimes |{\rm vac}\> + |\g\> \otimes b^\dagger(\xi_t) |{\rm vac}\> ,
\end{equation}
where the function $a(t)$ satisfies the following non-local equation	
	\begin{equation}\label{dot-a}
		\i \dot{a}(t) = \omega_0 a(t) + \int_0^t G(t-s) a(s)\;\mathrm{d}s, \;  \ \ a(0)=1 ,
	\end{equation}
with the memory kernel $G(t) = - i\int \mathrm{d} \omega\;|f(\omega)|^2 \e^{-\i \omega t}$, and  the boson profile $\xi_t(\omega)$ reads as follows

\begin{equation}\label{}
  \xi_t(\omega) = -i \int_0^t e^{-i\omega(t-s)} f(\omega) a(s) ds .
\end{equation}
The reduced dynamics has the  following form
\begin{equation}\label{eq:amplitude}
		\Lambda_t(\rho)=\begin{pmatrix}
			|a(t)|^2\,\rho_{\e\e} & a(t)\rho_{\e\g}\\
			a(t)^*\rho_{\g\e}& \rho_{\g\g} - |a(t)|^2\rho_{\e\e}
		\end{pmatrix} ,
\end{equation}
and hence it defines an amplitude-damping channel for any $t>0$. By construction one has $|a(t)|\leq 1$ and $a(0)=1$. For the map  (\ref{eq:amplitude}) positivity and complete positivity coincide.

\begin{Proposition} The dynamical map (\ref{eq:amplitude}) is CP-divisible if and only if $\frac{d}{dt} |a(t)|\leq 0$ for all $t \geq 0$.
\end{Proposition}
The corresponding time-local generator has the following form

\begin{equation}\label{}
  \mathcal{L}_t(\rho) = - \frac{i}{2} \varepsilon(t) [\sigma_z,\rho] + \gamma(t) \Big(\sigma_- \rho \sigma_+ - \frac 12 \{\sigma_+\sigma_-,\rho\} \Big) ,
\end{equation}
with

\begin{equation}\label{}
  \varepsilon(t) =  - 2{\rm Im} \frac{\dot{a}(t)}{a(t)} \ , \ \ \ \  \gamma(t) = -2{\rm Re} \frac{\dot{a}(t)}{a(t)} \ ,
\end{equation}
and CP-divisibility is equivalent to $\gamma(t) \geq 0$. Interestingly, for the amplitude damping evolution model CP-divisibility, P-divisibility and BLP condition of no information backflow coincide.

\subsection{Generalized amplitude-damping evolution}   \label{SUB-AD}

The above qubit model can be generalized as follows \cite{garrawy,Davide-1}: consider a quantum system living in the Hilbert space $\mathcal{H}_{\rm S}= \mathcal{H}_{\mathrm{e}} \oplus \mathcal{H}_{\mathrm{g}}$, with $\dim \mathcal{H}_{\mathrm{e}}=n$ and $\dim \mathcal{H}_{\mathrm{g}}=1$. Here $\mathcal{H}_{\mathrm{e}}$ corresponds to an $n$-dimensional excited sector, whereas $\mathcal{H}_{\mathrm{g}}$ corresponds to 1-dimensional ground sector spanned by $|\mathrm{g}\>$. The system is coupled to an $r$-mode bosonic bath ($r\leq n$), with the total system-bath Hamiltonian given by
	\begin{equation}\label{Hn}
		\mathbf{H} = H_{\mathrm{e}} \otimes \oper_{\rm E} + \oper_{\rm S} \otimes \sum_{j=1}^r \int \mathrm{d}\omega\;\omega b^\dagger_j(\omega) b_j(\omega) + H_{\rm int} ,
	\end{equation}
where $H_{\e}$ is the free Hamiltonian of the excited sector of the system, and the interaction Hamiltonian  reads
	\begin{equation}\label{hnint}
		H_{\rm int} = \sum_{j=1}^r\int \mathrm{d}\omega\; \Big( f_j(\omega) |\mathrm{g}\>\<\bm{\beta}_j| \otimes b^\dagger_{j}(\omega) + f_j^*(\omega) |\bm{\beta}_j\>\<\mathrm{g}| \otimes b_j(\omega) \Big) ,
	\end{equation}
with $|\bm{\beta}_1\>,\dots,|\bm{\beta}_r\> \in \mathcal{H}_{\e}$.  The bath  creation and annihilation operators $b^\dagger_j(\omega)$ and $b_j(\omega)$ satisfy the standard canonical commutation relations: $[b_i(\omega),b_{j}(\omega')]=0$ and $[b_i(\omega),b^\dag_{j}(\omega')]=\delta_{ij}\delta(\omega-\omega')$.

In the single-excitation sector the above model is solvable: starting from $\ket{\Psi(0)}=\ket{\psi_\e}\otimes\ket{\vac}$, with $\ket{\psi_\e}\in\mathcal{H}_{\e}$, one finds
	\begin{equation}
		\ket{\Psi(t)}=\mathbb{A}(t)\ket{\psi_\e}\otimes\ket{\vac}+\ket{\g}\otimes \bigotimes_{j=1}^r b_j^\dag(\xi_j(t))\ket{\vac_j},
	\end{equation}
where $|\vac\> = \bigoplus_j |\vac_j\>$. The operator $\mathbb{A}(t)$ satisfies a generalization of Eq.~\eqref{dot-a}:
\begin{equation}\label{eq:mathbba}
	\i\dot{\mathbb{A}}(t)=H_\e \mathbb{A}(t)+\int_0^t\mathbb{G}(t-s)\mathbb{A}(s)\;\mathrm{d}s, \;  \ \ \mathbb{A}(0)=\oper_{\rm e},
\end{equation}
with the memory kernel

\begin{equation}\label{}
\mathbb{G}(t) = -\i\sum_{\ell=1}^r\int\mathrm{d}\omega\;|f_\ell(\omega)|^2\e^{-\i\omega t}\ket{\bm{\beta}_\ell}\!\bra{\bm{\beta}_\ell} .
\end{equation}
Using a natural representation of the system density operator:
	\begin{equation}\label{}
		\rho = \left( \begin{array}{cc} \hat{\rho}_\e & |\mathbf{w}\> \\  \< \mathbf{w}| & \rho_\mathrm{g} \end{array} \right) ,
	\end{equation}
one finds the following formula for the reduced evolution \cite{Davide-1}
	\begin{equation}\label{MAD}
		\Lambda_t(\rho) = \left( \begin{array}{cc} \mathbb{A}(t)\hat{\rho}_\e \mathbb{A}^\dagger(t) & \mathbb{A}(t)|\mathbf{w}\> \\  \< \mathbf{w}|\mathbb{A}^\dagger(t) & \rho_\mathrm{g}(t) \end{array} \right) ,
	\end{equation}
where the population of the ground state reads $\rho_\mathrm{g}(t) = 1 -  {\rm Tr}( \mathbb{A}(t) \hat{\rho}_\e \mathbb{A}^\dagger(t))$. The operator $\mathbb{A}(t) : \mathcal{H}_\e \to \mathcal{H}_\e$ provides a multi-level generalization of the function $a(t)$ which appears in Eq.~\eqref{eq:amplitude}. Other generalizations of the amplitude-damping channel were recently analyzed in \cite{giovannetti,wilde}. The generalized amplitude damping map $\Lambda_t$ defined by (\ref{MAD}) is completely positive if and only if the operator norm $\| \mathbb{A}(t)\|_\infty \leq 1$. Again positivity and complete positivity coincide.

Let $\mathbb{A}(t) = e^{-\mathbb{B}(t)}$. Then $\| \mathbb{A}(t)\|_\infty \leq 1$ if and only if

\begin{equation}\label{}
  \mathbb{B}(t) + \mathbb{B}^\dagger(t) \geq 0 .
\end{equation}

\begin{Proposition} The generalized amplitude damping dynamical map $\{\Lambda_t\}_{t\geq 0}$ defined by (\ref{MAD}) is CP-divisible  if and only if

\begin{equation}\label{}
  \mathbb{L}(t) + \mathbb{L}^\dagger(t) \leq 0 ,
\end{equation}
where $ \mathbb{L}(t)= \dot{\mathbb{A}}(t) \mathbb{A}^{-1}(t)$. Moreover, CP-divisibility and P-divisibility coincide.
\end{Proposition}
One finds the following formula for the corresponding time-local generator

\begin{equation}\label{}
  \mathcal{L}_t(\rho) = \dot{\Lambda}_t\Lambda^{-1}_t(\rho) =  \left( \begin{array}{cc} \mathbb{L}(t)\hat{\rho}_\e + \hat{\rho}_\e\mathbb{L}^\dagger(t) & \mathbb{L}(t)|\mathbf{w}\> \\  \< \mathbf{w}|\mathbb{L}^\dagger(t) & -{\rm Tr}(\mathbb{L}(t)\hat{\rho}_\e + \hat{\rho}_\e\mathbb{L}^\dagger(t)) \end{array} \right)
\end{equation}
If $\mathbb{L}(t) = -i \mathbb{H}(t) - \mathbb{G}(t)$, then

\begin{equation}\label{}
   \mathcal{L}_t(\rho) = -i[\mathbb{H}(t),\rho] + |\mathrm{g}\>\<\mathrm{g}| {\rm Tr}(\mathbb{G}(t) \rho) - \frac 12 \{\mathbb{G}(t),\rho\} .
\end{equation}
The evolution is CP-divisible if $\mathbb{G}(t) \geq 0$ for all $t \geq 0$.

\subsection{Dephasing evolution}    \label{SUB-PD}

Let us  consider the following Hamiltonian of a $d$-level system $S$ coupled to an environment \cite{Open1,Open2,DEC1,DEC2,Ekert-Palma,Alicki-dephasing}

\begin{equation}\label{}
  \mathbf{H} = H_S \otimes \oper_E + \oper_S \otimes H_E + \sum_{k=0}^{d-1} P_k \otimes B_k ,
\end{equation}
where the system's Hamiltonian $H_S = \sum_k E_k P_k$, $P_k= |k\>\<k|$, and $B_k \in \mathrm{L}(\mathcal{H}_E)$.  One finds

\begin{equation}\label{}
  \mathbf{H} = \sum_{k=1}^d P_k \otimes H_k ,
\end{equation}
where the environmental operators $H_k$ are defined by $\, H_k = E_k \oper_E + H_E + B_k$. Computing the unitary operator

\begin{equation}\label{}
  \mathbf{U}_t = e^{-i \mathbf{H}t} = \sum_{k=1}^d P_k \otimes e^{-i H_k t} ,
\end{equation}
one eventually arrives at the reduced evolution of the $d$-level system which is represented via the Schur product

\begin{equation}\label{Schur}
  \Lambda_t(\rho) = \sum_{k,l=0}^{d-1} D_{kl}(t) P_k \rho P_l .
\end{equation}
 with

\begin{equation}\label{Dkl}
  D_{kl}(t) = {\rm Tr}\Big( e^{-i H_k t} \rho_E e^{i H_l t} \Big) = e^{-i(E_k - E_l)}\, {\rm Tr}\Big( e^{-i (H_E+B_k) t} \rho_E e^{i (H_E+B_l) t} \Big) .
\end{equation}
Observe that $D_{kk}(t) = 1$, and initially $D_{kl}(t=0)=1$.  This matrix is positive definite. Indeed, to prove it one needs to show that

\begin{equation}\label{}
  \sum_{k,l=1}^d D_{kl}(t) z_k z^*_l \geq 0 ,
\end{equation}
for arbitrary vector $z=(z_1,\ldots,z_d) \in \mathbb{C}^d$. One has $\sum_{k,l=1}^d D_{kl}(t) z_k z^*_l = \Tr (V \rho V^\dagger) \geq 0$,
with $V = \sum_{k=1}^{d} z_k e^{-i H_k t}$.

\begin{Proposition} The dynamical map (\ref{Schur}) is CP-divisible if and only if the matrix

\begin{equation}\label{}
  D_{kl}(t,s) = D_{kl}(t)/D_{kl}(s) ,
\end{equation}
is positive definite for any $t \geq s$. P- and CP-divisibility coincide.
\end{Proposition}
Markovian semigroups defined by (\ref{Schur}) were recently analyzd in \cite{Fabio-Hadamard}.

In the qubit case such dephasing evolution

\begin{equation}\label{eq:channel}
		\Lambda_t(\rho)=\begin{pmatrix}
			\rho_{00}&\rho_{01}\,D(t)\\
			\rho_{10}\,D^*(t)&\rho_{11}
		\end{pmatrix}, \ \ \ \ \ D(t) := D_{01}(t) = {\rm Tr}\Big( e^{-i H_0 t} \rho_E e^{i H_1 t} \Big) ,
	\end{equation}
may be realized via the following spin-boson Hamiltonian

\begin{equation}\label{}
  H_S = \int\mathrm{d}\omega\;\omega\,b^\dag_\omega b_\omega   \ , \ \ \ \ \ H_{\rm int} = \sigma_z \otimes (b^\dagger(f) + b(f^*)) ,
\end{equation}
where

\begin{equation}\label{}
  b(f^*) := \int d\omega\, f^*(\omega)b_\omega \ , \ \ \  b^\dagger(f) := \int d\omega\, f(\omega)b^\dagger_\omega .
\end{equation}
Assuming that $\rho_E = |\vac\>\<\vac|$ the decoherence function $D(t)$ reads

\begin{equation}\label{eq:dephasing-sb}
		D(t)=\Braket{\vac|\e^{i tH_1}\e^{- i tH_0}|\vac} .
	\end{equation}
Let us introduce a Weyl operator
\begin{equation}
		W(g)=\exp\left\{b^\dag(g)-b(g^*)\right\},
\end{equation}
where $g(\omega)$ is an arbitrary (square integrable) complex function. Weyl operators provide a (projective) representation of the Weyl algebra

\begin{equation}\label{eq:weyl_composition}
		W(g)W(h)=\exp\left\{-i{\rm Im} \int\mathrm{d}\omega\:g(\omega)^*h(\omega)\right\}W(g+h) .
	\end{equation}
Moreover, one finds
	\begin{equation}\label{eq:weyl_average}
		\Braket{\vac|W(g)|\vac}=\exp\left\{-\frac{1}{2}\int\mathrm{d}\omega\:|g(\omega)|^2\right\} ,
	\end{equation}
together with

	\begin{equation}\label{eq:weyl-rotation}
		e^{- i tH_{\rm B}}W(g)e^{ i tH_{\rm B}}=W(g_t),
	\end{equation}
where $g_t(\omega)=\e^{- i\omega t}g(\omega)$. Assuming that a form factor $f(\omega)$ satisfies $	\int\mathrm{d}\omega\;{|f(\omega)|^2}/{\omega}<\infty$ and  	$\int\mathrm{d}\omega\;{|f(\omega)|^2}/{\omega^2}<\infty$, one proves \cite{Alicki-dephasing}

\begin{eqnarray}
		\W{{f}/{\omega}}H_0\Wdag{{f}/{\omega}}&=&\tilde{E}_0+H_{\rm E};\\
		\Wdag{{f}/{\omega}}H_1\W{{f}/{\omega}}&=&\tilde{E}_1+H_{\rm E},
	\end{eqnarray}
where the renormalized energies read
\begin{equation}
		\tilde{E}_{k}= E_{k}-\int\mathrm{d}\omega\;{|f(\omega)|^2}/{\omega},\qquad	k=0,1.
\end{equation}
Note, that $\Delta E = \tilde{E}_{1} - \tilde{E}_{0} = {E}_{1} - {E}_{0}$. Finally, one computes $D(t)$ \cite{Open1,Ekert-Palma,Davide-2}

\begin{equation}
			D(t)=\e^{- i(E_0-E_1)t}\exp\left\{-4\int\mathrm{d}\omega\;\frac{|f(\omega)|^2}{\omega^2}(1-\cos\omega t)\right\}.
\end{equation}
Condition for CP-divisibility reduces to

\begin{equation}\label{}
  \frac{d}{dt} |D(t)| \leq 0 ,
\end{equation}
and hence the form factor $f(\omega)$ has to satisfy the following condition

\begin{equation}\label{}
  \int\mathrm{d}\omega\;\frac{|f(\omega)|^2}{\omega}\, \sin\omega t \, \leq \, 0 ,
\end{equation}
for any $t \geq 0$.

In the multi-level case let us consider

\begin{equation}
		\mathbf{H}=H_{\rm S}\otimes\oper_{\rm E} + \oper_{\rm S}\otimes H_{\rm E} + \sum_{j=0}^{d-1}\ketbra{j}{j}\otimes\left(b(f_j)+b^\dag(f_j)\right) ,
	\end{equation}
with $H_{\rm S}=\sum_j E_j\ketbra{j}{j}$, $H_{\rm E}$ the free boson Hamiltonian, and  $f_0,\dots,f_{d-1}$ are square-integrable coupling functions. Similar analysis as in the qubit case leads to \cite{Davide-2}

\begin{equation}\label{eq:dephasing_gsb}
			D_{j\ell}(t) = \e^{-i(E_j- E_\ell)t} e^{i\theta_{j\ell}(t)}\exp\left\{-\int\mathrm{d}\omega\; \left(|f_j(\omega)-f_\ell(\omega)|^2\right)\frac{1-\cos\omega t}{\omega^2}\right\} ,
		\end{equation}
		where the extra phases are defined as follows
		\begin{equation}\label{eq:theta}
			\theta_{j\ell}(t)=2{\rm Im} \int\mathrm{d}\omega\;f_\ell(\omega)^*f_j(\omega)\frac{1-\cos\omega t}{\omega^2} + \int\mathrm{d}\omega\,\left(|f_j(\omega)|^2-|f_\ell(\omega)|^2\right)\left(\frac{t}{\omega}-\frac{\sin\omega t}{\omega^2}\right) .
		\end{equation}

To find the corresponding time-local generator $\mathcal{L}_t$ one computes $ \mathcal{L}_t = \dot{\Lambda}_t \Lambda_t^{-1}$ and finds

\begin{equation}\label{L-deco}
  \mathcal{L}_t(\rho) = \sum_{k,l=0}^{d-1} L_{kl}(t) P_k \rho\, P_l ,
\end{equation}
where

\begin{equation}\label{}
  L_{kl}(t) = \frac{\dot{D}_{kl}(t)}{D_{kl}(t)} .
\end{equation}
Note, that the generator still does not have the canonical GKLS form. In order to find the corresponding canonical form let us define $d-1$ traceless Hermitian matrices

\begin{equation}\label{Sl}
  S_l = \frac{1}{\sqrt{l(l+1)}} \left( \sum_{k=0}^{l-1} P_k - l\, P_{l} \right)   \ , \ \ l=1,\ldots,d-1 .
\end{equation}
They generate a maximal commutative subgroup of $SU(d)$ and satisfy $\Tr (S_k S_l) = \delta_{kl}$. Clearly, any $P_k$ can be represented via

\begin{equation}\label{PS}
  P_k = \frac 1d \oper + \sum_{m=1}^d a_{km} S_m ,
\end{equation}
where the matrix $a_{kl}$ is defined via

\begin{equation}\label{}
  a_{kl} = {\rm Tr}(P_k S_l) = \frac{1}{\sqrt{l(l+1)}} \left( \sum_{k=0}^{l} \delta_{kl} - l\, \delta_{k,l+1} \right) .
\end{equation}
Note, that condition $P_k P_l = \delta_{kl} P_l$, implies for $k \neq l$

\begin{equation}\label{SSS}
  \frac{1}{d^2} \oper + \frac 1d \sum_{m=0}^{d-1} (a_{km} + a_{lm}) S_m + \sum_{m,n=1}^d a_{km} a_{ln} S_m S_n = 0 .
\end{equation}
Inserting (\ref{PS}) into (\ref{L-deco}) and using (\ref{SSS}) one finds the following formula for the time-local generator

\begin{equation}\label{LL-can}
  \mathcal{L}_t(\rho) = -i[H(t),\rho] + \sum_{m,n=0}^{d-1} K_{mn}(t) \left( S_m \rho S_n - \frac 12 \{ S_n S_m,\rho\} \right) ,
\end{equation}
where the time-dependent Hamiltonian $H(t)$, and the matrix $K_{mn}(t)$ are defined as follows

\begin{equation}\label{}
  H(t) =  - \frac 1d \sum_{k,l,m=0}^{d-1} [{\rm Im}\,L_{kl}(t)] a_{km} S_m ,
\end{equation}
and

\begin{equation}\label{}
  K_{mn}(t) = \sum_{k,l=0}^{d-1} L_{kl}(t) a_{km} a_{ln} .
\end{equation}
Formula (\ref{LL-can}) provides the canonical representation of the time-local generator of pure decoherence. It is clear that the dephasing evolution is CP-divisible if and only if the matrix $[K_{mn}(t)]$ is positive definite for all $t \geq 0$.

\subsection{Dephasing from stochastic Hamiltonians}

Interestingly, the dephasing dynamical map can be realized by averaging a stochastic unitary evolution. Consider a stochastic Hamiltonian $\, H(t) = \sum_{k=1}^d \varepsilon_k(t)  P_k$, where

\begin{equation}\label{}
  \varepsilon_k(t) = E_k + \xi_k(t) ,
\end{equation}
and $\xi_k(t)$ is a real Gaussian noise satisfying

\begin{equation}\label{}
  \<\!\< \xi_k(t) \>\!\> = 0 \ , \ \ \  \<\!\< \xi_k(t)\xi_l(s) \>\!\> = \delta_{kl}\gamma_k(t-s) ,
\end{equation}
where $\<\< ... \>\>$ denotes the noise average. Define a dynamical map via

\begin{equation}\label{}
  \Lambda_t(\rho) := \<\!\<\, U_\xi(t) \rho\, U_\xi^\dagger(t) \, \>\!\> ,
\end{equation}
where $U_\xi(t)$ is a stochastic unitary operator $U_\xi(t) = e^{-i \int_0^t H(\tau) d\tau}$. One finds

\begin{equation}\label{}
   \Lambda_t(\rho) = \sum_{k,l=0}^{d-1} D_{kl}(t) P_k \rho P_l ,
\end{equation}
with the decoherence matrix

\begin{equation}\label{}
  D_{kl}(t) = e^{-i(E_k-E_l)t} \, \exp\left(- \frac 12 \int_0^t ds \int_0^t du [\gamma_k(s-u) + \gamma_l(s-u)] \right)  ,
\end{equation}
where we used the well known property of the Gaussian noise

\begin{equation}\label{}
  \<\!\<\, e^{-i \int_0^t \xi_k(s) ds} \, \>\!\> = e^{- \frac 12 \<\!\<\, [ \int_0^t \xi_k(s) ds ]^2 \, \>\!\> } .
\end{equation}
Quantum dynamics generated by stochastic Hamiltonian was analyzed in great details in \cite{Kropf}. Note, that such evolution is a convex combination of unitary evolutions \cite{RU} and is always unital (see also \cite{NO}).

%\end{document}

\section{Implications of divisibility}   \label{IMPLICATIONS}

Divisibility is a mathematical property of a dynamical map. It does guarantee the existence of a propagator $\{V_{t,s}\}_{t\geq s}$ such that for any $t \geq s$ one has $\Lambda_t = V_{t,s} \Lambda_s$. In particular a dynamical map $\{\Lambda_t\}_{t \geq 0}$ is Markovian if and only if the corresponding propagator is CPTP.    In this section we review the most important physical implications of divisibility. It turns out that CP-divisibility (and often also P-divisibility) implies monotonic behaviour of various quantities, like e.g. guessing probability in the state discrimination process, monotonicity of important entropic quantities (generalization of the von Neumann relative entropy), and various correlations. It is, therefore, clear that violation of such  monotonic behaviour provides a witness of non-Markovianity and the very presence of memory effects.

\subsection{State discrimination}

Given two states $\rho_1$ and $\rho_2$ one defines the corresponding distinguishability (cf. discussion in section \ref{SUB-BLP})

\begin{equation}\label{a}
  D(\rho_1,\rho_2) = \frac 12  \|\rho_1 - \rho_2\|_1 .
\end{equation}
More general discrimination problem involves an ensemble of two states $\rho_1$ and $\rho_2$ with probabilities $p_1$ and $p_2$, respectively. In this more general scenario the distinguishability is defined by $ \| p_1\rho_1 - p_2 \rho_2\|_1$ and it reduces to (\ref{a}) whenever $p_1=p_2=1/2$. Any P-divisible dynamical map $\{\Lambda_t\}_{t \geq 0}$  implies \cite{Angel}

\begin{equation}\label{}
  \frac{d}{dt} \|  \Lambda_t(p_1\rho_1 - p_2 \rho_2) \|_1  \leq 0 ,
\end{equation}
for all pairs $\{\rho_1,p_1\},\{\rho_2,p_2\}$.

Consider now an ensemble of $N$ states $\rho_k$ prepared with probability $p_k$, i.e. a set $\mathcal{E} =\{p_k,\rho_k\}_{k=1}^N$. To discriminate states within $\mathcal{E}$  with minimal averaged error (or to maximize the success probability for the
guess) one has to devise an appropriate measurement scenario. The maximum success probability (called also a guessing probability) is defined as follows \cite{Helstrom}

\begin{equation}\label{}
  \mathrm{P}_{\rm guess}(\mathcal{E}) := \max \sum_{k} p_k {\rm Tr}(P_k \rho_k) ,
\end{equation}
where the maximum is over all POVMs $\{P_k\}$ defined on the systems Hilbert space $\mathcal{H}$. Note, that if $\mathcal{E}$ consists of two members $\mathcal{E}=\{p_1,\rho_1;p_2,\rho_2\}$, then due to Helstrom  \cite{Helstrom} (see also \cite{JUNU} for the review)

\begin{equation}\label{}
     \mathrm{P}_{\rm guess}(\mathcal{E}) = \frac 12 \Big( 1 + \| p_1 \rho_1 - p_2 \rho_2 \|_1 \Big) .
\end{equation}
A guessing probability can not increase during P-divisible evolution.

\begin{Proposition} For any P-divisible dynamical map $\{\Lambda_t\}_{t \geq 0}$ and any ensemble $\mathcal{E}$

\begin{equation}\label{P-mon}
  \frac{d}{dt} \mathrm{P}_{\rm guess}(\mathcal{E}(t)) \leq 0 ,
\end{equation}
where $\mathcal{E}(t) = \{p_k,\Lambda_t(\rho_k)\}_{k=1}^N$.
\end{Proposition}
Hence for any P-divisible (i.e. weakly non-Markovian evolution \cite{Sabrina}) our ability to discriminate states from a given ensemble decreases in the course of time.

Similar problem was considered in  \cite{datta} in the case of discrete dynamical maps. Let  $\{\Lambda_n\}_{n\geq 0}$ be a discrete dynamical map, that is, for any $n=0,1,2,\ldots$

\begin{equation}\label{}
  \Lambda_n : \mathrm{L}(H_S) \to \mathrm{L}(\mathcal{H}_n) ,
\end{equation}
is CPTP, where $\mathcal{H}_S,\mathcal{H}_1,\ldots$ are finite dimensional Hilbert spaces of dimensions $d_S,d_1,d_2, \ldots$, respectively. Moreover, we assume that $\Lambda_0 : \mathrm{L}(H_S)\to \mathrm{L}(H_S)$ is an identity map. $\{\Lambda_n\}_{n\geq 0}$ is CP-divisible if

\begin{equation}\label{}
  \Lambda_j = V_{j,i} \Lambda_i ,
\end{equation}
and $V_{j,i} : \mathrm{L}(\mathcal{H}_i) \to \mathrm{L}(\mathcal{H}_j)$ is CPTP for any $j > i$. Note, that divisibility, i.e. the existence of propagators $V_{t,s}$ is guaranteed if $\,   {\rm Ker} \Lambda_j \supset  {\rm Ker} \Lambda_i$ for $j > i$.

\begin{Definition} A time discrete dynamical map $\{\Lambda_n\}_{n\geq 0}$ is information decreasing if and only if for any $j > i $ one has
  \begin{equation}\label{PjPi}
     \mathrm{P}_{\rm guess}(\mathcal{E}_j) \leq  \mathrm{P}_{\rm guess}(\mathcal{E}_i) ,
  \end{equation}
where $\mathcal{E}_n =\{p_x,\Lambda_n(\rho_x)\}_x$. Finally, $\{\Lambda_n\}_{n\geq 0}$ is completely information decreasing if and only if ${\rm id}_S \otimes \Lambda_n$ is information decreasing (${\rm id}_S$ stands for the identity map on $\mathrm{L}(\mathcal{H}_S)$).
\end{Definition}
The main result of \cite{datta} consists in the following

\begin{Theorem} Time discrete dynamical map $\{\Lambda_n\}_{n\geq 0}$ is CP-divisible if and only if it is completely information decreasing.
\end{Theorem}

\subsection{Distinguishability of quantum channels}

Consider an ensemble of quantum channels $\mathcal{F}=\{p_k,\Phi_k\}$, where $\Phi_k : \LH \to \LH$. Define the corresponding guessing probability via

\begin{equation}\label{PI}
  {\rm P}_{\rm guess}(\mathcal{F}) = \sup_\rho {\rm P}_{\rm guess}(\mathcal{E}(\rho)) ,
\end{equation}
where $\mathcal{E}(\rho) = \{p_k,\Phi_k(\rho)\}$. The above strategy allows to use only states $\rho$ of the system living in $\HH$. Now, if one can use also states of a composite system living in $\mathbb{C}^k \otimes \HH$ (i.e. the system is coupled to $k$-dimensional ancilla), then the efficiency of discrimination may be increased \cite{Piani}. In this case (\ref{PI}) can be generalized as follows

\begin{equation}\label{PII}
  {\rm P}^{(k)}_{\rm guess}(\mathcal{F}) = \sup_{\rho_{SA}} {\rm P}_{\rm guess}(\mathcal{E}(\rho_{SA})) ,
\end{equation}
where the supremum is performed over all system-ancilla states. It is clear that

\begin{equation}\label{}
   {\rm P}_{\rm guess}(\mathcal{F}) =  {\rm P}^{(1)}_{\rm guess}(\mathcal{F}) \leq    {\rm P}^{(2)}_{\rm guess}(\mathcal{F})  \leq \ldots \leq  {\rm P}^{(d)}_{\rm guess}(\mathcal{F})  ,
\end{equation}
where $d = {\rm dim}\HH$.

\begin{Proposition} If the dynamical map $\{\Lambda_t\}_{t \geq 0}$ is $k$-divisible, then

\begin{equation}\label{}
  \frac{d}{dt}  {\rm P}^{(k)}_{\rm guess}(\mathcal{F}(t)) \leq 0 ,
\end{equation}
for any ensemble of quantum channels $\mathcal{F}$, where $\mathcal{F}(t) = \{p_k,\Lambda_t\, \Phi_k\}$.
\end{Proposition}
The above scheme simplifies if $\mathcal{F} = \{p_1,\Phi_1;p_2,\Phi_2\}$. In this case

\begin{equation}\label{}
  {\rm P}^{(k)}_{\rm guess}(\mathcal{F}) = \frac 12 \Big( 1 + \| p_1 \Phi_1 - p_2 \Phi_2\|^{(k)}_1 \Big)
\end{equation}
where we introduced a norm of a linear map $\Phi : \LH \to \LH$

\begin{equation}\label{}
  \| \Phi \|_1^{(k)} = \sum_{\rho_{SA}} \| [{\rm id}_k\otimes \Phi](\rho_{SA})\|_1 .
\end{equation}
For $k=d$ one recovers a celebrated diamond norm \cite{Kitaev} (cf. also \cite{Watrous})

\begin{equation}\label{}
  \| \Phi\|_\diamond :=  \| \Phi \|_1^{(d)} .
\end{equation}
Using the diamond norm one arrives at the following result \cite{Junu-DC}

\begin{Proposition} An invertible dynamical map $\{\Lambda_t\}_{t \geq 0}$ is CP-divisible if and only if

\begin{equation}\label{}
  \frac{d}{dt} \| p_1 \Lambda_t \, \Phi_1  - p_2 \Lambda_t \, \Phi_2 \|_\diamond \leq 0  ,
\end{equation}
for any  $\mathcal{F} = \{p_1,\Phi_1;p_2,\Phi_2\}$.
\end{Proposition}

\subsection{Entropic relations}

Divisible dynamical maps enjoy several monotonicity properties for important entropic quantities. One of the basic quantities is a relative entropy defined in (\ref{Relative-S}) (cf. \cite{Wehrl,O-P,TOM} for the detailed review).
%
%\begin{equation}\label{Relative}
%  S(\rho|\!|\sigma) = \left\{ \begin{array}{ll} {\rm Tr}[\rho(\log\rho - \log\sigma])\ , & \ \mbox{if}\ {\rm supp}\, \rho \subseteq {\rm supp}\, %\sigma \\ + \infty \ , & \ \mbox{otherwise} \end{array} \right. \ .
%\end{equation}
Relative entropy provides an important tool to discriminate between two states. Moreover,  it plays an essential role in quantum information theory \cite{QIT} and quantum thermodynamics (cf. section   \ref{SUB-QS}). It displays the following monotonicity property

\begin{Proposition} If the dynamical map $\{\Lambda_t\}_{t\geq 0}$ is P-divisible, then
  \begin{equation}\label{}
    \frac{d}{dt} S(\Lambda_t({\rho})|\!|\Lambda_t({\sigma}))\leq 0,
  \end{equation}
for any ${\rho}$ and ${\sigma}$ living in $\TTH$.
\end{Proposition}
The von Neumann entropy $S(\rho)$, contrary to relative entropy, in general is not monotonic. It is well known that $S(\Phi(\rho)) \geq S(\rho)$ for all $\rho$ if and only if the map $\Phi$ is bistochastic (PTP) \cite{ALBERTI}. Hence,   if the unital dynamical map $\{\Lambda_t\}_{t\geq 0}$ is P-divisible, then \cite{JPB-nasza,Paolo,Paolo-arxiv}

\begin{equation}\label{S>}
  \frac{d}{dt}S(\Lambda_t(\rho)) \geq 0 .
\end{equation}
For non unital maps the above inequality does not hold. However, one can derive the following bound \cite{Das-Wilde}

\begin{Proposition} If the dynamical map $\{\Lambda_t\}_{t\geq 0}$ is CP-divisible, then
  \begin{equation}\label{}
    \frac{d}{dt} S(\Lambda_t({\rho}))\geq - {\rm Tr}(\Pi_t \mathcal{L}^\ddag(\rho_t)) ,
  \end{equation}
where $\Pi_t$ is the projection onto the support of the state $\rho_t = \Lambda_t(\rho)$.
\end{Proposition}
In particular when $\rho_t > 0$ , then $\Pi_t = \oper$ and hence

\begin{equation}\label{}
    \frac{d}{dt} S(\Lambda_t({\rho}))\geq - {\rm Tr}(\mathcal{L}^\ddag(\rho_t)) .
\end{equation}
Using the diagonal representation

\begin{equation}\label{}
  \mathcal{L}^\ddag_t(X) = i[H(t),X] + \sum_k \gamma_k(t) \Big( L^\dagger_k(t) X L_k(t) - \frac 12 \{ L_k^\dagger(t) L_k(t),X\}\Big) ,
\end{equation}
one finds \cite{Fabio-Heide} (cf. also \cite{Abe}) ${\rm Tr}(\mathcal{L}^\ddag(\rho)) = \sum_k \gamma_k(t) {\rm Tr}( [L_k(t),L^\dagger_k(t)]\rho)$ and hence

\begin{equation}\label{}
 \frac{d}{dt} S(\Lambda_t({\rho})) \geq  \sum_k \gamma_k(t) {\rm Tr}( [L^\dagger_k(t),L_k(t)]\rho_t) .
\end{equation}

The original von Neumann relative entropy (\ref{Relative-S}) was generalized to a family of relative  quantum entropies -- so called R\'enyi-$\alpha$ divergences -- defined via \cite{O-P}
\begin{equation}\label{}
  S_\alpha(\rho|\!|\sigma) = \frac{1}{\alpha -1} \log {\rm Tr}[\rho^\alpha \sigma^{1-\alpha}] ,
\end{equation}
for $\alpha \in (0,1) \cup (1,+\infty)$. One recovers the standard relative entropy (\ref{Relative-S}) in the limit $\lim_{\alpha \to 1}  D_\alpha(\rho|\!|\sigma) =  D(\rho|\!|\sigma)$. R\'enyi-$\alpha$ divergence is monotonic under CPTP maps (i.e.
satisfies the data processing inequality) for $\alpha \in (0,1) \cup (1,2]$ \cite{O-P,Hiai} (and in the $\alpha \to 1$ limit).
 Moreover, it is known that it is monotonic under positive trace-preserving maps in the limit for $\alpha \to 0$ and for $\alpha \to 1$ (cf. \cite{Reeb}). Recently, a new class of R\'enyi-$\alpha$ divergences was introduced \cite{Tom,WWY}. These {\em sandwiched R\'enyi divergences} and are defined by

\begin{equation}\label{Sand}
   \widetilde{S}_\alpha(\rho|\!|\sigma) =  \frac{1}{\alpha-1} \log \left( {\rm Tr}\left[ \left( \sigma^{\frac{1-\alpha}{2\alpha}} \rho \, \sigma^{\frac{1-\alpha}{2\alpha}} \right)^\alpha \right] \right) ,
\end{equation}
if $\,{\rm supp}\, \rho \subseteq {\rm supp}\, \sigma$, and equals $+\infty$ otherwise. Clearly, if $\rho$ and $\sigma$ commute, then $ \widetilde{S}_\alpha(\rho|\!|\sigma) =  {S}_\alpha(\rho|\!|\sigma)$.  For an arbitrary quantum channel $\mathcal{E}$ the sandwiched R\'enyi relative entropy of order $\alpha$ satisfies the data processing inequality \cite{Lieb-RE,Tom,Beigi,Ogawa}
\begin{equation}\label{Data-CP}
  \widetilde{S}_\alpha(\mathcal{E}(\rho)|\!|\mathcal{E}(\sigma))    \leq \widetilde{S}_\alpha(\rho|\!|\sigma)  ,
\end{equation}
for $\alpha \in [\frac 12,1) \cup (1,\infty)$. Moreover
\begin{equation}\label{}
  \widetilde{S}_{1/2}(\rho|\!|\sigma) = -2 \log F(\rho,\sigma) ,
\end{equation}
where

\begin{equation}\label{}
  F(\rho,\sigma) = \Big({\rm Tr} \sqrt{ \sqrt{\rho} \sigma \sqrt{\rho}} \Big)^2 ,
\end{equation}
denotes the Uhlmann fidelity \cite{Uhlmann-F,Jozsa}. The monotonicity property of relative entropies w.r.t. to CP- and P-divisible maps may be summarized as follows:

\begin{tcolorbox}
\begin{Proposition}  If the dynamical map $\{\Lambda_t\}_{t\geq 0}$ is CP-divisible, then
  \begin{equation}\label{}
    \frac{d}{dt} S_\alpha(\Lambda_t({\rho})|\!|\Lambda_t({\sigma}))\leq 0, \ \ \ \alpha \in (0,1) \cup (1,2] ,
  \end{equation}
and

\begin{equation}\label{}
  \frac{d}{dt}  \widetilde{S}_\alpha(\Lambda_t(\rho) |\!| \Lambda_t(\sigma)) \leq 0 , \ \ \ \alpha \in \{\frac 12\} \cup (1,\infty) .
\end{equation}
If the dynamical map $\{\Lambda_t\}_{t\geq 0}$ is P-divisible, then
 \begin{equation}\label{}
    \frac{d}{dt} S_\alpha(\Lambda_t({\rho})|\!|\Lambda_t({\sigma})) \leq 0, \ \ \  \alpha \in \{0,1,2\} .
  \end{equation}
\end{Proposition}
\end{tcolorbox}
In particular for a CP-divisible dynamics one has

\begin{equation}\label{}
  \frac{d}{dt}  F(\Lambda_t(\rho),\Lambda_t(\sigma)) \geq 0 ,
\end{equation}
for any pair of states $\rho$ and $\sigma$ \cite{Usha}.

Note, that an original formula for the relative entropy (\ref{Relative-S}) is not bounded and may diverge even in finite dimensions. An interesting proposal for regularized version was provided in \cite{Ade1,Ade2}. Telescopic relative entropy \cite{Ade1} or quantum skew divergence \cite{Ade2} is defined as follows

\begin{equation}\label{}
  S_\mu(\rho|\!| \sigma) = \log(1/\mu)^{-1} S(\rho|\!| \mu \rho + (1-\mu)\sigma) ,
\end{equation}
with $\mu \in (0,1)$ (actually, the classical counterpart was introduced in \cite{Lee} under the name `skew divergence'). The quantum sqew divergence is bounded and it satisfies

\begin{equation}\label{}
  0 \leq    S_\mu(\rho|\!| \sigma)   \leq 1 .
\end{equation}
Moreover, $ S_\mu(\rho|\!| \sigma)=0$ only when $\rho=\sigma$, $ S_\mu(\rho|\!| \sigma)=1$ and $\rho \perp \sigma$. For Markovian CP-divisible evolution one finds

\begin{equation}\label{}
  \frac{d}{dt}  S_\mu(\Lambda_t(\rho)|\!| \Lambda_t(\sigma)) \leq 0 .
\end{equation}
Quantum skew divergence (telescopic relative entropy) in connection to non-Markovian evolution was recently analyzed in \cite{Bassano-2021}.

\subsection{Divisibility vs. Riemannian metrics}

A convex set of density operators may be equipped with a Riemannian metric which allows to define a distance between any pair of density  operators \cite{KAROL} (see also \cite{Amari-1,Amari-2}). Denote by $\mathcal{S}_{>0}(\HH)$ a subset of faithful states $\rho > 0$.

\begin{Definition} A Riemannian metric $\mathbf{\mathbf{g}}$ on $\mathcal{S}_{>0}(\HH)$ is monotone (or contractive) if for any quantum channel $\mathcal{E} : \LH \to \LH$

\begin{equation}\label{}
  \mathbf{g}_{\mathcal{E}(\rho)}(\mathcal{E}(A),\mathcal{E}(A)) \leq \mathbf{g}_\rho(A,A) ,
\end{equation}
 for any state $\rho$ and any Hermitian traceless operator $A$.
\end{Definition}
Geometrically, a Hermitian traceless operator $A$ may be interpreted as a tangent vector to $\mathcal{S}_{>0}(\HH)$ attached at $\rho$. The above formula states that the length of $A$ attached at $\rho$ is never smaller than the length of $\mathcal{E}(A)$ attached at $\mathcal{E}(\rho)$.
It is well known that contrary to the classical case where there is essentially a unique monotone Riemannian metric defined on a probability simplex (so called Fisher metric) in quantum case monotone Riemannian metrics are characterized as follows \cite{Cencov,Morozova,Petz,Sudar,Ruskai}

\begin{equation}\label{}
  \mathbf{g}_\rho(A,B) = {\rm Tr}(A \mathbf{K}^{-1}_\rho(B)) ,
\end{equation}
where the operator $\mathbf{K}_\rho : \LH \to \LH$ is defined via

\begin{equation}\label{}
  \mathbf{K}_\rho = k(L_\rho \,R_\rho) R_\rho ,
\end{equation}
with $L_\rho(A) = \rho A$,  $R_\rho(A) = A \rho $, and $k : \mathbb{R}_+ \to \mathbb{R}$ is an operator monotone function satisfying

\begin{equation}\label{k(t)}
  k(t) = t k(t^{-1}) ,
\end{equation}
together with $k(1)=1$. Recall, that $f: \mathbb{R}_+ \to \mathbb{R}_+$ is an operator monotone if for any $A \geq B \geq 0$ one has $f(A) \geq f(B)$. There is one to one correspondence between monotone Riemannian metrics and so-called Morozova-\v{C}encov function $k(t)$  \cite{Cencov,Morozova,Petz}.

%\begin{tcolorbox}
\begin{Proposition} If $\{\Lambda_t\}_{t\geq 0}$ is CP-divisible, then
\begin{equation}\label{}
  \frac{d}{dt}  \mathbf{g}_{\Lambda_t(\rho)}(\Lambda_t(A),\Lambda_t(A)) \leq 0 ,
\end{equation}
for any monotone Riemannian metric $\mathbf{g}$.
\end{Proposition}
%\end{tcolorbox}
Given a metric $\mathbf{g}$ one can define a geodesic distance between any two density operators

\begin{equation}\label{}
  D(\rho,\sigma) = \inf_{\gamma(t)} \int_0^1 \sqrt{ \mathbf{g}_{\gamma(t)}(\dot{\gamma}(t),\dot{\gamma}(t)) } dt ,
\end{equation}
where the infimum is over all trajectories $\gamma(t) \in \mathcal{S}_{>0}(\HH)$ connecting $\rho$ and $\sigma$, i.e. $\gamma(0) = \rho$ and $\gamma(1)=\sigma$. If the evolution is CP-divisible, then $\,  \frac{d}{dt} D(\Lambda_t(\rho),\Lambda_t(\sigma)) \leq 0\,$
for any pair of initial states $\rho$ and $\sigma$.

%\begin{figure} \label{}
%\begin{center}
%\hspace*{.4cm}
%\includegraphics[width=11cm]{rys_3a.pdf}
%\caption{(Color online) For CP-divisible evolution the distance between $\rho_t$ and $\sigma_t$ monotonically decreases. } \label{3a}
%\end{center}
%\end{figure}

\begin{Example} Unfortunately, it is very hard to find a geodesic distance for an arbitrary Morozova-\v{C}encov function $k(t)$. Analytic expressions for the distance are known only for $k(t) = \frac{2}{1+t}$  corresponding to Bures distance \cite{Bures,Uhlmann-1,Uhlmann-2,KAROL}

\begin{equation}\label{}
  D_{\rm B}(\rho,\sigma) = \sqrt{2(1- F(\rho,\sigma))} ,
\end{equation}
or, equivalently, Bures angle
\begin{equation}\label{}
  \ell_{\rm B}(\rho,\sigma) = \arccos[\sqrt{F(\rho,\sigma)}] ,
\end{equation}
and for $k(t)= \frac 14 (\sqrt{t}+1)$ corresponding to so called Wigner-Yanase angle \cite{Isola}

\begin{equation}\label{}
  \ell_{\rm WY}(\rho,\sigma) = \arccos[A(\rho,\sigma)] ,
\end{equation}
where $A(\rho,\sigma) = {\rm Tr}(\sqrt{\rho} \sqrt{\sigma})$ is known as quantum affinity (cf. the recent reviews \cite{Adesso,Jarzyna}).
\end{Example}

\begin{Remark} Note, that a trace distance
\begin{equation}\label{}
  D_{\rm Tr}(\rho,\sigma) = \| \rho - \sigma\|_1 ,
\end{equation}
is monotone but does not arise from the Riemannian metric. On the other hand the Hilbert-Schmidt distance

\begin{equation}\label{}
  D_{\rm HS}(\rho,\sigma) = \Big({\rm Tr}(\rho - \sigma)^2 \Big)^{1/2}  ,
\end{equation}
is Riemannian but not monotone. Hilbert-Schmidt distance was used recently \cite{Lo-Franco} to witness non-Markovian evolution via so called Hilbert Schmidt speed (cf. also \cite{Smerzi}).
\end{Remark}

%Interestingly, author of \cite{Ruskai} shown that monotonicity property of the Riemannian metric defined via function $k(t)$ satisfying %(\ref{k(t)}) may be infer  from the following result

%\begin{Proposition} For any pair of states $\rho$, $\sigma$, and any quantum channel $\mathcal{E}$
%
%\begin{equation}\label{Ruskai}
%  {\rm Tr}\Big( A^\dagger \frac{1}{R_\sigma + \lambda L_\rho}(A) \Big)\, \geq  \, {\rm Tr}\Big( \mathcal{E}(A^\dagger) %\frac{1}{R_{\mathcal{E}(\sigma)} + \lambda L_{\mathcal{E}(\rho)}}(\mathcal{E}(A)) \Big) ,
%\end{equation}
%with $\lambda > 0$ and arbitrary $A \in \LH$.
%\end{Proposition}

%\begin{Cor} If $\{\Lambda_t\}_{t\geq 0}$ is CP-divisible, then
%  \begin{equation}\label{}
%        \frac{d}{dt}  {\rm Tr}\Big( \Lambda_t(A^\dagger) \frac{1}{R_{\Lambda_t(\sigma)} + \lambda L_{\Lambda_t(\rho)}}(\Lambda_t(A)) \Big) \, %\leq \, 0 ,
%\end{equation}
%\end{Cor}
%Actually, inequality (\ref{Ruskai}) holds whenever $\mathcal{E}^\ddag$ is a unital Schwarz map. Hence, the above Corollary is also true for %Schwarz divisible dynamical maps.

In the qubit case the following formula for the Bures metric was derived in \cite{Hubner,Dittmann}

\begin{equation}\label{Hubner}
  D_B(\rho,\rho+d\rho)^2 = \frac 12 {\rm Tr}(d \rho \, d\rho) + (d \sqrt{\rho} )^2 .
\end{equation}
Moreover, using Bloch vector parametrization $\rho_k = \frac 12 (\oper + \mathbf{x}_k \cdot \boldsymbol\sigma)$ one finds \cite{Hubner}

\begin{equation}\label{}
  D_{\rm B}^2(\rho_1,\rho_2) = 2 - \sqrt{2} \sqrt{ 1 + \mathbf{x}_1 \mathbf{x}_2 +  \sqrt{1- |\mathbf{x}_1|^2} \sqrt{1 -|\mathbf{x}_2|^2} } .
\end{equation}
Monotone Riemannian metrics were recently analyzed in  \cite{Zanardi-m} in connection to symmetries and conservation laws of Markovian evolution.

%Non-Markovianity of continuous-variable Gaussian dynamical maps was analyzed in terms of Bures distance in \cite{Bures-Gaussian}.

\subsection{Divisibility and  information measures}

\subsubsection{Mutual information}

Mutual information  \cite{QIT,Wilde,Hayashi} serves  as a measure of total correlations encoded into a composite state $\rho_{AB}$

\begin{equation}\label{}
  I(\rho_{AB}) = S(\rho_A) + S(\rho_B) - S(\rho_{AB}) ,
\end{equation}
where $\rho_A$ and $\rho_B$ are marginal states of the composite state $\rho_{AB}$. Totally uncorrelated state $\rho_{AB} = \rho_A \otimes \rho_B$ results in $I(\rho_{AB}) =0$. Mutual information satisfies the data processing inequality \cite{Wilde,QIT}

\begin{equation}\label{}
  I([\mathcal{E} \otimes {\rm id}_B](\rho_{AB})) \leq  I(\rho_{AB}) ,
\end{equation}
for any quantum channel $\mathcal{E} : \mathrm{L}(\HH_A) \to \mathrm{L}(\HH_A)$. One has therefore \cite{Luo}

\begin{Cor} If $\{\Lambda_t\}_{t\geq 0}$ is CP-divisible, then

\begin{equation}\label{}
  \frac{d}{dt} I([\Lambda_t \otimes {\rm id}_A](\rho^{SA})) \leq 0 ,
\end{equation}
for any $\rho^{SA}$ living in $\mathcal{H}_S \otimes \HH_A$ and arbitrary ancillary Hilbert space $\mathcal{H}_A$.
\end{Cor}

\begin{Example} Consider qubit evolution generated by

\begin{equation}\label{}
  \mathcal{L}_t(\rho)= \frac 12 \gamma(t)(\sigma_z \rho \sigma_z - \rho) .
\end{equation}
Let $\mathcal{H}_A = \HH = \mathbb{C}^2$, and let $\rho^{SA} = |\psi^+_2\>\<\psi^+_2|$ be a maximally entangled state. One finds

\begin{equation}\label{}
  \rho^{SA}_t = \frac 12 \left( \begin{array}{cc|cc}
                         1 & 0 & 0 &  e^{-\Gamma(t)} \\
                         0 & 0 & 0 & 0 \\ \hline
                         0 & 0 & 0 & 0 \\
                         e^{-\Gamma(t)} & 0 & 0 & 1
                       \end{array} \right) ,
\end{equation}
with $\Gamma(t) = \int_0^t \gamma(\tau)d\tau$. Simple calculation \cite{Luo} leads to

\begin{equation}\label{}
  \frac{d}{dt} I(\rho^{SA}_t) = - \gamma(t) e^{\Gamma(t)} \log_2 \frac{1+ e^{\Gamma(t)}}{1-e^{\Gamma(t)}} ,
\end{equation}
and hence $\frac{d}{dt} I(\rho^{SA}_t) \leq 0$ if and only if $\gamma(t) \geq 0$.
\end{Example}

\subsubsection{Fisher information}

Quantum Fisher information  is one of the most important quantities for quantum metrology and quantum estimation theory \cite{Wootters-F,Giovanetti,est-1}. Its inverse  provides the lower bound of the error of the estimation through the quantum Cram\'er-Rao bound \cite{Helstrom,Holevo,est-1,est-2,est-3,est-4}. Quantum Fisher information is a direct generalization of the corresponding classical concept. Suppose that a probability vector $p_k(x)$ depends upon a real parameter $x$ and the goal is to estimate the value of $x$ out of the result of measurements. The classical Fisher information reads

\begin{equation}\label{Fisher-cl}
  F(x) = \sum_k p_k(x) \left( \frac{\partial_x p_k(x)}{p_k(x)} \right)^2 ,
\end{equation}
and in the case of several parameters $\mathbf{x}=(x^1,\ldots,x^r)$ it generalizes to the Fisher-Rao metric

\begin{equation}\label{FR-cl}
  F_{ab}(\mathbf{x}) = \sum_k \frac{\partial_a p_k(\mathbf{x}) \partial_b p_k(\mathbf{x})}{ p_k(\mathbf{x})} ,
\end{equation}
with $\partial_a = \partial/\partial x^a$.  The Fisher-Rao metric characterizes the inverse variance of an unbiased estimator $X^a$   \cite{Helstrom}

\begin{equation}\label{}
  \< X^a X^b \> - \< X^a\>\<X^b \> \geq  \Delta^{ab}_{\rm CR} := F^{ab} ,
\end{equation}
where $F^{ab}$ is the inverse of $F_{ab}$. The quantum analog of (\ref{Fisher-cl}) is defined by

\begin{equation}\label{}
  \mathcal{F}(x) = {\rm Tr}(\rho(x) L^2 ) ,
\end{equation}
where now $\rho(x)$ is a quantum state depending on a parameter $x \in \mathbb{R}$, and $L$ denotes so called  symmetric logarithmic derivative defined via

\begin{equation}\label{}
  \partial_x\rho(x) = \frac 12 (L \rho(x) + \rho(x)L) .
\end{equation}
If $\rho(x) = \sum_k p_k(x) |k(x)\>\<k(x)|$ is a spectral decomposition of $\rho(x)$, then

\begin{equation}\label{}
  \mathcal{F}(x) = \sum_k \frac{(\partial_x p_k(x))^2}{p_k(x)} + 2 \sum_{i\neq j} \frac{(p_i(x) - p_j(x))^2}{p_i(x) + p_j(x)} |\< i(x)|\partial_x j(x)\>|^2 .
\end{equation}
Hence $ \mathcal{F}(x)$ apart from the classical part corresponding to probabilities $p_k(x)$ contains purely quantum correction which takes into account $x$-dependence of eigenvectors $|k(x)\>$. Quantum Fisher information is directly related to the Bures distance \cite{est-2}

\begin{equation}\label{}
  D_{\rm B}(\rho(x),\rho(x+dx))^2 = \frac 14 \mathcal{F}(x) dx^2 .
\end{equation}
If the $x$-dependence is characterized via

\begin{equation}\label{}
  \rho(x) = e^{-i A x} \rho e^{i A x} ,
\end{equation}
with Hermitian operator $A$, then one defines  \cite{Helstrom,Holevo} quantum Fisher information $\mathcal{F}[\rho,A]$ of a state $\rho$ with respect to an observable $A$

\begin{equation}\label{}
\mathcal{F}(x) \equiv \mathcal{F}[\rho,A] = 2 \sum_{k\neq l} \frac{(p_k - p_l)^2}{p_k + p_l} |\< k(x)|A|l(x)\>|^2 .
\end{equation}
Note, that a classical part disappears since $p_k$ does not depend on $x$. The above scenario can be generalized to the case of several parameters $\mathbf{x}=(x_1,\ldots,x_r)$. One defines \cite{Helstrom,Holevo} a quantum Fisher information matrix

\begin{equation}\label{F-ij}
  \mathcal{F}_{\mu\nu}(\mathbf{x}) = \frac 12 {\rm Tr}(\rho(\mathbf{x}) \{L_\mu,L_\nu\}) ,
\end{equation}
where $L_\mu$ denotes  symmetric logarithmic derivative for the parameter $x^\mu$ defined via

\begin{equation}\label{}
  \partial_\mu \rho(\mathbf{x}) = \frac 12 (L_\mu \rho(\mathbf{x}) + \rho(\mathbf{x}) L_\mu ) .
\end{equation}
The diagonal elements of $\mathcal{F}_{\mu\nu}$

\begin{equation}\label{}
  \mathcal{F}_{\mu\mu}(\mathbf{x}) = {\rm Tr}(\rho(\mathbf{x}) L_\mu^2) ,
\end{equation}
define quantum Fisher information for a parameter $x_\mu$. One proves   \cite{Sun} the following

%\begin{tcolorbox}
\begin{Proposition}
Suppose that $\rho_t(\mathbf{x})$ satisfies time-local master equation

\begin{equation}\label{}
  \dot{\rho}_t(\mathbf{x}) = - i[H(t),\rho_t(\mathbf{x})] + \sum_k \gamma_k(t) \Big( V_k(t) \rho_t(\mathbf{x}) V^\dagger_k(t) - \frac 12 \{ V^\dagger_k(t)  V_k(t),\rho_t(\mathbf{x})\} \Big) .
\end{equation}
Then

\begin{equation}\label{}
  \frac{\partial}{\partial t} \mathcal{F}_{\mu\mu}(\mathbf{x}) = - \sum_k \gamma_k(t) {\rm Tr}\Big( \rho_t(\mathbf{x}) [L_\mu,V_k(t)]^\dagger [L_\mu,V_k(t)] \Big) .
\end{equation}
\end{Proposition}
%\end{tcolorbox}

\begin{Cor} If the evolution $\{\Lambda_t\}_{t \geq 0}$ is CP-divisible,  then the quantum Fisher information flow $ \frac{\partial}{\partial t} \mathcal{F}_{\mu\mu} \leq 0$ for $\mu=1,\ldots,r$.
\end{Cor}

Assuming that $\rho(\mathbf{x})$ is of full rank and using the spectral representation $\rho(\mathbf{x}) = \sum_k p_k(\mathbf{x}) |k(\mathbf{x})\>\<k(\mathbf{x})|$  one finds

\begin{equation}\label{}
  \mathcal{F}_{\mu\nu} = \sum_{i=1}^d \frac{\partial_\mu p_i \partial_\nu p_i}{p_i} + 2 \sum_{i \neq j} \frac{(p_i-p_j)^2}{p_i+p_j} {\rm Re}\, \<i|\partial_\mu |j\>\<j|\partial_\nu|i\>
\end{equation}
where we skipped dependence on $\mathbf{x}$.  Note, that the first term $ \sum_{i=1}^d p_i^{-1} \partial_\mu p_i \partial_\nu p_i$ reproduces the classical Fisher information and hence the second term provides purely quantum correction.
It turns out (cf. \cite{KAROL}) that quantum Fisher information matrix $\mathcal{F}_{\mu\nu}$ defines a Riemannian metric on the parameter space $(x_1,\ldots,x_r)$ (known also as a quantum geometric tensor \cite{Book}). This Riemannian metric is again directly related to the Bures metric on the space of quantum states

%(cf. Figure \ref{3b})

\begin{equation}\label{}
  D^2_{\rm B}(\rho(\mathbf{x}),\rho(\mathbf{x}+d\mathbf{x})) = \frac 14 \sum_{\mu,\nu}\mathcal{F}_{\mu\nu}(x) dx^\mu dx^\nu .
\end{equation}

%\begin{figure} \label{}
%\begin{center}
%\hspace*{.4cm}
%\includegraphics[width=13cm]{rys_3b.pdf}
%\caption{(Color online) A mapping from parameter space $\mathcal{M}$ to the space of states. The distance measured via quantum Fisher metric on %$\mathcal{M}$ corresponds to the distance measured by the Bures metric on $\mathcal{S}(\HH)$.    } \label{3b}
%\end{center}
%\end{figure}

\begin{Example} Consider again a qubit dephasing governed by a time local generator $\mathcal{L}_t(\rho) = \frac 12 \gamma(t)(\sigma_z \rho \sigma_z - \rho)$. A 1-parameter $x = \theta$ family of qubit states

\begin{equation}\label{}
  \rho(\theta) = e^{i \theta \sigma_z /2} \rho e^{- i \theta \sigma_z /2} = \left( \begin{array}{cc} \rho_{00} &\rho_{01} e^{i \theta} \\ \rho_{10} e^{-i \theta} & \rho_{11} \end{array} \right) ,
\end{equation}
gives rise to

\begin{equation}\label{}
  \mathcal{F}(\theta) d\theta^2 = 4 D^2_{\rm B}(\rho(\theta),\rho(\theta + d\theta)) .
\end{equation}
Using (\ref{Hubner}) one finds

\begin{equation}\label{}
   \mathcal{F}(\theta) = 4 |\rho_{01}|^2 ,
\end{equation}
and hence quantum Fisher information does not depend on the parameter $\theta$.  Now, if $\rho_t(\theta)$ satisfies $\partial_t\rho_t(\theta) = \mathcal{L}_t(\rho_t(\theta))$, then $\mathcal{F}(\theta) = 4 |\rho_{01}| e^{- \Gamma(t)}$, and hence $\partial_t \mathcal{F}(\theta) \leq 0$ if and only if $\gamma(t) \geq 0$ which provides a necessary and sufficient condition for Markovianity. Since the variance of any unbiased estimation  $\theta$ satisfies the quantum Cramer-Rao bound \cite{Helstrom}

\begin{equation}\label{}
  (\Delta \theta)^2 \geq \frac{1}{ \mathcal{F}(\theta)}  ,
\end{equation}
it is, therefore,  clear that in the course of time the estimation error (variance) increases whenever the evolution is Markovian.

\end{Example}
Authors of \cite{inter-power} introduced the concept of quantum interferometric power which is defined in terms of the minimal quantum Fisher
information obtained by local unitary evolution of one part of the system. It provides another interesting quantity based on quantum Fisher information which displays monotonic behaviour in the course of Markovian evolution.

\subsubsection{Wigner-Yanase-Dyson skew information }

Wigner-Yanase skew information \cite{Wigner-Yanase} of a state $\rho$ w.r.t. an observable $X=X^\dagger$ is defined by

\begin{equation}\label{}
  I^{\rm WY}(\rho,X) := - \frac 12 {\rm Tr}[\sqrt{\rho},X]^2 .
\end{equation}
The original motivation \cite{Wigner-Yanase} was a quantification of  the information contained in quantum state $\rho$  with respect to a conserved additive quantity represented by an observable $X$. This information-theoretic quantity was later generalized by Dyson into a family called
Wigner-Yanase-Dyson skew information

\begin{equation}\label{}
  I_p^{\rm WYD}(\rho,X) := - \frac{1}{p(1-p)} {\rm Tr}([{\rho}^p,X][\rho^{1-p},X]) ,
\end{equation}
for arbitrary 0 < p < 1.  The convexity of $I_{1/2}^{\rm WYD}$ was
already proved by Wigner and Yanase \cite{Wigner-Yanase}. The celebrated Wigner-Yanase-Dyson conjecture that $I_p^{\rm WYD}$ is convex in $\rho$ for any $p\in (0,1)$ was proved by Lieb \cite{Lieb} (ten year after original proposal by Wigner and Yanase \cite{Wigner-Yanase}). Recently, the Wigner-Yanase-Dyson skew information has been recognized as an interesting resource measures for the resource theory of asymmetry   (cf. \cite{Takagi}).  Interestingly, Wigner-Yanase information is directly related to Wigner-Yanase distance \cite{Isola}

\begin{equation}\label{}
  I^{\rm WY}(\rho,X) = 4\mathbf{g}^{\rm WY}_\rho([\rho,X],[\rho,X]) .
\end{equation}
If $\rho =|\psi\>\<\psi|$ is pure, then

\begin{equation}\label{}
  I^{\rm WY}(\rho,X) = \<\psi|X^2|\psi\> - \<\psi|X|\psi\>^2 ,
\end{equation}
is nothing but the variance of $X$. Wigner-Yanase-Dyson skew information enjoys the following property \cite{O-P,PETZ}

\begin{Proposition} For any quantum channel $\mathcal{E} : \LH \to \LH$

\begin{equation}\label{}
  I_p^{\rm WYD}(\mathcal{E}(\rho),X)  \leq  I_p^{\rm WYD}(\rho,\mathcal{E}^\ddag(X)) .
\end{equation}
for $p \in (0,1)$.
\end{Proposition}
Hence, if $\rho$ is an invariant state $\mathcal{E}(\rho)=\rho$, then

\begin{equation}\label{}
  I_p^{\rm WYD}(\rho,X)  \leq  I_p^{\rm WYD}(\rho,\mathcal{E}^\ddag(X)) .
\end{equation}

%\begin{tcolorbox}
\begin{Cor}[\cite{nasza-JPD}] If the evolution $\{\Lambda_t\}_{t \geq 0}$ is CP-divisible and $\Lambda_t(\rho_{\rm ss})=\rho_{\rm ss}$, then

%\begin{equation}\label{}
%  I_p^{\rm WYD}(\Lambda_t(\rho),X)  \leq  I_p^{\rm WYD}(\rho,\Lambda_t^\ddag(X)) .
%\end{equation}
%for $p \in (0,1)$. If $\rho_{\rm ss}$ is time independent invariant state
\begin{equation}\label{WYD-inv}
   \frac{d}{dt} I_p^{\rm WYD}(\rho_{\rm ss},\Lambda_t^\ddag(X)) \leq 0 ,
\end{equation}
for any $p \in (0,1)$.
\end{Cor}
%\end{tcolorbox}
In particular if $\rho_{\rm ss} = |\psi\>\<\psi|$, then (\ref{WYD-inv}) reduces to

\begin{equation}\label{}
  \frac{d}{dt} \Big( \<\psi| X_t^2 |\psi\> - \<\psi|X_t|\psi\>^2 \Big) \geq 0 ,
\end{equation}
where $X_t = \Lambda_t^\ddag(X)$ is a time evolved observable.

%\begin{Example}

%\end{Example}

\subsubsection{Divisibility vs. capacity of quantum channels}

A classical channel is represented by a stochastic matrix and there is one-to-one correspondence between classical channels and conditional probabilities.  Given a classical channel $T$ and the input state $\mathbf{p}$ one may introduce mutual information of `$T$' given an input $\mathbf{p}$ as follows
\begin{equation}\label{IpT}
  I(\mathbf{p},T) = S(\mathbf{p}) + S(T \mathbf{p}) - S(\mathbf{P})  ,
\end{equation}
where $\mathbf{P}$ denotes a joint state, that is, $P_{ij} = T_{ji} p^X_i$. Clearly,

\begin{equation}\label{}
  I(\mathbf{p},T) = I(\mathbf{P}) = S(\mathbf{p}) + S(\mathbf{q}) - S(\mathbf{P}) ,
\end{equation}
where $\mathbf{q}$ is a (second) marginal of $\mathbf{P}$, i.e. $q_i = \sum_j P_{ij} = \sum_j T_{ij} p_j$. One of the most important characteristics of the classical channel -- Shannon's capacity -- is defined in terms of mutual information as follows
\begin{equation}\label{}
  \mathcal{C}(T) := \sup_{\mathbf{p}} I(\mathbf{p},T) ,
\end{equation}
where the supremum is taken with respect to all input probability distributions $\mathbf{p}$  \cite{QIT,Holevo-73,Holevo,Wilde}.

\begin{Proposition} If $\{T(t)\}_{t \geq 0}$ is P-divisible classical dynamical map, then $\,   \frac{d}{dt} \mathcal{C}(T(t)) \leq 0$.
\end{Proposition}

Consider now a quantum channel $\Phi : \mathrm{L}(\HH_A) \to \mathrm{L}(\HH_B)$.  Being completely positive and trace-preserving map  it  gives rise to the environmental  representation \cite{Paulsen,Stormer}
\begin{equation}\label{}
  \Phi(\rho_A) = {\rm Tr}_E V \rho_A V^\dagger ,
\end{equation}
where $V : \mathcal{H}_A \to \mathcal{H}_B \otimes \mathcal{H}_E$ is a linear isometry, and $\mathcal{H}_E$ may be interpreted as an environmental Hilbert space. This representation allows to define a complementary channel $\Phi^c : \mathrm{L}(\mathcal{H}_A) \to \mathrm{L}(\mathcal{H}_E)$
\begin{equation}\label{compl}
  \Phi^c(\rho_A) := {\rm Tr}_B V \rho_A V^\dagger .
\end{equation}
Finally, complementary channel gives rise to the entropy exchange $S(\rho_A,\Phi)$ defined by
\begin{equation}\label{}
  S(\rho_A,\Phi) = S(\rho_E) .
\end{equation}
with $\rho_E = \Phi^c(\rho_A)$. The entropy exchange $S(\rho_A,\Phi)$  characterizes the information exchange between the original system $A$ and the environment during the process described by the quantum channel $\Phi$. It is a function of the input state $\rho_A$ and the channel itself and it was proved \cite{Schu} that $S(\rho_A,\Phi)$  does not depend on the particular environmental representation. The quantum analog of (\ref{IpT}) reads:  given an input state $\rho_A$ and a channel $\Phi$ one defines \cite{Holevo,O-P}
\begin{equation}\label{I2}
  I(\rho_A,\Phi) := S(\rho_A) + S(\rho_B) - S(\rho_E) ,
\end{equation}
where $\rho_B = \Phi(\rho_A)$ is an output state. Equivalently,
\begin{equation}\label{}
  I(\rho_A,\Phi) = S(\rho_A) + S(\Phi(\rho_A)) - S(\Phi^c(\rho_A)) .
\end{equation}
Note, that in the quantum case the above definition of  mutual information $ I(\rho_A,\Phi)$ does not in general coincide with $I(\rho_{AB}) = S(\rho_A) + S(\rho_B) - S(\rho_{AB})$ since the compound (joint) state $\rho_{AB}$ is not uniquely defined by the channel $\rho$ and the initial state \cite{Taka}. However, $ I(\rho_A,\Phi)$  also enjoys the data  processing inequality \cite{Wilde}. Using $ I(\rho_A,\Phi)$ one define coherent information \cite{Schu,Wilde}

\begin{equation}\label{}
  I_c(\rho_A,\Phi) :=  S(\Phi(\rho_A)) - I(\rho_A,\Phi) .
\end{equation}

\begin{Proposition} If $\{\Lambda_t\}_{t \geq 0}$ is a CP-divisible dynamical map, then

\begin{equation}\label{}
\frac{d}{dt}  I(\rho,\Lambda_t) \leq 0 , \ \ \ \  \frac{d}{dt}  I_c(\rho,\Lambda_t) \leq 0 ,
\end{equation}
for any initial state $\rho$.
\end{Proposition}
Finally, $I(\rho,\Phi)$ and $I_c(\rho,\Phi)$ allow to introduce capacity of the channel \cite{Wilde,Holevo,QIT}: entanglement assisted capacity

\begin{equation}\label{}
  C_{ea}(\Phi) := \sup_\rho I(\rho,\Phi) ,
\end{equation}
and quantum capacity for a single use of the channel
\begin{equation}\label{}
  Q(\Phi) := \sup_\rho I_c(\rho,\Phi) .
\end{equation}
The first quantity characterizes the amount of
classical information that can be transmitted along a quantum
channel when a sender and receiver to share an unlimited
amount of entanglement. The second provides the limit to the rate at which quantum
information can be reliably sent by the channel. One proves \cite{Bogna-capacity}

\begin{Proposition} If $\{\Lambda_t\}_{t \geq 0}$ is a CP-divisible dynamical map, then

\begin{equation}\label{}
\frac{d}{dt}  C_{ea}(\Lambda_t) \leq 0 , \ \ \ \  \frac{d}{dt}  Q(\Lambda_t) \leq 0 .
\end{equation}
\end{Proposition}
Hence, during a Markovian evolution represented by a family of channels $\{\Lambda_t\}_{t\geq 0}$, the capacity of $\Lambda_t$ decreases in the course of time. It is, therefore, clear that non-Markovian memory effects can temporally increase the capacity of $\Lambda_t$.

\subsubsection{Quantum entanglement and other correlations}

The original approach to non-Markovian evolution based on the concept of CP-divisibility \cite{RHP} was linked to evolution of quantum entanglement: if $\mathcal{E}$ is an entanglement measure then CP-divisible map $\{\Lambda_t\}_{t \geq 0}$ implies

\begin{equation}\label{}
  \frac{d}{dt} \mathcal{E}({\rm id} \otimes \Lambda_t)(\rho)) \leq 0 ,
\end{equation}
for any initial state $\rho$. Hence during the Markovian evolution the entanglement is monotonically decreasing. The connection between quantum entanglement and non-Markovian evolution was already analyzed by many authors (cf. reviews \cite{NM1,NM2,NM3}). Recently, an interesting approach based on so called negativity was proposed in \cite{Janek} in the 3-partite scenario. Similar monotonicity property holds for quantum discord \cite{discord-1,discord-2,discord-3}. The evolution  of correlations under non-Markovian evolution was analyzed recently in \cite{Acin-cor,Acin-cor2} (see also \cite{Johansson})  in analogy to an information backflow \cite{BLP} a distinct correlation backflow allows to witness non-Markovian dynamics (violation of CP-divisibility) \cite{Acin-cor,Acin-cor2}.

Another interesting notion closely related to quantum entanglement is quantum steering (cf. recent review  \cite{steering-Guhne}). Authors of
\cite{steering} shown that so called temporal steering weight (see \cite{steering-0} for more information) decreases monotonically whenever the evolution (of one party) is CP-divisible.

\subsection{Volume of accessible states}

Divisibility of a dynamical map $\{\Lambda_t\}_{t \geq 0}$ implies an interesting property of the volume of accessible states. Denote by $V_0$ the volume of entire set of states $\mathcal{S}(\HH)$. Using Bloch representation $\rho = \frac 1d \oper + \sum_{k=1}^{d^2-1} x_k \tau_k$ a set of states represents compact convex subset of $\mathbb{R}^{d^2-1}$ (generalization of 3-dimensional Bloch ball). The evolution of generalized Bloch vector $\mathbf{x} \in \mathbb{R}^{d^2-1}$ is governed by

\begin{equation}\label{}
  \mathbf{x}(t) = \Delta(t) \mathbf{x}(0) + \mathbf{r}(t) ,
\end{equation}
where $\Delta_{kl}(t) = {\rm Tr}(\tau_k \Lambda_t(\tau_l))$. Let $V(t)$ denote a volume of a subset of states accessible via the dynamical map $\Lambda_t$ at time $t>0$. Clearly $V(t) = |{\rm det}\,\Delta(t)| V_0$. Moreover

\begin{equation}\label{}
  {\rm det}\,\Delta(t) = {\rm det}\,\Lambda_t = {\rm det}\,\widehat{\Lambda}_t .
\end{equation}

\begin{Proposition}[\cite{Wolf-Mar2}] Let $\Phi : \LH \to \LH$ be a positive trace-preserving map. Then

\begin{equation}\label{}
  |{\rm det}\, \Phi| \leq 1 .
\end{equation}
Moreover,    $|{\rm det}\, \Phi| = 1$ if and only if $\Phi(X) = UXU^\dagger$ or $\Phi(X) = UX^TU^\dagger$, with some unitary operator $U$.
\end{Proposition}
Using this property one arrives at \cite{Volume}

\begin{Proposition} \label{PROP-VOL} If $\{\Lambda_t\}_{t \geq 0}$ is P-divisible, then

\begin{equation}\label{VOL}
  \frac{d}{dt} V(t) \leq 0
\end{equation}
for any $t \geq 0$.
\end{Proposition}

\begin{Example} For a qubit evolution governed by the time local master equation (\ref{sss}) one finds

\begin{equation}\label{}
  |{\rm det}\,\Delta(t)| = e^{-\Gamma(t)} ,
\end{equation}
with $\Gamma(t) = \sum_{k=1}^3 \Gamma_k(t)$, and hence condition (\ref{VOL})   implies

\begin{equation}\label{}
  \gamma_1(t) + \gamma_2(t) + \gamma_3(t) \geq 0 ,
\end{equation}
which is much weaker than condition for P-divisibility $\gamma_i(t) + \gamma_j(t) \geq 0$.
For the phase covariant evolution governed by (\ref{L+-t}) condition (\ref{VOL}) is equivalent to

\begin{equation}\label{}
  \gamma_+(t) + \gamma_-(t) + 2 \gamma_z(t) \geq 0 ,
\end{equation}
which is weaker than BLP condition

\begin{equation}\label{}
  \gamma_+(t) + \gamma_-(t) \geq 0 \ , \ \ \
  \gamma_+(t) + \gamma_-(t) + 4 \gamma_z(t) \geq 0 ,
\end{equation}
cf. (\ref{ggg-2}).

\end{Example}
Proposition \ref{PROP-VOL} states that for any P-divisible map the volume of accessible states monotonically shrinks in time. It does not, however, tell us about the shape of the body $\mathbf{B}_t$ of accessible states at time $t > 0$. Clearly $\mathbf{B}_t \subset \mathbf{B}_0 = \mathcal{S}(\HH)$. However, it is not clear how $\mathbf{B}_t$ and $\mathbf{B}_s$  are related for $t > s$ \cite{Chiara}.

%%%%%%%%%%%%%%%%%%%%%%%%%
%%%%%%%%%%%%%%%%%%%%%%%%%%%

\section{Beyond divisible maps}  \label{Beyond-d}

CP-divisible dynamical maps are completely characterized by the corresponding time-local generators. Indeed, an invertible dynamical map $\{\Lambda_t\}_{t \geq 0}$  is CP-divisible if and only if the $\mathcal{L}_t = \dot{\Lambda}_t \Lambda_t^{-1}$ is a
GKLS generator for all $t \geq 0$. Dynamical map which is not CP-divisible still satisfies time-local master equation.
However, $\dot{\Lambda}_t \Lambda_t^{-1}$ is no longer of GKLS form. A set of time-dependent GKLS generators is convex. It is no longer true for a set of all admissible generators. Essentially, $\mathcal{L}_t$ is admissible if $\mathcal{T} \exp(\int_0^t \mathcal{L}_\tau d\tau)$ is CPTP for all $t \geq 0$ and hence the problem of characterization of admissible generators is equivalent to characterization of all dynamical maps. In this section we consider two simple scenario to construct dynamical maps without referring to the corresponding time-local generators.

\subsection{Exponential representation and Magnus expansion}   \label{MAGNUS}

Consider a dynamical map defined as follows

\begin{equation}\label{LW}
  \Lambda_t = e^{\mathbf{L}_t} ,
\end{equation}
where $\mathbf{L}_t$ is a GKLS generator for all $t\geq 0$. It should be clear that not every dynamical map allows for such exponential representation, i.e. in general in (\ref{LW}) the operator $\mathbf{L}_t$ needs not be of GKLS form. By construction $\Lambda_t$ is CPTP. The initial condition $\Lambda_{t=0}={\rm id}$ implies $\mathbf{L}_{t=0}=0$, and hence it is convenient to introduce $\{\mathcal{M}_t\}_{t\geq 0}$ such that

\begin{equation}\label{}
  \mathbf{L}_t = \int_0^t \mathcal{M}_\tau d\tau ,
\end{equation}
that is, $\dot{\mathbf{L}}_t = \mathcal{M}_t$. It should be stressed that $\mathbf{L}_t$ is not a generator of $\Lambda_t$.  To find $\mathcal{L}_t = \dot{\Lambda}_t \Lambda_t^{-1} $ one uses well known formula \cite{Wilcox}

\begin{equation}\label{}
  \frac{d}{dt} e^{\mathbf{L}_t} = \int_0^1 e^{s \mathbf{L}_t} \mathcal{M}_t e^{(1-s) \mathbf{L}_t} ds ,
\end{equation}
and finds the following relations between $\mathcal{L}_t$ and $\mathcal{M}_t$

\begin{equation}\label{}
  \mathcal{L}_t =  \int_0^1 e^{s \mathbb{L}_t} \mathcal{M}_t e^{-s \mathbb{L}_t} ds .
\end{equation}
One has, therefore, two equivalent representations of $\{\Lambda_t\}_{t \geq 0}$

%\begin{tcolorbox}
\begin{equation}\label{}
\mathcal{T} \exp\left( \int_0^t \mathcal{L}_\tau d\tau\right)  = \Lambda_t = \exp\left( \int_0^t \mathcal{M}_\tau d\tau \right) ,
\end{equation}
%\end{tcolorbox}
for all $t \geq 0$. Equivalently,

%\begin{tcolorbox}
\begin{equation}\label{}
\mathcal{L}_t =  \dot{\Lambda}_t \Lambda_t^{-1} \ , \ \ \ \ \mathcal{M}_t = \frac{d}{dt} \log \Lambda_t .
\end{equation}
%\end{tcolorbox}
Note, that in the commutative case, i.e. $[\Lambda_t,\Lambda_s]=0$, one has $\mathcal{L}_t = \mathcal{M}_t$. In general, however, they are different. It should be stressed, that in general neither $\mathcal{L}_t$ nor $\mathcal{M}_t$ is of GKLS form.

Conversely, given $\mathcal{L}_t$ one finds the following Magnus expansion \cite{Magnus} for $\mathcal{M}_t$

\begin{equation}\label{}
  \mathcal{M}_n = \sum_{n=1}^\infty \mathcal{M}^{(n)}_t ,
\end{equation}
where the first three terms of this series read
\begin{eqnarray*}\label{}
  \mathcal{M}^{(1)}_t &=& \mathcal{L}_t , \\
  \mathcal{M}^{(2)}_t &=& \int_0^t d\tau [\mathcal{L}_t,\mathcal{L}_\tau] , \\
  \mathcal{M}^{(3)}_t &=& \int_0^t dt_1 \int_0^{t_1} dt_2 \Big( [\mathcal{L}_t,[\mathcal{L}_{t_1},\mathcal{L}_{t_2}] +  [\mathcal{L}_{t_2},[\mathcal{L}_{t_1},\mathcal{L}_{t}] \Big)  .
\end{eqnarray*}
For more details about the technique of Magnus expansion cf. \cite{M-review,M-review2}. Clearly, in the commutative case one has $\mathcal{M}_t = \mathcal{M}^{(1)}_t = \mathcal{L}_t$. We stress that whenever $\mathcal{L}_t$ has GKLS form then $\{\Lambda_t\}_{t \geq 0}$ is CP-divisible and whenever $\mathbf{L}_t$ has GKLS form then $\{\Lambda_t\}_{t \geq 0}$ is CPTP but in general it is not CP-divisible.

\subsection{Qubit evolution: a case study}% Lie algebraic methods}

To illustrate the connection between $\mathcal{L}_t$ and $\mathcal{M}_t$ let us consider a qubit evolution with

\begin{equation}\label{}
  \mathcal{M}_t = a_1(t) \mathcal{L}_+ + a_2(t) \mathcal{L}_- ,
\end{equation}
with $\mathcal{L}_\pm$ defined in (\ref{L+-}). One has therefore
\begin{equation}\label{}
    \mathbf{L}_t =  A_1(t) \mathcal{L}_+ + A_2(t) \mathcal{L}_- \ ,
\end{equation}
where $A_k(t) = \int_0^t a_k(u) du$. $\mathbf{L}_t$ has GKLS form if and only if $A_k(t) \geq 0$ for all $t \geq 0$. To compute the time-local generator

\begin{equation}\label{LMt}
  \mathcal{L}_t =  \int_0^1 e^{s \mathbf{L}_t} \mathcal{M}_t e^{-s \mathbf{L}_t} ds ,
\end{equation}
we shall use  the well-known Baker-Campbell-Hausdorff (BCH) formula
\begin{equation}\label{}
    e^X Y e^{-X} = Y + [X,Y] + \frac{1}{2!} [X,[X,Y]] + \ldots  = \sum_{k=0}^\infty \frac{1}{k!}\, {\rm ad}_X^k Y\ ,
\end{equation}
where ${\rm ad}_X Y := [X,Y]$ and ${\rm ad}^k_X Y := [X,{\rm ad}^{k-1}_X Y]$. One easily finds

\begin{equation}\label{}
  [\mathcal{L}_+,\mathcal{L}_-]= \mathcal{L}_+ - \mathcal{L}_- ,
\end{equation}
and hence simple algebra leads to

\begin{eqnarray*}
  {\rm ad}_{\mathbf{L}_t}^k \mathcal{L}_+  &=& (-1)^k A^{k-1}(t) A_2(t) ( \mathcal{L}_- - \mathcal{L}_+)\ , \\
   {\rm ad}_{\mathbf{L}_t}^k \mathcal{L}_-  &=& (-1)^k A^{k-1}(t) A_1(t) ( \mathcal{L}_+ - \mathcal{L}_-) \ ,
\end{eqnarray*}
where we introduced $A(t)=A_1(t) + A_2(t)$. After some algebra one arrives at

\begin{eqnarray}\label{ZL1Z}
    e^{s \mathbf{L}_t} \mathcal{L}_+  e^{-s \mathbf{L}_t} =   \left( 1 - \frac{A_2(t)}{A(t)}\big[1-e^{-sA(t)}\big] \right) \mathcal{L}_+ + \frac{A_2(t)}{A(t)}[1-e^{-sA(t)}] \mathcal{L}_-\ ,
\end{eqnarray}
and
\begin{equation}\label{ZL2Z}
    e^{s \mathbf{L}_t} \mathcal{L}_-  e^{-s \mathbf{L}_t} = \left( 1 - \frac{A_1(t)}{A(t)}[1-e^{-sA(t)}] \right) \mathcal{L}_- + \frac{A_1(t)}{A(t)}[1-e^{-sA(t)}] \mathcal{L}_+\ .
\end{equation}
Finally, formula (\ref{LMt}) gives

\begin{equation}\label{}
  \mathcal{L}_t = \gamma_1(t) \mathcal{L}_+ + \gamma_2(t) \mathcal{L}_- ,
\end{equation}
where the functions $\gamma_1(t)$ and $\gamma_2(t)$ are defined by
\begin{eqnarray}
  \gamma_1(t) = a_1(t) - f(t) \ ,  \ \ \   \gamma_2(t) = a_2(t) + f(t) \ ,
\end{eqnarray}
and the function $f(t)$ reads
\begin{equation}\label{}
    f(t) =  \frac{a_1(t) A_2(t) - a_2(t) A_1(t)}{A(t)} \, e^{-A(t)}\ .
\end{equation}
Hence, the time-local generator $\mathcal{L}_t$ has exactly the same structure as $\mathcal{M}_t$ but with functions $a_k(t)$ replaced by $\gamma_k(t)$. Note, however, that $\gamma_k(t)$ needs not be positive and hence in general $\mathcal{L}_t$ is not GKLS generator meaning that the corresponding dynamical map is non-Markovian.

\subsection{Cumulant expansion}

There is a natural scheme to define a dynamical map in terms of $\mathbf{L}_t$ instead of $\mathcal{L}_t$. Consider the reduced evolution in the interaction picture

\begin{equation}\label{red-1}
  \tilde{\Lambda}_t(\rho) = {\rm Tr}_E \tilde{U}_t \rho \otimes \rho_E \tilde{U}^\dagger_t ,
\end{equation}
where $\tilde{U}_t = \mathcal{T} \exp(-i \lambda \int_0^t \tilde{H}_{\rm int}(\tau)d\tau)$. Following \cite{Alicki-1989} consider the following cumulant expansion of $\tilde{\Lambda}_t$

\begin{equation}\label{}
  \tilde{\Lambda}_{t} = \exp\Big( \sum_{n=1}^\infty \lambda^n K^{(n)}_t \Big) .
   %= {\rm id} + \lambda K^{(1)}  + \lambda^2 \Big(  K^{(2)}(\Delta t) + \frac 12 (K^{(1)}(\Delta t))^2 \Big) + O(\lambda^3) .
\end{equation}
Assuming ${\rm Tr}(B_\alpha \rho_E)=0$ one finds $K^{(1)}  = 0$. Moreover, the second cumulant reads

\begin{eqnarray}\label{}
  K^{(2)}_t(\rho) &=& \int_0^t d\tau \sum_{\omega,\omega'} \sum_{\alpha,\beta} \Big\{ -i[\tilde{s}_{\alpha\beta}(\omega,\omega',s)\, A^\dagger_\alpha(\omega)A_\beta(\omega'),\rho] \nonumber \\
   &+& \tilde{\gamma}_{\alpha\beta}(\omega,\omega',s)\Big(  A_\beta(\omega') \rho A^\dagger_\alpha(\omega) - \frac 12 \{  A^\dagger_\alpha(\omega)A_\beta(\omega'),\rho\} \Big) \Big\} ,
\end{eqnarray}
where

\begin{equation}\label{}
  \tilde{s}_{\alpha\beta}(\omega,\omega',s) = e^{i(\omega-\omega')s}\, {s}_{\alpha\beta}(\omega,\omega',s) \ , \ \ \
   \tilde{\gamma}_{\alpha\beta}(\omega,\omega',s) = e^{i(\omega-\omega')s}\, {\gamma}_{\alpha\beta}(\omega,\omega',s) ,
\end{equation}
and

\begin{equation}\label{}
  {\gamma}_{\alpha\beta}(\omega,\omega',s) = \tilde{h}_{\alpha\beta}(\omega',s) + \tilde{h}^*_{\beta\alpha}(\omega,s) \ , \ \ \
  {s}_{\alpha\beta}(\omega,\omega',s) = \frac{1}{2i}\Big( \tilde{h}_{\alpha\beta}(\omega',s) - \tilde{h}^*_{\beta\alpha}(\omega,s)\Big) \ ,
\end{equation}
together with

\begin{equation}\label{}
  \tilde{h}_{\alpha\beta}(\omega,t) = \int_0^t e^{i\omega \tau} h_{\alpha\beta}(\tau) d\tau ,
\end{equation}
and $ h_{\alpha\beta}(\tau) = {\rm Tr}(B_\alpha(\tau)B_\beta \rho_E)$ is a two-point correlation function. One proves \cite{Alicki-1989} that the matrix

\begin{equation}\label{}
  \int_0^t {\gamma}_{\alpha\beta}(\omega,\omega',s) ds
\end{equation}
is positive definite and hence $ K^{(2)}_t$ is a legitimate GKLS generator. If the coupling constant $\lambda$ is sufficiently small then the following dynamical map

\begin{equation}\label{Al-Riv}
  \tilde{\Lambda}_t =  e^{ K^{(2)}_t} ,
\end{equation}
provides a good approximation to the original reduced dynamics (\ref{red-1}). Similar result was derived recently in \cite{Angel-refined}. Rivas called the dynamical map  (\ref{Al-Riv}) a refined weak coupling limit. It should be stressed that (\ref{Al-Riv}) is not a semigroup and in general it is non-Markovian, i.e. CP-divisibility is violated \cite{Angel-refined}. Clearly,  $ K^{(2)}_t$ plays the role of $\mathbf{L}_t$. Interestingly in the limit $t \to \infty$

\begin{equation}\label{}
  \lim_{t\to  \infty} \frac 1t  K^{(2)}_t = \mathbf{L}_{\rm D} ,
\end{equation}
one recovers a Davies generator in the weak coupling limit (cf. section \ref{sub-refined}).

\subsection{Mixing dynamical maps}

Suppose that $\{\Lambda^{(k)}_t\}_{t\geq 0}$ are reduced evolutions for $k=1,\ldots,N$. It is clear that any convex combination

\begin{equation}\label{}
  \Lambda_t = p_1 \Lambda^{(1)}_t + \ldots + p_N \Lambda^{(N)}_t ,
\end{equation}
defines a legitimate dynamical map. There is a natural microscopic representation of such a mixture given a microscopic model for each  dynamical map $\{\Lambda^{(k)}_t\}_{t\geq 0}$ \cite{PRA-Saverio,MIX}. Let

\begin{equation}\label{}
  \Lambda^{(k)}_t(\rho_S) = {\rm Tr}_{E_k} \Big(U^{(k)}_t \rho_S \otimes \rho_{E_k} U^{(k)\dagger}_t\Big) ,
\end{equation}
where $U^{(k)}_t = e^{-i H_k t}$ and $H_k$ is a Hamiltonian acting on $\mathcal{H}_S \otimes \mathcal{H}_{E_k}$ for $k=1,\ldots,N$. Now, we couple a system $S$ to $N$ environments and $N$-dimensional ancilla which results in the following total Hilbert space

\begin{equation}\label{}
  \mathcal{H}_{\rm total} = \mathcal{H}_S \otimes \mathcal{H}_E \otimes \mathcal{H}_A \ , \ \ \ \mathcal{H}_E = \mathcal{H}_{E_1} \otimes \ldots \otimes \mathcal{H}_{E_N} ,
\end{equation}
and ${\rm dim}\mathcal{H}_A = N$. One defines a total Hamiltonian

\begin{equation}\label{}
  \mathcal{H}_{\rm total} = H_1 \otimes \Pi_1 + \ldots + H_N \otimes \Pi_N ,
\end{equation}
where $\Pi_k = |k\>\<k|$ and $\{|k\>\}_{k=1}^N$ defines an orthonormal basis in $\mathcal{H}_A$. Taking
as initial state

\begin{equation}\label{}
  \rho_{\rm total} = \rho_S \otimes \rho_E  \otimes \rho_A , \ \ \ \rho_E = \rho_{E_1} \otimes \ldots \otimes \rho_{E_N} ,
\end{equation}
where  the ancilla state has the following form

\begin{equation}\label{}
  \rho_A = p_1 \Pi_1 + \ldots + p_N \Pi_N ,
\end{equation}
one immediately finds \cite{PRA-Saverio,MIX}

\begin{equation}\label{}
  \Lambda_t(\rho_S) = {\rm Tr}_E {\rm Tr}_A \Big( e^{-i H_{\rm total}t} \rho_S \otimes \rho_E  \otimes \rho_A e^{-i H_{\rm total}t} \Big) =  p_1 \Lambda^{(1)}_t(\rho_S) + \ldots + p_N \Lambda^{(N)}_t(\rho_S) .
\end{equation}
Interestingly, mixing does not preserve divisibility \cite{MIX} which means that in general mixing Markovian dynamical maps the resulting map is non-Markovian. In particular applying Markovian approximation separately in each $\mathcal{H}_S \otimes \mathcal{H}_i$ (e.g. weak coupling limit) one obtains a convex combination of Markovian semigroups

\begin{equation}\label{}
  \Lambda_t = p_1 e^{t \mathcal{L}_1} + \ldots + p_N  e^{t \mathcal{L}_N} ,
\end{equation}
with the corresponding generators $\mathcal{L}_1,\ldots,\mathcal{L}_N$.

\begin{Example} Consider a convex combination of Markovian semigroups \cite{Nina}

\begin{equation}\label{}
  \Lambda_t := p_1 e^{\mathcal{L}_1 t} +  p_2 e^{\mathcal{L}_2 t} +  p_3 e^{\mathcal{L}_3 t} ,
\end{equation}
with $\mathcal{L}_k(\rho) = \frac 12(\sigma_k \rho\sigma_k - \rho)$.  The corresponding time-local generator is of the form (\ref{sss}) with

\begin{equation}\label{}
  \left( \begin{array}{c} \gamma_1(t) \\ \gamma_2(t) \\ \gamma_3(t) \end{array} \right) = \left( \begin{array}{ccc} 1 & -1 & -1 \\ -1 & 1 & -1 \\ -1 & -1 & 1 \end{array} \right) \left( \begin{array}{c} \mu_1(t) \\ \mu_2(t) \\ \mu_3(t) \end{array} \right) ,
\end{equation}
with

\begin{equation}\label{}
  \mu_k(t) = \frac{1-p_k}{1-p_k - e^{2t}p_k} \, \ \ \ k=1,2,3 .
\end{equation}
Note, that for $p_1=p_2=\frac 12$ and $p_3=0$, one finds $\gamma_1(t)=\gamma_2(t)=1$ and $\gamma_3(t)=  - \tanh t $ which reproduce {\em eternally non-Markovian} evolution from the previous example. One easily checks that for any probability vector $(p_1,p_2,p_3)$ the corresponding evolution satisfies (\ref{ggg}) and hence the evolution is P-divisible \cite{Nina}. However, only for a fraction of probability vectors the evolution is CP-divisible.
\end{Example}

\begin{Example} Consider the generalized Pauli like evolution being the following convex combination of Markovian semigroups (cf. section \ref{SUB-Pauli})

\begin{equation}\label{}
  \Lambda_t = \sum_{\alpha=1}^{d+1} x_\alpha e^{\kappa t \mathcal{L}_\alpha}  = e^{- \kappa t} {\rm id} + (1- e^{- \kappa t}) \sum_{\alpha=1}^{d+1} x_\alpha  \Phi_\alpha,
\end{equation}
with $\kappa >0$, $(x_1,\ldots,x_{d+1})$ a probability distribution, and $\mathcal{L}_\alpha = \Phi_\alpha - {\rm id}$, with $\Phi_\alpha$ defined in (\ref{Phi-a}).
One finds \cite{Kasia-2016,Kasia-JPA} $\Lambda_t = p_0(t) {\rm id} + \sum_{\alpha=1}^{d+1} p_\alpha(t)  \Phi_\alpha$ with

\begin{equation}\label{}
  p_0(t) = \frac 1d (1+[d-1] e^{-\kappa t}) , \ \ \ p_\alpha(t) = \frac{d-1}{d} (1 - e^{-\kappa t}) x_\alpha .
\end{equation}
Moreover, $\Lambda_t$ satisfies time-local master equation with $\mathcal{L}_t = \sum_{\alpha=1}^{d+1} \gamma_\alpha(t) \mathcal{L}_\alpha$ and time dependent rates

\begin{equation}\label{}
  \gamma_\alpha(t) = - \frac{ (1-x_\alpha) \kappa}{1+ (e^{\kappa t}-1)x_\alpha} + \frac{\kappa}{d}   \sum_{\beta=1}^{d+1} \frac{ 1-x_\beta}{1+ (e^{\kappa t}-1)x_\beta} .
\end{equation}
Consider a special case corresponding to $x_1=x_2 = \frac 12$ and $x_\alpha=0$ for $\alpha > 2$. It gives rise  to the following transition rates \cite{Kasia-JPA}

\begin{equation}\label{}
  \gamma_1(t) = \gamma_2(t) = \frac{\kappa}{d} \frac{ d-1+e^{-\kappa t}}{1+ e^{-\kappa t}} , \ \ \
  \gamma_\alpha(t) = - \frac{\kappa}{d} \tanh \frac{\kappa t}{2} ,
\end{equation}
for $\alpha > 2$. Note, that $\gamma_1(t) = \gamma_2(t) >0$, however, $\gamma_\alpha(t) <0$ for $\alpha > 2$. Recalling that each $\gamma_\alpha$ enters the time-local generator with multiplicity $d-1$ we have  a time local generator with $(d-1)^2$ rates being negative for  $t > 0$. Actually, we do not know a generator with a larger number of {\em eternally} negative rates.
\end{Example}
The analysis of convex combination of Markovian semigroups was further developed in \cite{MIX-2,MIX-3}. One may also investigate the inverse problem: is it possible to obtain Markovian semigroup by mixing non-Markovian dynamical maps. Interestingly, to realize Markovian semigroup one needs to mix non-invertible maps. Several results were reported in \cite{MIX-4,MIX-4a,MIX-5,MIX-6}.

%%%%%%%%%%%%%%%%%%%%%%%%%%%%%%%%%%%
%%%%%%%%%%%%%%%%%%%%%%%%%%%%%%%%%%

\section{Markovianity beyond dynamical maps -- quantum regression}    \label{REGRESSION}

\subsection{Classical Markov process}

It should be stressed that the concept of Markovianity based on the Definition \ref{MARKOV} is not fully compatible  with the original notion of Markovianity in the theory of  classical  stochastic processes. Recall, that a classical stochastic process $\{x(t)\}_{t \geq t_0}$ is characterized by an infinite hierarchy of $n$-point probability distributions $p(x_n,t_n;x_{n-1},t_{n-1};\ldots,x_0,t_0)$. The process is Markovian \cite{Gardiner,Kampen} if the  conditional probabilities satisfy the following defining property

\begin{equation}\label{MC}
  p(x_n,t_n|x_{n-1},t_{n-1};\ldots,x_0,t_0) = p(x_n,t_n|x_{n-1},t_{n-1}) ,
\end{equation}
for any sequence $t_n \geq t_{n-1}\geq \ldots \geq t_0$. Using the definition of conditional probability (Bayes rule)

\begin{equation}\label{MC1}
  p(x_n,t_n|x_{n-1},t_{n-1};\ldots,x_0,t_0) := \frac{ p(x_n,t_n;x_{n-1},t_{n-1};\ldots,x_0,t_0) }{ p(x_{n-1},t_{n-1};x_{n-2},t_{n-2};\ldots,x_0,t_0)}  ,
\end{equation}
formula (\ref{MC1}) implies

\begin{equation}\label{MC2}
  p(x_n,t_n;x_{n-1},t_{n-1};\ldots,x_0,t_0) = p(x_n,t_n|x_{n-1},t_{n-1})\,  p(x_{n-1},t_{n-1}|x_{n-2},t_{n-2}) \, \ldots \,  p(x_1,t_1|x_{0},t_{0}) \, p(x_0,t_0) .
\end{equation}
Now, for any classical observable (function $A : \mathcal{X} \to \mathbb{R}$) one defines its average

\begin{equation}\label{}
  \< A(t) \> = \sum_{x \in \mathcal{X}}  A(x) p(x,t) ,
\end{equation}
and similarly given a set of observables $\{A_0,A_1,\ldots,A_n\}$ one defines the following correlation function

\begin{equation}\label{}
  \< A_n(t_n) A_{n-1}(t_{n-1}) \ldots A_1(t_1) A_0(t_0)  \> := \sum_{x_n,\ldots,x_0 \in \mathcal{X}}  A_n(x_n) \ldots A_0(x_0)  p(x_n,t_n;x_{n-1},t_{n-1};\ldots,x_0,t_0)  .
\end{equation}
Markov property of the process may be formulated as follows
\begin{tcolorbox}
A classical process $\{x(t)\}_{t\geq t_0}$ is Markovian if and only if for any set  of observables $\{A_0,A_1,\ldots,A_n\}$

\begin{eqnarray}\label{CRF}
 && \< A_n(t_n) A_{n-1}(t_{n-1}) \ldots A_1(t_1) A_0(t_0)  \> =  \\  && \sum_{x_n,\ldots,x_0 \in \mathcal{X}}  A_n(x_n) \ldots A_0(x_0) p(x_n,t_n|x_{n-1},t_{n-1})\,  p(x_{n-1},t_{n-1}|x_{n-2},t_{n-2}) \, \ldots \,  p(x_1,t_1|x_{0},t_{0}) \, p(x_0,t_0) \nonumber ,
\end{eqnarray}
where $t_n \geq \ldots \geq t_0$.

\end{tcolorbox}
One calls (\ref{CRF}) a {\em classical regression formula} \cite{NM4}. P-divisibility of the classical dynamical map $ T(t) = T(t,s) T(s)$ is defined in terms of 2-point conditional probability, i.e. $T_{ij}(t,s)=p(x_i,t|x_j,s)$ satisfies Chapman-Kolmogorov equation

\begin{equation}\label{}
  T(t,s)T(s,u) = T(t,u) ,
\end{equation}
for $t>s>u$. However, $T(t,s)$ does not provide any information about $n$-point correlations which are critical to decide about the Markovianity of the process  \cite{Hanggi-1,Hanggi-2}. Therefore, having an access to the dynamical map $T(t)$ only, one can not use the original definition of Markovianity of the stochastic process \cite{NM4}. Moreover, there are classical non-Markovian processes for which Chapman-Kolmogorov equation holds.

\begin{Example} Following \cite{Rosenblatt} consider $m$ state process $x_n \in \{0,1,\ldots,m-1\}$ such that

\begin{equation}\label{}
  p(x_{n+2},t_{n+2}|x_{n+1},t_{n+1};x_{n},t_{n}) = \frac 1m \Big( 1 - \cos\Big[\frac{2\pi}{m}(2x_{n+2}-x_{n+1}-x_n) \Big] \Big) .
\end{equation}
Clearly, the process is non-Markovian since the above conditional probability does depend upon $x_n$. However

\begin{equation}\label{}
  p(x_{n+2},t_{n+2}|x_{n+1},t_{n+1})  = \sum_{x_n=0}^{m-1}  p(x_{n+2},t_{n+2}|x_{n+1},t_{n+1};x_{n},t_{n})  = \frac 1m ,
\end{equation}
and it satisfies Chapman-Kolmogorov equation.
\end{Example}

\subsection{Quantum regression formula}

In the quantum case the problem is even more subtle since to  define a  joint probability one needs to perform a series of measurements which would disturb the state of the system and affect the subsequent outcomes. The problem of a proper definition of a quantum stochastic process was already addressed  by Lindblad \cite{L-stoch} and then independently developed by Accardi, Frigerio and Lewis in a series of papers \cite{Q-stoch2,Q-stoch3,Q-stoch4}. The detailed exposition of this approach is beyond the scope of this report. Interestingly, one may derive a quantum analog of (\ref{CRF}) --- a {\em quantum regression formula} \cite{Gardiner,Open1,NM4,Carmichael-QR} which may be considered as a quantum analog of Markovianity condition for the quantum process.

To define an evolution of an open system one needs the unitary evolution $\mathbb{U}_{t}$ of the `system + environment' and the initial environmental state $\rho_E$ at $t_0$. Consider two families of system's operators $\{A_0,A_1,\ldots,A_n\}$ and $\{B_0,B_1,\ldots,B_n\}$ and let

\begin{equation}\label{}
  A_k(t) = \mathbb{U}^\ddag_{t} (A_k \otimes \oper_E) \ , \ \ \ \  B_k(t) = \mathbb{U}^\ddag_{t} (B_k \otimes \oper_E) ,
\end{equation}
be the corresponding Heisenberg picture evolution.  Assuming initial factorization $\rho_{SE}(0) = \rho_S \otimes \rho_E$ and  introducing a set of maps $C_k : \LH \to \LH$

\begin{equation}\label{}
  C_k(X) = B_k X A_k ,
\end{equation}
one arrives at

\begin{tcolorbox}
\begin{Definition} An open quantum system evolution defined by $\{ \mathbb{U}_{t,t_0},\rho_E\}$ satisfies  quantum egression formula if and only if the time-ordered correlation functions satisfy

\begin{equation}\label{QRF}
% {\rm Tr}_{S+E} \Big( \mathbb{C}_n \mathbb{U}_{t_n - t_{n-1}} \mathbb{C}_{n-1} \ldots \mathbb{C}_{t_1} \mathbb{U}_{t_1}(\rho_S \otimes \rho_E) %\Big)  =
  \< A_1(t_1) \ldots A_n(t_n) B_n(t_n) \ldots B_1(t_1) \>_{SE} =
{\rm Tr}_S \Big( C_n  V_{t_n,t_{n-1}} C_{n-1} V_{t_{n-1},t_{n-2}} \ldots C_1 \Lambda_{t_1,t_0} (\rho_S) \Big) ,
\end{equation}
for any $t_n \geq \ldots t_1 \geq 0$, and any sets of operators $\{A_1,\ldots,A_n,B_1,\ldots,B_n\}$, where  $\< X_{SE} \>_{SE} = {\rm Tr}(X_{SE} \rho_S \otimes \rho_E)$.
\end{Definition}
\end{tcolorbox}
In (\ref{QRF}) $V_{t,s} := {\Lambda}_{t} {\Lambda}^{-1}_{s}$ and $\Lambda_t$ is a reduced evolution of the system, i.e. $\Lambda_t(\rho_S) = {\rm Tr}_E(\mathbb{U}_t(\rho_S \otimes \rho_E))$. Such formula was considered recently in \cite{Andrea-Bassano,Ilia}. The above formulation of quantum regression was generalized in \cite{NM4}: Li, Hall and Wiseman proposed to replace the propagator $V_{t,s}$ by the family of maps $\widetilde{\Lambda}_{t,s}$ defined via

\begin{equation}\label{}
   \widetilde{\Lambda}_{t,s}(\rho_S) = {\rm Tr}_E \Big( \mathbb{U}_{t-s}(\rho_S \otimes \mathbb{W}_s(\rho_E)) ,
\end{equation}
where $\mathbb{W}_s$ represents some unitary evolution of the environment (cf. \cite{NM4} for more details).  In full analogy with a classical case they proposed the following

\begin{Definition} A quantum evolution defined by $\{ \mathbb{U}_{t},\rho_E\}$  is Markovian if and only if it satisfies (general) quantum regression formula.
\end{Definition}

Note, that

\begin{equation}\label{}
   \< A_1(t_1) \ldots A_n(t_n) B_n(t_n) \ldots B_1(t_1) \>_{SE} = {\rm Tr}_{S+E} \Big( \mathbb{C}_n \mathbb{U}_{t_n - t_{n-1}} \mathbb{C}_{n-1} \ldots \mathbb{C}_{t_1} \mathbb{U}_{t_1}(\rho_S \otimes \rho_E) \Big) .
\end{equation}
It was proved \cite{Dumcke-WCL} that in the weak coupling (van Hove) limit

\begin{equation}\label{QRF-3}
  {\rm Tr}_{S+E} \Big( \mathbb{C}_n \mathbb{U}_{t_n - t_{n-1}} \mathbb{C}_{n-1} \ldots \mathbb{C}_{t_1} \mathbb{U}_{t_1}(\rho_S \otimes \rho_E) \Big)   \longrightarrow {\rm Tr}_S \Big( C_n e^{(t_n-t_{n-1}) \mathcal{L}} C_{n-1} e^{(t_{n-1}-t_{n-2}) \mathcal{L}} \ldots C_1 e^{t_1 \mathcal{L}}(\rho_S) \Big) ,
\end{equation}
where $\mathcal{L}$ stands for the weak coupling limit Davies generator. The same holds for the  singular coupling limit \cite{Dumcke-WCL} as well.

Interestingly, this formulation is very close to the  original formulation of Markovian quantum process due to Lindblad \cite{L-stoch}, and Accardi, Lewis and Frigerio \cite{Q-stoch2,Q-stoch3,Q-stoch4}. Moreover, it is essentially equivalent to the recent approach based on quantum process tensor \cite{Modi1,Modi2,Modi3,Modi3a} (cf. also recent tutorial \cite{Modi4}). This approach was further applied in  \cite{Andrea1,Andrea2,Filippov-Modi} in the study of open quantum systems dynamics.

%et us observe that when $\mathbb{U}_{t,s} = \mathbb{U}_{t-s}$ and the initial state $\rho_E$ satisfies $\mathbb{W}_t(\rho_E) = \rho_E$, then

%\begin{equation}\label{}
%   \widetilde{\Lambda}_{t,s}(\rho) = {\rm Tr}_E \Big( \mathbb{U}_{t-s}(\rho \otimes \rho_E) \Big) = \Lambda_{t-s} ,
%\end{equation}
%and hence (\ref{QRF}) reduces to

%\begin{equation}\label{QRF-2}
%  \< A_1(t_1) \ldots A_n(t_n) B_n(t_n) \ldots B_1(t_1) \>_{SE} = {\rm Tr}_S \Big( C_n {\Lambda}_{t_n-t_{n-1}} C_{n-1} {\Lambda}_{t_{n-1}-t_{n-2}} %\ldots C_1 {\Lambda}_{t_1}(\rho_S) \Big) .
%\end{equation}
This discussion clearly shows that having an access to the whole process in $\HH \otimes \HH_E$ one immediately finds that CP-divisibility is only a necessary condition for Markovianity of the quantum process defined via (\ref{QRF}). However, CP-divisibility provides a perfect {\em intrinsic} characterization of Markovianity in terms of the system's dynamical map.

%In what follows we assume that the evolution is entirely characterized in terms of $\{\Lambda_t\}_{t \geq 0}$ and whenever we use the very term %{\em Markovian} it means that $\{\Lambda_t\}_{t \geq 0}$ is CP-divisible.

%%%%%%%%%%%%%%%%%%%%%%%%%%
%%%%%%%%%%%%%%%%%%%%%%%%%%%%%

%\subsection{Quantum regression for paradigmatic models}

It was proved recently  \cite{Davide-1,Davide-2}  that both amplitude damping model and phase damping model considered in sections \ref{SUB-AD} and \ref{SUB-PD}, satisfy quantum regression theorem when the form factors $f_j(\omega)$ are flat (do not depend upon $\omega$) and the initial state of the environment is the vacuum of the boson field.

\subsection{Conditional past-future correlation}

A related concept of quantum Markovianity based on three subsequent measurements was recently proposed by Budini \cite{Budini1,Budini2,Budini3}. The starting point is an observation that for a classical Markovian stochastic process past (at time $t_x$) and future (at time $t_z$) events become statistically independent when conditioned on a given state at the present time $t_y$. Hence if the observation of the process at $t_x < t_y < t_z$ gives three outcomes $x,y,z$, then for a Markov process one obviously finds

\begin{equation}\label{}
  p(z,y,x) = p(z|y)p(y|x) p(x) ,
\end{equation}
where $p(x)$ is the probability of the first event at $t_x$, and $p(y|x)$ is a conditional probability of the event $y$ at $t_y$ provided one observed an event $x$ at $t_x$ (similarly for $p(z|y)$). Simple application of the Bayes rule leads to

\begin{equation}\label{}
  p(z,x|y) = \frac{p(z,y,x)}{p(y)} = \frac{p(z,y,x)}{p(y)} = \frac{p(z|y,x)p(y,x)}{p(y)} = p(z|y,x) p(x|y) ,
\end{equation}
and hence for a Markovian process

\begin{equation}\label{ppp}
  p(z,x|y) = p(z|y) p(x|y) ,
\end{equation}
due to $p(z|y,x) = p(z|y)$. Formula (\ref{ppp}) shows that events $z$ and $x$ (in the future and in the past) are statistically independent given an even $y$ (at present). The  retrodicted conditional probability $p(x|y)$ may be computed as follows

\begin{equation}\label{}
  p(x|y) = p(y|x) \frac{p(x)}{p(y)} \ , \ \ \ p(y) = \sum_{x'} p(y,x') = \sum_{x'} p(y|x')p(x') .
\end{equation}
Violation of (\ref{ppp}) provides a clear witness that a process is non-Markovian. Given a classical observable $O$ one defines \cite{Budini1} the conditional past-future correlation

\begin{equation}\label{CPF}
 C_{\rm pf}:= \sum_{z,x} [p(z,x|y) - p(z|y)p(x|y)]O_z O_x ,
\end{equation}
where $O_z$ denotes the value of $O$ in the state `$z$' (and similarly for $O_x$). Whenever $C_{\rm pf} \neq 0$ for some $O$, then the process is non-Markovian.

Consider now quantum evolution governed by the unitary map $\mathbb{U}_t$ acting on $\mathrm{L}(\HH_S \otimes \HH_E)$.  Let $\{\Pi_x\}$, $\{\Pi_y\}$, and $\{\Pi_z\}$ be projective measurement satisfying

$$   \sum_x \Pi_x = \oper_S , \ \ \ \sum_z \Pi_z = \oper_S , \ \ \ \sum_z \Pi_z = \oper_S . $$
Let $\rho_{SE}(0)$ be an initial `system + environment' state. After the first measurement $\Pi_x$ at $t_x$ one obtains

\begin{equation}\label{}
  \rho^x_{SE} = \frac{\Pi_x \otimes \oper_E \mathbb{U}_{t_x}(\rho_{SE}(0)) \Pi_x \otimes \oper_E }{{\rm Tr}_{SE} [ \Pi_x \otimes \oper_E \mathbb{U}_{t_x}(\rho_{SE}(0)) ]} = \Pi_x \otimes \rho^x_E .
\end{equation}
A second measurement $\Pi_y$ at $t_y$ leads to

\begin{equation}\label{}
  \rho^{y}_{SE} = \frac{\Pi_y \otimes \oper_E \mathbb{U}_{t_y - t_x}(\rho^x_{SE}) \Pi_y \otimes \oper_E }{{\rm Tr}_{SE} [ \Pi_y \otimes \oper_E \mathbb{U}_{t_y- t_x}(\rho^x_{SE})]} = \Pi_y \otimes \rho^{yx}_E ,
\end{equation}
and

\begin{equation}\label{}
  p(y|x) = {\rm Tr}_{SE} [\Pi_y \otimes \oper_E \mathbb{U}_{t_y- t_x}(\rho^x_{SE})] .
\end{equation}
Finally, a third measurement $\Pi_z$ at $t_z$ results in

\begin{equation}\label{}
  \rho^{z}_{SE} = \frac{\Pi_z \otimes \oper_E \mathbb{U}_{t_z - t_y}(\rho^{yx}_{SE}) \Pi_z \otimes \oper_E }{{\rm Tr}_{SE} [ \Pi_z \otimes \oper_E \mathbb{U}_{t_z- t_y}(\rho^{yx}_{SE})]} = \Pi_z \otimes \rho^{zyx}_E ,
\end{equation}
and

\begin{equation}\label{}
  p(z|y,x) = {\rm Tr}_{SE} [\Pi_z \otimes \oper_E \mathbb{U}_{t_z- t_y}(\rho^{y}_{SE})] .
\end{equation}
These data allow to compute  conditional past-future correlation $C_{\rm pf}$ given an observable $O=O^\dagger$.
In \cite{Budini2,Budini3} the detailed analysis for amplitude damping and phase damping models is provided.

\section{Memory kernel approach}   \label{SEC-MEMORY}

The reduced evolution $\Lambda_{t}(\rho) = {\rm Tr}_E \Big( \mathbb{U}_{t}(\rho \otimes \rho_E) \Big)$ satisfies time non-local memory kernel master equation

\begin{equation}\label{MK-eq}
  \dot{\Lambda}_{t} = \int_{0}^t \mathcal{K}_{t-\tau} \Lambda_{\tau} d\tau , \ \ \ \Lambda_{0} = {\rm id} ,
\end{equation}
where the kernel $\mathcal{K}_t$ is defined via the total system-environment Hamiltonian (cf.  section \ref{Reduced}). Since the structure of $\mathcal{K}_t$ is highly nontrivial and requires the full knowledge of the unitary evolution of the composite system $\mathbb{U}_{t}$ the dynamical problem defined by (\ref{MK-eq}) is rather untractable. It would be interesting to characterize properties of $\mathcal{K}_t$ which guarantee that the solution $\Lambda_{t}$ defines a legitimate CPTP dynamical map.  As already observed in \cite{Stenholm} this problem is very hard. Actually,  we show that it involves an infinite hierarchy of nontrivial conditions. In this section we characterize a few classes of admissible kernels which already found interesting applications in the literature. Essentially, the problem we address in this section may be formulated as follows: {\em how to construct an admissible memory kernel}? It should be stressed that the fact that the evolution is governed by a memory kernel master equation does not necessarily mean that it is non-Markovian. However, conditions for Markovianity (i.e. CP-divisibility) are not easy to formulate in terms of $\mathcal{K}_t$. For another related concept of quantum channels with memory cf. \cite{Mancini}. The memory kernel master equations were analyzed by several authors (cf. e.g. \cite{Mem1,Mem1a,Mem1b,Mem2,Mem3,Vacchini-NJP,Mem4,Wege-1}). Another approach based on Feshbach projection technique was proposed in \cite{PRL-2013} (see also \cite{PRL-Walter}).

%In this section we consider only time homogeneous case (cf. Section \ref{SEC-V}).

\subsection{General structure}

Applying the Laplace transform (LT) to (\ref{MK-eq}) one obtains the following equation

\begin{equation}\label{}
  s\widetilde{\Lambda}(s) - {\rm id} = \widetilde{\mathcal{K}}(s) \widetilde{\Lambda}(s) ,
\end{equation}
where the LT is defined as $\,\widetilde{\Lambda}(s) := \int_0^\infty e^{-s t} \Lambda_t dt\,$ and similarly for $\widetilde{\mathcal{K}}(s)$. It gives rise to the  following simple relation in the LT domain
\begin{equation}\label{}
  \widetilde{\Lambda}(s) = \frac{1}{s - \widetilde{\mathcal{K}}(s) } .
\end{equation}
Clearly, the kernel $\mathcal{K}_t$ is physically legitimate if and only if the inverse transform of $ \widetilde{\Lambda}(s) = (s - \widetilde{\mathcal{K}}(s))^{-1}$ is CPTP. Let us observe that if $\Lambda_t$ is CP then

\begin{equation}\label{}
  (-1)^n \frac{d^n}{ds^n} \widetilde{\Lambda}(s) = \int_0^\infty t^n e^{-s t} \Lambda_t dt ,
\end{equation}
is CP for all $s > 0$ and $n=1,2,\ldots$.  Note, that this property is very similar to the well known property of completely monotonic (CM) functions \cite{Bernstein}.

\begin{Definition} A function $g : [0,\infty) \to \mathbb{R}$ is completely monotonic if

\begin{equation}\label{}
  (-1)^n \frac{d^n}{ds^n} f(s) \geq 0 ,
\end{equation}
for all $n=1,2,\ldots$, and $s > 0$.
\end{Definition}
These functions are fully characterized by the following

\begin{Theorem}[Bernstein] A function  $g : [0,\infty) \to \mathbb{R}$ is completely monotonic if and only if

\begin{equation}\label{}
  g(s) = \int_0^\infty e^{-s t} f(t) dt ,
\end{equation}
where $f(t) \geq 0$, that is, $g(s)$ is a Laplace transform of a non-negative function $f(t)$.
\end{Theorem}

\begin{Definition} We call an operator function $R(s) : \LH \to \LH$ for $s > 0$  completely monotonic CP if

\begin{equation}\label{}
  (-1)^n \frac{d^n}{ds^n}  R(s)  ,
\end{equation}
is completely positive for $n=1,2,\ldots$, and $s > 0$.
\end{Definition}

\begin{Theorem} An operator function $R(s)$ is completely monotonic (completely) positive if and only if

\begin{equation}\label{}
  R(s) = \int_0^\infty e^{-s t} F_t dt ,
\end{equation}
and $F_t : \LH \to \LH$ is a  (completely) positive map for all $t \geq 0$.
\end{Theorem}
Proof: let $F_t$ be the inverse transform of $R(s)$. Clearly, if $F_t$ (completely) positive, then $R(s)$ is completely monotonic  (completely) positive. Suppose, that $R(s)$ is completely monotonic positive (the proof for complete positivity is analogous). Now, for any pair of rank-1 projectors $P$ and $Q$ define two functions

\begin{equation}\label{}
  g(s) = {\rm Tr}(P R(s)(Q))  \ , \ \ \ f(t) = {\rm Tr}(P F_t(Q)) ,
\end{equation}
related via $g(s) = \int_0^\infty e^{-st} f(t) dt$. Clearly, $g(s)$ is completely monotonic and hence Bernstein theorem implies that $f(t) \geq 0$. Since, for any pair of projectors ${\rm Tr}(P F_t(Q)) \geq 0$ one concludes that $F_t$ is a positive map for $t \geq 0$. \hfill $\Box$

This analysis leads us to the following
%\begin{tcolorbox}

\begin{Theorem} A memory kernel $\mathcal{K}_t$ is physically legitimate if and only if $(s - \widetilde{\mathcal{K}}(s))^{-1}$ is completely monotonic CP.
\end{Theorem}

%\end{tcolorbox}
It clearly shows that characterisation of physically admissible memory kernel is rather untractable since it requires infinite hierarchy of conditions for its Laplace transform  $\widetilde{\mathcal{K}}(s)$.

\begin{Example} To illustrate the intricate relation between time-local and memory kernel description let us consider a simple qubit evolution

\begin{equation}\label{}
  \Lambda_t(\rho) = \left( \begin{array}{cc} \rho_{00} & \lambda(t) \rho_{01} \\ \lambda(t) \rho_{10} & \rho_{11} \end{array} \right) ,
\end{equation}
where $\lambda(t)$ is a (real) decoherence function such that $\lambda(0)=1$ and $|\lambda(t)| \leq 1$. This evolution may be characterized either by time-local generator $\mathcal{L}_t$ or memory kernel $\mathcal{K}_t$

\begin{equation}\label{}
  \mathcal{L}_t(\rho) = \frac 12 \gamma(t) ( \sigma_z \rho \sigma_z - \rho) \ , \ \ \ \  \mathcal{K}_t(\rho) = \frac 12 \kappa(t) ( \sigma_z \rho \sigma_z - \rho) \ ,
\end{equation}
where

\begin{equation}\label{}
\gamma(t) = - \dot{\lambda}(t)/\lambda(t) \ , \ \ \ \ \widetilde{\kappa}(s) = \frac{ s\widetilde{\lambda}(s) - 1}{\widetilde{\lambda}(s)} .
\end{equation}
Markovianity condition $\gamma(t) \geq 0$ is, however, not very transparent in terms of $\kappa(t)$. Note, that for highly non-Markovian  evolution $\lambda(t) = \cos t$ one finds highly singular $\gamma(t) = \tan t$ and perfectly regular kernel with $\kappa(t)=1$ for $t \geq 0$.
It shows that singularity of the time-local description might be completely removed in the memory kernel approach. For more examples of memory kernels for qubit evolution cf. \cite{sabrina-kernel1,sabrina-kernel2}.

\end{Example}

Note, that

\begin{equation}\label{}
  \frac{1}{s - \widetilde{\mathcal{K}}(s)} = \frac{1}{s}\,  \frac{1}{{\rm id} - \widetilde{\mathbb{K}}(s)} ,
\end{equation}
where $\widetilde{\mathbb{K}}(s) = \frac{1}{s} \widetilde{\mathcal{K}}(s)$, and hence it leads to the following expansion

\begin{equation}\label{x1}
  \frac{1}{s - \widetilde{\mathcal{K}}(s)} = \frac{1}{s}\,  \Big( {\rm id} +  \widetilde{\mathbb{K}}(s) +  \widetilde{\mathbb{K}}^2(s) +  \widetilde{\mathbb{K}}^3(s) +\ldots \Big) .
\end{equation}
Passing to the time domain one finds the following expansion for the dynamical map

%\begin{tcolorbox}
\begin{eqnarray}\label{x2}
  \Lambda_t & =& {\rm id} + \int_0^t \mathbb{K}_\tau d\tau + \int_0^t \mathbb{K}_\tau \ast \mathbb{K}_\tau d\tau + \int_0^t \mathbb{K}_\tau \ast \mathbb{K}_\tau\ast \mathbb{K}_\tau d\tau + \ldots \\
  &=& {\rm id} +  \int_0^t \mathbb{K}_\tau d\tau + \int_0^t dt_1 \int_0^{t_1} dt_2 \mathbb{K}_{t_1-t_2}  \mathbb{K}_{t_2} + \int_0^t dt_1 \int_0^{t_1} dt_2 \int_0^{t_2} dt_3 \mathbb{K}_{t_1-t_2}  \mathbb{K}_{t_2-t_3} \mathbb{K}_{t_3} + \ldots .  \nonumber
\end{eqnarray}
%\end{tcolorbox}
The above expansion should be compared with the corresponding Dyson expansion in terms of a time-local generator $\mathcal{L}_t$

%\begin{tcolorbox}
\begin{equation}\label{x3}
  \mathcal{T} \exp\left( \int_0^t \mathcal{L}_\tau d\tau \right) = {\rm id} +  \int_0^t \mathcal{L}_\tau d\tau + \int_0^t dt_1 \int_0^{t_1} dt_2 \mathcal{L}_{t_1}  \mathcal{L}_{t_2} + \int_0^t dt_1 \int_0^{t_1} dt_2 \int_0^{t_2} dt_3 \mathcal{L}_{t_1}  \mathcal{L}_{t_2} \mathcal{L}_{t_3} + \ldots .
\end{equation}
%\end{tcolorbox}
Note, that if $\mathcal{L}_t$ is GKLS for all $t \geq 0$, then (\ref{x3}) provides a CP evolution. However, it is no longer true for (\ref{x2}): even if $\mathbb{K}_t$ is GKLS for all $t \geq 0$ formula (\ref{x2}) does not in general provide CP map $\Lambda_t$.
%Clearly, complete positivity of (\ref{x1}) is only necessary (but not sufficient) for complete positivity of (\ref{x2}).
In the case of a semigroup evolution the problem dramatically simplifies. Since $\mathcal{K}_t = \delta(t) \mathcal{L}$ one finds

\begin{equation}\label{}
  \frac{1}{s - \widetilde{\mathcal{K}}(s)} = \frac{1}{s- \mathcal{L}} = R(s,\mathcal{L}) ,
\end{equation}
where $R(s,\mathcal{L})$ denotes the resolvent operator. Note, that

\begin{equation}\label{}
  (-1)^n \frac{d^n}{ds^n}  R(s,\mathcal{L})  = {n!}\,  R(s,\mathcal{L})^{n+1}  ,
\end{equation}
and hence $R(s,\mathcal{L})$ is completely monotonic CP if and only if $R(s,\mathcal{L})$ is CP, that is, the infinite hierarchy of condition reduces to a single condition for the resolvent $R(s,\mathcal{L})$. In this case $\mathbb{K}_t = \mathcal{L}$ and hence (\ref{x2}) reduces to

\begin{equation}\label{}
  \Lambda_t = {\rm id} + \int_0^t d\tau \mathcal{L}  + \int_0^t  \tau d\tau \mathcal{L}^2 + \int_0^t  \tau^2 d\tau \mathcal{L}^3 + \ldots = {\rm id} + \mathcal{L}t + \frac{(\mathcal{L}t)^2}{2} + \frac{(\mathcal{L}t)^3}{3!} + \ldots ,
\end{equation}
which recovers $e^{\mathcal{L} t}$. Hence for a semigroup  both expansions (\ref{x2}) and (\ref{x3}) coincide. This makes the Markovian semigroup exceptional.

\subsection{Sufficient conditions for legitimate memory kernels}

Since the problem of finding necessary and sufficient conditions for memory kernels giving rise to legitimate dynamical maps is rather untractable let us consider a particular construction of $\mathcal{K}_t$ which provides a direct generalization of GKLS generator \cite{DC-AK}. Consider a GKLS generator

\begin{equation}\label{}
  \mathcal{L}(\rho) = -i[H,\rho] + \Phi(\rho)- \frac 12 \{ \Phi^\ddag(\oper),\rho\} .
\end{equation}
It can be rewritten as $\,\mathcal{L} =  \Phi - Z$, where

\begin{equation}\label{}
  Z(\rho) = C \rho + \rho C^\dagger ,
\end{equation}
and $C = iH + \frac 12  \Phi^\ddag(\oper)$. One has

\begin{equation}\label{}
  \frac{1}{s- \mathcal{L}} = \frac{1}{s+Z} \frac{1}{{\rm id} - \Phi (s+Z)^{-1}} ,
\end{equation}
and hence defining a map

\begin{equation}\label{}
  \widetilde{N}(s) := \frac{1}{s+Z} ,
\end{equation}
one obtains the following representation for the dynamical map in the LT domain

\begin{equation}\label{}
  \widetilde{\Lambda}(s) = \frac{1}{s- \mathcal{L}} =   \widetilde{N}(s) \frac{1}{1- \Phi  \widetilde{N}(s)} =  \widetilde{N}(s) \Big( {\rm id} + \Phi \widetilde{N}(s) + \Phi \widetilde{N}(s)\Phi \widetilde{N}(s) + \ldots \Big) ,
\end{equation}
provided the expansion is convergent which is guaranteed requiring $\| \Phi  \widetilde{N}(s) \|_1 < 1$. Going back to the time domain one arrives at the following representation

\begin{equation}\label{N-Phi-N}
  \Lambda_t = N_t + N_t \ast \Phi N_t +  N_t \ast \Phi N_t \ast \Phi N_t + \ldots ,
\end{equation}
where $F_t \ast G_t = \int_0^t F_{t-\tau} G_\tau d\tau$ denotes the convolution. Note, that the map $N_t$

\begin{equation}\label{}
  N_t(\rho) = e^{-Z t} \rho = e^{- Ct} \rho e^{- C^\dagger t} ,
\end{equation}
is  completely positive but not trace-preserving. Since $\Phi$ is completely positive the following map $\, Q_t :=\Phi N_t\,$ is completely positive as well. Hence, one arrives at the following representation

\begin{equation}\label{N-Q}
  \Lambda_t = N_t + N_t \ast Q_t +  N_t \ast Q_t \ast Q_t + \ldots ,
\end{equation}
which by construction is completely positive (each term in this  series is completely positive). It is, therefore, clear, that the role of $Q_t$ is to restore trace-preservation which is violated by the map $N_t$. This representation has a clear interpretation in terms of quantum jumps: $N_t$ represents deterministic completely positive evolution and a completely positive map $\Phi$ represents a stochastic quantum jump. Each term  in (\ref{N-Phi-N}) represents a process with a fixed number of jumps and the evolution between jumps is realized via the map $N_t$. Taking into account all terms one eventually ends up with completely positive trace-preserving map $\Lambda_t$. The above representation  of the dynamical semigroup  allows for the following generalization \cite{DC-AK, DC-AK-2016}

%\begin{tcolorbox}
\begin{Theorem}
Let $\{N_t\}_{t \geq 0}$ be a family of completely positive maps  such that $N_0 = {\rm id}$. Let $\{Q_t\}_{t \geq 0}$ be a family of completely positive maps such that $\| \widetilde{Q}(s) \|_1 < 1 $, and

\begin{equation}\label{normal}
  {\rm Tr}[\dot{N}_t(\rho) + Q_t(\rho)] = 0 ,
\end{equation}
for all $\rho \in \TTH$, or equivalently

\begin{equation}\label{}
  \dot{N}_t^\ddag(\oper) = Q_t^\ddag(\oper) .
\end{equation}
Then representation (\ref{N-Q}) provides a legitimate dynamical map.

\end{Theorem}
%\end{tcolorbox}
Proof: condition $\| \widetilde{Q}(s) \|_1 < 1 $ guarantees the converges of the series and clearly the map $\Lambda_t$ is completely positive. Note, that condition (\ref{normal}) does guarantee that $\Lambda_t$ is trace-preserving. Indeed, passing to the LT domain this condition reads as follows

\begin{equation}\label{normal1}
  {\rm Tr}[(s \widetilde{N}(s) - {\rm id} + \widetilde{Q}(s))\rho] = 0 .
\end{equation}
Now, $  {\rm Tr}\widetilde{\Lambda}(s)[{\rm id} - \widetilde{Q}(s)](\rho) = {\rm Tr} \widetilde{N}(s)(\rho)$, and hence using
 ${\rm Tr}[\widetilde{\Lambda}(s)\rho] = \frac 1s {\rm Tr}\rho$ the result is proved. \hfill $\Box$

How the above representation fits the memory kernel master equation (\ref{MK-eq})? Let us represent the kernel via

\begin{equation}\label{}
  \mathcal{K}_t = \Phi_t - Z_t ,
\end{equation}
such that ${\rm Tr}[\mathcal{K}_t(\rho)] =0$, that is, ${\rm Tr}[\Phi_t(\rho)] = {\rm Tr}[Z_t(\rho)]$ for all $\rho \in \TTH$. Assuming that $N_t$ is a solution to

\begin{equation}\label{}
  \dot{N}_t = - \int_0^t Z_{t-\tau} N_\tau d\tau \ , \ \ \ \ N_0 = {\rm id} ,
\end{equation}
and introducing

\begin{equation}\label{}
  Q_t = \Phi_t \ast N_t ,
\end{equation}
one finds the solution to (\ref{MK-eq}) in the form (\ref{N-Q}). Hence, if $\Phi_t \ast N_t$ is completely positive for all $t \geq 0$, then $\Lambda_t$ defines a legitimate dynamical map. Note, that

\begin{equation}\label{}
  \widetilde{N}(s) \frac{1}{s -\widetilde{\Phi}(s)\widetilde{N}(s) } = \frac{1}{s -\widetilde{N}(s)\widetilde{\Phi}(s) }  \widetilde{N}(s) ,
\end{equation}
and hence introducing another map

\begin{equation}\label{}
  \mathcal{P}_t = N_t \ast \Phi_t ,
\end{equation}
one finds an equivalent representation of the dynamical map

\begin{equation}\label{}
  \Lambda_t = N_t + \mathcal{P}_t \ast N_t +  \mathcal{P}_t \ast \mathcal{P}_t \ast N_t +  \ldots .
\end{equation}
The above analysis provides a direct construction of a large class of admissible memory kernels.

\begin{tcolorbox}
How to construct an admissible kernel $\mathcal{K}_t = \Phi_t - Z_t$:

\begin{itemize}
  \item let $\{N_t\}_{t \geq 0}$ be  completely positive with $N_0 = {\rm id}$. Define $Z_t$ such that $\widetilde{Z}(s) = \widetilde{N}(s)^{-1} - s\, {\rm id}$,
  \item let $\{\Phi_t\}_{t \geq 0} $ be a map such that 1)  $\Phi_t^\ddag(\oper) = Z_t^\ddag(\oper)$, 2) either $Q_t = \Phi_t \ast N_t$ or $\mathcal{P}_t = N_t \ast \Phi_t$ is completely positive, and 3) either $\| \widetilde{Q}(s)\|_1 < 1 $ or $\|  \widetilde{\mathcal{P}}(s)\|_1 < 1 $,
\end{itemize}
then $\mathcal{K}_t = \Phi_t - Z_t$ is a legitimate memory kernel. The solution to (\ref{MK-eq}) defines the following map

\begin{eqnarray}\label{}
  \Lambda_t &=& N_t + N_t \ast \Phi_t \ast  N_t +  N_t \ast \Phi_t \ast N_t \ast \Phi_t \ast N_t + \ldots , \nonumber \\
  &=& N_t + N_t \ast Q_t +  N_t \ast Q_t \ast Q_t + \ldots , \\
  &=& N_t + \mathcal{P}_t \ast N_t +  \mathcal{P}_t \ast \mathcal{P}_t \ast N_t +  \ldots , \nonumber
\end{eqnarray}
which is  completely positive and trace-preserving.

\end{tcolorbox}

\begin{Cor} Suppose that $\{N_t\}_{t \geq 0}$ is completely positive and trace non-increasing, i.e. $\frac{d}{dt} N_t^\ddag(\oper) \leq 0$, then there exists completely positive $\{Q_t\}_{t\geq 0}$ satisfying $Q_t^\ddag(\oper) = - \dot{N}_t^\ddag(\oper)$ such that (\ref{N-Q}) defines a legitimate dynamical map.
\end{Cor}
Summarizing, generalization of Markovian semigroup consists in

\begin{itemize}
  \item replacing a  completely positive and trace non-increasing semigroup $N_t = e^{- Z t}$ by an arbitrary completely positive and trace non-increasing map $\{N_t\}_{t \geq 0}$,
  \item replacing completely positive map $\Phi N_t$ by a convolution $\Phi_t \ast N_t$.

\end{itemize}
Note that in the case of semigroup $\Phi$ has to be completely positive. However, in general it is sufficient that $\Phi_t \ast N_t$ is CP. The key difference is that for a semigroup $N_t^{-1} = e^{Z t}$ is still completely positive and hence $\Phi = Q_t N_t^{-1}$ has to be completely positive. It is no longer true for a general map $N_t$.

\subsection{New form of dynamical equations}

The above analysis shows that there is a large class of quantum dynamical maps $\{\Lambda_t\}_{t \geq 0}$ which is characterized be a pair of maps: completely positive  $\{ N_t\}_{t \geq 0}$ satisfying $N_0={\rm id}$ and completely positive  $\{Q_t\}_{t \geq 0}$. These two maps are not independent and have to satisfy a normalization condition (\ref{normal}). In \cite{DC-AK-2016}) we called such pair of maps $\{N_t,Q_t\}$ a {\em legitimate pair}. This class contains a Markovian semigroup $\{N_t=e^{-Zt},Q_t = \Phi N_t\}$ and  gives rise to a lot of new dynamical maps beyond Markovian semigroups. Since the dynamical map is characterized by a pair $\{N_t,Q_t\}$ or, equivalently, $\{N_t,\mathcal{P}_t\}$, it would be interesting to formulate a dynamical equation for the map $\{\Lambda_t\}_{t \geq 0}$ in terms  of $\{N_t,Q_t\}$ or $\{N_t,\mathcal{P}_t\}$.

%\begin{tcolorbox}

\begin{Proposition} The  dynamical map $\{\Lambda_t\}_{t \geq 0}$ satisfies

\begin{equation}\label{}
  \dot{\Lambda}_t = \int_0^t \mathbf{K}_{t - \tau} \Lambda_\tau d\tau + \dot{N}_t ,
\end{equation}
where the new kernel $K_t$ is defined via the LT as follows

\begin{equation}\label{}
  \widetilde{\mathbf{K}}(s) = s \widetilde{N}(s) \widetilde{Q}(s) \widetilde{N}^{-1}(s) = s \widetilde{\mathcal{P}}(s) .
\end{equation}
\end{Proposition}

%\end{tcolorbox}
The proof follows immediately in the LT domain

\begin{equation}\label{}
  s \widetilde{\Lambda}(s) - {\rm id} = \widetilde{\mathbf{K}}(s)\widetilde{\Lambda}(s) + s\widetilde{N}(s) - {\rm id}
\end{equation}
and hence

\begin{equation}\label{}
  \widetilde{\Lambda}(s) = \frac{1}{{\rm id} - \frac 1s \widetilde{\mathbf{K}}(s)} \widetilde{N}(s) = \frac{1}{{\rm id} - \widetilde{\mathcal{P}}(s)} \widetilde{N}(s) .
\end{equation}
Note, that in the time domain $\mathbf{K}_t = \dot{\mathcal{P}}_t + \delta(t) \mathcal{P}_0$ implies

\begin{equation}\label{}
   \int_0^t K_{t - \tau} \Lambda_\tau d\tau = \mathcal{P}_0 +  \int_0^t \partial_t \mathcal{P}_{t - \tau} \Lambda_\tau d\tau ,
\end{equation}
which leads the following equation

\begin{equation}\label{}
 \dot{\Lambda}_t = \mathcal{P}_0 \Lambda_t +  \int_0^t \partial_t \mathcal{P}_{t - \tau} \Lambda_\tau d\tau + \dot{N}_t ,
\end{equation}
with three characteristic terms:  time-independent local generator $\mathcal{P}_0$,  memory kernel part $\dot{\mathcal{P}}_t \ast \Lambda_t$, and inhomogeneous term $\dot{N}_t$. Using

\begin{equation}\label{}
  \int_0^t \partial_t \mathcal{P}_{t - \tau} \Lambda_\tau d\tau = - \int_0^t \partial_\tau \mathcal{P}_{t - \tau} \Lambda_\tau d\tau =
  - \mathcal{P}_0 \Lambda_t + \mathcal{P}_t + \int_0^t \mathcal{P}_{t - \tau} \dot{\Lambda}_\tau d\tau ,
\end{equation}
one gets another equivalent   form of the dynamical equation

\begin{equation}\label{PALMA-like}
 \dot{\Lambda}_t =   \int_0^t \mathcal{P}_{t - \tau} \dot{\Lambda}_\tau d\tau + \mathcal{P}_t + \dot{N}_t .
\end{equation}

\begin{Example} \label{EX-Palma} Authors of \cite{Palma-2013} derived the following master equation for the evolution of the system's density operator $\rho_t$:

\begin{equation}\label{Palma-2013}
  \dot{\rho}_t = \Gamma \int_0^t e^{-\Gamma(t-\tau)} \mathcal{F}_{t-\tau} \dot{\rho}_\tau d\tau + e^{-\Gamma t} \mathcal{F}_t \rho_0 ,
\end{equation}
where $\{\mathcal{F}_t\}_{t \geq 0}$ is a dynamical map satisfying $\mathcal{F}_0 = {\rm id}$, and $\Gamma > 0$. Note, that (\ref{Palma-2013}) has exactly the form of (\ref{PALMA-like}) with

\begin{equation}\label{}
  N_t = e^{-\Gamma t} \mathcal{F}_t , \ \ \ \mathcal{P}_t = {Q}_t = \Gamma e^{-\Gamma t} \mathcal{F}_t .
\end{equation}

\end{Example}
Interestingly, a set of legitimate pairs $\{N_t,Q_t\}$ satisfies the following properties:

\begin{enumerate}
  \item convexity: if $\{N_t,Q_t\}$ and $\{N'_t,Q'_t\}$ are legitimate pairs, then $\{N,t,Q_t\}$ with

$$   N_t = p N_t + p' N_t' \ , \ \ \ Q_t = p Q_t + p' Q'_t , $$
and $p+p'=1$, provides a legitimate pair,

  \item reduction: if $\{\mathbf{N}_t,\mathbf{Q}_t\}$ is a legitimate pair acting in $\mathcal{H} \otimes \mathcal{H}_E$, then

$$   N_t(\rho) := {\rm Tr}_E \mathbf{N}_t( \rho \otimes \omega_E) \ , \ \ \ Q_t(\rho) := {\rm Tr}_E \mathbf{Q}_t( \rho\otimes \omega_E) , $$
is a legitimate pair ($\omega_E$ is a fixed density operator in $\mathcal{H}_E$),

\item gauge transformations: if $\{N_t,Q_t\}$ is legitimate pair, then

$$ N'_t = \mathcal{F}_t N_t \ , \ \ \ Q'_t = \mathcal{G}_t Q_t , $$
where $\{\mathcal{F}_t\}_{t \geq 0}$ is an arbitrary dynamical map, and  $\{\mathcal{G}_t\}_{t \geq 0}$ is an arbitrary family of  CPTP maps, defines a new (gauge transformed) legitimate pair.
\end{enumerate}
Note, that in Example \ref{EX-Palma} a pair $\{N_t,Q_t\}$ is gauge transformed from $\{ e^{-\Gamma t} {\rm id}, \Gamma e^{-\Gamma t} {\rm id}\}$.

\subsection{Classical semi-Markov evolution}

The above analysis can be applied in the classical case as well. A classical memory kernel master equation for the stochastic matrix $T_{kl}(t)$ has the following structure

\begin{equation}\label{semi-class}
  \partial_t T_{kl}(t) = \int_0^t \sum_m \Big[ W_{km}(t-\tau) T_{ml}(\tau) - W_{mk} (t-\tau) T_{kl}(\tau) \Big] d\tau ,
\end{equation}
that is, the classical memory kernel reads

\begin{equation}\label{}
  K_{mn}(t) = W_{mn}(t) - D_{mn}(t) \  \ , \ \ \ \ \ D_{mn}(t) = \delta_{mn} \sum_k W_{kn}(t) .
\end{equation}
An interesting example of a classical process beyond Markovian semigroup is provided by so called  semi-Markov evolution of a probability vector $\mathbf{p}(t)$. Such evolution is  uniquely determined by a semi-Markov matrix $q_{mn}(t) \geq 0$  such that $\int_0^t q_{mn}(\tau) d\tau$
defines the probability of a jump from a state $n$ to a state $m$ at time $t$ (provided that the system stays at state $n$ at $t=0$).
Assuming, that a jump from any state $n$ eventually occurs one has the following normalization condition

\begin{equation}\label{}
  \lim_{t \to \infty}  \int_0^t \sum_{m=1}^d q_{mn}(\tau) d\tau  = 1 .
\end{equation}
The functions $f_n(t) := \sum_m q_{mn}(t)$ may be, therefore, interpreted as waiting time distribution and hence one may define  the  corresponding survival probability via

\begin{equation}\label{}
  g_n(t) = 1 - \int_0^tf_n(\tau) d\tau .
\end{equation}
Defining the diagonal matrix $\,  n_{ij}(t) := \delta_{ij} g_j (t)$, one finally obtains the following classical dynamical map

\begin{equation}\label{}
  T(t) = n(t) + (n \ast q)(t) + (n \ast q \ast q)(t)+ \ldots ,
\end{equation}
or, equivalently, in the LT domain

\begin{equation}\label{}
  \widetilde{T}(s) = \widetilde{n}(s) \frac{1}{\oper - \widetilde{q}(s)} .
\end{equation}
Such evolution provides a solution to the classical memory kernel master equation with

\begin{equation}\label{W-D}
  \widetilde{W}_{mn}(s) := \frac{\widetilde{q}_{mn}(s)}{\widetilde{g}_n(s)} , \ \ \ \ \
  \widetilde{D}_{mn}(s) = \delta_{mn}\, \frac{1-s \widetilde{g}_n(s)}{\widetilde{g}_n(s)} .
\end{equation}
Note, that the map $n(t)$ is a solution to

\begin{equation}\label{}
  \dot{n}(t) = - \int_0^t D(t-\tau) n(\tau) d\tau \ , \ \ \ n(0) = \oper .
\end{equation}

A typical example of a semi-Markov matrix may be constructed as follows

\begin{equation}\label{}
  q_{mn}(t) = \Pi_{mn} f_n(t) ,
\end{equation}
where $\Pi_{mn}$ is a stochastic matrix (classical channel). The functions (waiting time distributions) $f_n : \mathbb{R}_+ \to \mathbb{R}$ satisfy: $f_n(t) \geq 0$ and  $\int_0^\infty f_n(\tau) d \tau = 1$. One finds

\begin{equation}
  \widetilde{W}_{mn}(s) = \Pi_{mn}  \frac{ s \widetilde{f}_n(s)}{1 -  \widetilde{f}_n(s)} , \ \ \ \
  \widetilde{D}_{mn}(s) = \delta_{mn}\, \frac{\widetilde{f}_n(s)}{\widetilde{g}_n(s)}  =  \delta_{mn}\, \frac{s\widetilde{f}_n(s)}{1-\widetilde{f}_n(s)} .
\end{equation}
Finally,  the memory kernel reads as follows

\begin{equation}\label{}
  \widetilde{K}_{mn}(s) = \widetilde{W}_{mn}(s) - \widetilde{D}_{mn}(s) = (\Pi_{mn} - \delta_{mn}) \widetilde{k}_n(s) ,
\end{equation}
with

\begin{equation}\label{}
  \widetilde{k}_n(s) = \frac{s\widetilde{f}_n(s)}{1-\widetilde{f}_n(s)} .
\end{equation}
Equivalently, the functions  $f_n(t)$ and $g_n(t)$ satisfy the following relation $\,  f_n = k_n \ast g_n$. Note, that one recovers Markovian semigroups taking waiting time distributions

\begin{equation}\label{}
  f_n(t) = \gamma_n e^{- \gamma_n t} ,
\end{equation}
with $\gamma_n > 0$. Indeed,  $k_n(t) = \delta(t) \gamma_n$, and hence

\begin{equation}\label{}
  {K}_{mn}(t) = {W}_{mn}(t) - {D}_{mn}(t) = \delta(t)(\Pi_{mn} - \delta_{mn}) \gamma_n ,
\end{equation}
defines a standard Kolmogorov generator.

%\begin{Example}

Consider the simplest scenario corresponding to $f_n(t) = f(t)$ for $n=1,\ldots,d$. In this case

\begin{equation}\label{}
  n_{ij}(t) = g(t) \delta_{ij} \  , \ \ \ \ q_{ij}(t) = f(t) \Pi_{ij} ,
\end{equation}
with $g(t) = 1 - \int_0^t f(\tau) d\tau$. Hence, the memory kernel reads

\begin{equation}\label{}
  K(t) = k(t) (\Pi - \oper) ,
\end{equation}
with

\begin{equation}\label{}
  \widetilde{k}(s) = \frac{s\widetilde{f}(s)}{1-\widetilde{f}(s)} .
\end{equation}
In the LT domain

\begin{equation}\label{}
  \widetilde{T}(s) = \frac{1}{s - \widetilde{K}(s)} = \frac{\widetilde{g}(s)}{\oper - \widetilde{f}(s) \Pi }\ .
\end{equation}
Consider $\Pi$ satisfying $\Pi^2 = \Pi$, i.e. $\Pi$ defines a projector. In this case one easily finds the solution of the memory kernel master equation

\begin{equation}\label{}
  T(t) = n(t) + \int_0^t q(\tau)d\tau = \left(1- \int_0^t  f(\tau)d\tau \right) \oper +  \int_0^t f(\tau)d\tau \, \Pi .
\end{equation}
Let us observe that $T(t)$ satisfies time-local master equation

\begin{equation}\label{}
  \dot{T}(t) = L(t) T(t) ,
\end{equation}
with the following generator

\begin{equation}\label{}
  L(t) = \gamma(t) (\Pi - \oper) ,
\end{equation}
where

\begin{equation}\label{}
  \gamma(t) = \frac{f(t)}{g(t)} = \frac{f(t)}{ 1 - \int_0^t f(\tau) d\tau } .
\end{equation}
One has, therefore, the following suggestive relation between time-local and memory kernel description
%\begin{tcolorbox}

\begin{equation}\label{}
  \widetilde{K}(s) =  \frac{ \widetilde{f}(s)}{ \widetilde{g}(s) }\, (\Pi - \oper) \ \ \ \ \longleftrightarrow \    \ \ \ \   L(t) = \frac{f(t)}{g(t)} \, (\Pi - \oper) \ .
\end{equation}

%\end{tcolorbox}

 Interestingly, in this case semi-Markov evolution, i.e. $f(t) \geq 0$, is also Markovian, i.e. $\gamma(t) \geq 0$. Note, however, that  $T(t)$ defines a legitimate classical dynamical map whenever  $\int_0^t f(\tau) d\tau  \geq 0$. Hence, whenever $0 \leq \int_0^t f(\tau) d\tau  \leq 1$ but $f(t) \ngeqslant 0$, then the evolution $T(t)$ is no longer semi-Markov. Take for example $f(t) = \gamma \sin(\gamma t)$. One gets for $t \geq 0$:

\begin{equation}\label{}
  k(t) = \gamma^2  , \ \ \ \  \frac{f(t)}{g(t)} = \gamma \tan(\gamma t) ,
\end{equation}
that is, this evolution which is no longer semi-Markov is governed by a constant memory kernel $K(t) = \gamma^2(\Pi- \oper)$ and highly singular time-local generator $L(t) = \gamma \tan(\gamma t)(\Pi-\oper)$.

A second interesting case corresponds to $\Pi^2 = \oper$. One finds

\begin{equation}\label{}
  {T}(t) = \frac{1+ \widetilde{q}(t)}{2}\oper +  \frac{1 - \widetilde{q}(t)}{2} \Pi ,
\end{equation}
where

\begin{equation}\label{}
  \widetilde{q}(s) = \frac 1s \frac{1-\widetilde{f}(s)}{1+ \widetilde{f}(s)} .
\end{equation}
In this case one may have semi-Markov evolution which is no longer Markovian \cite{Vacchini-NJP}. One easily finds the corresponding time-local generator

\begin{equation}\label{}
  L(t) =  -\frac{\dot{q}(t)}{q(t)} (\Pi- \oper) .
\end{equation}
Clearly, taking $f_0(t) = \gamma e^{-\gamma t}$ one obtains $q(t) = e^{- 2 \gamma t}$ giving rise to Markovian semigroup. However, taking

\begin{equation}\label{}
  f(t) = (f_0 \ast f_0)(t) = \gamma^2 t e^{- \gamma t}
\end{equation}
one gets

\begin{equation}\label{}
  q(t) = e^{- \gamma t} \Big( \cos(\gamma t) + \sin(\gamma t) \Big) ,
\end{equation}
and hence

\begin{equation}\label{}
  -\frac{\dot{q}(t)}{q(t)}=  \frac{2\gamma}{1+ \tan(\gamma t)} ,
\end{equation}
which is no longer positive and even highly singular. One has, therefore, a perfectly  regular memory kernel $K(t)$ and highly singular time local generator $L(t)$ giving rise to semi-Markov evolution which is not even P-divisible (strongly non-Markovian):

\begin{equation}\label{}
  {K}(t) =  \gamma^2 e^{-2 \gamma t} (\Pi - \oper) \  \ \ \ \longleftrightarrow \ \ \ \  L(t) =   \frac{2\gamma}{1+ \tan(\gamma t)}\, (\Pi - \oper) \ .
\end{equation}
The intricate relation between time-local master and memory kernel master equations were recently analyzed in \cite{Nina-2020,Nina-2020a}.

%\end{Example}

\subsection{Quantum semi-Markov evolution}

The classical semi-Markov evolution
\begin{equation}\label{}
  n(t) = g(t) \oper \ , \ \ \ q(t) = f(t) \Pi .
\end{equation}
may be immediately generalized for the quantum case as follows

\begin{equation}\label{}
  N_t = g(t) {\rm id} \ , \ \ \ Q_t = f(t) \mathcal{E} ,
\end{equation}
where $\mathcal{E} : \TTH \to \TTH$ is an arbitrary quantum channel. Moreover, exploiting a gauge freedom   we arrive at the following general family of legitimate pairs $\{N_t,Q_t\}$
%\begin{tcolorbox}
\begin{equation}\label{}
  N_t = g(t) \mathcal{G}_t \ , \ \ \ Q_t = f(t) \mathcal{E} \mathcal{F}_t  ,
\end{equation}
or
\begin{equation}\label{}
  N_t = g(t) \mathcal{G}_t \ , \ \ \ \mathcal{P}_t = f(t) \mathcal{F}_t \mathcal{E} ,
\end{equation}
%\end{tcolorbox}
where $\{\mathcal{G}_t\}_{t \geq 0}$ is a dynamical map and $\{\mathcal{F}_t\}_{t \geq 0}$ is a family of CPTP maps. One arrives at the following representation   \cite{Bassano-PRL,Bassano-Sci}

%\begin{tcolorbox}
\begin{equation}\label{}
  \Lambda_t = g(t) \mathcal{G}_t + \sum_{n=1}^\infty \int_0^t dt_1 \int_0^{t_1} dt_2 \ldots \int_0^{t_{n-1}}dt_n \, p(t;t_1,\ldots,t_n) \, \mathcal{G}_{t-t_1} \mathcal{E} \mathcal{F}_{t_1-t_2} \mathcal{E} \ldots \mathcal{E}\mathcal{F}_{t_{n-1}-t_n}\mathcal{E}\mathcal{F}_{t_n}
\end{equation}
or, equivalently
\begin{equation}\label{}
  \Lambda_t = g(t) \mathcal{G}_t + \sum_{n=1}^\infty \int_0^t dt_1 \int_0^{t_1} dt_2 \ldots \int_0^{t_{n-1}}dt_n \,q(t;t_1,\ldots,t_n) \, \mathcal{F}_{t-t_1} \mathcal{E} \mathcal{F}_{t_1-t_2} \mathcal{E} \ldots \mathcal{E}\mathcal{F}_{t_{n-1}-t_n}\mathcal{E}\mathcal{G}_{t_n}
\end{equation}
%\end{tcolorbox}
where
\begin{equation}\label{}
   p(t;t_1,\ldots,t_n) = g(t-t_1)f(t_1-t_2) \ldots f(t_{n-1}-t_n) f(t_n) ,
\end{equation}
and
\begin{equation}\label{}
  q(t;t_1,\ldots,t_n) = f(t-t_1)f(t_1-t_2) \ldots f(t_{n-1}-t_n) g(t_n) .
\end{equation}

%The above representations provide trajectory description of the quantum evolution

\begin{Example} Consider $  N_t = g(t) \mathcal{G}_t$ and $\mathcal{P}_t = f(t) \mathcal{F}_t \mathcal{E}$, with

\begin{equation}\label{}
  \mathcal{F}_t = \mathcal{G}_t = e^{\mathcal{L}t} ,
\end{equation}
for some fixed GKLS generator $\mathcal{L}$. One finds

\begin{equation}\label{}
  \widetilde{N}(s) = \widetilde{g}(s-\mathcal{L}) \ , \ \ \ \widetilde{\mathcal{P}}(s) = \widetilde{f}(s-\mathcal{L}) \, \mathcal{E} ,
\end{equation}
and hence

\begin{equation}\label{}
  \widetilde{\Phi}(s) = \widetilde{N}^{-1}(s) \widetilde{\mathcal{P}}(s) = \frac{\widetilde{f}(s-\mathcal{L})}{\widetilde{g}(s-\mathcal{L})}\, \mathcal{E} ,
\end{equation}
together with

\begin{equation}\label{}
  - \widetilde{Z}(s) = s\,{\rm id} - \widetilde{N}^{-1}(s) = \frac{s\widetilde{f}(s-\mathcal{L})- {\rm id}}{\widetilde{g}(s-\mathcal{L})} = \mathcal{L} -  \frac{\widetilde{f}(s-\mathcal{L})}{\widetilde{g}(s-\mathcal{L})} .
\end{equation}
Finally, one arrives at the following memory kernel

\begin{equation}\label{}
  \widetilde{\mathcal{K}}(s) =  \mathcal{L} + \frac{\widetilde{f}(s-\mathcal{L})}{\widetilde{g}(s-\mathcal{L})} \, (\mathcal{E} - {\rm id}) ,
\end{equation}
giving rise to the following master equation

\begin{equation}\label{}
  \dot{\Lambda}_t = \mathcal{L}\Lambda_t + \int_0^t d\tau k(\tau) e^{\tau \mathcal{L}} (\mathcal{E} - {\rm id}) \Lambda_{t-\tau} ,
\end{equation}
where the memory function $k(t)$ is defined via

\begin{equation}\label{k-f-g}
  \widetilde{k}(s) = \frac{\widetilde{f}(s)}{\widetilde{g}(s)}  .
\end{equation}
In particular for $\mathcal{L} = \gamma(\mathcal{E} - {\rm id})$ it gives rise to

\begin{equation}\label{}
  \dot{\Lambda}_t = \mathcal{L}\Lambda_t + \frac{1}{\gamma}\int_0^t d\tau k(\tau) e^{\tau\mathcal{L}} \mathcal{L}\Lambda_{t-\tau} .
\end{equation}
Let us observe that the non-local term reproduces a post-Markovian master equation analyzed in \cite{Lidar-post}. Interestingly, the authors of \cite{Lidar-post} observed that in general the memory function $k(t)$ does not lead to legitimate dynamical map. According to our construction $k(t)$ is no longer arbitrary  but is defined by (\ref{k-f-g}).

\end{Example}

\subsection{A hybrid approach}

Let us  consider a hybrid approach \cite{Chaos} in which the dynamics of the density operator satisfies the following condition: evolution of populations is independent of evolutions of coherences. Such evolution is a characteristic feature of the Davies GKLS generator. Let us assume
that a quantum dephasing process is governed by a time-local generator $\mathcal{L}_t$, whereas the `classical process' describing the evolution of populations is governed by a classical semi-Markov generator

\begin{equation}\label{hybrid}
  \dot{\Lambda}_t = \mathcal{L}_t \Lambda_t + \int_0^t \mathcal{K}_{t-\tau} \Lambda_\tau d\tau ,
\end{equation}
where

\begin{equation}\label{}
  \mathcal{L}_t(\rho) = -i[H(t),\rho] + \sum_{k,l} D_{kl}(t) \Big(  |k\rangle \langle k| \rho |l\rangle \langle l| - \frac 12 \delta_{kl} \{  |k\rangle \langle k| ,\rho\} \Big) ,
\end{equation}
with $H(t) = \sum_k E_k(t)|k\>\<k|$, and the kernel

\begin{equation}\label{}
   \mathcal{K}_t(\rho) = \sum_{k\neq l} W_{kl}(t) \Big( |k\rangle \langle l| \rho |l\rangle \langle k|  - \frac 12 \{ |l\rangle \langle l|,\rho\} \Big) ,
\end{equation}
where $W_{kl}(t)$ is constructed out of semi-Markov matrix $q_{kl}(t)$ via (\ref{W-D}). Note, that both $\mathcal{L}_t$ and $\mathcal{K}_t$ are covariant (cf. section \ref{SEC-COV}), i.e. satisfy

\begin{equation}\label{}
  \mathbb{U}_\mathbf{x} \, \mathcal{L}_t = \mathcal{L}_t \, \mathbb{U}_\mathbf{x} \ , \ \ \ \ \
  \mathbb{U}_\mathbf{x} \, \mathcal{K}_t = \mathcal{K}_t \, \mathbb{U}_\mathbf{x} ,
\end{equation}
where $\mathbb{U}_\mathbf{x}(\rho) = U_\mathbf{x} \rho U^\dagger_\mathbf{x}$, and $U_\mathbf{x} = \sum_k e^{i x_k} |k\>\<k|$. The structure of (\ref{hybrid}) implies that  the populations $\rho_{kk}(t)$ are only affected by $ \mathcal{K}_t$ but not by $\mathcal{L}_t$. Hence $p_k(t) = \rho_{kk}(t)$ evolves according to a classical semi-Markov evolution. Now, the time evolution of coherences  reads

\begin{equation}\label{}
  \rho_{kl}(t) =  \lambda_{kl}(t) \exp\left( -i \int_0^t(E_k(\tau) - E_l(\tau))d\tau \right) \exp\left( - \int_0^t D_{kl}(\tau) d\tau \right)   ,
\end{equation}
and $\lambda_{kl}(t)$ satisfy

\begin{equation}\label{}
  \dot{\lambda}_{kl}(t) = - \frac 12 \int_0^t (w_k(t-\tau) + w_l(t-\tau)) \lambda_{kl}(\tau) d\tau , \ \ \ \ \lambda_{kl}(0) = 1 ,
\end{equation}
where $w_{l}(t) = \sum_k W_{kl}(t)$. Hence, coherences are affected both by time-local generator  $\mathcal{L}_t$ and the memory kernel $\mathcal{K}_t$. Note, that the dynamical map may be represented as a composition of three mutually commuting maps:

\begin{equation}\label{}
  \Lambda_t = \mathbb{U}_t \, \Phi_t^{\rm dec} \, \Phi_t^{\rm diss} ,
\end{equation}
where $ \mathbb{U}_t = e^{-i \int_0^t H(u)du} \rho e^{i \int_0^t H(u)du}$, pure decoherence map

\begin{equation}\label{}
  \Phi^{\rm dec}_t \rho = \sum_k  |k\rangle \langle k| \rho |k\rangle \langle k| +  \sum_{k\neq l} e^{- \int_0^t D_{kl}(u)du}   |k\rangle \langle k| \rho |l\rangle \langle l| ,
\end{equation}
and

\begin{equation}\label{}
  \Phi^{\rm diss}_t \rho = \sum_{k,l} {T}_{kl}(t)  |k\rangle \langle l| \rho |l\rangle \langle k| + \sum_{k\neq l} \lambda_{kl}(t)  |k\rangle \langle k| \rho |l\rangle \langle l| ,
\end{equation}
where $T_{kl}(t)$ solves the classical master equation (\ref{semi-class}).

\begin{Proposition}[\cite{Chaos}] The map $\{\Lambda_t\}_{t \geq 0}$ is completely positive if and only if

\begin{equation}\label{}
  \Phi^{\rm dec}_t[\mathbf{C}(t)]\geq 0 ,
\end{equation}
where

\begin{equation}\label{C}
  \mathbf{C}_{kk}(t) := {T}_{kk}(t) \ , \ \ \ \mathbf{C}_{kl}(t) := \lambda_{kl}(t) , \ \ (k \neq l) .
\end{equation}

\end{Proposition}
Hence, by an appropriate engineering of decoherence rates $D_{kl}(t)$ we can control complete positivity of $\Lambda_t$. Similar approach was proposed in \cite{Zanardi} where on top of a Markovian generator $\mathcal{L}$ authors consider `noise' generated through an appropriate memory kernel. It is shown that   engineering the kernel allows to reduce the overall noise (e.g. one can increase the channel fidelity). This shows that a hybrid approach might have also interesting applications.

\section{Concluding remarks}  \label{CONCLUSIONS}

In this report we analyzed quantum evolution represented by a quantum dynamical map, i.e. a family of completely positive trace-preserving maps $\{\Lambda\}_{t \geq 0}$. In the simplest scenario such evolution is defined by a dynamical semigroup which is generated by the celebrated GKLS generator. Such evolution is realized whenever appropriate Markovian approximations are justified. In particular, whenever the system-environment interaction is sufficiently weak and there is a separation of characteristic system's and environment's time scales, then Markovian semigroup provides a good approximation for the true (reduced) system's evolution.  Interestingly, Markovian semigroups are perfectly consistent with universal laws of thermodynamics. In general, however, one is forced to analyze dynamical maps which are not semigroups.

Here, I stressed that elegant mathematical concepts, like e.g. complete positivity, provide a unifying scheme for the description of quantum evolution. General quantum evolution was discussed using  two approaches based on time-local generators and non-local memory kernels of Nakajima-Zwanzig form. Time-local approach enables one to introduce a natural notion of Markovian evolution which is represented by CP-divisible dynamical map. Divisibility property allows to factorize a map as a composition of propagators $V_{t,s}$ from time `$s$' to time `$t$'. It turns out that CP-divisible maps display several monotonicity properties of  characteristic quantities like relative entropy, fidelity and entanglement measures. CP-divisible evolution can be essentially characterized by time-dependent GKLS generators.
However, true non-Markovian dynamics breaks the CP-divisibility. A characteristic feature of quantum evolution beyond Markovian regime is, therefore, violation of some of the above mentioned monotonicity properties. The most evident sign of non-Markovian memory effect is the information backflow from the environment back to the system. This is a typical effect observed for the majority of non-Markovian evolutions.  However, as already stressed in \cite{NM4}, Markovianity is highly context dependent and the authors of \cite{NM4} presented the intricate hierarchy of different concepts.
In particular, having an access to the total system-environment unitary evolution the very concept of Markovianity might be considerably refined. In this case the property known as quantum regression formula provides the most natural definition which reduces to the standard definition of a classical stochastic process in the commutative (classical) scenario. Now, CP-divisibility provides only a necessary condition for quantum regression to hold and even a dynamical semigroup might violate quantum regression formula and hence display non-Markovian effects. It should be stressed, however, that CP-divisibility provides the most natural definition of system's Markovian evolution when only intrinsic properties of the system are taken into account. I hope that this presentation supplements existing reviews \cite{NM1,NM2,NM3} which were focused more on the analysis of non-Markovian behaviour of many physical systems.

There are still several interesting topics which deserve further analysis. In particular the structure of time-local generators beyond CP-divisible, i.e. Markovian, scenario is poorly understood. The same applies for the memory kernel approach which seems to be well understood only in the semi-Markov case. Moreover, it would be interesting to search for more connections between these two approaches.
In recent years we have faced the rapid development in the field of quantum thermodynamics which is
intimately connected to the quantum theory of open systems  and strongly
influenced by the ideas and methods of quantum information

There is an evident connection between the theory of open quantum systems and quantum thermodynamics which attracts  considerable attention these days. Open system dynamics is perfectly consistent with basic  laws of thermodynamics in the weak coupling regime. However, in the strong coupling case there are still several open problems which deserve more detailed analysis. Moreover, the thermalization processes are still not fully understood. Even for the Markovian semigroup in the weak coupling limit, governed by Davies GKLS master equation, the system thermalizes to the Gibbs state w.r.t. the system's Hamiltonian. However, it is rather evident that thermalization should  lead to the state being the reduction of a  true Gibbs state of the total system. Such issues are currently intensively discussed.

The notion of a quantum channel might be generalized to so-called quantum superchannel, i.e. a completely positive map which maps quantum channels to quantum channels (cf. the recent review \cite{SUPER}). It would be interesting to generalize the presented  analysis for the superchannel scenario. In the recent paper \cite{SUPER-L} the generalization of GKLS structure for a semigroup of superchannels was derived. It seems quite natural to consider superchannel evolution beyond semigroup case as well.

In this report I deal with finite dimensional case only. Using more powerful mathematical tools (infinite dimensional operator algebras) one may provide the corresponding analysis in the general infinite dimensional scenario. I hope that the presented report might be helpful for students and researchers to clarify existing concepts, mathematical notions, intricate connections between them and/or to start the research of open quantum systems.

\section*{Acknowledgments}

I am grateful to Andrzej Kossakowski for many years of inspiring collaboration in the field of open quantum systems and his scientific and human support.
%I thank to anonymous referees for a number of valuable suggestions. 
I thank Saverio Pascazio, Fabio Benatti, Roberto Floreanini, Beppe Marmo, Kavan Modi, Sabrina Maniscalco, Jyrki Piilo, \'Angel Rivas, Andrea Smirne, Gen Kimura, Kimmo Louma, Karol \.Zyczkowski, Sergey Denisov, Wojciech Tarnowski, Sagnik Chakraborty, Davide Lonigro for useful discussions and collaboration on several problems related to open quantum systems. This report was partially supported by the National Science Center project 2018/30/A/ST2/00837.

%I thank Dorota Chru\'sci\'nska for preparing figures.
%I also thank many colleagues with whom I had a pleasure to discuss several issues of open quantum systems:  S. Pascazio, S. Maniscalco,  J. %Piilo, \'A. Rivas, F. Benatti, G. Kimura, K. \.Zyczkowski, S. Denisov, K. Modi, A. Smirne, S. Huelga, B. Vacchini, B. Marmo and F. Petruccione.

\newlist{abbrv}{itemize}{1}
\setlist[abbrv,1]{label=,labelwidth=1in,align=parleft,itemsep=0.1\baselineskip,leftmargin=!}

\section*{List of abbreviations}

\addcontentsline{toc}{section}{List of abbreviations}

\begin{abbrv}

\item[BLP]    Breuer-Laine-Piilo

\item[CM]			Completely Monotone

\item[CP]			Completely Positive

\item[CPTP]         Completely Positive Trace-Preserving

\item[CPU]      Completely Positive Unital

\item[GKLS]         Gorini-Kossakowski-Lindblad-Sudarshan

\item[LT] Laplace Transform

\item[MUBs]      Mutually Unbiased Bases

\item[NZ]   Nakajima-Zwanzig

\item[PTP]        Positive Trace-Preserving

\item[RHP]     Rivas-Huelga-Plenio

\item[WYD]    Wigner-Yanase-Dyson

\item[${\rm id}$]  identity map

\item[$\|X\|_1$] trace norm

\item[$\|X\|_\infty$] operator norm

\item[$\mathrm{L}(\HH)$]  vector space of linear operators $\HH \to \HH$

\item[$\mathcal{T}(\HH)$]   Banach space $(\mathrm{L}(\HH),\|\ \|_1)$

\item[$\mathcal{B}(\HH)$]   an algebra of bounded operators  $(\mathrm{L}(\HH),\|\ \|_\infty)$

\item[$(X,Y)_{\rm HS}$]   Hilbert-Schmidt inner product $(X,Y)_{\rm HS} := {\rm Tr}(X^\dagger Y)$

\item[$\Phi^\ddag$]  a dual map $(\Phi(X),Y)_{\rm HS} = (X,\Phi^\ddag(Y))_{\rm HS} $

\item[$\rho_{\rm ss}$] steady state

\item[$\rho_\beta$] a thermal state at the inverse temperature $\beta$

\item[$(X,Y)_{\rm ss}$] an inner product w.r.t. $\rho_{\rm ss}$, i.e.  $(X,Y)_{\rm ss}:= {\rm Tr}(\rho_{\rm ss} X^\dagger Y)$

\item[$U_t$]     Unitary operator in the Hilbert space

\item[$\mathbb{U}_t$]   Unitary map $\mathbb{U}_t(X) = U_t X U_t^\dagger$

\item[$V_{t,s}$] a propagator $V_{t,s} := \Lambda_t \Lambda_s^{-1}$

\end{abbrv}

%\begin{acronym}
%        \acro{cp}[CP]{Completely Positive}
       % \acro{ann}[ANN]{Artificial Neural Network}
%\end{acronym}

% bibliography
%\phantomsection
\addcontentsline{toc}{section}{References}
%\section*{References}

%\bibliographystyle{elsarticle-num}

%\bibliographystyle{plain} % We choose the "plain" reference style
%\bibliography{refs} % Entries are in the refs.bib file

\end{document}